\PassOptionsToClass{Review}{sagej}
\documentclass[sageh,times]{sagej}

\usepackage{iftex}
\ifPDFTeX
  \usepackage[T1]{fontenc}
\else
  \usepackage{fontspec}
\fi

\usepackage{array}
\usepackage{tabularx}
\usepackage{etoolbox}

\newenvironment{widetable}%
  {\par\addvspace{\intextsep}\leavevmode\noindent}%
  {\par\addvspace{\intextsep}}

\newenvironment{apptable}%
  {\par\addvspace{\intextsep}\noindent}%
  {\par\addvspace{\intextsep}}

\newcommand{\tblwidth}{\dimexpr\linewidth-4pt\relax}

\newcolumntype{L}{>{\raggedright\arraybackslash}X}
\newcolumntype{R}{>{\raggedleft\arraybackslash}X}

\newcommand{\breakableunderscore}{\renewcommand{\_}{\textunderscore\allowbreak}}
\AtBeginEnvironment{tabular}{\breakableunderscore}
\AtBeginEnvironment{tabularx}{\breakableunderscore}
\usepackage{xltabular}
\AtBeginEnvironment{xltabular}{\breakableunderscore}
\AtBeginEnvironment{xltabular}{}
\newcommand{\ttbreak}[1]{{\breakableunderscore\texttt{#1}}}
\newcommand{\stripsafebreak}{}
\usepackage{microtype}
\usepackage[hidelinks]{hyperref}

\AtBeginDocument{\pagestyle{plain}}

\newenvironment{sagestatement}[1]%
  {\subsection*{\normalsize\sagesf\bfseries #1}\begin{refsize}\noindent}%
  {\end{refsize}}
\newcommand{\statementshead}[1]{\section*{\normalsize\sagesf\bfseries #1}}

\def\volumeyear{2026}

\begin{document}

\runninghead{Nishizawa et al.}

\title{Instruction-Aware Transfer of GPU Implementation Crossovers:
a Tensor-Product Discontinuous Galerkin
Operator on GB200, H100, and A100}

\author{Seiya Nishizawa\affilnum{1}, Yuta Kawai\affilnum{1} and Xuanzhengbo Ren\affilnum{2}}

\affiliation{\affilnum{1}RIKEN Center for Computational Science, Japan\\
\affilnum{2}Information Technology Center, The University of Tokyo, Japan}

\corrauth{Seiya Nishizawa, RIKEN Center for Computational Science,
7-1-26 Minatojima-minami-machi, Chuo-ku, Kobe, Hyogo 650-0047, Japan}

\email{s-nishizawa@riken.jp}

\begin{abstract}
We compare up to six implementations of an FP64 tensor-product discontinuous Galerkin operator at $p = 7$--$1023$ on GB200, H100, and A100 with fixed numerics, input, and code.
The fastest path is machine dependent: at $p = 511$, a fused-epilogue GEMM path is 5.0\% faster than a fused Tensor Core kernel on GB200 and 6.9\% slower on A100.
Nominal-peak scaling reproduces the A100 ranking but fails on H100, where the cost of FP64 arithmetic in the profiler-defined FP64 budget depends on the instruction.
We therefore price each issued instruction at its architecture-dependent cost in that budget, using counter profiles from one or more calibration GPUs and one derived calibration factor per kernel, with no fit to the target.
A retrospective leave-one-GPU-out evaluation, holding out target timings and counters but not the model form, transfers the $p \ge 511$ path times with 1.9\% median error at the common MMA shape, and the direction of the crossover at the H100 shape.
Restricting the streaming multiprocessors of one A100, with nothing fitted, reproduces the A100-to-GB200 reversal.
At production-relevant orders the model does not resolve the fastest fused path; the result identifies the crossover mechanism.
\end{abstract}

\keywords{Discontinuous Galerkin, high-order methods, GPU, Tensor Core,
kernel fusion, GEMM, roofline model, machine balance, instruction-aware transfer}

\maketitle

\section{Introduction}
\label{sec:introduction}
High-order discontinuous Galerkin (DG) discretizations are attractive for atmospheric and fluid dynamics.
Their element-local data dependencies suit graphics processing units (GPUs), and their arithmetic intensity grows with polynomial order.
On tensor-product elements, the dominant cost is a sequence of one-dimensional contractions along each coordinate direction.
Therefore, the operator never forms a dense matrix of the full three-dimensional elements.

Two implementation strategies are common.
Element-local \emph{fused} kernels carry the flux evaluation, the contractions, and the surface lift through registers and shared memory.
In these kernels, no intermediate field reaches global memory.
\emph{Library} implementations pass each contraction to a dense general matrix multiplication (GEMM).
The optimized mainloop of the GEMM reaches a larger fraction of the arithmetic roof than hand-written code normally does.
These implementations also store intermediate fields in global memory.
The usual argument for the choice is arithmetic intensity.
At high order the contraction dominates and the library should win.
At low order the traffic dominates and fusion should win.
Hardware dependence by itself is unsurprising.
The open question is whether a crossover that is measured for a fixed operator can be transferred to another architecture without timing the candidate implementations there.

This transfer question remains open for three reasons.
Existing comparisons generally do not separate the two strategies.
In one study, the unit of comparison is a whole code that a different group tuned \citep{fischer2020}.
In another study, the comparison is not made, because the GEMMs are small and irregular at the shapes of that study and the operator needs fusion that the libraries do not offer \citep{tu2026}.
The best-performing choice is known to move between machines.
However, no study that Table~\ref{tab:positioning} summarizes compares implementation families across architectures at a fixed operator, input, and code base.
In addition, no such study predicts the ranking on a target architecture without timing the paths there.
Roofline arguments are made at the level of a kernel or of an implementation \citep{swirydowicz2019,cao2026}.
These arguments do not establish whether the binding resource, and the fraction of its roof that a kernel reaches, are preserved when the architecture changes (Section~\ref{sec:roofline-its-granularity}).

We extract the three-dimensional advection operator of a production DG dynamical core into a benchmark that preserves the array-based interface and the numerics of the operator.
We implement the operator along six paths that are organized as pairs.
Three paths are fused kernels, and three paths are built from library GEMMs.
Each pair changes one implementation choice only.
This choice is the matrix multiply-accumulate (MMA, written \texttt{mma} in instruction names) instruction, the per-GEMM library/mainloop choice, or the epilogue fusion (Section~\ref{sec:operator-implementation-paths}).
We measure them on three GPU systems (GB200, H100, A100) over polynomial degrees $p = 7$ to 1023.
At each order, we hold the numerics and the input fixed across implementations.
The scope is a mechanism and architecture-transfer study of implementation crossovers.
This study does not present a practical autotuner for production-order DG.

\begin{figure*}[tp]
\centering
\includegraphics[width=\linewidth]{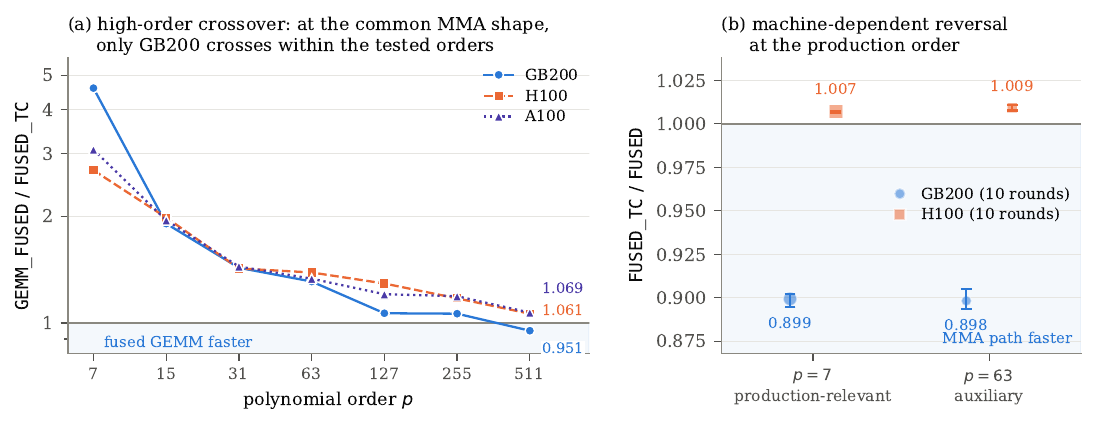}
\caption{
  Machine-dependent changes in the fastest implementation of the same DG operator.
  (a) Ratio $\texttt{GEMM\_FUSED}/\texttt{FUSED\_TC}$ against polynomial order $p$ in the common-shape comparison, on a logarithmic vertical axis.
  (b) Ratio $\texttt{FUSED\_TC}/\texttt{FUSED}$ on GB200 and H100.
  Section~\ref{sec:operator-implementation-paths} defines the names in these ratios.
  A ratio below 1 means that the implementation that is named first is the faster one.
  Points are medians of interleaved runs of the step-loop wall time per RK stage (Section~\ref{sec:timing-protocol}).
  Error bars in (b) are the min--max range over the ten rounds on each machine.}
\label{fig:f1}
\end{figure*}

The implementation crossover is not fixed by the operator alone.
With the same operator, numerics, input, and benchmark code base, at $p = 511$ the fused-epilogue path that uses CUDA Templates for Linear Algebra Subroutines and Solvers (CUTLASS) requires 5.0\% less step-loop time than the fully fused Tensor Core kernel on GB200.
On A100, however, the same path requires 6.9\% more step-loop time (Figure~\ref{fig:f1}(a)).
Here CUDA is NVIDIA's Compute Unified Device Architecture GPU programming platform, and the step-loop time is the wall time per Runge--Kutta (RK) stage of the step loop.
The MMA shape is held at \texttt{m8n8k4} on every machine here, and we call this the common-shape comparison.
Section~\ref{sec:architecture-dependent-reversal} also defines the preferred-shape comparison, in which H100 selects a wider shape.
A small but repeatably resolved machine-dependent ordering is already present at the production-relevant order $p = 7$.
At this order, the fused Tensor Core kernel is 10.1\% faster than the fused CUDA Core kernel on GB200 and 0.7\% slower on H100.
The larger crossover, and its prediction, are demonstrated at high order.

Timing each path on the target architecture is not always an option, because the target may not be available yet.
For example, an architecture is selected before procurement, or a port is prepared before access to the system is granted.
However, the workload and the toolchain that compiles for the target are available in such cases.
We propose a performance model for this situation, which requires no performance measurement on the target.
Its inputs are the capacities in the target datasheet and the MMA shape that a cross-compiled binary for the target issues.
The model also requires the workload to have been profiled with hardware counters on one or more architectures that are available (Section~\ref{sec:subpipe-model}).
Our three contributions are the following.
\begin{itemize}
\item We measure a machine-dependent crossover for a fixed operator, numerics, input, and code base.
By a machine-dependent crossover we mean a crossover between architecture--software-stack combinations.
We show that nominal-peak scaling does not transfer this crossover (Sections~\ref{sec:why-optimum-moves-order} and \ref{sec:instruction-aware-model}).

\item We show retrospectively that, once the instruction-pricing form is specified, its parameters transfer a workload that was profiled on other architectures to a target.
This transfer uses no target timings or counters for parameter instantiation.
The transfer is quantitative for the kernels that we compile ourselves, and it is conditional on the dispatched kernel family for vendor-library kernels (Sections~\ref{sec:instruction-priced-subpipe} and \ref{sec:predictive-validation}).

\item We provide an orthogonal perturbation that is consistent with the machine-balance explanation.
In this experiment, we restrict the active streaming multiprocessors (SMs) of one A100.
The ranking crosses 1 and agrees with GB200 at matched nominal dynamic random-access memory (DRAM) bytes per active-SM cycle (Section~\ref{sec:balance-controlling-variable}).
\end{itemize}

The conceptual distinction in the second contribution is that work that is counted in floating-point operations (FLOP) is not architecture invariant for these kernels.
The same FLOP count can expand into instructions with different execution costs.
On the three architectures that are tested here, the issued-instruction demand becomes transferable when each instruction is priced by the empirical relation that Section~\ref{sec:instruction-priced-subpipe} establishes.

\section{Background and Related Work}
\label{sec:background-related-work}

\subsection{Tensor-product DG operator}
\label{sec:tensor-product-dg-operator}
We consider the nodal discontinuous Galerkin discretization of three-dimensional linear advection, $\partial q/\partial t + \nabla \cdot (q\,\mathbf{u}) = 0$, of a scalar $q$.
The velocity is $\mathbf{u}=(u,v,w)$, the Cartesian coordinates are $(x,y,z)$, and the time is $t$.
The domain is a periodic unit cube that is partitioned into hexahedral elements.
The basis is a tensor-product Lagrange basis on Legendre--Gauss--Lobatto nodes.
With $N_q = p + 1$ nodes per direction and $N_e$ elements in the mesh, each element carries $N_p = N_q^3$ volume nodes and $N_{\mathrm{fp}} = 6 N_q^2$ face nodes.
The tendency at every volume node is
\begin{equation}
\frac{\partial q}{\partial t} = - ( E_x D_x(q u) + E_y D_y(q v) + E_z D_z(q w) + L f^{*} ),
\end{equation}
where $D_x$, $D_y$, and $D_z$ denote the one-dimensional differentiation matrix $D$ when it is applied along the first, second, and third index of the element tensor, respectively.
The symbols $E_x$, $E_y$, $E_z$ are the per-node metric factors, and $L$ lifts the surface flux difference $f^{*}$ from the $6 N_q^2$ face nodes onto the volume nodes.
On an affine hexahedron the lift is separable, so the operator stores a $(N_q, 6)$ one-dimensional lift and not a dense $(N_q^3, 6 N_q^2)$ matrix.

The surface term is evaluated on all six faces from the interior (M) and exterior (P) traces of $q$ and $\mathbf{u}$.
In this strong form, $f^{*}$ is the numerical flux minus the interior flux.
With the normal velocities $V_M = \mathbf{u}_M \cdot \mathbf{n}$ and $V_P = \mathbf{u}_P \cdot \mathbf{n}$ for the outward unit normal $\mathbf{n}$ of each face node, this difference for the local Lax--Friedrichs numerical flux is
\begin{equation}
\begin{aligned}
f^{*} &= \tfrac{1}{2} F \bigl( q_P V_P - q_M V_M - \alpha (q_P - q_M) \bigr), \\
\alpha &= \tfrac{1}{2} \lvert V_P + V_M \rvert,
\end{aligned}
\end{equation}
where $F$ is the per-face-node surface-to-volume scale factor and $\alpha$ is the local dissipation speed.
The code uses this form of $\alpha$.
The more common choice for the local Lax--Friedrichs flux is $\alpha = \max(\lvert V_M \rvert, \lvert V_P \rvert)$.
For the constant velocity of the benchmark input, the two choices give the same value.

The three contractions cost $6 N_q N_p$ floating-point operations per element against $O(N_p)$ words of element state.
Therefore, the arithmetic intensity of the volume term grows linearly in $N_q$.
In contrast, the ratio of face nodes to volume nodes is $N_{\mathrm{fp}}/N_p = 6/N_q$.
Throughout, $q$, $u$, $v$, $w$, and the per-direction $E=(E_x,E_y,E_z)$ (\texttt{Escale} in the code) are pointwise fields.
The normal $\mathbf{n}$ (\texttt{normal\_fn}) and $F$ (\texttt{Fscale}) are per-face-node, per-element fields.
The benchmark input prescribes a constant velocity, but no implementation exploits this.

For atmospheric turbulence in implicit large-eddy simulation, DG at eleventh order resolves shorter wavelengths than third- and seventh-order finite-volume schemes at comparable degrees of freedom.
In addition, DG at seventh and eleventh order costs less in total than the third-order finite-volume core in both implicit and explicit large-eddy simulation \citep{kawai2026}.
We therefore take $p = 7$ to 11 as the production-relevant range.

\subsection{GPU implementations of the tensor-product operator}
\label{sec:gpu-implementations-tensor-product}
The tensor-product contraction (sum factorization) dominates nodal DG operators.
Its cost argument is well known \citep{orszag1980,deville2002}.
On GPUs the contraction has been optimized since the earliest ports \citep{kloeckner2009,swirydowicz2019}.
The contraction has also been optimized in Nek5000 and NekRS \citep{fischer2022}, the Modular Finite Element Methods (MFEM) library \citep{anderson2021}, the matrix-free framework of deal.II \citep{kronbichler2012,kronbichler2019}, and libParanumal \citep{swirydowicz2019}.
Several of these codes share a common element-operator interface \citep{brown2021,kolev2021}.
The Center for Efficient Exascale Discretizations (CEED) bake-off problems \citep{fischer2020} compared three such codes on a CPU platform and a fourth on V100 GPUs.
Table~\ref{tab:positioning} places these and the following studies on the comparison axes of this paper.
The table also records what each study held fixed while comparing.

No sharp boundary separates the two strategies that Section~\ref{sec:introduction} defines.
For tensor bases, \citet{beams2020} build the contraction from device-level batched GEMMs that are fused inside one kernel.
For non-tensor bases, they select among standard batched Basic Linear Algebra Subprograms (BLAS) calls.
The library side of this trade-off uses the vendor and template GEMM libraries \citep{nvidia-cublas,thakkar2023}.
The $x$ volume GEMM of this operator has a reduction dimension $K = N_q$ and output columns $N = N_q^2 N_e$ (Section~\ref{sec:library-paths-gemm-gemm}).
This GEMM is very long and thin, and it therefore differs from the many-small-matrix range that batched BLAS targets \citep{abdelfattah2021}.
Epilogue fusion is a feature of the template libraries \citep{thakkar2023}.
This feature enables a matrix multiplication to be fused with the operations that follow it, in its epilogue \citep{thakkar2023,shah2024}.

The use of matrix engines for numerical computation is an established line of work \citep{markidis2018,haidar2018,ootomo2022}.
For a matrix-free symmetric interior penalty Galerkin (SIPG) DG Laplacian on A100, \citet{cui2024} shows that the gain of FP64 Tensor Cores depends on a conflict-free shared-memory layout for the MMA fragments.
This layout nearly halves the shared-memory wavefronts.
The layout also removes most of the memory-pipeline stalls of the kernel that uses the same instructions without the layout.
The $2.3\times$ gain that they report is measured on A100, whose FP64 Tensor Core peak is twice its CUDA Core peak.
On AMD MI250X, kernels that are built from the FP64 matrix-core instruction give the highest throughput for the CEED bake-off kernels at the higher polynomial degrees \citep{kolev2023}.
The stated reason is the register-level operand access of the matrix cores, not the arithmetic peak.
Both results place the gain in data movement.
We adopt the layout of \citet{cui2024} in the Tensor Core path of this paper.
\citet{tu2026} program FP64 Tensor Cores for MFEM operators and combine this with kernel fusion.
They report up to $2\times$ performance gains on GH200 and GB200 at scale.
On CPU matrix engines, \citet{wang2026} report a $1.6\times$ gain against the $4\times$ peak advantage.

\subsection{Positioning}
\label{sec:positioning}
Table~\ref{tab:positioning} shows the gap that this paper fills.
Studies that report more than one GPU architecture do so for one implementation.
Studies that compare implementations do so on one architecture, or at the level of complete codes \citep{fischer2020}.
A machine-dependent optimum is the starting assumption of autotuning and of performance portability \citep{bell2009,sedaghati2015,medina2014,brown2021}.
However, none of the work that we cite in this area reports the ranking of two implementations on a machine that it did not measure.

The closer comparison for the model of Section~\ref{sec:instruction-aware-model} is the prediction of GPU performance for a machine on which the workload was not measured.
This prediction problem is not new.
\citet{guo2015} predict the time of a sparse matrix-vector kernel on a target GPU from its measurement on a reference GPU.
\citet{ardalani2015} predict the GPU speed-up of a program from a run of its CPU code through a learned map.
\citet{vanlanker2024} project a kernel that was measured on one NVIDIA GPU onto another.
They use a hierarchical roofline and a comparison of the assembly instructions that the compiler emits for the two GPUs, and their mean errors are 5.9 to 20.3\% on three mini-applications.
The last of these studies is the closest in purpose to Section~\ref{sec:subpipe-model}, which also reads the instructions of the binary that is compiled for the target.
The new element of this paper is not the transfer of a profiled workload to an unseen machine but the ranking of implementation families on that machine.
\citet{jarmusch2026} build analytical models of B200 and MI300A from microbenchmarks of each architecture.
They predict the time of one microbenchmark kernel at a time, with mean errors of 1.3\% on B200 and, after a per-case calibration, 0.09\% on MI300A.
They reuse the model structure on H200 and MI250X after they update the bandwidth and cache parameters, and their application-level errors on these machines are much larger.
Their models characterize an architecture and price its arithmetic by a throughput per precision.
The model of this paper differs in what it predicts, which is the ranking of qualitatively different implementations of one operator.
Our model also differs in what it prices, which is the instruction that each implementation issues.

\citet{konstantinidis2017} build a roofline from device microbenchmarks and hardware-counter profiles of a kernel.
\citet{stevens2020} count operations symbolically and rank variants of three computations, one of which is a discontinuous Galerkin differentiation operator.
Both studies take their machine-specific parameters from measurements on the machine that they predict.
\citet{lee2025}, \citet{muralidharan2016}, and \citet{hochgraf2025} forecast latency, select code variants, or reuse GEMM variants on a GPU that was not among the GPUs on which their predictors were trained or tuned.
All three studies learn from a set of measured GPUs.
The model of Section~\ref{sec:subpipe-model} takes no measurement from the target for the paths that this study compiles.
This model prices the arithmetic capacity by the share of a counter-defined FP64 execution budget that the issued instruction uses, and not by a nominal machine peak (Section~\ref{sec:instruction-priced-subpipe}).
The reason is that Section~\ref{sec:does-accounting-transfer-a100} shows that the nominal FP64 Tensor Core peak fails on H100 for these kernels.
The implementation families that we rank are very different programs, such as an element-local fused kernel against a vendor GEMM.
These families also differ in the resource that binds them.
Therefore, we rank them by a mechanism and not by a learned map from device features to a choice.

\begin{table*}[tp]
\centering
\footnotesize
\caption{
  Positioning of the studies that Section~\ref{sec:gpu-implementations-tensor-product} discusses on the comparison axes that this paper uses, together with what each study held fixed while comparing.
  A dash indicates that the cited study does not report the corresponding comparison, not that the comparison is absent from the literature.
  BP and BK denote bake-off problems and kernels, respectively, DOF denotes degrees of freedom, and MAGMA denotes Matrix Algebra on GPU and Multicore Architectures.
  Entries, including the reported hardware, are taken from the cited work itself.}
\label{tab:positioning}
\setlength{\tabcolsep}{3pt}
\begin{tabularx}{\tblwidth}{@{}>{\hsize=.91\hsize}L>{\hsize=.83\hsize}L>{\hsize=.95\hsize}L>{\hsize=.74\hsize}L>{\hsize=.83\hsize}L>{\hsize=1.74\hsize}L@{}}
\toprule
Study & Fused / matrix-free & Compared with a library GEMM & Matrix engine & GPU architectures reported & What was held fixed while comparing \\
\midrule
Kl\"ockner et al. 2009 & yes & no & no & one & -- \\
\'Swirydowicz et al. 2019 & yes & no & no & one (P100) & operator and machine; optimization sequence compared, with an empirically calibrated roofline \\
Fischer et al. 2020 (bake-off) & yes & no & no & one (V100) for one code; the other three codes are on a CPU platform & problem (BP/BK) and DOF count; codes, numerics, tuning teams, and platforms differ; the comparison is between whole codes, not along a single axis \\
Beams et al. 2020 & yes (fused device-level GEMM) & yes (selection among batch-BLAS calls) & no & one (V100) & machine and code base; the GEMM call is varied per basis \\
Cui 2024 & yes & no & yes, Tensor Core & one (A100) & machine; wide-MMA against MMA, and the conflict-free layout, fixed one at a time \\
Kolev et al. 2023, Sec. 2.1 & yes & no & yes, matrix core & one (MI250X) & machine; order swept, N = 1 to 15, and the vector and matrix-core implementations compared \\
Kolev et al. 2023, Sec. 2.2 & yes (MAGMA non-tensor backend) & yes (against standard BLAS) & no & two (A100, MI250X) & backend (MAGMA) and problem; the blocking parameter \texttt{nb} is tuned separately on each GPU by an offline sweep \\
Cao et al. 2026 & yes & no (the baseline is the Nek axhelm kernel) & yes, FP64 Tensor Core on A100 & two (NVIDIA A100, Hygon K100) & operator and code base; recompute against load of the geometric factors, with the Nek axhelm kernel as the baseline \\
Tu et al. 2026 & yes & explicitly rejected (small, irregular GEMMs; fusion needed) & yes, FP64 Tensor Core & two (GH200, GB200) & everything but the machine: the same implementation is run on both \\
Wang et al. 2026 & yes (CPU) & no & yes, CPU matrix engine & -- (CPU) & operator and machine; staged optimization \\
This paper & yes & yes & yes, FP64 Tensor Core & three (GB200, H100, A100) & code base, numerics, input (\texttt{NeX/NeY/NeZ}, \texttt{PolyOrder}, \texttt{dt}, \texttt{nstep}), and the surrounding kernels and their launch schedule, fixed by the path contract; three ablation axes \\
\bottomrule
\end{tabularx}
\end{table*}

\subsection{Roofline and the level at which it is applied}
\label{sec:roofline-its-granularity}
The comparisons in this paper are interpreted with roofline arguments \citep{williams2009}.
These arguments have been applied to this operator at the level of a kernel or of a complete implementation \citep{swirydowicz2019,cao2026}.
Roofline arguments have also been placed against several roofs, level by level of the memory hierarchy \citep{yang2020} or against instruction throughput \citep{ding2019}.
We examine whether the fraction of a roof that a path reaches, and the resource that binds the path, are preserved when the architecture changes (Section~\ref{sec:instruction-aware-model} and Section~\ref{sup:roofline-placement-using-measured} of the supplemental material).
Section, table, and figure numbers with the prefix S refer to the supplemental material.
This examination requires that the arithmetic roof is selected according to the instruction that a path issues.
The reason is that the FP64 Tensor Core peak is twice the CUDA Core peak on H100 and A100 but equal to it on GB200 (Table~\ref{tab:machines}).
Scalable Computing for Advanced Library and Environment (SCALE) is a library and model for regional atmospheric simulation \citep{nishizawa2015,sato2015}.
For the discontinuous Galerkin core of SCALE on CPUs, \citet{ren2026} find roofline and execution-cache-memory predictions unsuitable in the arithmetic-intensive range.
They estimate parameters of their analytical model with a learned model on the machine that they predict.
On GPUs we re-price the arithmetic roof instead.
The model of Section~\ref{sec:instruction-aware-model} is fitted to no target measurement at all.

\section{Operator and Implementation Paths}
\label{sec:operator-implementation-paths}

\subsection{Extracted benchmark}
\label{sec:extracted-benchmark}
SCALE-DG is the discontinuous Galerkin dynamical core that is under development for SCALE.
\citet{kawai2026} and \citet{ren2026} study this core.
The three-dimensional advection kernels of SCALE-DG are published as a standalone extraction \citep{kawai-extraction}.
This extraction removes the Message Passing Interface (MPI) decomposition, mesh management, input and output, and physical parameterizations of the full model.
The extraction keeps the DG kernels and their memory-access patterns, including the indirect face addressing.
The six implementation paths of Table~\ref{tab:paths} were developed for this study and are released separately \citep{thiswork-code}.

The benchmark retains the original numerics and the array-based interface of Section~\ref{sec:tensor-product-dg-operator}.
The benchmark keeps only the tendency evaluation and the time integration, which is a three-stage third-order strong-stability-preserving RK scheme.
Therefore, the tendency is evaluated three times per step.
Every run occupies a single GPU without MPI, and the periodic halo is a device-side gather within that GPU.

The domain is the unit cube at every order, so an increase of the element count refines the mesh and does not enlarge the domain.
Orders $p = 7$ through $255$ are run at a fixed $256^3$ volume degrees of freedom, i.e., $N_e = (256 / N_q)^3$.
From $p = 511$ upward, those orders run at $N_e = 1$, i.e., the degree-of-freedom count grows as $N_q^3$.
The order coverage differs among the machines because of memory capacity (Table~\ref{tab:machines}).
The payload at $p = 1023$ reaches 180 GiB and fits only in a GB200.
A100 supports orders only up to $p = 575$.

\subsection{Overview of the six paths}
\label{sec:paths-overview}
All six paths compute the same tendency with the same numerics.
They differ in how the computation is divided into GPU kernels and in the instruction that performs the multiply-add operations of the contractions.
The paths form two families of three paths, which correspond to the two strategies of Section~\ref{sec:introduction} (Table~\ref{tab:paths}).

The first family is the fused family: \texttt{FUSED}, \texttt{FUSED\_TC}, and \texttt{FUSED\_DFMA}.
In these paths, hand-written kernels read the state of an element from global memory.
The kernels then compute the volume fluxes, the three derivatives, and the surface term in shared memory and registers.
No volume-sized intermediate field is written to global memory, and only the tendency is written back.
\texttt{FUSED} performs the contractions on the CUDA Cores, and \texttt{FUSED\_TC} performs them with the FP64 Tensor Core instruction.
\texttt{FUSED\_DFMA} is a control.
It is built from the source of \texttt{FUSED\_TC}, but it uses ordinary double-precision multiply-add instructions in place of the Tensor Core instruction.

The second family is the library family: \texttt{GEMM}, \texttt{GEMM\_CUTE}, and \texttt{GEMM\_FUSED}.
In these paths, each RK stage is a sequence of separate kernels.
One kernel computes the volume fluxes, and three GEMMs then compute the three derivatives.
Another kernel computes the numerical flux on the faces.
A final kernel applies the metric weights and the separable surface lift, and it assembles the tendency.
We call this kernel the lift and assembly kernel.
The kernels pass their results to each other through global memory.
\texttt{GEMM} uses cuBLAS, NVIDIA's CUDA BLAS library, for the three GEMMs.
\texttt{GEMM\_CUTE} uses CUTLASS in place of cuBLAS for the GEMMs where CUTLASS is faster.
\texttt{GEMM\_FUSED} keeps the GEMM mainloops of \texttt{GEMM\_CUTE}, but it does the work of the lift and assembly kernel at the end of the last GEMM, which is called its epilogue.
Therefore, \texttt{GEMM\_FUSED} runs one kernel fewer and does not write the last derivative to global memory.

In short, the fused family keeps the intermediate data on chip.
The library family moves the intermediate data through global memory, but it performs the contractions with tuned library GEMMs.
Within each family, pairs of paths differ in one choice only (Section~\ref{sec:controlled-one-axis-comparisons}).
The following subsections describe each path in detail.

\subsection{Fused CUDA Core path (FUSED)}
\label{sec:fused-cuda-core-path}
\texttt{FUSED} stages the differentiation matrix, the pointwise volume fluxes, and the numerical fluxes of the six faces in shared memory.
It performs the contractions as length-$N_q$ inner products in registers.
The per-order grouping of the directions over kernels and the block and thread tiling are given in Section~\ref{sup:fused-per-order-structure} of the supplemental material.
It is implemented only at orders $p \le 255$.

\subsection{Fused Tensor Core path (FUSED\_TC) and the iso-schedule ablation (FUSED\_DFMA)}
\label{sec:fused-tensor-core-path}
The Tensor Core path replaces the inner products with the FP64 Tensor Core instruction \texttt{mma.sync.aligned.m8n8k4.f64}.
The operand fragments of the instruction require their own data layout.
Therefore, the shared-memory panels, the chunking of the contraction, and the register budget differ from the CUDA Core path.
\texttt{FUSED\_DFMA} separates the instruction from the schedule that is built around it.
In this build, double-precision fused multiply-add (DFMA) instructions replace the double-precision matrix multiply-accumulate (DMMA) instruction.
This build keeps the same fragment layout, thread counts, tiles, register budget, and summation order.
The pair $\texttt{FUSED\_DFMA} \leftrightarrow \texttt{FUSED\_TC}$ therefore differs in the instruction alone.
The ratio of the two paths is the value of the MMA instruction at a fixed schedule (Figure~\ref{fig:f3}(a)).
Beyond $p = 511$, the only fused paths are \texttt{FUSED\_TC} and \texttt{FUSED\_DFMA} at $p = 767$.
No fused path is implemented at $p = 575$ or $p = 1023$.

\subsection{Library paths (GEMM, GEMM\_CUTE, GEMM\_FUSED)}
\label{sec:library-paths-gemm-gemm}
The three volume GEMMs of the library paths have the following shapes.
The $x$ GEMM has $M = K = N_q$ and $N = (N_p / N_q)\, N_e$.
Here $M$, $N$, and $K$ are the output rows, the output columns, and the reduction depth, respectively, and they are distinct from the trace labels M and P above.
The $y$ GEMM has $M = N = K = N_q$ and a batch count of $N_q N_e$.
The $z$ GEMM has $M = N_q^2$, $N = K = N_q$, and a batch count of $N_e$.

The three paths differ on two axes, and only one axis moves at a time.
\texttt{GEMM\_CUTE} differs from \texttt{GEMM} only in the library that runs each volume GEMM.
\texttt{GEMM\_FUSED} keeps the mainloops, tiles, and per-GEMM library assignment of \texttt{GEMM\_CUTE}.
It applies two epilogue fusions, the carry and the fold.
The carry is the fusion that Section~\ref{sec:paths-overview} describes.
It moves the work of the lift and assembly kernel into the epilogue of the last GEMM, and we call this GEMM the carrier.
The fold is enabled from $N_q = 64$ upward.
The fold applies the \texttt{Escale} metric weighting inside the $x$ and $y$ volume GEMMs, and it lets the $y$ GEMM consume the $x$ derivative directly.
As a result, the carrier reads two arrays instead of five.
Fusion has a scheduling penalty.
Without fusion, the three volume GEMMs overlap on concurrent streams.
With fusion, the carrier, and the $y$ GEMM where the fold is enabled, cannot start until the derivatives that they read are complete.
The per-order library assignment and the choice of the carrier were swept (Section~\ref{sup:swept-choices-library-paths} of the supplemental material).
This sweep ensures that the library axis and the fusion axis do not move together.

\subsection{Controlled one-axis comparisons}
\label{sec:controlled-one-axis-comparisons}
The six paths of Table~\ref{tab:paths} support the three paired comparisons that its last column names: the instruction, the library, and the epilogue fusion.
Each pair moves one implementation choice and holds the rest fixed.
We call what a pair holds fixed its path contract.
The fusion pair, \texttt{GEMM\_CUTE} against \texttt{GEMM\_FUSED}, also gives up the volume-GEMM concurrency.
Therefore, at the orders where the unfused control overlaps volume GEMMs, this difference underestimates the value of the epilogue fusions themselves.
For this reason, the difference is a bound and not a direct comparison (Section~\ref{sec:why-optimum-moves-order}).
The cross-strategy comparison that gives the main claim of the paper, \texttt{GEMM\_FUSED} against \texttt{FUSED\_TC}, is not one of these pairs.
Those two production paths differ in nearly everything.
Therefore, we interpret their difference through the one-axis comparisons and through the kernel-level model of Section~\ref{sec:instruction-aware-model}, and not as the effect of one choice.
Optimizations that were implemented, measured on the complete path, and not adopted are recorded in Section~\ref{sup:rejected-optimizations} of the supplemental material.

\begin{table*}[tp]
\centering
\footnotesize
\caption{
  Implementation paths.
  The last column names the controlled comparison in which each path participates, with the panel letters of Figure~\ref{fig:f3}.}
\label{tab:paths}
\noindent\begin{tabularx}{\tblwidth}{@{}lLLLL@{}}
\toprule
Path & Volume contraction & Surface lift \& assembly & Role & Controlled comparison \\
\midrule
\texttt{FUSED} & hand-written shared-memory panels, length-$N_q$ inner products & fused & CUDA Core production path & --- \\
\texttt{FUSED\_TC} & \texttt{mma.sync.m8n8k4} (FP64 Tensor Core, TC) & fused & Tensor Core production path & (a) instruction \\
\texttt{FUSED\_DFMA} & fused multiply-add (FMA) on the same fragment layout, same source & fused & iso-schedule ablation of the instruction & (a) instruction \\
\texttt{GEMM} & cuBLAS, all three GEMMs & separate kernel & library baseline & (b) library \\
\texttt{GEMM\_CUTE} & CUTLASS-based, with per-GEMM library assignment; unweighted epilogues & separate kernel & unfused control for the CUTLASS path & (b) library, (c) fusion \\
\texttt{GEMM\_FUSED} & CUTLASS-based, mainloops and library assignment identical to \texttt{GEMM\_CUTE} & carry (fused into the epilogue of the last GEMM) & fused-epilogue production path & (c) fusion \\
\bottomrule
\end{tabularx}
\end{table*}

\section{Experimental Methodology}
\label{sec:experimental-methodology}

\subsection{Machines}
\label{sec:machines}
All measurements use one whole GPU of each of the three systems of Table~\ref{tab:machines}.
These systems are a Grace-Blackwell GB200, an H100 SXM 94 GB, and an A100 SXM 40 GB.
The one exception is the experiment of Section~\ref{sec:balance-controlling-variable}, which restricts the SMs that are available to the benchmark on one GPU.
Each site provides its own production software stack, including the cuBLAS library that dispatches the kernels in Table~\ref{tab:cublas}.
Therefore, the machine axis varies the GPU architecture together with its stack, and the effects of the two are not separated here.
The instruction-expansion evidence isolates an architectural mechanism, but the path-level crossover does not separate architecture from stack.

\begin{table*}[tp]
\centering
\footnotesize
\caption{
  Hardware and software configuration of the three systems.}
\label{tab:machines}
\noindent\begin{tabularx}{\tblwidth}{@{}LLLL@{}}
\toprule
 & GB200 & H100 & A100 \\
\midrule
Site / scheduler & RIKYU / Slurm & TSUBAME 4 / Grid Engine (SGE, Sun Grid Engine) & Wisteria/BDEC-01 Aquarius / Fujitsu Technical Computing Suite (TCS) \\
Host processor & NVIDIA Grace (AArch64, \texttt{-tp neoverse-v2}) & x86-64 AMD Zen 4 (\texttt{-tp znver4}) & Intel Xeon Platinum 8360Y, Ice Lake (\texttt{-tp icelake-server}) \\
Process / MPI & single process, no MPI & same & same \\
Compute capability & 10.0 & 9.0 & 8.0 \\
Streaming multiprocessors & 152 & 132 & 108 \\
SM boost clock & 2062 MHz & 1980 MHz & 1410 MHz \\
Memory capacity & 185 GiB & 93.6 GiB & 40 GiB \\
FP64 CUDA Core peak & 40.1 TFLOP/s & 33.5 TFLOP/s & 9.7 TFLOP/s \\
FP64 Tensor Core peak & 40.1 TFLOP/s ($1.0\times$) & 66.9 TFLOP/s ($2.0\times$) & 19.5 TFLOP/s ($2.0\times$) \\
High-bandwidth memory (HBM) peak & 7.9 TB/s & 2.396 TB/s & 1.555 TB/s \\
Driver & 580.173.02 & 595.71.05 & 535.54.03 \\
nvfortran / nvcc & 26.3-0 / 13.1.115 & 26.1-0 / 13.1.80 & 25.9-0 / 12.9.86 \\
Build flags & \texttt{-gpu=cc100} / \texttt{-arch=sm\_100} & \texttt{-gpu=cc90} / \texttt{-arch=sm\_90} & \texttt{-gpu=cc80} / \texttt{-arch=sm\_80} \\
\bottomrule
\end{tabularx}
\end{table*}

\subsection{Timing protocol}
\label{sec:timing-protocol}
Each run reports three timing quantities.
They are the CUDA-event device time of the tendency computation, the step-loop wall time per RK stage, and the end-to-end wall time per step (\texttt{Main}).
The step-loop time per stage is the primary metric of Section~\ref{sec:performance-crossovers}.
The device time serves the kernel-resolved analysis of Section~\ref{sec:instruction-aware-model}.
The end-to-end time is a cross-check (Section~\ref{sup:agreement-three-timing-measures} of the supplemental material).
The step-loop interval includes the RK update and the halo exchange, but it excludes an in-loop diagnostic (Section~\ref{sup:difference-between-stage-device} of the supplemental material).
An absolute device time per stage is the CUDA-event sum divided by the number of tendency calls that it covers (Section~\ref{sup:delayed-cuda-event-harvesting} of the supplemental material).

At each order, all implementation paths run from one namelist in which only \texttt{DqdtKernel\_Type} changes.
Therefore, the problem and timing configuration are identical across paths (Section~\ref{sup:benchmark-configurations} of the supplemental material).
CUDA graphs were disabled for all timings that this paper reports.
The reason is that the timed configuration should launch its kernels in the same way as the profiled configuration.
The kernel-resolved analysis of the profiled configuration (Section~\ref{sec:instruction-aware-model}) needs one profiler record per kernel launch.
We also ran with the graph enabled at $p = 7$ and 15 on the three machines.
In these runs, the graph removes a fixed cost per launch of 0.5 to 3.5\% of the end-to-end time per step.
No ratio of two paths crosses 1 in any of the 90 pairs of the six cells, and the fastest path of every cell is unchanged (Section~\ref{sup:graph-replay} of the supplemental material).
Paths are executed in turn in interleaved rounds with one run per path per round.
There are six rounds at $p \le 511$ and three at $p \ge 575$.
Dedicated within-job comparisons state their own round counts.
We report the median and the median absolute deviation over those runs.
Failed runs are excluded from the medians and are listed in the measurement record (Section~\ref{sup:data-provenance-reproducibility} of the supplemental material).

\subsection{Profiling protocol}
\label{sec:profiling-protocol}
We profile with Nsight Systems and Nsight Compute in that graph-disabled configuration.
Section~\ref{sup:profiler-invocation} of the supplemental material gives the collected metrics and the exact invocation.
Nsight Compute locks the SM clock well below the production boost clock.
This lower clock can change the apparent bottleneck (Section~\ref{sup:profiler-measurements-effect-clock} of the supplemental material).
We therefore use Nsight Compute to identify mechanisms, and we base optimization decisions on wall-clock measurements from interleaved runs.
The profiles that decide which resource binds are taken with unlocked clocks.

\subsection{Numerical validation}
\label{sec:numerical-validation}
Correctness is checked at two levels.
First, the complete owned tendency field is compared point by point against a reference path.
In this comparison, the velocity and metric coefficients vary from point to point on purpose, because the constant-velocity benchmark alone does not test the full interface (Section~\ref{sup:definition-point-varying-coefficients} of the supplemental material).
The reference is the unfused CUDA Fortran baseline \texttt{CUDAFORTRAN\_SPLIT} at $p \le 127$, and \texttt{GEMM} above that order.
This baseline is not among the six measured paths.
At the fixed-degree-of-freedom orders, we divide the largest absolute difference over the field by the largest absolute value of the reference field.
This quantity is between bit-identity and $10^{-14}$ for the library paths, and it is at most $2.1 \times 10^{-13}$ for the fused element-local paths on all three architectures.
At the single-element orders $p \ge 511$, the same quantity against \texttt{GEMM} is $6.4 \times 10^{-15}$ or better in every combination of machine and order that fits in memory.
The largest absolute value of the reference field is 8.55 at every order, so the largest absolute difference is $5.5 \times 10^{-14}$.
These combinations cover every single-element cell of the performance evaluation (Table~\ref{tab:a-validation-coverage} in Section~\ref{sup:numerical-validation-coverage} of the supplemental material).
On H100, the two MMA shapes produce bit-identical fields on both CUTLASS paths at $p = 511$, 575, and 767, so the shape affects performance only.

Second, the constant-velocity problem is integrated in time and compared with its exact solution.
All paths agree at the round-off level, and the error shows the expected third-order temporal convergence down to the round-off floor (Table~\ref{tab:validation} in Section~\ref{sup:numerical-validation-coverage} of the supplemental material).

\subsection{Sources of variation and claim threshold}
\label{sec:sources-variation-claim-threshold}
Compute-node-to-compute-node variation within one machine is approximately 0.1--0.6\%.
A 2.5-year compiler-version difference produces a median change of approximately 0.4\%, and the worst observed case is 3.2\% (Table~\ref{tab:variation} in Section~\ref{sup:separating-compiler-code-generation} of the supplemental material).
For comparisons assembled across compute nodes or measurement campaigns with the same compiler and executable, we therefore do not interpret differences below 1\%.
The 3.2\% outlier occurs for \texttt{FUSED\_DFMA} across compiler versions, so 1\% is not a claim threshold for such comparisons.
A difference of up to 3.2\% across compiler versions is treated as unresolved unless it is reproduced with a common build or by paired evidence.
For paired comparisons within one job on the same compute node with the same executable, these sources of variation cancel.
For these comparisons, we regard the local ordering as resolved only when the repeated-run ranges do not overlap.
The repeated-run range is the minimum-to-maximum interval of the runs of one path in one job.
At $p \ge 575$, where only three runs estimate the range, the 1\% same-build cross-campaign criterion is the one that applies.
Sections~\ref{sup:data-provenance-reproducibility} and~\ref{sup:provenance} of the supplemental material give the campaign, job, compute node, and source revision behind every figure and table of the main text.

\section{Performance Crossovers}
\label{sec:performance-crossovers}

\subsection{Performance landscape}
\label{sec:performance-landscape}
\begin{figure*}[tp]
\centering
\includegraphics[width=0.62\linewidth]{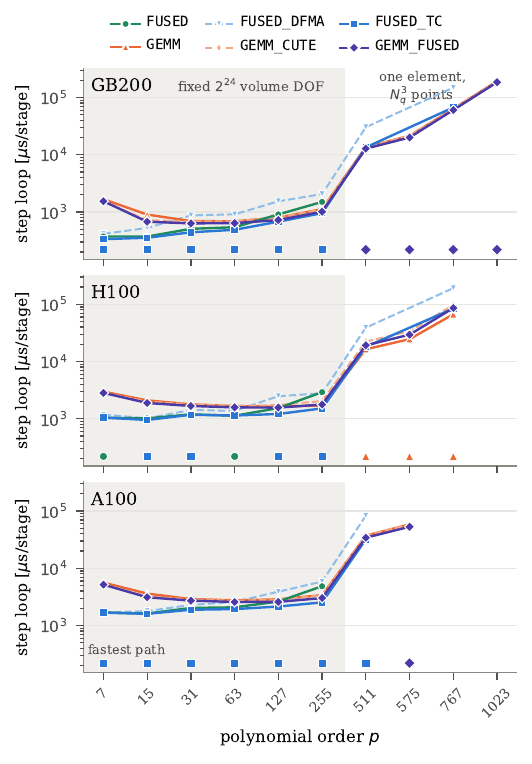}
\caption{
  Performance landscape: step-loop time per RK stage (logarithmic) for the six implementation paths of Table~\ref{tab:paths} on GB200, H100, and A100.
  Each point is the median over interleaved rounds under identical problem and timing configurations across paths.
  In the shaded region, $p \le 255$, the volume degree-of-freedom count is fixed at $256^3$.
  At larger orders a single element is used, so absolute times across this boundary mix order and problem size.
  The markers along the bottom of each panel indicate the fastest path at each order under the repeated-run range criterion of Section~\ref{sec:sources-variation-claim-threshold}.}
\label{fig:f2}
\end{figure*}

Figure~\ref{fig:f2} shows the step-loop time of the implementation paths of Section~\ref{sec:operator-implementation-paths} over the polynomial orders that each machine can run (Section~\ref{sec:extracted-benchmark}).
The path that minimizes this time changes with order on GB200 and H100, and at several orders it differs among the machines.
Therefore, no single path is optimal across orders and architectures.
On A100, the fused Tensor Core path is fastest at every order for which a fused path is implemented (Section~\ref{sec:fused-tensor-core-path}).

On GB200, in CUDA-device time, the fastest path reaches 14.0\% of the 40.1 TFLOP/s FP64 arithmetic roof and 80.0\% of the 7.9 TB/s HBM roof at $p = 7$.
At $p = 255$, at the same $256^3$ volume points, the fastest path reaches 75.1\% and 15.9\% of the same two roofs.
Therefore, the order moves the operator from a bandwidth-dominated toward a compute-dominated range.

\subsection{Order-dependent crossover}
\label{sec:order-dependent-crossover}
Figure~\ref{fig:f1}(a) plots the ratio $\texttt{GEMM\_FUSED}/\texttt{FUSED\_TC}$ against $p$.
On GB200, the ratio decreases steadily from 4.60 at $p = 7$ to 0.951 at $p = 511$.
The ratio crosses 1 between $p = 255$ and $p = 511$, so the ordering of the two main strategies reverses with order.
The large ratio at low order has a simple reason.
At $p = 7$, a volume contraction applies an $8 \times 8$ matrix along one index of the element tensor, so the reduction depth is 8.
This depth is too shallow to pay back the fixed cost of a dense GEMM.

Above $p = 511$, the fused paths exist only at $p = 767$, so Figure~\ref{fig:f1}(a) ends at $p = 511$.
At $p = 767$ on GB200, \texttt{FUSED\_TC} and \texttt{FUSED\_DFMA} take 67.1 and 148.8 ms per stage, against 60.5 ms for \texttt{GEMM\_FUSED}.
Therefore, $\texttt{GEMM\_FUSED}/\texttt{FUSED\_TC}$ is 0.902, and $\texttt{GEMM}/\texttt{FUSED\_TC}$ falls from 1.014 at $p = 511$ to 0.946.
All three library paths are faster than the fused kernel.
In CUDA-device time from $p = 511$ to 767, \texttt{FUSED\_TC} grows by a factor of 5.06, which equals $(768/512)^4$ for the leading $N_q^4$ term.
Over the same interval, \texttt{GEMM\_FUSED} and \texttt{GEMM} grow by 4.80 and 4.72.
Therefore, the library paths gain efficiency with order, and the fused path does not.
The one-axis ratios stay flat over the same interval.

\subsection{Machine-dependent reversal}
\label{sec:architecture-dependent-reversal}
The crossover in Figure~\ref{fig:f1}(a) does not occur at the same order on every machine.
At $p = 511$, the path definitions, input, and numerics are the same on all machines, and the MMA shape is the default $8\times 8\times 4$.
Under these conditions, \texttt{GEMM\_FUSED} requires 5.0\% less step-loop time than \texttt{FUSED\_TC} on GB200, but 6.1\% more on H100 and 6.9\% more on A100.
Only GB200 crosses 1 within the tested orders at that shape.
The sign of the reversal is the same under all three timing measures of Section~\ref{sec:timing-protocol} (Section~\ref{sup:agreement-three-timing-measures} of the supplemental material).

Two comparisons are kept separate throughout the paper.
The \emph{common-shape} comparison holds the MMA shape at $8\times 8\times 4$ on all three architectures.
Figure~\ref{fig:f1}(a) plots this comparison, and the accounting of Section~\ref{sec:instruction-aware-model} is built on it.
The \emph{preferred-shape} comparison lets each architecture use the shape that its production dispatch selects.
This comparison gives what a user of the code obtains.
The two comparisons differ on H100 alone, and there the difference changes the winner.
With the $16\times 8\times 4$ shape that H100 selects at high order, $\texttt{GEMM\_FUSED}/\texttt{FUSED\_TC}$ is 0.870 instead of 1.061.
With this shape, \texttt{GEMM\_FUSED} is the fastest path on H100 at $p = 511$ and $p = 575$.
Table~\ref{tab:two-shapes-p511} collects the $p = 511$ values of this ratio in the two comparisons and under the two timers of Section~\ref{sec:timing-protocol}.
Every claim about H100 names the comparison in which it is made.
The two comparisons have different roles.
The common-shape comparison establishes the mechanism by which the ranking transfers between architectures (Sections~\ref{sec:instruction-aware-model} and~\ref{sec:predictive-validation}).
The preferred-shape comparison tests the consequence of that mechanism for the configuration that a user runs (Section~\ref{sec:operational-predictions}).

\begin{table}[tp]
\centering
\footnotesize
\caption{
  $\texttt{GEMM\_FUSED}/\texttt{FUSED\_TC}$ at $p = 511$ in the two comparisons, by timer.
  A value below 1 means that the fused-epilogue library path is faster.
  The common shape is \texttt{m8n8k4} on every machine. The preferred shape differs from it on H100 alone (\texttt{m16n8k4}), so the other two columns repeat.
  Step-loop time per stage is the metric of this section. The device time per stage is the metric of Sections~\ref{sec:instruction-aware-model} and~\ref{sec:predictive-validation}.
  Table~\ref{tab:a-p511-ratio} of the supplemental material lists the campaign of every value.}
\label{tab:two-shapes-p511}
\begin{tabular}{@{}llrrr@{}}
\toprule
Comparison & Timer & GB200 & H100 & A100 \\
\midrule
common shape & step loop & 0.951 & 1.061 & 1.069 \\
common shape & device & 0.948 & 1.069 & 1.068 \\
preferred shape & step loop & 0.951 & 0.870 & 1.069 \\
preferred shape & device & 0.948 & 0.854 & 1.068 \\
\bottomrule
\end{tabular}
\end{table}

The shape is a choice on H100 only.
The reason is that the assembler splits the wider shapes into the same instruction sequence on GB200, and the wider shapes are unavailable on A100 (Section~\ref{sec:instruction-priced-subpipe}).
In addition, the shape is a parameter of the CUTLASS paths only.

At the production-relevant order $p = 7$, the fused Tensor Core kernel and the fused CUDA Core kernel swap places between GB200 and H100 (Figure~\ref{fig:f1}(b)).
The Tensor Core path saves 10.1\% on GB200, whereas on H100 the CUDA Core path is ahead by 0.7\%.
The same ordering holds at $p = 63$, where the ratio is 0.898 on GB200 and 1.009 on H100.
This pairing is not a one-axis comparison, but it ranks the two production kernels within each machine.
The H100 ordering passes the criterion of Section~\ref{sec:sources-variation-claim-threshold}, because the ranges over ten interleaved rounds do not overlap.
A re-measurement on another H100 node with a separately built executable reproduces this ordering (Section~\ref{sup:data-provenance-reproducibility} of the supplemental material).

The $p = 255$ and $p = 511$ cases differ in problem size as well as in order.
However, an element-count sweep at fixed order moves $\texttt{GEMM\_FUSED}/\texttt{FUSED\_TC}$ by less than 1\% when the problem size grows by a factor of 27 at $p = 255$ and 8 at $p = 511$.
Therefore, the problem size does not explain the crossover (Section~\ref{sup:element-count-sweep} of the supplemental material).

\subsection{Controlled one-axis evidence}
\label{sec:one-axis-evidence}
\begin{figure*}[tp]
\centering
\includegraphics[width=\linewidth]{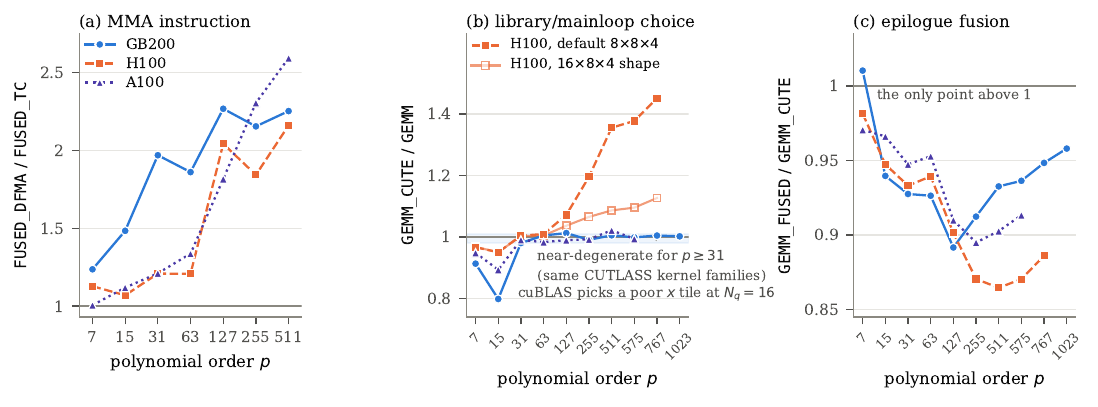}
\caption{
  One-axis ablations of three implementation choices: (a) the inner-product instruction, $\texttt{FUSED\_DFMA}/\texttt{FUSED\_TC}$; (b) the per-GEMM library/mainloop choice, $\texttt{GEMM\_CUTE}/\texttt{GEMM}$; and (c) epilogue fusion (the carry and the fold), $\texttt{GEMM\_FUSED}/\texttt{GEMM\_CUTE}$. The ratios in (c) include the concurrency that is lost when one volume GEMM is the fusion carrier.
  Each pair holds fixed the written contract of Section~\ref{sec:controlled-one-axis-comparisons}.
  Values above 1 indicate that the numerator is slower.
  In (b), the two H100 series are the two MMA shapes of Section~\ref{sec:architecture-dependent-reversal}.
  All panels are step-loop time per stage. The values are medians over six interleaved rounds from the campaign of Figure~\ref{fig:f1}(a).}
\label{fig:f3}
\end{figure*}

The crossovers above compare complete paths that differ in more than one way.
Therefore, we also report the three one-axis comparisons of Section~\ref{sec:controlled-one-axis-comparisons} (Figure~\ref{fig:f3}).
Appendix~\ref{app:one-axis-mechanism} gives the counter-level mechanism of each axis.

On the MMA axis on GB200, the replacement of the Tensor Core DMMA instructions with DFMA requires 23.7\% more time at $p = 7$ and 86.2\% more at $p = 63$.
On H100, the same replacement requires 12.8\% and 20.8\% more time.
The gain is larger on GB200, where the nominal roofs of the two instructions are equal, than on H100, where they differ by a factor of two.
The counters explain this: the memory supply limits the MMA gain on H100 and not on GB200 (Table~\ref{tab:mma}).
The ratios are an upper bound on the value of the MMA instruction.
The reason is that the DFMA version runs a warp shape that is suitable for the Tensor Core version (Appendix~\ref{app:one-axis-mechanism}).

On the library axis, the two sides are nearly identical on GB200 and A100.
The reason is that cuBLAS on these machines dispatches CUTLASS kernels of the same families and tile shapes (Table~\ref{tab:cublas}, Appendix~\ref{app:one-axis-mechanism}).
The axis compares two different implementations only on H100.
There, the ratio is 1.36--1.45 at $p \ge 511$ with the default MMA shape, and it is 1.04--1.13 at $p \ge 127$ with the shape that H100 prefers.

On the fusion axis on GB200, fusion requires 1.0\% more time at $N_q = 8 (p = 7)$.
Fusion saves a maximum of 10.8\% at $N_q = 128 (p = 127)$, and the saving falls to 4.2\% at $N_q = 1024 (p = 1023)$.
Only the decreasing branch above $N_q = 128$ follows the direction of the $1/N_q$ scaling of Section~\ref{sec:why-optimum-moves-order}.
Below this point, the fusion absorbs only the lift and assembly kernel (Section~\ref{sup:low-order-fusion-axis} of the supplemental material).

\subsection{Why the optimum moves with order}
\label{sec:why-optimum-moves-order}
The reversal of \texttt{FUSED\_TC} and \texttt{GEMM\_FUSED} between $p = 255$ and $p = 511$ on GB200 results from two competing effects.
The \emph{full-fusion saving} is the work that the full fusion of \texttt{FUSED\_TC} removes.
Throughout the paper, this saving is the separate \texttt{volume\_flux} kernel that \texttt{GEMM\_FUSED} executes in every stage and that \texttt{FUSED\_TC} never launches.
The carry and the fold of \texttt{GEMM\_FUSED} (Section~\ref{sec:library-paths-gemm-gemm}) are a different fusion.
This fusion is included in the \texttt{GEMM\_FUSED} side and is not the full-fusion saving.
The \emph{contraction-kernel gap} is the time by which the three contraction kernels of \texttt{FUSED\_TC} exceed the three library GEMMs of \texttt{GEMM\_FUSED}.

At the two orders on either side of the crossover, we record the individual contractions of both paths (Section~\ref{sup:low-order-fusion-axis} of the supplemental material).
Relative to the CUDA-device time per stage of \texttt{GEMM\_FUSED}, the full-fusion saving is 13.9\% at $p = 255$ and 8.5\% at $p = 511$.
At the same two orders, the contraction-kernel gap is 8.5\% and 14.0\%.
The difference between the two paths is the time of \texttt{FUSED\_TC} minus the time of \texttt{GEMM\_FUSED}.
Therefore, a positive value means that \texttt{GEMM\_FUSED} is faster.
The gap minus the saving is the attributed difference.
The attributed differences, $-$5.3\% and +5.5\%, reproduce the sign of the measured $-$6.4\% and +5.5\%.

\begin{figure*}[tp]
\centering
\includegraphics[width=\linewidth]{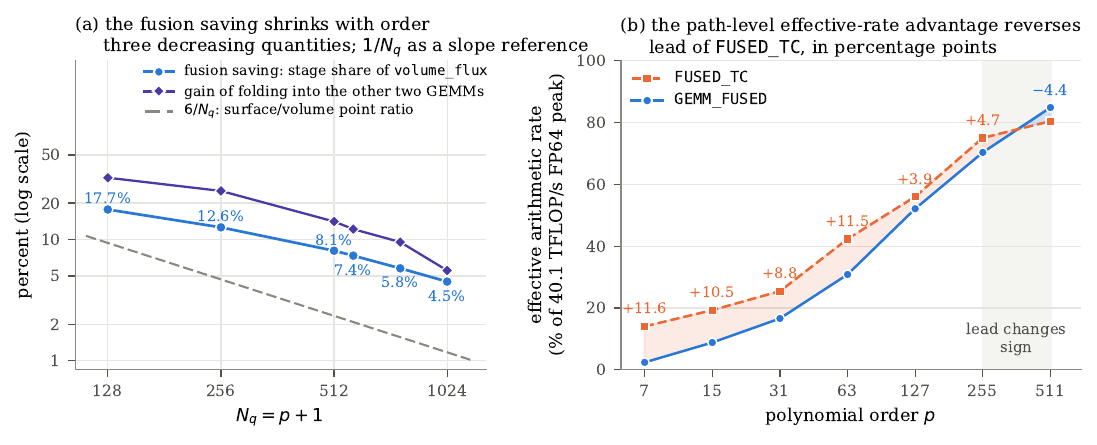}
\caption{
  On GB200, the benefit of fusion decreases with polynomial order and the effective-rate advantage of the fused Tensor Core implementation changes sign.
  (a) Three quantities against $N_q = p+1$ on logarithmic axes, so that a quantity that is proportional to $1/N_q$ appears as a line of slope $-1$. Circles: the full-fusion saving, the share of the steady-state stage period that \texttt{volume\_flux} takes.
  Diamonds: the gain of the fold of Section~\ref{sec:library-paths-gemm-gemm} against the unfolded configuration. Both sides use the same four-stage carrier, and overlap is disabled.
  Dashed: the surface-to-volume point ratio $6/N_q$.
  The plotted quantities are the full-fusion saving and the fold, not the carry, and they are not additive.
  (b) Effective arithmetic rate of \texttt{FUSED\_TC} and \texttt{GEMM\_FUSED} against $p$, normalized by the 40.1 TFLOP/s FP64 peak of GB200, from $p = 7$ through $p = 511$. The \texttt{FUSED\_TC} advantage is labeled in percentage points.
  All measurements are from one frozen executable (Section~\ref{sup:data-provenance-reproducibility} of the supplemental material).}
\label{fig:f4}
\end{figure*}

The full-fusion saving shrinks because the removed \texttt{volume\_flux} kernel performs $O(1)$ work per volume point, whereas each tensor contraction performs $O(N_q)$ work per volume point.
Therefore, the share of this kernel in a contraction-dominated stage is expected to decrease approximately as $1/N_q$, like the surface-to-volume point ratio $6/N_q$.
On the production \texttt{GEMM\_FUSED} path, the fraction of each RK stage that \texttt{volume\_flux} occupies falls from 17.7\% at $N_q = 128$ to 4.5\% at $N_q = 1024$ (Figure~\ref{fig:f4}(a)).
The removed traffic of the fold is likewise $O(1)$ per volume point, and the gain of the fold falls with that share.
However, neither series follows $1/N_q$ exactly (Section~\ref{sup:low-order-fusion-axis} of the supplemental material).
Therefore, we read $1/N_q$ as the direction of the decay and not as a rate.

The contraction-kernel gap moves in the opposite direction.
We use the conventional GEMM count of $2 N_q^4$ operations per contraction and the 40.1 TFLOP/s FP64 roof of GB200.
Against this roof, the three CUTLASS volume GEMMs of \texttt{GEMM\_FUSED} lead the three hand-written contraction kernels of \texttt{FUSED\_TC} by 7.9 percentage points of the roof at $N_q = 256$ and by 12.4 percentage points at $N_q = 512$ (Section~\ref{sup:low-order-fusion-axis} of the supplemental material).
The hand-written kernels also execute the fluxes and the epilogue, which their operation count omits.
In addition, the contractions were profiled at these two orders only.
Therefore, the gap is an upper bound on the difference in mainloop efficiency, and the measurements do not establish that the gap widens with order.
The concurrency that fusion gives up changes the time by at most 0.9\% (Section~\ref{sup:low-order-fusion-axis} of the supplemental material).

Figure~\ref{fig:f4}(b) summarizes the two effects through the effective arithmetic rate of each path, which is normalized by the same roof.
The library path becomes faster than the fused path between $N_q = 256$ and $N_q = 512$.
This change happens at the same place as the wall-time crossover in Figure~\ref{fig:f1}(a).
The one further measured point, $p = 767$, continues the trend (Section~\ref{sec:order-dependent-crossover}).

\section{An Instruction-Aware Cross-Architecture Performance Model}
\label{sec:instruction-aware-model}

\subsection{Model at a glance}
\label{sec:model-at-a-glance}
Write the time of kernel $k$ on machine $m$ in SM cycles as
\begin{equation}
T(k,m) = \eta(k)\,\max_{r} D_r(k,m).
\label{eq:model}
\end{equation}

The demands $D_r$ are in SM cycles.
The index $r$ runs over the counter-defined FP64 budget, L1, level-2 cache (L2), DRAM, instruction issue, the arithmetic logic unit (ALU) and FP32 FMA pipes, and shared-memory wavefronts.
The dimensionless calibration-architecture-profiled factor $\eta$ is the ratio of the elapsed SM cycles of the kernel on the calibration machine to its largest demand there.
A value $\eta = 1$ therefore denotes a kernel that runs exactly at its modeled roof.
Here and throughout, $\eta(k)$ denotes a value that is measured on a specified calibration machine.
This value is not a machine-independent property of the kernel.
When more than one calibration machine is available, the machine is chosen per kernel by the rule of Section~\ref{sec:subpipe-model}.
We do not write the calibration machine as an index of $\eta$.
The prediction in time divides the cycle count of Equation~\ref{eq:model} by the boost clock $f_m$ of the target (Table~\ref{tab:machines-per-sm-cycle}),
\begin{equation}
\hat t(k,m) = \frac{\eta(k)\,\max_{r} D_r(k,m)}{f_m},
\label{eq:model-time}
\end{equation}
and the path total is the sum of $\hat t$ over the kernels of the path.
Every kernel instance is profiled on a calibration machine at the same order and launch configuration.
The factor $\eta$ is carried from there, and it is not refitted on the target.
Therefore the model predicts a new machine and not a new workload (Appendix~\ref{app:model-contract}).
Two properties are used throughout.
First, the architecture enters only as capacities per SM cycle (Table~\ref{tab:machines-per-sm-cycle}) and as the instruction expansion that the compiler performs for it.
Second, $\eta$ is the only quantity that is read from a timing.
Sections~\ref{sec:does-accounting-transfer-a100} and~\ref{sec:instruction-priced-subpipe} derive the form of the demands from measurements.
Section~\ref{sec:subpipe-model} states the complete model.
Appendix~\ref{app:model-contract} states the inputs, the output, and the range of the model as a contract (Table~\ref{tab:contract}).

\subsection{Nominal roofs reproduce A100 and fail on H100}
\label{sec:does-accounting-transfer-a100}
The model is built kernel by kernel.
Therefore, from here on and in Section~\ref{sec:predictive-validation}, ratios are stated in CUDA-device time per stage unless the production timer is named explicitly.
The $p = 511$ reversal of Section~\ref{sec:architecture-dependent-reversal} is then $\texttt{GEMM\_FUSED}/\texttt{FUSED\_TC} = 0.948$ on GB200 and 1.068 on A100 (Table~\ref{tab:two-shapes-p511}).

We first tested whether nominal hardware roofs transfer this reversal.
The simplest such model assigns each path one arithmetic roof and one bandwidth roof.
It constrains the A100-to-GB200 time ratio of a path to lie between the ratio of the two arithmetic roofs and the ratio of the two bandwidth roofs.
For the CUDA Core paths, this interval is approximately 4.1 to 5.1.
Eleven of the thirteen measured CUDA Core path ratios lie outside this interval, so the measurements reject the model.
The measurements show that the resource that a kernel comes closest to saturating is not preserved across architectures (Section~\ref{sup:path-level-two-roof} of the supplemental material).
We therefore decomposed each path at $p = 511$ into its DRAM-saturated kernels and the remaining kernels, from a profile on GB200 at production clocks.
We scaled the two components by the ratio of HBM peaks and by the ratio of nominal FP64 Tensor Core peaks, and we fitted no coefficient.
The decomposition, the classification threshold (70\% of peak DRAM throughput), and its sensitivity are given in Section~\ref{sup:nominal-roof-accounting} of the supplemental material.
Two rival explanations that measurement rejects are given in Section~\ref{sup:rejected-hypotheses} of the supplemental material.

This accounting reproduces the GB200-to-A100 reversal.
For $\texttt{GEMM\_FUSED}/\texttt{FUSED\_TC}$ it predicts 1.075 on A100 from the GB200 decomposition, against the measured CUDA-device ratio of 1.068.
The mechanism is the relative cost of the surrounding work that fusion removes.
\texttt{GEMM\_FUSED} retains DRAM-saturated work, which is 9\% of its tendency call on GB200.
This work becomes relatively more expensive on A100.
In contrast, \texttt{FUSED\_TC} has no such work.

The same procedure does not transfer to H100.
With the H100 nominal peak ratios, the procedure gives $\texttt{GEMM\_FUSED}/\texttt{FUSED\_TC} = 1.34$, against a measured 1.069 at the common instruction shape.
The procedure also gives $\texttt{GEMM}/\texttt{FUSED\_TC} = 1.812$ against a measured 0.894.
The model therefore misses not only the magnitude but also the direction of a ranking (Section~\ref{sup:nominal-roof-failure-h100} of the supplemental material).

The measurements identify the immediate source of the failure.
The fraction of the nominal Tensor Core peak that a mainloop-dominated kernel reaches is not preserved across architectures.
At $p = 511$ and the common shape, the CUTLASS volume GEMMs of \texttt{GEMM\_FUSED} reach 92.7\% of the GB200 Tensor Core roof but only 45.2\% of the H100 roof (Section~\ref{sup:roofline-crossover-depends-architecture} of the supplemental material).
The nominal Tensor Core peak is therefore not the arithmetic capacity that this kernel can actually use.
The next section shows that the coefficient that is needed for H100 is not identifiable from the calibration data of GB200 and A100 within this scaling model.

\subsection{Instruction-dependent pricing resolves the nominal-roof failure}
\label{sec:instruction-priced-subpipe}
The idea of this section is simple.
The nominal FP64 Tensor Core FLOP/s is not the arithmetic capacity that transfers between these machines.
The quantity that transfers is the cost of each issued DMMA instruction in the FP64 execution budget that the profiler defines.
The shape of the issued instruction sets how much of that budget one unit of arithmetic consumes.
The rest of this section derives this pricing from the counters.

Under the per-SM-cycle normalization of Table~\ref{tab:machines-per-sm-cycle}, GB200 and A100 have the same FP64 Tensor Core throughput per SM cycle to within 0.2\%.
Therefore, on that pair, a roof that is proportional to the Tensor Core throughput is numerically the same as a roof that is proportional to SM cycles.
No ratio of two measured times can separate these two roofs.
Only H100 separates them, by a factor of two.
A calibration set of GB200 and A100 therefore moves the CUDA Core FP64 axis and the DRAM axis (by factors of 2.01 and 2.47) but not the Tensor Core axis.
As a result, the coefficient that decides the H100 ranking is not identifiable from this calibration set within the scaling model.
The failure of Section~\ref{sec:does-accounting-transfer-a100} is thus evidence that this calibration set does not vary the coordinate that matters, and not evidence that a roofline model cannot transfer.
Because we add no parameter that is fitted to the target, that coordinate must be fixed from a quantity that belongs to the kernel.
That quantity is the issued instruction.
The form of this pricing was chosen after the H100 failure (Section~\ref{sec:experimental-scope-limitations}).

\begin{table*}[tp]
\centering
\footnotesize
\caption{
  Machine capacities normalized per SM cycle, the form in which the architecture enters the model of Section~\ref{sec:subpipe-model}.
  B denotes bytes and SM-cycle one SM clock cycle.
  The FP64 Tensor Core column is the nominal peak under this normalization. Section~\ref{sec:instruction-priced-subpipe} replaces this peak with the FLOP that one DMMA instruction of the issued shape carries.
  The SM$\times$clock column is the product of the SM count and the boost clock.
  The ratio rows are per SM cycle. Their product with the SM$\times$clock ratio is the total peak ratio that constrains a ratio of two measured times.}
\label{tab:machines-per-sm-cycle}
\begin{tabularx}{\tblwidth}{@{}>{\hsize=1.15\hsize}L>{\hsize=.65\hsize}R>{\hsize=.80\hsize}R>{\hsize=1.10\hsize}R>{\hsize=1.15\hsize}R>{\hsize=1.15\hsize}R>{\hsize=1.00\hsize}R@{}}
\toprule
 & SMs & Boost (GHz) & SM$\times$clock (SM$\cdot$GHz) & FP64 Tensor Core (FLOP/SM-cycle) & FP64 CUDA Core (FLOP/SM-cycle) & HBM (B/SM-cycle) \\
\midrule
GB200 & 152 & 2.062 & 313.4 & 127.9 & 127.9 & 25.21 \\
A100 & 108 & 1.410 & 152.3 & 128.1 & 63.7 & 10.21 \\
H100 & 132 & 1.980 & 261.4 & 256.0 & 128.2 & 9.17 \\
\midrule
GB200-to-A100 ratio & & & 2.06 & 1.00 & 2.01 & 2.47 \\
GB200-to-H100 ratio & & & 1.20 & 0.50 & 1.00 & 2.75 \\
\bottomrule
\end{tabularx}
\end{table*}

Under the normalization of Table~\ref{tab:machines-per-sm-cycle}, H100 offers twice the FP64 Tensor Core throughput per SM cycle of GB200.
However, that advantage is available only through the wider DMMA instructions that H100 executes as one instruction.
A kernel that continues to issue the \texttt{m8n8k4} shape is charged four cycles per instruction in the counter-defined FP64 budget on H100 as on GB200.
Therefore this shape delivers 128 FLOP per budget cycle on H100 as well, and not the nominal factor-of-two gain.
The counter-defined arithmetic demand is therefore set by the issued instruction together with its architecture-dependent price, and not by the nominal Tensor Core peak alone.
As a result, a model that transfers must price that instruction on each machine.

On all three architectures, Nsight Compute reports the activity of the FP64 CUDA Core pipeline and of the FP64 Tensor Core pipeline through one counter, \texttt{sm\_\_pipe\_shared\_cycles\_active}.
We model the activity that this counter reports as one counter-defined FP64 budget.
We do not claim that a single physical unit stands behind the counter.
The profiler documentation describes its pipelines as physical or logical, and it does not fix which.
The documentation also states that pipeline metric names and exposed pipeline terms vary across GPU architectures \citep{nvidia-ncu-guide}.
The model requires only that both instruction classes contribute to the same counter-defined activity at a fixed cost per instruction.
Measurement establishes that this holds on each of the three machines, though the documentation does not establish it.
The measured counter agrees with the sum of the two instruction classes, and each class has a fixed cost per instruction.
The comparison covers every kernel that issues a DMMA instruction on the three machines.
Section~\ref{sup:closure} of the supplemental material tabulates it kernel by kernel.
Because both sides of the comparison are profiler counters, it is an empirical counter relationship and not an independent measurement of a hardware mechanism.
The evidence that this pricing is needed for the transfer is the floor ablation of Table~\ref{tab:floor-ablation} below.
At $p = 511$ on GB200 at production clocks, one Tensor Core instruction contributes 4.01 to 4.02 cycles to that budget in the nine contraction kernels of \texttt{GEMM}, \texttt{GEMM\_FUSED}, and \texttt{FUSED\_TC}.
This value is obtained after the FP64 CUDA Core instructions of those kernels are subtracted at the price that is given below.
These instructions take at most 2.4\% of the counter.
The predicted cost is 4.00, the 512 FLOP of one \texttt{DMMA.8x8x4} divided by the 127.9 FLOP per SM cycle of the machine.
The path \texttt{FUSED} at $p = 255$ issues no Tensor Core instruction at all.
This path measures 0.51 cycles per FP64 CUDA Core warp instruction, against 0.50 that is predicted in the same way from 64 FLOP.
Both instruction classes are thus priced by their own FLOP count divided by the same 128 FLOP per SM cycle.
The counter balance of kernel $k$ on machine $m$ then closes as
\begin{equation}
S(k,m) = c_{\mathrm{tc}}(k,m)\,N_{\mathrm{DMMA}}(k,m) + c_{\mathrm{cc}}(m)\,N_{\mathrm{fp64}}(k),
\label{eq:subpipe}
\end{equation}
\begin{equation}
c_{\mathrm{tc}}(k,m) = \max\!\left(4,\ \frac{F_{\mathrm{DMMA}}(k,m)}{C_{\mathrm{tc},m}}\right),
\qquad
c_{\mathrm{cc}}(m) = \frac{64}{C_{\mathrm{cc},m}},
\label{eq:subpipe-prices}
\end{equation}
where $S$ is the FP64-budget cycle demand that is summed over all SMs of the machine.
The quantities $c_{\mathrm{tc}}$ and $c_{\mathrm{cc}}$ are the cycle costs per Tensor Core instruction and per CUDA Core instruction, respectively.
$N_{\mathrm{fp64}}(k)$ is the executed count of warp-wide DFMA instructions, which the calibration counter \texttt{sm\_\_inst\_executed\_pipe\_fp64} reports.
$N_{\mathrm{DMMA}}(k,m)$ is the executed count of DMMA instructions of the shape that the binary that is compiled for $m$ issues.
The profiler counter \texttt{sm\_\_inst\_executed\_pipe\_tensor} reports that count only on a machine that is run.
Therefore this counter is not an input of a prediction that does not run the target.
The model multiplies the calibration value of the same counter by the ratio of the FLOP per DMMA instruction of the calibration shape to that of the target shape, and it uses the result as the count on the target.
This scaling holds fixed the FLOP that the kernel executes through DMMA.
$F_{\mathrm{DMMA}}(k,m)$ is the FLOP that one DMMA instruction of the shape carries on the machine $m$.
$C_{\mathrm{tc},m}$ and $C_{\mathrm{cc},m}$ are the FP64 Tensor Core and CUDA Core throughputs per SM cycle of Table~\ref{tab:machines-per-sm-cycle}, respectively.
The floor of four cycles is the measured minimum contribution of one DMMA instruction to the FP64 budget.
This minimum is 3.96 to 4.02 cycles in the 33 kernels that issue DMMA at $p = 511$ on the three machines (Section~\ref{sup:closure} of the supplemental material).
On GB200 and A100, $F_{\mathrm{DMMA}}/C_{\mathrm{tc},m}$ is 4.00 for the only shape that they issue (Table~\ref{tab:dmma-prices}).
Therefore the counters of these two machines cannot separate the floor from the FLOP price.
The two-branch form of Equation~\ref{eq:subpipe-prices} is identified from the H100 counters alone.
There, we subtract the CUDA Core share and use the SM clock that the profiler records.
Then one \texttt{DMMA.8x8x4} contributes 4.00 to 4.03 cycles to the budget at $p \ge 511$, where the floor binds.
One \texttt{DMMA.16x8x4} contributes 3.99 to 4.09 cycles at $p = 511$, where the floor and the FLOP price are equal.
One \texttt{DMMA.16x8x8} of the cuBLAS kernels contributes 7.93 to 8.04 cycles, where the FLOP price alone applies.
This value was read before the FLOP branch of Equation~\ref{eq:subpipe-prices} took its present form.
Therefore the cuBLAS totals of H100 do not test that branch as a prediction that was made in advance.
Every shape carries a power-of-two FLOP count, and $N_{\mathrm{DMMA}}$ is rescaled by the FLOP ratio between shapes.
Therefore $c_{\mathrm{tc}} N_{\mathrm{DMMA}}$ depends on the shape only through the floor (Section~\ref{sup:closure} of the supplemental material).

\begin{table}[tp]
\centering
\footnotesize
\caption{
  Cycle cost per DMMA instruction in the counter-defined FP64 budget, by machine and shape (Equation~\ref{eq:subpipe-prices}).
  $F_{\mathrm{DMMA}}$ is the FLOP that one instruction carries, and $C_{\mathrm{tc},m}$ is the FP64 Tensor Core capacity per SM cycle of Table~\ref{tab:machines-per-sm-cycle}.
  The price $c_{\mathrm{tc}}$ is the larger of the four-cycle floor and their ratio. Among the three machines studied, the floor binds only on H100.}
\label{tab:dmma-prices}
\begin{tabular}{@{}llrrr@{}}
\toprule
Machine & Shape & $F_{\mathrm{DMMA}}$ & $F_{\mathrm{DMMA}}/C_{\mathrm{tc},m}$ & $c_{\mathrm{tc}}$ \\
\midrule
GB200 & \texttt{m8n8k4} & 512 & 4.003 & 4.003 \\
A100 & \texttt{m8n8k4} & 512 & 3.997 & 4.000 \\
H100 & \texttt{m8n8k4} & 512 & 2.000 & 4.000 \\
H100 & \texttt{m16n8k4} & 1024 & 4.000 & 4.000 \\
H100 & \texttt{m16n8k8} & 2048 & 8.000 & 8.000 \\
H100 & \texttt{m16n8k16} & 4096 & 16.000 & 16.000 \\
\bottomrule
\end{tabular}
\end{table}

The architecture enters Equation~\ref{eq:subpipe} only through the DMMA shapes that it holds in one instruction.
In the native GPU assembly (SASS), \texttt{m8n8k4} is one \texttt{DMMA.8x8x4} on \texttt{sm\_90} and on \texttt{sm\_100} alike.
In contrast, each wider shape is one instruction on \texttt{sm\_90} and a sequence of two to eight \texttt{DMMA.8x8x4} on \texttt{sm\_100}.
Section~\ref{sup:dmma-expansion-check} of the supplemental material lists and checks the expansions.
The FLOP that one DMMA instruction carries for the shape that the compiled code issues is thus the kernel-side quantity that fixes the coordinate that the calibration set leaves undetermined.

Two parts of Equations~\ref{eq:subpipe} and~\ref{eq:subpipe-prices} can be isolated on the same inputs (Table~\ref{tab:floor-ablation}).
In each case, $\eta$ is recalibrated under the same rule for the choice of the calibration machine.
Without the floor, $c_{\mathrm{tc}} = F_{\mathrm{DMMA}}/C_{\mathrm{tc},m}$ on every machine, and 42 of the 140 path totals of Section~\ref{sec:logo-three-directions} change.
On A100 the change is below 0.1 percentage points, because the price there is 4.00 with or without the floor.
On H100 the narrow shape then costs two cycles and not four.
Therefore every H100 path that issues \texttt{DMMA.8x8x4} is underpredicted by 30 to 40\% at $p \ge 511$.
In addition, $\texttt{GEMM}/\texttt{FUSED\_TC}$ reverses at $p = 511$ and at $p = 767$, because the cuBLAS kernels of \texttt{GEMM} issue the wide shape.
The floor does not touch the price of this wide shape.
The three GB200 totals at $p = 767$ are calibrated from H100 alone, because A100 has no measurement at that order.
Therefore the recalibrated $\eta$ carries the halved H100 price to GB200 and overpredicts these three totals by 66 to 84\%.
The floor is therefore the part of the price that makes the transfer to H100 work.
When the sum of Equation~\ref{eq:subpipe} is replaced by $\max(c_{\mathrm{tc}} N_{\mathrm{DMMA}},\ c_{\mathrm{cc}} N_{\mathrm{fp64}})$, the two classes become two separate resources in the maximum of Equation~\ref{eq:model}.
However, the CUDA Core term is at most 7.9\% of $S$ in the 154 profiled kernels that the FP64 budget binds and that issue both classes.
Therefore the binding resource of no kernel changes.
Fifty-six path totals change, by 0.1\% at the median and 5.4\% at most.
The median errors at $p \ge 511$ move by about one percentage point at most.
One scored cell of Section~\ref{sec:signal-to-noise-rule} changes sign below the screening threshold ($\texttt{FUSED}/\texttt{FUSED\_TC}$ on H100 at $p = 127$).
Because of this change, the screen selects 101 machine-axis cells and is correct on 95, against 100 and 94.
The sum is therefore the form in which the counter closes (Section~\ref{sup:closure} of the supplemental material).
The transfer rests on the price per instruction and its floor.

\begin{table*}[tp]
\centering
\footnotesize
\caption{
  The effect of the four-cycle floor of Equation~\ref{eq:subpipe-prices} and of the additive form of Equation~\ref{eq:subpipe} on the leave-one-GPU-out transfer, with every other input and rule held fixed.
  The second column is the model of this paper. The third column is the same model with $c_{\mathrm{tc}} = F_{\mathrm{DMMA}}/C_{\mathrm{tc},m}$. The fifth column is the same model in which the sum of Equation~\ref{eq:subpipe} is replaced by the larger of its two terms. Both variants are recalibrated.
  Errors are median absolute path-total errors over the $p \ge 511$ cells of one held-out machine (twelve on GB200 and H100, eight on A100). The two ratio rows give the predicted $\texttt{GEMM}/\texttt{FUSED\_TC}$ on held-out H100 at the common shape, against the measured device-time ratio.}
\label{tab:floor-ablation}
\begin{tabular}{@{}lrrrr@{}}
\toprule
 & With floor & Without floor & Measured & Max form \\
\midrule
H100, median error at $p \ge 511$ & 1.7\% & 33.4\% & & 1.7\% \\
GB200, median error at $p \ge 511$ & 1.3\% & 4.7\% & & 1.3\% \\
A100, median error at $p \ge 511$ & 5.4\% & 5.5\% & & 6.6\% \\
GB200, three paths at $p = 767$ & $-0.9$ to $+1.2$\% & $+66$ to $+84$\% & & $-0.9$ to $+1.2$\% \\
H100, $\texttt{GEMM}/\texttt{FUSED\_TC}$, $p = 511$ & 0.888 & 1.441 & 0.919 & 0.887 \\
H100, $\texttt{GEMM}/\texttt{FUSED\_TC}$, $p = 767$ & 0.732 & 1.228 & 0.756 & 0.732 \\
\bottomrule
\end{tabular}
\end{table*}

\subsection{The proposed model}
\label{sec:subpipe-model}
We call the resource that reaches the maximum in Equation~\ref{eq:model} the \emph{binding resource} of the kernel.
On the model side, the binding resource is the resource whose demand sets the prediction.
On the measurement side, it is the resource of largest measured utilization, against which the model side is scored.
Neither of the two is a bottleneck whose cause has been established (Figure~\ref{fig:a11a}(b) in Section~\ref{sup:path-level-two-roof} of the supplemental material).
The agreement rates that we report below are the agreement between those two quantities.
Every demand of Equation~\ref{eq:model} is a counter sum over the whole kernel divided by the product of a capacity per SM cycle and the SM count of the machine.
Therefore $D_r$ is in SM cycles throughout.
For the two demands whose capacity differs between the machines,
\begin{equation}
\begin{aligned}
D_{\mathrm{FP64}}(k,m) &= \frac{S(k,m)}{N_{\mathrm{SM},m}}, \\
D_{\mathrm{DRAM}}(k,m) &= \frac{B_{\mathrm{DRAM}}(k)}{C_{\mathrm{DRAM},m}\,N_{\mathrm{SM},m}},
\end{aligned}
\label{eq:demands}
\end{equation}
where $S$ is the priced FP64-budget demand of Equation~\ref{eq:subpipe}, and $B_{\mathrm{DRAM}}(k)$ is the DRAM byte count of the kernel that the profiler reports on the calibration machine.
The quantities $C_{\mathrm{DRAM},m}$ and $N_{\mathrm{SM},m}$ are the HBM bytes per SM cycle and the SM count of Table~\ref{tab:machines-per-sm-cycle}, respectively.
The remaining demands have the same form, $D_r = X_r(k) / (C_r\,N_{\mathrm{SM},m})$, where $X_r$ is the counter sum and $C_r$ the capacity per SM cycle.
The transfer contract treats every non-DMMA counter numerator for the same kernel instance as transferable from the selected calibration machine, unless a transformation is stated explicitly.
These numerators include $N_{\mathrm{fp64}}$, DRAM bytes, L1 wavefronts, issue, ALU, FMA, and L2 activity.
Only $N_{\mathrm{DMMA}}$ is rescaled for the target instruction shape.
This transfer is an assumption about the generated kernel instance and not a source-level invariant.
The three systems use different compiler and library versions (Table~\ref{tab:machines}).
Apart from the cross-machine DRAM-byte check below, these numerator counts are not independently guaranteed to be architecture invariant.
The direct matched-role audit in Section~\ref{sup:counter-numerator-portability} of the supplemental material finds exact transfer of $N_{\mathrm{fp64}}$ and close transfer of L1 wavefronts for the kernels that are compiled here.
However, the audit finds materially larger differences in issue, ALU, FMA, and L2 counts.
Those latter demands do not bind any matched pair except four L2 pairs.
The remaining capacities (L1 wavefronts, issue slots, ALU-pipe instructions, and L2 bytes per SM cycle) are taken as the same on all three machines.
The FMA-pipe capacity is two warp instructions per SM cycle on A100 and four on GB200 and H100.
This capacity binds no kernel.
Thus the FP64-budget pricing, DRAM capacity, and unused FMA capacity are the machine-dependent entries.
Table~\ref{tab:capacities} of the supplemental material gives the sources of these entries and checks them against the peaks that Nsight Compute implies through its utilization counters.
The DRAM demand carries the measured bytes of a kernel to a target with a different L2 capacity as though these bytes were a property of the kernel.
At path level, the traffic for a fixed path and order agrees across machines to within 17.5\% and usually to within a few percent (Section~\ref{sup:two-byte-counts-differ} of the supplemental material).

The elapsed SM cycles that define $\eta$ are the profiled duration of the kernel multiplied by the SM clock that the profiler records for that kernel.
The clock is unlocked when the profiles are taken (Section~\ref{sec:profiling-protocol}).
Because $\eta$ is defined from the largest demand, it absorbs the latency that the 12 to 18\% occupancy of these kernels exposes.
A value above one is the factor by which the elapsed time exceeds the modeled roof.
The inverse $1/\eta$ is the fraction of the roof that the kernel reaches.
We define this factor only after we take the maximum, so that the maximization stays free to select a different binding resource on the target machine.
We do not assert that $\eta$ is machine independent.
The transfer of this factor is the assumption of the model.
Section~\ref{sec:subpipe-model-negatives} measures the drift of $\eta$ between machines where that assumption fails.

The transfers of Section~\ref{sec:predictive-validation} calibrate on two machines.
The rule that selects one calibration machine is applied per kernel, not per path.
The prediction takes $\eta(k)$, together with the counter numerators, from the calibration machine on which the kernel appears and whose $\eta(k)$ lies closest to 1.
This machine is the machine on which the kernel sits closest to a roof.
The rule involves no target quantity in any individual transfer.
However, the rule was selected retrospectively after we compared alternatives on the same leave-one-GPU-out population.
The rule minimizes the worst error over all orders.
The mean is slightly better in one high-order worst case, and the larger-$\eta$ rule is better in the A100 median (Section~\ref{sup:calibration-set-choice} of the supplemental material).
When the rule is replaced by the larger $\eta$ or by the mean, the three directional medians at $p \ge 511$ change from 1.3/5.4/1.7\% to 4.1/3.6/3.5\% or 2.4/4.2/1.7\%, respectively.
Neither fixed alternative uniformly improves the three directions.

The profiles also show why L1 must appear in Equation~\ref{eq:model} even though it is not the roof on the calibration machine.
At $p = 511$ on GB200, the contraction kernels are bound by the counter-defined FP64 budget, at 81.9 to 95.5\% of elapsed cycles.
L1 is consistently second, at 33.8 to 58.4\% (Table~\ref{tab:kernels} of the supplemental material).
On H100 at the wide shape, the FP64-budget demand of those same kernels falls to 41 to 48\%, and L1 takes over as the roof.

The target contributes three kinds of input, none of which is a timing.
The first is the boost clock $f_m$ of Equation~\ref{eq:model-time}.
No clock is read on the target, so the difference between its boost clock and the clock at which it actually runs is part of the reported error.
When we weight the recorded clock by kernel duration, it is 99.6 to 99.8\% of the boost clock at $p \ge 511$ and 97.8 to 99.2\% at $p \le 255$ on the three machines.
Therefore this part accounts for at most 0.4\% of a path-total error in the observed high-order accuracy regime of Section~\ref{sec:signal-to-noise-rule} and for at most 2.2\% below it.

The second is the instruction expansion.
The instruction counts are executed counts that the profiler reports on the calibration machine.
For the kernels that we compile ourselves, the target contributes the DMMA shape that each kernel issues there.
This shape is read with \texttt{cuobjdump} from a binary that is cross-compiled for the target, so no access to the target is needed for these kernels.
When that shape differs from the calibration shape, $N_{\mathrm{DMMA}}$ is rescaled as in Section~\ref{sec:instruction-priced-subpipe}.
The volume GEMMs that the CUTLASS paths assign to cuBLAS (Section~\ref{sec:library-paths-gemm-gemm}) are treated like the roles of \texttt{GEMM} below.

The third input is the condition of application.
We call the list of kernels that a path launches on a machine, with their instruction families, the \emph{kernel list} of that path on that machine.
The model carries every kernel of the calibration kernel list to the target.
Therefore the condition of application of the model is that the target kernel list matches.
This means that the same kernel roles exist on the target and that their instruction family is known.
The cross-compiled binary does not give the MMA shape that cuBLAS chooses at run time.
Therefore, for the cuBLAS roles, the model conditionally prices the shape that gives the smallest demand among the shapes that the target supports (Equation~\ref{eq:instruction-enumeration} in Appendix~\ref{app:model-contract}).
The supported shapes can be enumerated from the instruction set or a cross-compiler without target access.
However, support does not imply that a vendor library will dispatch that implementation.
Therefore the library family remains a condition of application, and the shape selection is a model assumption and not a prediction of library policy.
The choice matters on H100, where the floor binds.
When the cuBLAS roles are priced with the calibration shape instead, five H100 \texttt{GEMM} totals leave their error band and four predicted signs reverse.
In contrast, the shape that cuBLAS actually dispatches changes no scored total (Appendix~\ref{app:model-contract}).
This shape is read from one execution of the target.

The cost of calibration is one calibration-architecture-profiled number $\eta$ per kernel instance (one kernel of one path at one order and launch configuration).
One leave-one-GPU-out direction of Section~\ref{sec:logo-three-directions} rests on 187 to 219 such instances.

\section{Leave-One-GPU-Out Evaluation of Parameter Transfer}
\label{sec:predictive-validation}
The scores of this section hold every parameter out of the target in three directions.
However, they were computed after all three machines had been measured.
The model form was motivated by the observed failure of nominal-peak scaling.
The per-kernel closest-$\eta$ calibration rule was selected by comparing alternatives on this same leave-one-GPU-out population.
The evaluation therefore tests parameter transfer under a retrospectively specified model, and it is not an out-of-sample validation of the four-cycle-floor model form.

\subsection{Leave-one-GPU-out in three directions}
\label{sec:logo-three-directions}
We calibrate on two of the three GPUs and compare against the third.
No timing of the third enters any parameter.
All three sweeps run the same benchmark body, and the clock lock is disabled.
The kernels that we compile ourselves use the $8\times 8\times 4$ MMA shape.
This common-shape evaluation is the control that establishes the parameter transfer.
Section~\ref{sec:operational-predictions} then tests the operational consequence of this transfer at the shape that H100 selects.

Table~\ref{tab:holdout-h100} shows the H100 direction at $p = 511$.
The three path totals are predicted to within 5.2\%.
The two cross-machine rank ratios both cross 1 between GB200 and H100, and both ratios are predicted on the correct side of 1.
In contrast, the nominal-roof scaling of Section~\ref{sec:does-accounting-transfer-a100} misses the magnitude of both ratios and the direction of the second ratio.
The first ratio is a hold-out of matching kernels, because both paths that enter it issue only \texttt{DMMA.8x8x4}.
In contrast, the cuBLAS kernels of \texttt{GEMM} in the second ratio differ between the two machines.
The mechanism is that of the A100 reversal, through a different pair of roofs.
The counter-defined FP64 budget does not shrink on H100 at $8\times 8\times 4$.
Therefore the contraction kernels of both paths grow by the SM-cycle ratio 1.20.
In contrast, the \texttt{volume\_flux} kernel, which only \texttt{GEMM\_FUSED} carries, is DRAM bound and grows by the HBM ratio 3.30 (Table~\ref{tab:holdout-h100}, last column).

\begin{table*}[tp]
\centering
\footnotesize
\caption{
  Leave-one-GPU-out prediction of H100 at $p = 511$, in CUDA-device time per tendency call, from a model that is calibrated on GB200 and A100 (Section~\ref{sec:subpipe-model}).
  No H100 timing enters any parameter.
  Errors are relative to the H100 measurement.
  The last two rows are the cross-machine rank ratios, for which the nominal-roof scaling of Section~\ref{sec:does-accounting-transfer-a100} gives 1.34 and 1.81.}
\label{tab:holdout-h100}
\begin{tabularx}{\tblwidth}{@{}>{\hsize=1.6\hsize}L>{\hsize=.75\hsize}R>{\hsize=.75\hsize}R>{\hsize=.75\hsize}R>{\hsize=.55\hsize}R>{\hsize=1.6\hsize}L@{}}
\toprule
Quantity & GB200 (ms) & H100 measured & H100 predicted & Error & Binding resource, GB200 to H100 \\
\midrule
\texttt{FUSED\_TC}, path total & 12.87 & 15.71 & 15.42 & $-1.9$\% & FP64 budget to FP64 budget \\
\texttt{GEMM\_FUSED}, path total & 12.20 & 17.24 & 16.91 & $-1.9$\% & FP64 budget to FP64 budget \\
\texttt{GEMM}, path total & 13.08 & 14.44 & 13.69 & $-5.2$\% & FP64 budget to FP64 budget \\
\midrule
\texttt{GEMM\_FUSED} mainloop & 3.63 & 4.37 & 4.34 & $-0.6$\% & FP64 budget to FP64 budget \\
\texttt{GEMM\_FUSED} \texttt{volume\_flux} & 1.04 & 3.40 & 3.41 & $+0.2$\% & DRAM to DRAM \\
\texttt{GEMM} \texttt{separable\_lift} & 1.15 & 3.51 & 3.58 & $+2.0$\% & DRAM to DRAM \\
\midrule
$\texttt{GEMM\_FUSED}/\texttt{FUSED\_TC}$ & 0.948 & 1.097 & 1.096 & $<0.1$\% & \\
$\texttt{GEMM}/\texttt{FUSED\_TC}$ & 1.016 & 0.919 & 0.888 & $-3.4$\% & \\
\bottomrule
\end{tabularx}
\end{table*}

The profiled-window H100 ratio 1.097 in Table~\ref{tab:holdout-h100} differs by 2.6\% from the base-sweep device-time value 1.069 in Table~\ref{tab:two-shapes-p511}.
Both values resolve the same ordering (Table~\ref{tab:a-p511-ratio} of the supplemental material lists the two campaigns).
The $<0.1\%$ value is therefore the model residual within that profiling campaign only, and we do not claim that it is reproducible below the 2.6\% cross-campaign difference.

\begin{figure*}[tp]
\centering
\includegraphics[width=\linewidth]{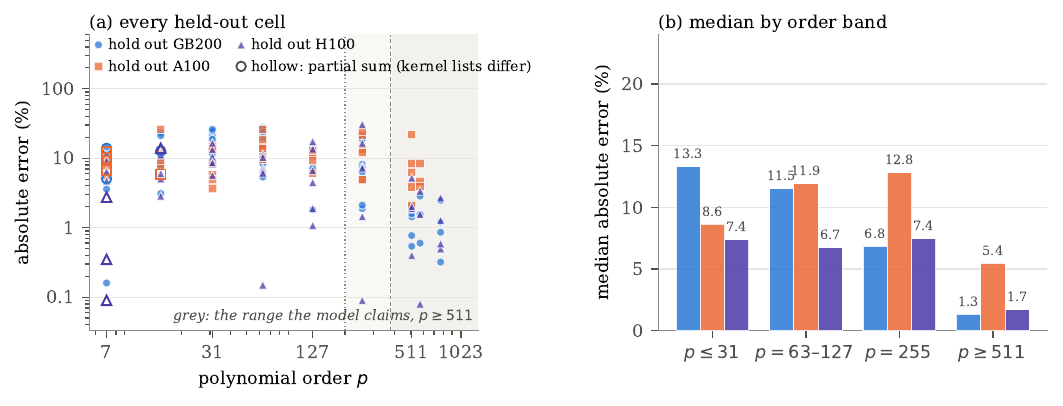}
\caption{
  Leave-one-GPU-out in all three directions. No timing of the held-out GPU enters any parameter.
  (a) Absolute error of every held-out cell against polynomial order.
  The twelve cells at $p \le 15$ whose kernel lists do not match the calibration machines are marked.
  (b) Median of those errors by order band, over every cell in (a).
  The transition order $p = 255$ is a band of its own (Section~\ref{sec:signal-to-noise-rule}).}
\label{fig:f5}
\end{figure*}

Over all orders, this gives 140 path-total cells, of which 128 have kernel lists that match between the calibration machines and the target.
In the other twelve cells, cuBLAS dispatches a different number of kernels of a different instruction family on the target.
These twelve cells are the three library paths at $p = 7$ and \texttt{GEMM} at $p = 15$ on every target.
Therefore the kernel lists of Section~\ref{sec:subpipe-model} do not match in these cells.
We score these cells on the kernels that the lists share.
We mark them in Figure~\ref{fig:f5}(a) and do not drop them, because the mismatch is itself the low-order limit of the transfer (Section~\ref{sec:subpipe-model-negatives}).
At $p \ge 511$ the median path-total error is 1.3, 5.4, and 1.7\% when GB200, A100, and H100, respectively, are held out (Figure~\ref{fig:f5}(b)).
At $p \le 31$, on the path totals whose kernel lists match, the corresponding errors are 16.6, 8.6, and 8.5\%, respectively.
These medians use the closest-$\eta$ calibration rule, which was selected retrospectively on this same population (Section~\ref{sec:subpipe-model}).
The two fixed alternatives are the mean and the larger $\eta$.
They give 2.4, 4.2, and 1.7\% and 4.1, 3.6, and 3.5\%, respectively.
Therefore the high-order accuracy does not depend on that selection by more than a few percentage points.
An error below 2\% appears in only two of the three directions (Figure~\ref{fig:f5}(a)).
The binding resource of the model agrees with the resource of largest measured utilization on the target for 88, 97, and 91\% of the kernels.
In the cells whose kernel lists match, the agreement is 96, 98, and 98\%.

The hardest direction depends on the polynomial order.
At $p \le 31$ the hardest direction holds out GB200.
The reason is that the two calibration machines, A100 and H100, lie closest together in HBM bytes per SM cycle, and GB200 differs most from them in this coordinate.
Inside the observed $p \ge 511$ regime, the hardest direction holds out A100.
In that direction, the model predicts a shorter time than is measured on all eight cells that the three machines share ($p = 511$ and 575).
A100 is also the worst source when it calibrates the other two machines (Section~\ref{sup:calibration-set-choice} of the supplemental material).
Therefore this error belongs to that machine and not to the direction of the transfer.

\subsection{Seven ratios, three directions}
\label{sec:seven-ratios}
An implementation is chosen on rankings and not on absolute times.
Of the fifteen unordered pairwise ratios of the six paths, we score seven.
Three of them are the controlled one-axis comparisons of Section~\ref{sec:controlled-one-axis-comparisons}.
The other four are rankings of paths that differ in more than one implementation choice.
One of these four is $\texttt{GEMM\_FUSED}/\texttt{FUSED\_TC}$, which is the ranking of Figure~\ref{fig:f1}(a).
At $p \ge 63$ in all three directions, the sign is correct in 83 of 92 cells (Table~\ref{tab:seven-ratios}).
The weakest ratio, $\texttt{GEMM\_CUTE}/\texttt{GEMM}$, has the smallest true difference of the seven and one of the two smallest errors.
Therefore there is almost nothing to predict on this ratio.
In contrast, $\texttt{FUSED}/\texttt{FUSED\_TC}$ is weak because its own error is large.
Section~\ref{sec:signal-to-noise-rule} makes that distinction quantitative.

\begin{table*}[tp]
\centering
\footnotesize
\caption{
  The seven ratios are scored on the sign of $\ln r$ at $p \ge 63$ in all three leave-one-GPU-out directions. Three of the seven ratios are the controlled one-axis comparisons. Here $r$ is the ratio of the time of the first named path to the time of the second.
  The true difference is the median $|\ln r_{\mathrm{meas}}|$. The error is the median $|\ln (r_{\mathrm{pred}} / r_{\mathrm{meas}})|$, where pred and meas denote predicted and measured, respectively.
  Both are quoted in percent, as $100\,|\ln r|$ (Section~\ref{sec:signal-to-noise-rule}).
  The MMA row is the iso-schedule ablation. Therefore its true difference of 81.4\% (a factor of 2.26 in time) is the time difference that the instruction produces on the schedule that was built for it (Section~\ref{sup:register-budget-cost-iso} of the supplemental material).
  The fusion row is written as $\texttt{GEMM\_CUTE}/\texttt{GEMM\_FUSED}$, which is the inverse of the orientation of Figure~\ref{fig:f3}(c). The scored sign, the true difference, and the error do not depend on the orientation.}
\label{tab:seven-ratios}
\begin{tabularx}{\tblwidth}{@{}>{\hsize=1.9\hsize}L>{\hsize=.6\hsize}R>{\hsize=.5\hsize}R>{\hsize=.7\hsize}R>{\hsize=.6\hsize}R>{\hsize=1.7\hsize}L@{}}
\toprule
ratio & correct & rate & true diff. & error & what it compares \\
\midrule
$\texttt{FUSED\_DFMA}/\texttt{FUSED\_TC}$ & 12/12 & 100\% & 81.4 & 11.8 & the MMA instruction (one-axis) \\
$\texttt{GEMM}/\texttt{FUSED\_TC}$ & 14/14 & 100\% & 23.0 & 3.9 & library against fused \\
$\texttt{GEMM\_CUTE}/\texttt{GEMM\_FUSED}$ & 17/17 & 100\% & 11.5 & 4.8 & epilogue fusion (one-axis) \\
$\texttt{GEMM\_FUSED}/\texttt{FUSED\_TC}$ & 13/14 & 93\% & 13.9 & 2.1 & the ranking of Figure~\ref{fig:f1}(a) \\
$\texttt{FUSED\_DFMA}/\texttt{FUSED}$ & 8/9 & 89\% & 34.1 & 6.8 & MMA fragment layout \\
$\texttt{FUSED}/\texttt{FUSED\_TC}$ & 8/9 & 89\% & 24.9 & 14.3 & CUDA Core against Tensor Core \\
$\texttt{GEMM\_CUTE}/\texttt{GEMM}$ & 11/17 & 65\% & 1.3 & 2.1 & the library (one-axis) \\
\midrule
total & 83/92 & 90\% & & & \\
\bottomrule
\end{tabularx}
\end{table*}

\subsection{Choosing the fastest path at the common shape}
\label{sec:fastest-path-selection}
A correct pairwise ranking is not yet a choice of implementation, because a user wants the fastest path and not a correct order of two slow ones.
We therefore also score the choice itself.
In every target-and-order cell, the path with the smallest predicted total is compared with the path with the smallest measured total.
Only the paths whose kernel lists match on the target (Section~\ref{sec:logo-three-directions}) enter this comparison.
The cells are those of the common-shape comparison (Section~\ref{sec:architecture-dependent-reversal}).
Therefore, on H100, the candidate set is not the one that a user of the code obtains.
The set that a user obtains is scored below.
This gives 26 cells with three to six paths each (Table~\ref{tab:fastest-path} of the supplemental material).
The predicted fastest path is the measured fastest path in 21 of the 26 cells.
The five wrong choices are all at $p \le 63$ on A100 and H100, and they cost 0.5 to 15.2\% over the measured fastest path.
The screen of Section~\ref{sec:signal-to-noise-rule} applies to the choice as well.
The screen selects a predicted fastest path only when the ratio of that path to every other path of the cell has a separation-to-error ratio $S/E \ge 1$.
For each of the fifteen pairs, the family error is formed in the same way as for the seven scored ratios.
The screen selects a fastest path in 15 cells, and that path is the measured fastest path in all 15.
The eleven cells below the threshold contain all five wrong choices.
The simpler rule takes the path that is fastest on a calibration machine at the same order.
This rule is correct in 30 of the 50 combinations of a target cell and a calibration machine.
In addition, 15 of its 20 wrong choices cost more than 5\% (Section~\ref{sup:fastest-path-selection} of the supplemental material).

On H100, the candidate set that a user obtains has the $16\times 8\times 4$ variants of \texttt{GEMM\_CUTE} and \texttt{GEMM\_FUSED} in place of the common-shape variants (Section~\ref{sec:architecture-dependent-reversal}).
Therefore we repeat the choice on the nine H100 cells with that set (Table~\ref{tab:fastest-path-preferred} of the supplemental material).
The predicted fastest path is the measured fastest path in seven of the nine cells.
The two misses are the $p = 7$ and $p = 15$ cells that are also wrong at the common shape.
No family error exists at the preferred shape (Section~\ref{sec:operational-predictions}).
Therefore the seven of nine is a count of correct choices and not of claimed ones.
At $p = 511$, 575, and 767, the shape changes the measured fastest path from \texttt{GEMM} to \texttt{GEMM\_FUSED}.
The retrospective hold-out calculation uses no H100 measurement for parameter instantiation, and it follows the change in all three cells.
However, at $p = 767$ the measured margin of 0.5 to 2.8\% does not separate the two paths.

\subsection{Operational predictions}
\label{sec:operational-predictions}
The model predicts CUDA-device time per tendency call in the profiled runs.
In contrast, a user times the production binary without a profiler (Section~\ref{sup:difference-between-stage-device} of the supplemental material).
We also score the rankings against the production CUDA-device time per stage.
The cells are the 88 cells at $p \ge 63$ of the seven ratios of Table~\ref{tab:seven-ratios} that the production sweep times on both sides.
Of these cells, 81 are correct.
The two measures agree on the sign in 87 of the 88.
On the other cell, both measures place the two paths within 0.2\% of each other.
No screened-in cell of Section~\ref{sec:signal-to-noise-rule} changes between the two measures.

At the preferred shape on H100, \texttt{GEMM\_FUSED} is faster than the same path at the $8\times 8\times 4$ shape, in step-loop time (Section~\ref{app:larger-mma-shape-h100} of the supplemental material).
The path is 5.5\% faster at $p = 255$, and 18 to 21\% faster at $p \ge 511$.
The preferred shape enters the model through the DMMA shape of the cross-compiled binary (Section~\ref{sec:subpipe-model}).
The model is scored on 20 path totals, with a median error of 8.1\%, and on twelve ranking cells, of which eleven have the correct sign (Appendix~\ref{app:crossover-map}).
At $p = 511$, the shape change turns $\texttt{GEMM\_FUSED}/\texttt{FUSED\_TC}$ from 1.069 at the common shape to 0.854 at the preferred shape.
The hold-out calculation reproduces the reversal without using H100 measurements for parameter instantiation.
Its prediction is 0.710 against a measured 0.860 in the profiled window.
That realized error of 19.2\% (as $100\,|\ln (r_{\mathrm{pred}} / r_{\mathrm{meas}})|$, Section~\ref{sec:signal-to-noise-rule}) is five times the family error of this ratio on the machine axis, 3.6\%.

Appendix~\ref{app:crossover-map} projects the whole curve $\texttt{GEMM\_FUSED}/\texttt{FUSED\_TC}$ against order onto the two other machines from the counters of one calibration machine.
With calibration on either machine, the crossing on H100 is placed between $p = 255$ and 511, which is where it is measured.
No crossing is produced on A100, where none is measured.
The two departures of the projection move no predicted cell across 1.

\section{Changing Machine Balance on One GPU}
\label{sec:balance-controlling-variable}
Sections~\ref{sec:instruction-aware-model} and~\ref{sec:predictive-validation} account for the reversal by a change in machine balance across machines.
However, machines differ in many coordinates at once (Section~\ref{sec:instruction-priced-subpipe}).
This section moves the main coordinate of that balance on one GPU.
This coordinate is the DRAM bytes that the machine supplies per active-SM cycle (Table~\ref{tab:machines-per-sm-cycle}).
The compiled program, the clock, and the per-SM capacities stay fixed.
This coordinate is not the whole-machine FLOP-to-byte ratio of a roofline diagram.
Throughout this section the coordinate is the nominal supply, the peak HBM bandwidth divided by the product of the active SM count and the clock.
The bandwidth that the memory system sustains with fewer active SMs was not measured directly.

\subsection{MPS as a way to change the balance}
\label{sec:mps-balance-knob}
Two ways to move this coordinate were considered.
Lowering the SM clock was rejected before the run.
A lower clock raises the DRAM bytes per SM cycle, but it also shrinks the unloaded latency in SM cycles by the same factor.
Therefore the experiment could not have said which of the two moved (Section~\ref{sup:balance-knob-measurements} of the supplemental material).
The CUDA Multi-Process Service (MPS) variable \texttt{CUDA\_MPS\_ACTIVE\_THREAD\_PERCENTAGE} reduces the SM resources that are exposed to the client.
The variable leaves the code, instruction set, compiler, cuBLAS kernel selection, clock, and nominal per-SM hardware capacities unchanged.
The principal coordinate that changes is therefore DRAM bytes per active-SM cycle.
Fewer active SMs also mean more waves per SM, which the wave control of Section~\ref{sec:wave-control} measures.
Fewer active SMs also mean fewer concurrent requests to the memory system and more L2 capacity per active SM, and these effects are not separated here.
Therefore the experiment shows that the ranking moves with the main balance coordinate.
The experiment does not show that this coordinate alone changed.

The obtained SM count is read from \texttt{\%smid} in the same job as the timing (Table~\ref{tab:balance-knob}).
The sweep uses the production timer, \texttt{Step loop per stage}.
The sweep covers the six requested SM fractions of Table~\ref{tab:balance-knob} (and, on A100, the supply-matched setting of Section~\ref{sec:same-reversal-one-machine}), the six paths of Table~\ref{tab:paths}, and $p = 63$, 255, and 511 on GB200 and A100.
H100 is absent because site policy does not allow MPS there.

\subsection{The same reversal on one machine, and the supply-matched point}
\label{sec:same-reversal-one-machine}
At $p = 511$ on A100, $\texttt{GEMM\_FUSED}/\texttt{FUSED\_TC}$ falls from 1.069 at the full machine to 0.934 at a quarter of the machine.
The ratio crosses 1 between 75 and 50\% (Figure~\ref{fig:f6}(a)).
This is the change of winner that Section~\ref{sec:architecture-dependent-reversal} reports between A100 and GB200, and it is produced here on one GPU with the same binary, the same cuBLAS, and the same clock.
At $p = 255$ both machines cross 1 along this axis (Figure~\ref{fig:f6}(b)).
Therefore three of the six combinations of machine and order cross.

If the ranking follows the balance, then equal balance on the two machines should bring the two rankings together.
A100 at 44 SMs matches the unrestricted GB200 in nominal DRAM bytes per active-SM cycle to 0.6\% (Table~\ref{tab:balance-knob}).
The two machines also agree in FP64 Tensor Core throughput per SM cycle to approximately 0.2\% (Section~\ref{sec:instruction-priced-subpipe}).
At that point, $\texttt{GEMM\_FUSED}/\texttt{FUSED\_TC}$ is 0.966 against 0.950 on GB200, a difference of 1.6\%.
In contrast, the unrestricted comparison differs by 12.6\%.
The point was measured and not interpolated.
Repeats of both sides on other compute nodes place the difference between 0.6 and 1.6\% (Section~\ref{sup:data-provenance-reproducibility} of the supplemental material).
The match is one of nominal supply and not of measured balance.
The sustained bandwidth, the request concurrency, and the L2 capacity per active SM of the restricted A100 were not measured.
Therefore we call this point the supply-matched point.
The match should not bring every ratio together, because the CUDA Core FP64 throughput per SM cycle still differs by a factor of two.
Section~\ref{sup:balance-knob-measurements} of the supplemental material reports the ranking cells on which the matched A100 still disagrees with GB200.

\subsection{The wave control}
\label{sec:wave-control}
The SM restriction also raises the number of waves that each SM executes.
Therefore we move the wave count alone through the element count $N_e$, which does not move the DRAM bytes per SM cycle at all.
Eight times the waves at a fixed SM count, from $N_e = 4^3$ to $8^3$, change $\texttt{GEMM\_FUSED}/\texttt{FUSED\_TC}$ by a relative 1.1 to 2.6\%.
In contrast, the cut from 100 to 25\% of the SMs at a fixed $N_e$ changes the ratio by 9.6 to 9.7\%.
The effect of the SM cut is therefore 3.6 to 8.8 times larger (Figure~\ref{fig:f6}(c)).
The control is taken at $p = 63$ on GB200.
At this point the binding resources differ from those at the $p = 511$ crossing on A100 (Sections~\ref{sec:why-optimum-moves-order} and \ref{sec:architecture-dependent-reversal}).
Therefore the control supports the balance reading of the crossing, but it does not bound the wave effect there (Section~\ref{sec:experimental-scope-limitations}).

\subsection{Predicting where the crossover lies on the balance axis}
\label{sec:predicting-balance-crossover}
\begin{figure*}[tp]
\centering
\includegraphics[width=\linewidth]{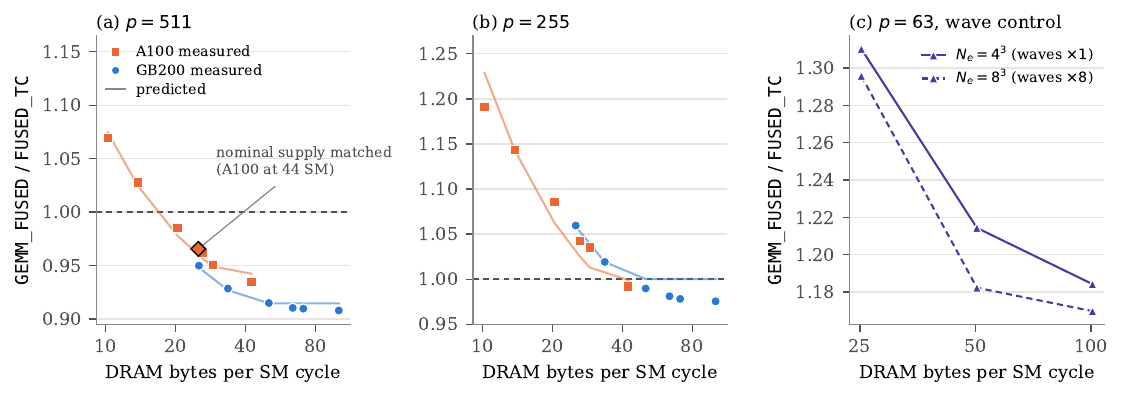}
\caption{
  Machine balance changed on one GPU.
  Markers are measured.
  Lines are predicted as in Section~\ref{sec:predicting-balance-crossover}, and nothing is fitted to this axis.
  (a) $\texttt{GEMM\_FUSED}/\texttt{FUSED\_TC}$ at $p = 511$.
  The diamond is the supply-matched point, A100 at 44 SMs (Section~\ref{sec:same-reversal-one-machine}).
  (b) The same at $p = 255$, where both machines cross.
  (c) The wave control at $p = 63$ on GB200 with $N_e = 4^3$ and $8^3$ (Section~\ref{sec:wave-control}).
  The sizes $N_e = 1$ and $2^3$ are excluded because their working sets, 42 and 336\,MB, respectively, are within a factor of three of the 126\,MB L2.
  Timings are \texttt{Step loop per stage}.
  The estimator of each cell is given in Section~\ref{sup:data-provenance-reproducibility} of the supplemental material.}
\label{fig:f6}
\end{figure*}

The model of Section~\ref{sec:subpipe-model} was never fitted to the balance axis.
Its form and counter inputs were committed before the restricted sweep (Section~\ref{sup:data-provenance-reproducibility} of the supplemental material).
Therefore the model can be scored on this axis.
The counters of a restricted machine come from the same machine at 100\%, and the target supplies only its SM count.
In Equation~\ref{eq:demands}, the term $N_{\mathrm{SM},m}$ is the active SM count.
The term $C_{\mathrm{DRAM},m}$ is multiplied by the ratio of the full to the active SM count, because the restriction removes SMs and not bandwidth.
At $p = 511$ on A100, every one of the six predicted points is within 0.9\% of the measurement.
The point where the ratio crosses 1 is placed between 75 and 50\%, which is where the measured crossing lies (Figure~\ref{fig:f6}(a)).
This agreement is indirect evidence for the nominal-supply assumption of Section~\ref{sec:mps-balance-knob}.
The DRAM-bound work of \texttt{GEMM\_FUSED} is predicted with the full bandwidth at every SM fraction.
Therefore a loss of sustained bandwidth under the restriction would appear as an error that grows along this axis, and such an error is not observed at this order.
The sensitivity of this check is limited, because the DRAM-bound work is a small share of the path.
Over the whole sweep of 90 points, the response curves are predicted with a median absolute error of 6.7\% and a 90th percentile of 19.0\%.
The misses concentrate at low order on A100.
There the model classifies every kernel as DRAM bound and predicts no response.
However, the measurement responds by a factor of 1.74 to 2.05.
This disagreement is the low-order limit of Section~\ref{sec:subpipe-model-negatives}, not a property of this axis.

\section{Range of Validity and Limitations}
\label{sec:trust-and-limits}

\subsection{Absolute times at high order and rankings outside that regime}
\label{sec:signal-to-noise-rule}
Absolute times transfer accurately at high order.
In the present data set, this observed high-order accuracy regime is $p \ge 511$.
At these orders there are 32 path totals.
They transfer with a median error of 1.9\%, and 31 of them are within 10\%.
The range is read from the measured transfers and not derived from $\eta$.
The reason is that no aggregation of the calibration-side $\eta$ over a path separates the good transfers from the bad (Section~\ref{sup:validity-band-selection} of the supplemental material).
This error uses the retrospectively selected closest-$\eta$ calibration rule.
The fixed alternatives give somewhat larger but comparable high-order errors (Section~\ref{sec:logo-three-directions}).
This range is not a prospectively identifiable validity domain.
The order alone does not establish accuracy on a new GPU before that target is measured.
The median path-total error is pooled over the three leave-one-GPU-out directions of Section~\ref{sec:logo-three-directions}.
It falls from 10.7\% at $p = 127$ to 7.5\% at $p = 255$ and to 2.0\% at $p = 511$.
We call $p = 255$ the transition order (Section~\ref{sup:model-departures} of the supplemental material).

Rankings remain useful outside that regime.
The reason is that a ranking is decided not by the size of the error but by its comparison with the difference that is ranked.
We state this as an observation on the present data sets and not as a decision rule.
For a pair of paths, the predicted separation is $|\ln r_{\mathrm{pred}}|$.
The family error is the median of $|\ln (r_{\mathrm{pred}} / r_{\mathrm{meas}})|$ for that pair on transfers in which the target machine takes no part.
Both are quoted in percent, as $100\,|\ln r|$.
Their ratio is the separation-to-error ratio $S/E$.
Where $S/E \ge 1$, which we call a \emph{screened-in} cell, the predicted sign is correct in 94 of 100 cells on the machine axis and in 156 of 167 cells on the balance axis.
Where $S/E < 1$, a sign would be correct in 25 of 37 and 15 of 23 cells.
The threshold was set after the three machines had been measured, and the family error is formed on a population of three GPUs.
The denominator is specific to each pair, because the family errors on the machine axis run from 1.3 to 29.6\%.
The same predicted difference is then large for one pair and small for another, and a single threshold on $|\ln r_{\mathrm{pred}}|$ would treat those pairs as if their errors were equal.
When we divide by the family error, the same comparison applies to every pair.
A selection of the same number of cells by $|\ln r_{\mathrm{pred}}|$ alone is correct in 95 of 100 and 160 of 167 cells, so the family-specific denominator adds no selectivity on these data.
Section~\ref{sup:sn-scoring} of the supplemental material gives the definition, the cell populations, the hit rate as a function of the threshold (Figure~\ref{fig:f7}), the misses by family, the cost of a wrong screened-in cell and the bootstrap intervals.

\subsection{At low order}
\label{sec:production-order-rankings}
A production dynamical core uses $p \simeq 7$ to 11 (Section~\ref{sec:gpu-implementations-tensor-product}).
This range lies outside the observed high-order accuracy regime for absolute times.
To sample this low-order regime and a nearby case, Table~\ref{tab:low-order-rankings} of the supplemental material scores the $S/E$ screen at $p = 7$ and 15 on the 24 rankings of path totals.
Eleven cells reach $S/E \ge 1$, and ten of the eleven are correct.
A forced sign would be right on eight of the thirteen that do not reach it.
The cross-strategy ranking $\texttt{GEMM\_FUSED}/\texttt{FUSED\_TC}$ and the epilogue axis reach the threshold in all of their cells, and the sign is correct in all of those cells.
The ranking $\texttt{FUSED}/\texttt{FUSED\_TC}$ has a median measured difference of 3.3\%.
None of its six cells reach the threshold ($S/E$ is 0.31 on GB200 and 0.06 on H100 at $p = 7$).
Therefore the $p = 7$ reversal of Figure~\ref{fig:f1}(b) is a case that the model does not rank.
The one wrong screened-in cell is $\texttt{FUSED\_DFMA}/\texttt{FUSED}$ on H100 at $p = 15$ (Section~\ref{sup:low-order-rankings} of the supplemental material).

\subsection{What the model does not reproduce}
\label{sec:subpipe-model-negatives}
Six disagreements between the model and the measurement mark limits of the model, and they fall into three groups.
Section~\ref{sup:model-departures} of the supplemental material gives the mechanism and the magnitude of each.
The first group is latency-bound kernels.
The element-boundary flux kernel is latency bound and not roof bound.
Therefore its transfer error spans $-57$ to $+29\%$ over the three directions.
A positive error means that the prediction is slower than the measurement.
The weight of this kernel is at most 1.2\% of a tendency call in the observed $p \ge 511$ regime.
The stall classes \texttt{lg\_throttle} and \texttt{barrier} follow no machine coordinate that we tried.
In \texttt{lg\_throttle}, the L1 queue for local and global memory operations is full.
In \texttt{barrier}, sibling warps of the same thread block have not yet reached a barrier.
The second group is kernels whose binding resource changes between machines, so that $\eta$ does not transfer.
This happens for the CUDA Core path \texttt{FUSED} at $p = 63$ and 255.
The same happens for the contraction kernels of the other paths at $p = 63$, which are L1 bound on GB200 and bound by DRAM or the FP64 budget on the other two machines.
It also happens for the CUTLASS paths at the preferred shape of H100, where the FP64-budget and L1 roofs are nearly tied.
Therefore the gain of the shape change is overestimated, although the direction and the roof of the gain are predicted in all sixteen cases.
The third group is information that the model does not receive.
At $p \le 15$, cuBLAS dispatches different instruction families on the three machines, and the library's choice is not an input.
When those cells are set aside, the low-order error is largely one factor per machine and order, the drift of $\eta$ between the machines.
That factor divides out of a ratio, which is why accurate rankings are observed over a wider range than accurate absolute times.

\subsection{Limitations and outlook}
\label{sec:experimental-scope-limitations}
The limits of the evidence fall into four groups.

The first group is the retrospective specification of the model.
The functional form, the four-cycle floor, and the calibration-machine rule were fixed after the three machines had been measured (Section~\ref{sec:predictive-validation}).
The ranking observation of Section~\ref{sec:signal-to-noise-rule} characterizes the present data sets and is not a validated decision rule.
A rule that chooses a single calibration machine from datasheet supply ratios is rejected in the form that we tested.
That form selects the machine with the smallest supply per unit of SM throughput, and on these three machines it selects A100.
When the rule uses only the DRAM bytes per SM cycle, the measurements follow that ordering.
However, three machines cannot separate this ordering from other orderings that agree with it by chance.
Therefore we do not claim that form (Section~\ref{sup:calibration-set-choice} of the supplemental material).

The second group is the machine population.
The machines are three GPUs of one vendor, one generation apart.
They share one counter set and one compiler toolchain, and they do not vary memory bandwidth and arithmetic throughput independently.
Each of them runs the software stack of its site.
The four-cycle DMMA floor is separated from FLOP pricing only by H100.
Therefore the transfer that is established here is conditional on the empirical pricing relation of Equation~\ref{eq:subpipe-prices}.
A fourth architecture could falsify that relation.
The high-order error in the A100 direction is one-sided (Section~\ref{sec:logo-three-directions}).
Therefore that error is a drift of $\eta$ that is specific to that machine, and no term of the demand model represents it.
The SM-restriction experiment moves its coordinate upward only, and it was run on two of the three machines.
Nsight Compute was not applied to MPS clients.
Therefore every kernel decomposition behind the balance axis is the unrestricted one.

The third group is the transferred inputs, which are assumptions and not measurements on the target.
The non-DMMA counter numerators of a generated kernel instance are assumed to be portable across architectures unless they are transformed explicitly.
The matched-role audit of Section~\ref{sup:counter-numerator-portability} of the supplemental material finds $N_{\mathrm{fp64}}$ identical and L1 wavefronts close for the kernels that are compiled here.
In contrast, issue, ALU, FMA, and L2 activity are less portable.
Most of those demands do not bind in these data, so the path accuracy does not validate their transferability independently.
The library kernel family that the model takes as an input at $p \le 15$ is a function of the architecture and of its software stack.
Nothing here establishes transfer to a different vendor.

The fourth group is the distance from a production configuration.
The crossover and its prediction are demonstrated at high order, which is a controlled parameter and not a production configuration.
At the production-relevant order $p = 7$, the machine-dependent ordering is small (Section~\ref{sec:architecture-dependent-reversal}).
The comparison that decides the fastest fused path at this order is below the screening threshold (Section~\ref{sec:production-order-rankings}).
The transfer error was not scored at a second element count.
Therefore the observed accuracy at $p \ge 511$ is a statement about the measured configuration with $N_e = 1$.
The element-count sweep of Section~\ref{sup:element-count-sweep} of the supplemental material separates order from problem size for the measured ranking only.
All six paths are implementations within one code base, and this is what makes the comparisons one-axis.
However, the results are not calibrated against an independently developed DG code, because the CEED bake-off operators carry no numerical flux.
The operator is linear advection on a periodic unit cube on one GPU without MPI.
For compressible Euler or the complete SCALE-DG model, the pointwise flux has far more arithmetic than the $O(1)$ per node here.
Therefore the size of the full-fusion saving and the crossover order move with the flux arithmetic.
However, the $1/N_q$ scaling of Section~\ref{sec:why-optimum-moves-order} is a ratio and still holds.
Time integration, inter-GPU communication, and physical parameterizations reduce a tendency-level speedup further in a complete model.
At $p = 7$ the tendency is 75--96\% of an RK stage (Section~\ref{sup:difference-between-stage-device} of the supplemental material).
FP64 emulation on reduced-precision matrix instructions \citep{ozaki2012,ootomo2024,ozaki2026,schwarz2026} is measured in Section~\ref{sup:emulated-ozaki-scheme-fp64} of the supplemental material as a reference only.
At the accuracy setting that reproduces the operator, no tested emulated path is faster than native FP64 for these matrix shapes.

\section{Conclusion}
\label{sec:conclusion}
We compared up to six implementations of one three-dimensional tensor-product discontinuous Galerkin advection operator on three GPU systems, GB200, H100, and A100, each with its site-specific software stack.
At each polynomial order, the numerics and the input were held fixed.
Three one-axis controls separate the matrix instruction, the vendor-library choice, and the epilogue fusion.
Four conclusions follow.
First, the better implementation of one operator changes between these architecture--software-stack combinations at a fixed problem.
Neither the element-local fused strategy nor the library GEMM strategy is always the best.
The winner changes with order on GB200 and H100, and at $p = 511$ it also changes between systems.
Second, the nominal peak does not transfer that change.
Third, an accounting that prices counter-defined FP64 activity per issued instruction, with no parameter that is fitted to the target, transfers the change quantitatively at high order and at the common instruction shape.
This holds in three retrospective leave-one-GPU-out directions and on a restriction of the active SMs of one A100 that moves the $p = 511$ ranking across 1.
At the preferred shape of H100, a wider MMA shape that does not exist on GB200 or A100, the same accounting transfers the direction of the ranking.
Outside the high-order regime, rankings remain useful where the predicted separation is large compared with the transfer errors observed elsewhere.
Fourth, the latency-bound range of a production dynamical core is still to be predicted.
At $p = 7$ and 15, the ratio that decides the fastest path, \texttt{FUSED}/\texttt{FUSED\_TC}, is below the screening threshold in all six cells, so the screen returns no fastest path there.
This result suggests that an explicit measure of memory-level parallelism is needed to predict those cells.

The model is a transfer model for a workload that has been profiled on one architecture, and not a general performance predictor.
It needs one profiled calibration factor per kernel instance, and it predicts a new machine and not a new workload.
The nominal-model failure motivated the form, and the leave-one-GPU-out scores selected the calibration-machine rule.
Therefore those scores establish retrospective parameter transfer and not prospective model discovery.
The ranking screen is an empirical characterization of the present data sets.

\begin{acks}
This research used computational resources of the supercomputer RIKYU provided by the RIKEN Center for Computational Science.
This work used computational resources TSUBAME4.0 supercomputer provided by Institute of Science Tokyo through Joint Usage/Research Center for Interdisciplinary Large-scale Information Infrastructures and High Performance Computing Infrastructure in Japan (Project ID: jh260024).
This research was conducted using the FUJITSU Supercomputer PRIMEHPC FX1000 and FUJITSU Server PRIMERGY GX2570 (Wisteria/BDEC-01) at the Information Technology Center, The University of Tokyo.

\emph{Use of generative artificial intelligence (AI).} Generative AI tools (Claude Code, Anthropic; ChatGPT, OpenAI; and Cursor, Anysphere) were used at multiple stages of this study. The first author (S.N.) used them for GPU code development and optimization, performance analysis, preparation of analysis and visualization scripts, exploration of possible explanations for observed performance behavior, and preparation and review of the manuscript.
The co-authors used them for review of the manuscript.

The tools were used interactively as part of the research process.
The first author formulated research questions, hypotheses, and optimization ideas and directed the AI tools to develop, analyze, or test specific possibilities.
The AI tools also proposed candidate hypotheses, optimization strategies, analysis approaches, and interpretations that were not necessarily initiated by the first author.
Ideas that arose from either the first author or the AI tools could motivate subsequent implementation, profiling, analysis, or additional experiments.

AI-generated suggestions and interpretations were treated as candidates for investigation and not as scientific evidence.
All reported performance results were obtained from actual executions on the hardware described in this paper.
Generated or modified code was checked for correctness and evaluated experimentally.
Proposed explanations were assessed against measured data, profiling results, source code, and relevant hardware documentation.
The first author directed the overall investigation, determined which questions and candidate explanations deserved further study, designed and evaluated the computational experiments, and integrated the resulting evidence.
The authors are responsible for the final scientific claims and conclusions.
\end{acks}

\statementshead{Statements and Declarations}

\begin{sagestatement}{Ethical considerations}
Not applicable.
This study involves no research on humans or animals.
\end{sagestatement}

\begin{sagestatement}{Consent to participate}
Not applicable.
\end{sagestatement}

\begin{sagestatement}{Consent for publication}
Not applicable.
\end{sagestatement}

\begin{dci}
The authors declared no potential conflicts of interest with respect to the research, authorship, and/or publication of this article.
\end{dci}

\begin{funding}
The authors disclosed receipt of the following financial support for the research, authorship, and/or publication of this article: This work was supported by the Japan Society for the Promotion of Science (JSPS) Grants-in-Aid for Scientific Research (KAKENHI; grant number JP25H01109).
This work was supported by MEXT Next-Generation Computational Science Global Reach \& Innovation Program (Generative AI Spec-Driven Breakthrough of SCALE Modeling) Japan Grant Number JPMXP1920260210.
\end{funding}

\begin{sagestatement}{Data availability}
The scripts that produce every figure and the inputs that those scripts read under \texttt{figs/data/} are archived at Zenodo \citep{thiswork-data}.
These inputs are the raw logs of the three-machine path sweep behind Figures~\ref{fig:f1}(a), \ref{fig:f2}, \ref{fig:f3}, and \ref{fig:a11a}(a) of the supplemental material, and Tables~\ref{tab:a-timing-gb200}--\ref{tab:a-timing-a100}.
They also include the logs of the balance campaign behind Figure~\ref{fig:f6} and Table~\ref{tab:balance-knob} of the supplemental material, which include the \texttt{\%smid} readout.
They also include the logs and configurations of the $p = 767$ fused job of Section~\ref{sec:order-dependent-crossover}.
Finally, they include the per-cell model output behind Figures~\ref{fig:f5} and~\ref{fig:f7}, Table~\ref{tab:seven-ratios}, and Table~\ref{tab:low-order-rankings} of the supplemental material.
The model of Section~\ref{sec:subpipe-model} reads per-kernel Nsight Compute counters.
These counters are included as \texttt{figs/data/subpipe\_inputs.csv} with a column definition, together with the job, kernel name, launch configuration, and the derived demands and calibration factor of every kernel instance.
Therefore Figures~\ref{fig:f5} and~\ref{fig:f7} can be rebuilt from the counters.
The raw Nsight Compute reports behind the remaining tables and panels are not included in the archive.
The provenance record (Section~\ref{sup:provenance} of the supplemental material) identifies them by scheduler job, and they are available from the corresponding author.
The benchmark itself is the extracted advection operator of Section~\ref{sec:extracted-benchmark}.
The extraction from which it derives is publicly available \citep{kawai-extraction}.
The six implementation paths that we measured here are released as a public fork of that extraction on the branch \texttt{feature/cuda} \citep{thiswork-code}.
The namelists of Table~\ref{tab:a-configs} and the build configuration for the paths are released together with them.
The provenance record maps every figure and table to the scheduler job, the compute node, and the source revision at which it was measured.
Those revisions are resolvable in that repository.
\end{sagestatement}

\appendix
\setcounter{section}{0}
\renewcommand{\thesection}{A\arabic{section}}
\counterwithin{figure}{section}
\renewcommand{\thefigure}{\thesection\alph{figure}}
\counterwithin{table}{section}
\renewcommand{\thetable}{\thesection\alph{table}}

\section{Mechanism of the one-axis ablations}
\label{app:one-axis-mechanism}
This appendix holds the Nsight Compute evidence behind the three one-axis ratios of Section~\ref{sec:one-axis-evidence}.

\subsection{The MMA axis}
We profiled four orders with Nsight Compute, $p = 15$ through $p = 255$.
At these orders, the Tensor Core implementation executes 2.3--5.0 times fewer dynamic instructions on GB200 and 2.4--5.4 times fewer on H100 (Table~\ref{tab:mma}).
At $p = 63$, the two H100 kernels move from 42--70\% to 75--89\% of DRAM throughput, whereas the two GB200 kernels reach only 26--41\%.
Within the clock qualifications of Table~\ref{tab:mma}, this is consistent with the memory supply truncating the MMA gain on H100, but not on GB200.
Beyond $p = 63$, the H100 ratio rises to approximately two, and the GB200 ratio remains near 2.2.
Therefore $p = 63$ is not the arithmetic-bound limit of this comparison.
The wall-time ratios of Figure~\ref{fig:f3}(a) are an upper bound on the value of the MMA instruction on this schedule.
The reason is that the DFMA version runs the warp shape that was chosen for the Tensor Core version.
The warp shape of the DFMA version alone can be retuned, outside the path definitions and outside production dispatch.
When it is retuned in this way, the DFMA version requires 28.3\% less time at $p = 31$ and 26.5\% less at $p = 127$ (Section~\ref{sup:register-budget-cost-iso} of the supplemental material).

\subsection{The library axis}
On GB200 the ratio $\texttt{GEMM\_CUTE}/\texttt{GEMM}$ stays within 2\% of 1 for $p \ge 31$, from 0.980 to 1.013, and A100 behaves similarly.
The reason is that cuBLAS on these machines dispatches CUTLASS-generated kernels from the same families and tile shapes as the explicit CUTLASS implementation (Table~\ref{tab:cublas}).
At the two lowest orders, GB200 leaves that band.
The reason is that the cuBLAS selection rule selects tiles that are wider than these contraction shapes, so predicates discard much of each tile.
At $N_q = 8$, the replacement of cuBLAS by the order-specific CUTLASS mainloops reduces the $x$ and $z$ GEMMs from 179.2 to 91.4~$\mu$s and from 159.3 to 92.0~$\mu$s per stage.
The per-kernel differences sum to 158.2~$\mu$s against a path difference of 155.5~$\mu$s.
This sum includes 3.1~$\mu$s from the $y$ GEMM, which is cuBLAS on both sides.
The CUTLASS FP64 multistage iterator cannot be built with an $N = 8$ tile either, but its smallest valid $N = 16$ tile halves the waste.
At $N_q = 16$, for example, cuBLAS selects a $32\mathord{\times}32$ tile for the $x$ GEMM that takes 355.0~$\mu$s.
This tile executes 3.66 times as many instructions as the 89.3~$\mu$s CUTLASS tile that is matched to $N = N_q$.
A100 also leaves the band at these orders, but its $N_q = 8$ cuBLAS path dispatches a SIMT double-precision kernel.
Therefore we do not explain that cell by the GB200 tile mechanism.
The near-identity of the two sides is not built into the axis.
\texttt{GEMM\_CUTE} assigns at most one of the three volume GEMMs to cuBLAS at any order, and at most orders it assigns none.
Therefore the comparison moves at least two of the three GEMMs everywhere.
On H100, at $p = 511$, the two paths move the same DRAM bytes at the same occupancy.
The explicit CUTLASS mainloop, however, issues 2.3--2.4 times as many instructions for the same arithmetic.

\subsection{The fusion axis in device time}
In CUDA-device time, the fusion axis of Figure~\ref{fig:f3}(c) is at most 1.8 percentage points lower than in step-loop time, and its sign never differs.
The rising branch below $N_q = 128$ is decomposed per kernel in Section~\ref{sup:low-order-fusion-axis} of the supplemental material.

\begin{table*}[tp]
\centering
\footnotesize
\caption{
  Mechanism of the MMA axis, \texttt{FUSED\_TC} against \texttt{FUSED\_DFMA}.
  Ratios are \texttt{FUSED\_TC} divided by \texttt{FUSED\_DFMA}, so a value below 1 is a speedup for the Tensor Core version.
  Arrows read \texttt{FUSED\_DFMA} to \texttt{FUSED\_TC}.
  Where two numbers appear, the order runs two distinct fused kernels ($xz$ and $y$), and both are given.
  At $p = 255$ the profiling window contains several invocations of one kernel, and the range is given.
  All entries are Nsight Compute measurements at the default locked clock of the profiler (\texttt{--set full}). \emph{Time ratio} is the ratio of the profiled kernel durations under that lock, not a wall-clock ratio.
  The wall-clock form of this axis is Figure~\ref{fig:f3}(a).
  The lock is a different fraction of the boost clock on each machine, and the SM-side utilizations are not clock-invariant. Therefore the percentages are within-machine mechanism indicators and not a machine-to-machine comparison of hardware utilization.}
\label{tab:mma}
\noindent\begin{tabularx}{\tblwidth}{@{}lRRRRRRR@{}}
\toprule
$p$ & instruction ratio, GB200 & instruction ratio, H100 & GB200 SM throughput [\%] & GB200 DRAM throughput [\%] & GB200 time ratio & H100 DRAM throughput [\%] & H100 time ratio \\
\midrule
15 & 0.434 & 0.414 & 58.8 $\rightarrow$ 25.3 & 22.0 $\rightarrow$ 40.4 & 0.543 & 80.1 $\rightarrow$ 86.3 & 0.883 \\
31 & 0.199 & 0.185 & 76.8 / 69.9 $\rightarrow$ 38.6 / 35.7 & 19.0 / 16.9 $\rightarrow$ 57.7 / 46.3 & 0.350 & 69.1 / 60.4 $\rightarrow$ 89.2 / 87.5 & 0.715 \\
63 & 0.271 & 0.320 & 74.3 / 75.7 $\rightarrow$ 53.1 / 57.2 & 16.8 / 11.0 $\rightarrow$ 40.8 / 26.2 & 0.415 & 69.6 / 42.2 $\rightarrow$ 88.7 / 74.7 & 0.674 \\
255 & 0.297 & 0.294 & 60.6--62.2 $\rightarrow$ 76.6--77.8 & 6.0--6.6 $\rightarrow$ 13.1--15.3 & 0.439 & 25.3--28.2 $\rightarrow$ 52.6--59.3 & 0.447 \\
\bottomrule
\end{tabularx}
\end{table*}

\begin{table*}[tp]
\centering
\footnotesize
\caption{
  Kernels that cuBLAS dispatches for the three volume GEMMs.
  The statement is restricted to the shapes that this benchmark calls and to the library versions of Table~\ref{tab:machines}. These shapes are FP64, \texttt{nn} (neither input matrix transposed), \texttt{align1} (single-element alignment), and the $M$, $N$, $K$ of the volume contractions.
  It is not a statement about cuBLAS in general.
  The last column is the measured \texttt{GEMM\_CUTE} to \texttt{GEMM} ratio of Figure~\ref{fig:f3}(b).}
\label{tab:cublas}
\begin{tabularx}{\tblwidth}{@{}>{\hsize=.5\hsize}L>{\hsize=1.25\hsize}L%
                             >{\hsize=1.25\hsize}L>{\hsize=1.0\hsize}L@{}}
\toprule
Machine & Kernels cuBLAS dispatches & Relation to our CUTLASS mainloop & Library axis \\
\midrule
GB200 (sm\_100) & \texttt{cutlass\_80\_tensorop\_d884gemm\_64x128\_16x3\_nn\_align1}, \texttt{...\_64x64\_16x4\_...}, \texttt{...\_64x32\_16x4\_...} & same family, same tiles, same \texttt{DMMA.8x8x4}; for the $x$ GEMM at $p = 255$ even the register count, grid, and block size agree & nearly identical: 0.980--1.013 for $p \ge 31$ \\
A100 (sm\_80) & \texttt{cutlass\_80\_tensorop\_d884gemm\_128x64\_16x3\_nn\_align1}, \texttt{...\_64x64\_16x4\_...}, \texttt{...\_64x32\_16x4\_...} & same family and tiles; at $p = 63$ all three GEMMs match & nearly identical \\
H100 (sm\_90) & \texttt{sm90\_xmma\_gemm\_f64f64\_f64f64\_f64\_nn\_n\_tilesize64x128x16\_stage3\_warpsize2x2x1\_tensor16x8x8\_execute\_kernel\_\_5x\_cublas}, and the same kernel at \texttt{tilesize128x64x16\_stage3} and \texttt{tilesize64x64x16\_stage4} & hand-written sm\_90 kernels that use \texttt{tensor16x8x8}, a third MMA shape that neither our default $8\times 8\times 4$ nor the H100-preferred $16\times 8\times 4$ uses & different: 1.36--1.45 at $p \ge 511$ (1.20 already at $p = 255$) with the default shape, 1.04--1.13 at $p \ge 127$ with the H100-preferred shape \\
\bottomrule
\end{tabularx}
\end{table*}

\section{Order-resolved crossover map}
\label{app:crossover-map}
Table~\ref{tab:reversal-map} projects the order-resolved ratio $\texttt{GEMM\_FUSED}/\texttt{FUSED\_TC}$ of Section~\ref{sec:operational-predictions} onto the two other machines from the counters of one calibration machine.
No measurement from the predicted machine is used for parameter instantiation.
As in Section~\ref{sec:predictive-validation}, the model form itself was specified after all three machines had been measured.
The crossings are those of Figure~\ref{fig:f1}(a), which reads them in step-loop time.
The non-tendency remainder that separates the two timers is common to the paths.
It moves every ratio toward 1 without a change of sign (Section~\ref{sup:difference-between-stage-device} of the supplemental material).
At $p = 7$, the two \texttt{GEMM\_FUSED} cells of A100 and the cell of H100 that is projected from A100 have kernel lists that do not match those of the calibration machine.
Therefore these cells are scored on the common kernels, as in Section~\ref{sec:logo-three-directions}.
The H100 measured value on that common set is 2.140 against the projection from A100.
At $p = 511$ the measured rows differ from a second measurement by 1.0\% on A100 and 0.8\% on H100.
No spread moves a cell across 1 (Section~\ref{sup:p511-ratio-provenance} of the supplemental material).

At the shape that H100 prefers, there is one ranking miss among the twelve cells of Section~\ref{sec:operational-predictions}.
This miss is the epilogue axis $\texttt{GEMM\_CUTE}/\texttt{GEMM\_FUSED}$ at $p = 127$, which is predicted at 0.988 against a measured 1.117.
At $p = 767$, the common-shape prediction of $\texttt{GEMM\_FUSED}/\texttt{FUSED\_TC}$ is 0.996 against a measured 1.014.
This is the only sign that this ratio gets wrong on the machine axis.
At the preferred shape, the prediction is 0.607 against a measured 0.753, and this sign is right.

Neither of the two failures in Table~\ref{tab:reversal-map} changes a sign.
The ratio at $p = 511$ on H100 is underestimated.
The projections from GB200 and from A100 are 0.709 and 0.718, respectively, against a measured 0.860.
This error is the overestimate of the MMA-shape gain at high order that Section~\ref{sec:subpipe-model-negatives} explains.
Both projections from GB200 carry the dip that GB200 shows at $p = 127$, which neither target shows.
At $p \le 127$, the two projections from GB200 coincide to three digits, although A100 and H100 differ in both coordinates of Table~\ref{tab:machines-per-sm-cycle}.
At those orders, the target enters the numerator and the denominator through the same factor, so the projected ratio is set by the calibration machine alone.
Those cells lie well above 1 on every machine, so the carried-over shape costs no ranking.

\begin{table*}[tp]
\centering
\footnotesize
\caption{
  The order-resolved crossover map, $\texttt{GEMM\_FUSED}/\texttt{FUSED\_TC}$ in CUDA-device time per tendency call.
  A value below 1 (bold) means that the fused-epilogue library path is the faster one.
  Predictions are retrospective projections of one calibration machine onto the others, and they use no measurement from the predicted machine for parameter instantiation.
  Within the tested orders, the measured ratio crosses 1 between $p = 255$ and $p = 511$ on GB200 and, at the preferred shape, on H100. At the common shape of Figure~\ref{fig:f1}(a), H100 does not cross. The measured ratio does not cross on A100. Both projections of H100 reproduce its crossing interval, the projection of A100 produces no crossing, and no projection introduces a crossing anywhere else.
  H100 rows, including the predictions, are at the $16\times 8\times 4$ shape that H100 prefers at high order.
  Orders $p \le 255$ lie below the observed high-order accuracy regime of Section~\ref{sec:signal-to-noise-rule}. Therefore the lower end of the measured crossing interval is at the transition order $p = 255$, and the upper end is inside that regime.}
\label{tab:reversal-map}
\begin{tabularx}{\tblwidth}{@{}>{\hsize=2.3\hsize}L>{\hsize=.81\hsize}R>{\hsize=.81\hsize}R>{\hsize=.81\hsize}R>{\hsize=.81\hsize}R>{\hsize=.81\hsize}R>{\hsize=.81\hsize}R>{\hsize=.81\hsize}R@{}}
\toprule
 & $p = 7$ & 15 & 31 & 63 & 127 & 255 & 511 \\
\midrule
GB200 measured (calibration) & 5.955 & 2.220 & 1.555 & 1.361 & 1.031 & 1.063 & \textbf{0.948} \\
\midrule
A100 measured & 2.168 & 2.271 & 1.579 & 1.408 & 1.257 & 1.237 & 1.079 \\
A100 predicted from GB200 & 2.201 & 2.062 & 1.415 & 1.356 & 1.181 & 1.264 & 1.055 \\
\midrule
H100 measured & 3.265 & 2.316 & 1.541 & 1.460 & 1.269 & 1.097 & \textbf{0.860} \\
H100 predicted from GB200 & 3.383 & 2.055 & 1.415 & 1.356 & 1.181 & 1.157 & \textbf{0.709} \\
H100 predicted from A100 & 2.159 & 2.252 & 1.565 & 1.408 & 1.245 & 1.123 & \textbf{0.718} \\
\bottomrule
\end{tabularx}
\end{table*}

\section{Contract of the model}
\label{app:model-contract}
Table~\ref{tab:contract} collects the inputs, the output, and the range of the model of Section~\ref{sec:subpipe-model}, and it names the section in which each entry is established.

The MMA shape of the cuBLAS roles is not in the cross-compiled binary (Section~\ref{sec:subpipe-model}), and its price is measured here.
Suppose that the cuBLAS roles are priced with the shape of the calibration machines instead of the shape that Equation~\ref{eq:instruction-enumeration} selects.
Then seven of the 140 path totals change, and four of the rankings of Section~\ref{sec:seven-ratios} that are built on them reverse sign.
Two of the four are on the cross-strategy ranking $\texttt{GEMM}/\texttt{FUSED\_TC}$.
At $p = 511$ its prediction moves from 0.888 to 1.306 against a measured 0.919, and at $p = 767$ from 0.732 to 1.146 against 0.756.
The conditional shape is chosen without running the target.
The set $\mathcal S_{k,m}$ of instruction shapes that a target kernel role $k$ supports is known from the instruction set or from a cross-compiler.
Every other demand is kept fixed, and the model chooses
\begin{equation}
\widehat T_{\mathrm{enum}}(P,m)=\min_{\boldsymbol{s}\in\mathcal S_m(P)}\sum_{k\in P}\frac{\eta(k)\max_r D_r(k,m;s_k)}{f_m},
\label{eq:instruction-enumeration}
\end{equation}
where $\boldsymbol{s}\in\mathcal S_m(P)$ assigns one shape $s_k\in\mathcal S_{k,m}$ to each selectable kernel role $k$ of path $P$, and $f_m$ is the boost clock of Equation~\ref{eq:model-time}.
The sum is over kernels, so the minimum is taken role by role.
The enumeration is over MMA shapes within one instruction family.
The family that the library selects at $p \le 15$ remains the condition of application of Section~\ref{sec:subpipe-model}.
At $p = 511$ on held-out H100, the enumeration covers \texttt{m8n8k4}, \texttt{m16n8k4}, \texttt{m16n8k8}, and \texttt{m16n8k16} for the three cuBLAS roles.
This enumeration separates the narrow shape, on which the floor binds, from the three wide shapes.
The wide shapes tie at 13.69~ms, because their instruction-count reduction and their per-instruction price cancel under Equation~\ref{eq:subpipe-prices}.
The shape that cuBLAS dispatches there, \texttt{m16n8k8}, is one of the three, and the measured path time is 14.44~ms (Table~\ref{tab:holdout-h100}).
When the target can be run, the dispatched shape can be read from the kernel name of one execution and used in place of the enumerated shape.
On the three machines here, the two give the same total in every scored cell.
On GB200 and A100, the enumerated shape is \texttt{m8n8k4}, which is also the dispatched one.
On H100, the roles for which cuBLAS dispatches a narrow shape at $p \ge 31$ are bound by DRAM under either price.
At $p \ge 127$, every cuBLAS role there dispatches a wide shape.
The scores of Section~\ref{sec:predictive-validation} therefore hold for the enumerated shape as well.

\begin{table*}[tp]
\centering
\footnotesize
\caption{
  The contract of the model of Section~\ref{sec:subpipe-model}: its inputs, its output, the range of evidence reported, and what falls outside it.
  Each entry is established in the section that is named against it.}
\label{tab:contract}
\begin{tabularx}{\tblwidth}{@{}>{\hsize=.55\hsize}L>{\hsize=1.45\hsize}L@{}}
\toprule
 & Content \\
\midrule
Calibration inputs & One Nsight Compute profiling sweep per calibration machine over the paths and orders of interest, which supplies for every kernel instance the instruction counts, DRAM bytes, L1 wavefronts, and elapsed SM cycles, and hence the demands $D_r$ and one calibration factor $\eta$ (187 to 219 instances per leave-one-GPU-out direction, Section~\ref{sec:subpipe-model}).
All non-DMMA counter numerators are assumed transferable for the same generated kernel instance; only the DMMA count is transformed for the target instruction shape (audit in Section~\ref{sup:counter-numerator-portability} of the supplemental material). \\
Target inputs & Capacities per SM cycle from the datasheet (Table~\ref{tab:machines-per-sm-cycle}); for kernels compiled here, the MMA instruction expansion read with \texttt{cuobjdump} from a binary cross-compiled for the target; for vendor-library roles alone, the supported instruction family and set of MMA shapes, from which Equation~\ref{eq:instruction-enumeration} conditionally selects the shape with the smallest demand.
No execution on the target is needed, but supported shapes do not predict vendor-library dispatch policy (Section~\ref{sec:subpipe-model}). \\
Not taken from the target & Any timing, any counter, and any parameter that is fitted to the target (the one cuBLAS role of the CUTLASS paths at $p = 7$ and 31 is treated as a library role, Section~\ref{sec:subpipe-model}).
When the target can be run, the library shape that is dispatched in one execution can replace the enumerated shape; on these three machines the two routes give the same total in every scored cell (this appendix). \\
Output & Equation~\ref{eq:model} in SM cycles per kernel instance, the same count divided by the boost clock of the target (Equation~\ref{eq:model-time}) in time, its sum over the kernels of a path, and the quotient of two path totals, which is the ranking on which an implementation is chosen. \\
Observed absolute-time accuracy & At $p \ge 511$ and the common instruction shape: median path-total errors of 1.3, 5.4, and 1.7\% in the three leave-one-GPU-out directions (Section~\ref{sec:logo-three-directions}); at the preferred shape of H100 the two shape-dependent CUTLASS paths err by 10.2 to 19.0\% (Section~\ref{sec:operational-predictions}).
This regime is drawn from the measured transfer errors, is not prospectively identifiable from $p$ or calibration-side $\eta$, and does not establish validity on a new GPU before target measurements; the production orders $p = 7$ to 11 lie outside it (Section~\ref{sec:signal-to-noise-rule}). \\
Empirically screened rankings & Where the predicted difference exceeds the leave-target-out error of that family of ratios: the sign is correct on 94 of 100 machine-axis cells and 156 of 167 balance-axis cells at the common shape, the screen returns no winner below the threshold, and a forced sign would be correct on 65 to 68\% of those cells (Section~\ref{sec:signal-to-noise-rule}).
When the screen is applied to the choice of the fastest path, it selects a path in 15 of 26 cells and matches the measured fastest path in all 15 (Section~\ref{sec:fastest-path-selection}); at the preferred shape of H100 no family error exists and correct signs are reported only as correct signs (Section~\ref{sec:operational-predictions}). \\
Outside the contract & A kernel instance that has been profiled on no machine, that is, a new order, implementation, or operator; the choice of kernel family that the library itself makes at $p \le 15$, which is an input and not a prediction; latency-bound kernels at low occupancy, for which the FP64-budget pricing has no counterpart (Section~\ref{sec:subpipe-model-negatives}); and vendors other than the measured one (Section~\ref{sec:experimental-scope-limitations}). \\
Class of evidence & The target contributes no fitted parameter in any of the three directions, but the nominal-model failure motivated the functional form and the same scores selected the per-kernel calibration-machine rule, so the machine axis tests retrospective parameter transfer and not prospective model discovery; both screening rates are empirical characterizations of the present data sets (Sections~\ref{sec:predictive-validation} and~\ref{sec:experimental-scope-limitations}). \\
\bottomrule
\end{tabularx}
\end{table*}

\begin{biogs}
\textbf{Seiya Nishizawa} received his Doctor of Philosophy (PhD) in geophysics from Kyoto University,
Japan, in 2005. After postdoctoral positions at the Research Institute for
Mathematical Sciences and the Department of Geophysics, Kyoto University, and at
the Center for Planetary Science, Kobe University, he was an assistant professor
at the Center for Planetary Science, Kobe University, from 2009 to 2012. In 2012
he joined the RIKEN Center for Computational Science (then the Advanced Institute
for Computational Science) as a research scientist, and since 2023 he has been a
senior research scientist in its Computational Climate Science Research Team. He
has published more than 40 peer-reviewed articles on atmospheric turbulence,
turbulence--cloud interaction, high-resolution atmospheric modeling, and
numerical methods for geophysical fluid dynamics. He has served as an editor and
steering committee member of \emph{Scientific Online Letters on the Atmosphere}
(SOLA).

\textbf{Yuta Kawai} received his PhD in science from Kobe University, Japan, in
2018. He has been with the RIKEN Center for Computational Science since 2015,
first as a junior research associate until 2018, then as a postdoctoral
researcher from 2019 to 2024, and as a technical scientist since 2024. He works
on the research, development, and practical application of SCALE-DG, an
atmospheric model with a discontinuous Galerkin dynamical core aimed at future
high-resolution atmospheric simulations, and, in collaboration with
computational scientists, on optimizing its performance on modern
high-performance computing systems. His research interests are in computational
meteorology.

\textbf{Xuanzhengbo Ren} received his Bachelor of Science (BS) in computer science from Harbin
Engineering University, China, in 2018, his Master of Science (MS) in computer software and theory
from the University of Chinese Academy of Sciences, China, in 2021, and
his PhD in informatics from Nagoya University, Japan, in 2026. From 2025 to 2026
he was a junior research associate with the Computational Climate Science
Research Team, RIKEN Center for Computational Science, and since 2026 he has
been a project assistant professor at the Information Technology Center, The
University of Tokyo. His research interests include high-performance computing
and software auto-tuning.
\end{biogs}

\clearpage
\setcounter{section}{0}
\counterwithout{figure}{section}
\counterwithout{table}{section}
\setcounter{figure}{0}
\setcounter{table}{0}
\renewcommand{\thesection}{S\arabic{section}}
\renewcommand{\thetable}{S\arabic{table}}
\renewcommand{\thefigure}{S\arabic{figure}}
\counterwithin*{equation}{section}
\renewcommand{\theequation}{\thesection.\arabic{equation}}
\renewcommand{\theHsection}{S.\arabic{section}}
\renewcommand{\theHtable}{S.\arabic{table}}
\renewcommand{\theHfigure}{S.\arabic{figure}}
\renewcommand{\theHequation}{S.\arabic{section}.\arabic{equation}}
\renewenvironment{apptable}%
  {\par\addvspace{\intextsep}\noindent\begin{minipage}{\linewidth}}%
  {\end{minipage}\par\addvspace{\intextsep}}

\section*{Supplemental material}

This document accompanies the main text.
Section~\ref{sup:rejected-optimizations} records the optimizations that were implemented, measured, and not adopted, with the measured effect and the mechanism that profiling indicates.
Section~\ref{sup:provenance} maps every measured figure and table of the main text and of this document to the scheduler job, the compute node, and the source revision from which it was taken.
It also reports the checks that relate the measurement campaigns to each other.
Sections~\ref{sup:low-order-fusion-axis}--\ref{sup:roofline-placement-using-measured} hold evidence that the main text cites but does not need in full.
These sections give the order dependence of the full-fusion saving, the element-count sweep, and the measurement details behind the profiling rules and claim thresholds.
They also give the nominal-roof two-component accounting that reproduces A100 and fails on H100, the closure of the counter-defined FP64 budget, the calibration-side choices, the disagreements between the model and the measurement, and the scoring of the empirical ranking screen.
Next, they give the cell-level tables of the screened rankings at low order and of the choice of the fastest path, the register-budget cost of the iso-schedule ablation, and the absolute timing values behind every figure.
Finally, they give the benchmark configurations, the implementation details of the paths and the validation coverage, the cost of emulated FP64 GEMM, and the roofline placement of every measured path.
Section, figure, table, and appendix numbers without the prefix S refer to the main text.

\section{Rejected optimizations}
\label{sup:rejected-optimizations}

This section records optimizations of the fused and CUTLASS paths that were implemented, measured, and not adopted.
For each optimization, it gives the measured effect and the mechanism that profiling indicates.
The section documents the tuning behind the paths of Table~\ref{tab:paths} of the main text.
Each entry shows that an optimization must be judged on the complete path.
Unless stated otherwise, timings are medians over interleaved rounds on an exclusively allocated GPU.
Each timing is compared against the immediately preceding implementation, which differs only in the stated change.
The ranges do not overlap unless noted.

\subsection{A wider z tile: faster on its own, slower in the production path}
\label{app:wider-z-tile-faster}
The $z$ volume GEMM in the CUTLASS paths uses a $64\times 32$ thread-block tile.
In contrast, cuBLAS selects $64\times 64$ for the same matrix shape.
A replacement of the tile in \texttt{VolumeGemmSet} changes both the unfused control path and the fused production path, because they share the same mainloop by design.
On GB200, however, the change has opposite effects on the two paths (Table~\ref{tab:a-wider-z-tile}).

\begin{apptable}
\captionof{table}{A wider $z$-GEMM thread-block tile on GB200. \texttt{Main} time per step with a $64\times 64$ tile for the $z$ volume GEMM, relative to the production $64\times 32$ tile at the same order. Negative values denote a speedup and positive values a slowdown. Eight interleaved rounds, four for $p \ge 575$. \texttt{GEMM} is the cuBLAS control, which this source change does not touch.}
\label{tab:a-wider-z-tile}
\noindent\begin{tabularx}{\tblwidth}{@{}lLLL@{}}
\toprule
$p$ & \texttt{GEMM\_CUTE} & \texttt{GEMM\_FUSED} & \texttt{GEMM} (control) \\
\midrule
63 & $-$1.2\% & +8.6\% & +0.04\% (overlapping) \\
127 & $-$1.0\% & +4.3\% & $-$0.04\% (overlapping) \\
255 & $-$0.7\% & +2.9\% & +0.03\% (overlapping) \\
511 & $-$0.4\% & +0.4\% & $-$0.01\% (overlapping) \\
575 & $-$0.4\% & +0.4\% & $<0.01$\% (overlapping) \\
767 & $-$0.3\% & +0.4\% & $<0.01$\% (overlapping) \\
\bottomrule
\end{tabularx}
\end{apptable}

The cuBLAS control is within $\pm$0.04\% with overlapping ranges, so it bounds run-to-run drift.
The wider tile improves the unfused path and degrades the fused one at every order.
Occupancy is unchanged (12.1 against 12.2\% at $p = 63$).
However, the SM throughput of the epilogue carrier falls from 71.0 to 59.2\% at $p = 127$, although its instruction count falls.
Also, at $p = 127$ the wider carrier begins to spill.
The halved block-level concurrency hurts the latency-sensitive fused carrier, whose epilogue also reads the other two derivatives and the face data.
The lighter unfused epilogue does not suffer from the halved concurrency.
Both paths share one mainloop by design, so the fused path decides the production tile.

\subsection{A larger MMA shape on H100}
\label{app:larger-mma-shape-h100}
On H100, $16\times 8\times 4$ is the best MMA shape at every measured order, whereas $16\times 8\times 8$ never wins.
Table~\ref{tab:a-mma-shape-h100} reports \texttt{Main} time per step relative to cuBLAS.

\begin{apptable}
\captionof{table}{MMA shape sweep on H100. \texttt{Main} time per step relative to cuBLAS for \texttt{GEMM\_CUTE} and \texttt{GEMM\_FUSED} at the three MMA shapes. Negative values indicate a speedup. The two highest-order measurements in this sweep contained occasional interference from other work on the machine, which only made runs slower, and we report the minimum observed time to remove it. The text below checks those rows against the interleaved medians of the base sweep.}
\label{tab:a-mma-shape-h100}
\noindent\begin{tabularx}{\tblwidth}{@{}lRRRRLL@{}}
\toprule
$p$ & \texttt{GEMM\_CUTE} $8\times 8\times 4$ & $16\times 8\times 4$ & $16\times 8\times 8$ & \texttt{GEMM\_FUSED} $8\times 8\times 4$ & $16\times 8\times 4$ & $16\times 8\times 8$ \\
\midrule
127 & +7.1\% & +3.7\% & +6.2\% & $-$3.4\% & $-$3.8\% & $-$1.1\% \\
255 & +19.6\% & +6.6\% & +11.6\% & +4.0\% & $-$1.5\% & +5.7\% \\
511 & +35.1\% & +8.6\% & +9.2\% & +16.5\% & $-$4.5\% & $<0.1$\% \\
575 & +37.3\% & +9.5\% & +20.6\% & +29.0\% & $-$2.7\% & +1.6\% \\
\bottomrule
\end{tabularx}
\end{apptable}

The result does not depend on the minimum estimator.
The base three-machine sweep (Section~\ref{sup:campaigns}) timed the $16\times 8\times 4$ variants of both CUTLASS paths on H100 in interleaved rounds.
Its medians relative to cuBLAS ($+8.6$ and $-4.0\%$ at $p = 511$, $+9.5$ and $-2.4\%$ at $p = 575$) agree with the table to within 0.5 points.
Every preferred-shape value that the main text quotes is taken from that base sweep and not from this table.

The register file is the limit.
For the $x$ GEMM at $p = 255$, $16\times 8\times 4$ raises the register count from 228 to 248 and lowers the instruction count from 68.2 to 56.2 million without spilling.
In contrast, $16\times 8\times 8$ reaches the 255-register limit, spills (13.66 million local-load sectors), and executes more instructions than $16\times 8\times 4$.

The shape is a parameter of the CUTLASS paths only.
\texttt{FUSED\_TC} issues \texttt{mma.sync.m8n8k4} from hand-written parallel thread execution (PTX) assembly.
The fragment layout of this PTX is shared with \texttt{FUSED\_DFMA}, the iso-schedule ablation of Section~\ref{sec:fused-tensor-core-path} of the main text.
Therefore, a wider shape would be a different kernel and would break that pair.
On GB200, where the kernel was developed, the assembler expands the wider shapes into the same \texttt{DMMA.8x8x4}.
A trial \texttt{m16n8k8} variant at $p = 7$ cost 24.4\% more time because of register pressure.
No variant that is native to H100 was built.

\subsection{Trading occupancy for register budget}
\label{app:trading-occupancy-register-budget}
At $p = 255$, neither a deeper Tensor Core chunk ($BK = 32$, reduction entries per chunk) nor a fused $x{+}y$ kernel fits the production register budget.
Both need a residency of two blocks per SM instead of three.
Table~\ref{tab:a-occupancy-budget} gives the cost of that budget on its own.

\begin{apptable}
\captionof{table}{Cost of the additional register budget at $p = 255$ on GB200. CUDA device time per Runge--Kutta stage for the production configuration and for the three variants that buy a larger budget. Each variant is labeled with its chunk depth, its resident blocks per SM, and its register count. The relative column is against the production configuration. The per-stage values are the CUDA-event sum divided by the $N + 1$ calls that it covers (Section~\ref{sup:delayed-cuda-event-harvesting}).}
\label{tab:a-occupancy-budget}
\noindent\begin{tabularx}{\tblwidth}{@{}LRR@{}}
\toprule
variant & microseconds per stage & relative \\
\midrule
production ($BK = 16$, 3 blocks/SM, 168 registers) & 868.4 & --- \\
2 blocks/SM alone (208 registers) & 953.4 & +9.8\% \\
$BK = 32$ with 2 blocks/SM (255 registers) & 963.9 & +11.0\% \\
$BK = 32$ with 3 blocks/SM (168 registers, 100 B spill) & 1140.6 & +31.4\% \\
\bottomrule
\end{tabularx}
\end{apptable}

A doubled $BK$ halves the chunk count, but the barrier-stall count falls only about 5\% and the instruction count rises 2.7\%.
Therefore, barrier synchronization is not the binding resource.
The same budget cost rejects the $x{+}y$ fusion without a full implementation.
The removed \texttt{dqdt} store and reload would save 14.6 $\mu$s per stage (1.7\%).
However, the residency that the fusion needs costs 85.0 $\mu$s.

The trade-off differs at $p = 127$.
At this order, a halved thread count in the $xz$ kernel gives 128 instead of 64 registers per thread.
The budget is spent on a wider warp tile together with a double-buffered chunk loop.
The pair is 4.3\% faster, but its parts are worth only 0.6 and 0.3\% alone.
The reason is that the prefetch of the double buffer covers a global-memory wait (the long-scoreboard stall falls from 11.37 to 2.91 warps per issue).
The best chunk depth depends on that prefetch.
Without a double buffer, a deeper chunk wins.
With a double buffer, depth 16 is 4.3\% faster than depth 64.
The budget matters only when the concurrency that it costs is exchanged for something that hides a latency on the critical path.

\subsection{Software prefetching into shared memory}
\label{app:software-prefetching-into-shared}
Software prefetching with \texttt{cp.async} made every measured configuration slower.
The slowdown was +6.9\% for the epilogue \texttt{Escale} data of the $p = 63$ $xz$ kernel, +2.3\% and +1.3\% for two variants of the $p = 127$ $xz$ kernel, and +1.1\% for face-side values at $p = 255$.
Profiling shows one mechanism.
The prefetch traffic competes for the combined L1 cache and texture-unit (L1/TEX) path with the panel staging of the GEMM mainloop.
In addition, variants that need an extra barrier lose overlap between the loads and the surrounding computation.

\subsection{An order-dependent optimization}
\label{app:order-dependent-optimization}
The addition of \texttt{\_\_restrict\_\_} qualifiers to the kernel arguments improves performance by 0.5\% at $p = 7$ and by 1.2\% at $p = 127$.
However, the addition produces no measurable improvement at $p = 255$.
At the latter order, the intended change in code generation is present (\texttt{LDG.E.64.CONSTANT}), with unchanged register usage and no spilling.
Nevertheless, twelve interleaved rounds give 1869.9 versus 1868.1 $\mu$s per stage, with overlapping timing ranges.
The reason is that at $p = 255$ the affected loads are no longer on the critical path.

\subsection{Bulk asynchronous copy with the Tensor Memory Accelerator (TMA)}
\label{app:bulk-asynchronous-copy-tma}
No reported path uses the Tensor Memory Accelerator.
Two applications were evaluated and rejected on measurements.
With the \texttt{sm\_100} target that we use, CuTe TMA support needs \texttt{sm\_100a} or \texttt{sm\_100f}.
TMA adds no bandwidth where bandwidth is the limit.
For the two transfer shapes that this operator generates, TMA and ordinary loads agree within 1.8\% in a twenty-run microbenchmark.
TMA removes the issue cost.
L1 load/store unit (LSU) wavefronts fall by 89\%, global load sectors and shared-store bank conflicts fall to zero, and registers per thread fall from 30 to 16.
The DRAM bytes are unchanged.
Therefore, TMA can help only where L1/TEX throughput, registers, or issue slots bind.
None of these resources binds in the $p = 255$ $z$-GEMM epilogue.
In that epilogue, the long-scoreboard stall is 11.3\% of the kernel against 36.6\% for the fixed-latency execution dependency.
The occupancy of the epilogue is fixed by shared memory and not by registers.
In the $p = 7$ fused Tensor Core kernel (91.9\% of peak L1/TEX throughput), all four TMA swizzles for 8-byte elements make the $x$ and $y$ contractions two-way bank conflicted.
The reason is that the swizzle folds node bit 4 onto address bit 1, where node bit 1 already lies.
As a result, the staged variant has a net +8 wavefronts per warp, and the L1 traffic that the transformation exists to reduce increases.

\subsection{Reductions that do not save time}
\label{app:reductions-do-not-pay}
Three variants achieve a large measured reduction and are nevertheless slower.
These variants show that a saving helps only where the saved resource binds.
They also show that at low occupancy the removed work may have been filling a latency.
Staging the M-side face gather of the $p = 7$ fused Tensor Core kernel in shared memory cuts global load sectors by 26\%.
However, the version without bank conflicts is 5.5\% slower.
The reason is that ALU-pipe instructions rise 52\% in a kernel where integer work is already 43\% of thread instructions at 96.7\% occupancy.
An algebraic simplification of the shared-memory swizzle removes 13\% of the native GPU assembly (SASS) in the $p = 63$ fused kernels with a bit-identical output.
Nevertheless, the simplified version is 2.4\% slower.
In contrast, the same change is 2.0\% faster at $p = 127$.
The kernel at that order runs at about 50\% occupancy.
The $p = 63$ $xz$ kernel has only 24\% occupancy, and there the removed integer work had filled gaps in the MMA dependency chain.
Moving the staging address computation out of the chunk loop of the $p = 255$ fused Tensor Core kernel removes 34\% of the loop-body instructions and adds 2.5\% to the stage time.
The reason is that the instruction count does not bind there (Section~\ref{app:trading-occupancy-register-budget}).

\section{Measurement provenance}
\label{sup:provenance}

This section maps every measured figure and table of the main text and of this document to the scheduler job, the compute node, and the source revision from which it was taken.
The main text describes the campaigns and their time order.
The identifiers are kept here so that the main text does not have to carry them.

Job numbers are Slurm on GB200 (RIKYU), Grid Engine (SGE) on H100 (TSUBAME 4.0), and Fujitsu TCS on A100 (Wisteria/BDEC-01 Aquarius).
Commits are abbreviated hashes in the benchmark repository, the public fork of the SCALE-DG kernel extraction on branch \texttt{feature/cuda} (\url{https://github.com/seiya/scale-dg-kernel-extraction}).
The history of that branch was published without rewriting, so every hash below resolves there.
Paths under \texttt{figs/data/} refer to the archived data set \citep{thiswork-data}.

\subsection{Floats of the main text}
\label{sup:floats-main}

Floats whose content is derived and not measured directly (Figures~\ref{fig:f5} and~\ref{fig:f7}, Tables~\ref{tab:seven-ratios}, \ref{tab:low-order-rankings}, and~\ref{tab:fastest-path}) are listed against the campaign from which their inputs come.
The files \texttt{figs/data/sn\_cells.csv} and \texttt{figs/data/logo\_path\_totals.csv} hold the per-cell model output itself.

\begin{widetable}
\captionof{table}{Scheduler jobs and source revisions behind the figures and tables of the main text. The numbers are job numbers, and the commits are abbreviated hashes in the benchmark repository.}
\label{tab:a-provenance-main}
\begin{xltabular}{\tblwidth}{@{}>{\raggedright\arraybackslash}p{0.27\tblwidth}X@{}}
\toprule
Float & Jobs and commits \\
\midrule
\endhead
\bottomrule
\endlastfoot
Figure~\ref{fig:f1} & (a) GB200 123614, H100 8708077, A100 9652100; (b) GB200 109215, H100 8586793 \\
Figure~\ref{fig:f2} & GB200 123614 (\texttt{3d6ce77}) plus the $p = 767$ fused-path points from 105229 (\texttt{22e7297}); H100 8708077 (\texttt{1cfb57e}) plus the $p = 767$ fused-path points from 8650257 (\texttt{8a8975c}); A100 9652100 (\texttt{1cfb57e}) \\
Figure~\ref{fig:f3} & GB200 123614, H100 8708077, A100 9652100 \\
Figure~\ref{fig:f4} & GB200 only, commit \texttt{cec10a6}, one frozen binary (SHA256 \texttt{e0514e20...6d02}): fusion-saving traces and effective-rate timings 109874 (compute node c190); four-stage, no-overlap fold-only A/B 110017 (compute node c055) \\
Figures~\ref{fig:f5} and~\ref{fig:f7}, Tables~\ref{tab:seven-ratios}, \ref{tab:low-order-rankings}, and~\ref{tab:fastest-path} & the instruction-pricing campaign (below): GB200 92507, 92904, and 96222; A100 9568133, 9569112, and 9579761; H100 8621140, 8621935, 8621936, and 8631948 \\
Figure~\ref{fig:f6}, Table~\ref{tab:balance-knob} & the balance campaign (below): GB200 96451 and A100 9580356; SM counts GB200 96586 and A100 9580732; supply-matched point A100 9624726, repeated by 9624364 and 9624517 (9580806 replaced), with the GB200 side repeated as 113977; wave control GB200 96631 \\
Table~\ref{tab:holdout-h100} & GB200 92507, A100 9568133, H100 8621140 \\
Table~\ref{tab:floor-ablation} & the instruction-pricing campaign, as Figure~\ref{fig:f5}; no new measurement, the same inputs priced without the four-cycle floor (\texttt{figs/data/logo\_path\_totals\_nofloor.csv}) \\
Figure~\ref{fig:a11a} & (a) GB200 123614, A100 9652100; (b) GB200 82944, A100 9523596 \\
Figure~\ref{fig:a11b} & (a) kernel decomposition, GB200 83053 and A100 9523829; (b) measured ratios, GB200 83462 and A100 9527108 \\
\end{xltabular}
\end{widetable}

\subsection{Floats of the appendices and of this document}

Table~\ref{tab:cublas} (cuBLAS kernel names) holds no timing and is not listed.
Table~\ref{tab:balance-knob} is listed with Figure~\ref{fig:f6} above, because both come from the balance campaign.

\begin{widetable}
\captionof{table}{Scheduler jobs and source revisions behind the figures and tables of the appendices of the main text and of this document. The entries follow the conventions of Table~\ref{tab:a-provenance-main}.}
\label{tab:a-provenance-sup}
\begin{xltabular}{\tblwidth}{@{}>{\raggedright\arraybackslash}p{0.27\tblwidth}X@{}}
\toprule
Float & Jobs and commits \\
\midrule
\endhead
\bottomrule
\endlastfoot
Table~\ref{tab:nesweep} & GB200 83260 and 83261, compute node c359, commit \texttt{baf69b2} \\
Table~\ref{tab:mma} & GB200 81659 and H100 8577233, both at the default locked clock of the profiler \\
Table~\ref{tab:reversal-map} & GB200 92904 and 96222, A100 9569112 and 9579761, H100 8621935, 8621936, and 8631948 \\
Table~\ref{tab:fastest-path-preferred} & \texttt{figs/data/logo\_path\_totals\_16x8x4.csv}; the shape-independent paths as Table~\ref{tab:fastest-path}; the $16\times 8\times 4$ totals of \texttt{GEMM\_CUTE} and \texttt{GEMM\_FUSED} on H100 from SGE 8621936 (eight orders), 8621412 ($p = 511$ and 575), and 8648736 ($p = 767$, all four paths of that cell); predictions calibrated on GB200 and A100 at $8\times 8\times 4$ \\
Table~\ref{tab:a-timing-gb200} & Slurm 123614, compute node c286, commit \texttt{3d6ce77}; raw log \texttt{figs/data/slurm-123614.out} \\
Table~\ref{tab:a-timing-h100} & SGE 8708077, compute node r17n2, commit \texttt{1cfb57e}; raw log \texttt{figs/data/dg\_h100\_paths.o8708077} \\
Table~\ref{tab:a-timing-a100} & TCS 9652100, compute node wa35, commit \texttt{1cfb57e}; raw log \texttt{figs/data/dg\_a100\_paths.9652100.out} \\
Table~\ref{tab:a-event-harvesting} & frozen GB200 executable of commit \texttt{e113535} \\
Clock-lock measurements (Section~\ref{sup:profiler-measurements-effect-clock}) & Slurm 84347 (compute node c398) and 84359 (compute node c184) on GB200, frozen executables, no source change \\
Compiler versus library separation (Section~\ref{sup:separating-compiler-code-generation}) & TCS 9524314, 9524414, and 9524473 on A100 \\
Table~\ref{tab:a-validation-coverage}, orders above $p = 255$ & SGE 8611479 on H100 ($p = 511$, 575, and 767, both MMA shapes); TCS 9555767 on A100 ($p = 511$) and TCS 9660271 on A100 ($p = 575$, the frozen executable of TCS 9652100); timing disabled \\
Table~\ref{tab:a-attained-fraction} & one campaign with \texttt{ncu --clock-control none}: GB200 Slurm 83214, H100 SGE 8585547 (compute node r18n9), A100 TCS 9525913 \\
Table~\ref{tab:a-ratio-by-shape} & the campaign above, and the later H100 job SGE 8602846 (compute node r13n6) that supplies the $8\times 8\times 4$ rows of H100, the production-clock row, and the second $16\times 8\times 4$ entry \\
Table~\ref{tab:a-cublas-vs-cute} & SGE 8611480 and 8611482 (one per shape), compute node r7n8, \texttt{ncu --clock-control none}; the separate $8\times 8\times 4$ \texttt{GEMM\_CUTE} profile of Section~\ref{sup:roofline-placement-using-measured} is SGE 8602523, compute node r21n9, commit \texttt{9d1dfd9}, \texttt{ncu --clock-control none} \\
Table~\ref{tab:binding}, Figure~\ref{fig:s8} & GB200 Slurm 83141, 83209, 83214, and 115157 ($p = 1023$, compute node c272, commit \texttt{7004e82}); H100 SGE 8585090 and 8585547; A100 TCS 9525666 and 9525913 \\
Table~\ref{tab:kernels} & Slurm 83053, compute node c310, 2 rounds, production clocks \\
Table~\ref{tab:variation} & compute node to compute node: GB200 Slurm 78678 (compute node c186, tree \texttt{232dd27}), 83431 (compute node c397), and 83261 (compute node c359, commit \texttt{baf69b2}); compiler version: the three A100 jobs of the compiler-versus-library separation above; paired within-job rounds: SGE 8586793 (compute node r14n2, commit \texttt{7a95be3}), the H100 side of Figure~\ref{fig:f1}(b) \\
Tables~\ref{tab:eta-strata} and~\ref{tab:validity-band-selection} & derived from the instruction-pricing campaign (below); no additional GPU run. Table~\ref{tab:validity-band-selection} is recomputed from the three calibration-side columns of \texttt{figs/data/logo\_path\_totals.csv} \\
Table~\ref{tab:a-device-stage-ratio} & derived from the raw logs of Tables~\ref{tab:a-timing-gb200}, \ref{tab:a-timing-h100}, and~\ref{tab:a-timing-a100} \\
Table~\ref{tab:a-validation-field} & the validation part of the base sweep jobs: GB200 Slurm 123614, H100 SGE 8708077, A100 TCS 9652100 \\
Table~\ref{tab:a-traffic-ratio} & the same \texttt{ncu --clock-control none} campaign as Table~\ref{tab:binding} and Figure~\ref{fig:s8}: GB200 Slurm 83141 (compute node c116, \texttt{357e2f8}), 83209 and 83214 (compute node c282, \texttt{a445aa3}), 115157 ($p = 1023$) \\
Table~\ref{tab:a-threshold-sweep} & re-aggregation of the Table~\ref{tab:kernels} profile (Slurm 83053, compute node c310, commit \texttt{e241570}); the measured ratios against which it is scored are GB200 83462 and A100 TCS 9527108 and 9555962 (compute node wa34) \\
Table~\ref{tab:a-mma-shape-h100} & SGE 8577787, compute node r9n1, commit \texttt{55bf94c}; the $p = 255$ register and instruction counts are the Nsight Compute part of the same job \\
Table~\ref{tab:a-wider-z-tile} & timing A/B Slurm 82569 and Nsight Compute 82586, both compute node c389; the $64\times 64$ variant was applied to the working tree at \texttt{d15a18c} and reverted without being committed, and the rejection is recorded in \texttt{51f6f96} \\
Table~\ref{tab:a-occupancy-budget} & A/B Slurm 78165, Nsight Compute 78172, and the fused $x{+}y$ ablation 78175, all compute node c186, working tree on \texttt{5fb1b8a}; the $p = 127$ double-buffer prefetch figures of the same subsection are Slurm 75886 and 75888 (compute node c184) and 75895 (compute node c188), working tree on \texttt{a060741} \\
Table~\ref{tab:a-closure} & \texttt{figs/data/subpipe\_closure\_summary.csv}, derived from \texttt{figs/data/subpipe\_inputs.csv}, that is, from the instruction-pricing campaign jobs listed below \\
Table~\ref{tab:a-counter-portability} & \texttt{figs/audit\_counter\_numerators.py} applied to \texttt{figs/data/subpipe\_inputs.csv}; no additional GPU run \\
Table~\ref{tab:a-measured-cost} & Slurm 78846, compute node c186, one job of twelve interleaved rounds, tree \texttt{c74e682} (source identical to \texttt{232dd27}) \\
Table~\ref{tab:a-spill-residency} & spill executions and stall reasons: Slurm 78612, compute node c186, commit \texttt{63803da}; the measured-change column is from 78846 \\
DFMA thread-shape retuning (28.3, 7.9, and 26.5\% at $p = 31$, 63, and 127; Section~\ref{sup:register-budget-cost-iso} and Section~\ref{sec:one-axis-evidence}) & Slurm 79899 (compute node c086, tree \texttt{07f983d}), 79344 (compute node c387, tree \texttt{8421463}), and 79799 (compute node c079, tree \texttt{1c81d60}); the nine slower \texttt{FUSED\_TC} variants are 79344 and 79350 \\
Table~\ref{tab:a-emulation-accuracy} & $p = 7$ rows: run on the GB200 login node without a scheduler job, tree \texttt{44b02a6}; the $p = 255$ and $p = 511$ extension in the text: Slurm 88634, compute node c181, commit \texttt{e19c834} \\
Table~\ref{tab:a-scheme-one-cost} & Slurm 83482, compute node c164, exclusive, tree \texttt{e113535} (log header \texttt{7a95be3}, same source) \\
Table~\ref{tab:a-vendor-emulation} & Slurm 84461, compute node c191, exclusive, tree \texttt{663df57}; the adaptive-mode $104.3\times$, the $p = 1023$ point, and the H100/A100 attempts (SGE 8599546, TCS 9545645, neither usable) are Slurm 86678, compute node c384, commit \texttt{094aabe}; the $p = 767$ bit-identity check is Slurm 88634 \\
\end{xltabular}
\end{widetable}

\subsection{Campaigns}
\label{sup:campaigns}

\subsubsection{Base three-machine path sweep}

The campaign behind Figures~\ref{fig:f1}(a), \ref{fig:f2}, \ref{fig:f3}, and~\ref{fig:a11a}(a) and the absolute-timing tables: GB200 Slurm 123614 (compute node c286, commit \texttt{3d6ce77}), H100 SGE 8708077 (compute node r17n2, commit \texttt{1cfb57e}), A100 TCS 9652100 (compute node wa35, commit \texttt{1cfb57e}).
The two commits differ in job scripts and reports only, so the three machines were measured from one benchmark source.
The executables report the number of tendency calls that the CUDA-event sum covers (Section~\ref{sup:delayed-cuda-event-harvesting}).

\subsubsection{Added $p = 767$ fused-path points (Figure~\ref{fig:f2})}

\begin{itemize}
\item GB200: Slurm 105229, compute node c395, commit \texttt{22e7297}, six interleaved rounds of five paths; logs, metadata, and generated configurations under \texttt{figs/data/p767\_fused\_gb200/}.
\item H100: SGE 8650257, compute node r3n5, commit \texttt{8a8975c}; two interleaved rounds in each of two profiler metric groups; values in \texttt{figs/data/h100\_p767\_fused\_8650257.csv}.
\end{itemize}

The source audit between \texttt{22e7297} and \texttt{3d6ce77} is reported in Section~\ref{sup:data-provenance-reproducibility}.
The only source files that differ are the three that hold the CUDA-event timer.
No kernel file is changed.

\subsubsection{Fusion axes (Figure~\ref{fig:f4})}

Commit \texttt{cec10a6}, one frozen executable (SHA256 \texttt{e0514e20...6d02}).
Slurm 109874 (compute node c190) supplies the six \texttt{volume\_flux} traces and all fourteen path-timing cells.
Slurm 110017 (compute node c055) reuses the identical executable for the fold-only axis.
\texttt{figs/data/fig4a\_uniform.csv} and \texttt{figs/data/fig4b\_uniform.csv} retain the unrounded measurements, the round ranges, and the job and binary provenance.
Both CSVs were regenerated from the same raw logs with upstream \texttt{analysis/summarize\_fig4\_uniform.py} at commit \texttt{1e8e9b2}.
This script divides the CUDA-event sum of each run by the number of tendency calls that it covers (\texttt{device\_calls\_summed}).
For the executables of this paper, this number is $3 \times$ steps + 1, because the delayed event read-back adds the last warm-up call to the sum.

\subsubsection{The CUDA-event sum covers one call more than the timed stages (Section~\ref{sup:delayed-cuda-event-harvesting})}

\begin{itemize}
\item Mechanism: \texttt{mod\_advect3d\_eq.f90} (\texttt{advect3d\_eq\_reset\_timers}, \texttt{setup\_advect3d\_eq\_finalize}) and \texttt{mod\_cuda\_dg\_kernels.cuf} (\texttt{dg\_ev\_harvest}) at every measurement commit from \texttt{e910708} to \texttt{cec10a6}.
The mechanism was corrected upstream in \texttt{1e8e9b2}.
This commit drops the pending measurement at the reset and prints \texttt{Device-time calls summed:}.
\item Direct check (the 1.0150 against 70/69 in Section~\ref{sup:delayed-cuda-event-harvesting}): the $p = 7$ \texttt{GEMM\_FUSED} timing configuration of Slurm 109874 was run three times with each of two executables, interleaved.
The two executables are the frozen \texttt{cec10a6} executable and the \texttt{1e8e9b2} executable, on a login-node GB200 (no Slurm job).
The sums are 0.102070 / 0.102030 / 0.102045 s against 0.100536 / 0.100536 / 0.100557 s.
Recorded in upstream \texttt{reports/measurement\_inventory.md} section 30.2.
\item The $p = 1023$ two-interval test (Table~\ref{tab:a-event-harvesting}) is the measurement listed under that table above; only its interpretation changed.
\end{itemize}

\subsubsection{Profiling of $p = 767$ and $p = 1023$}

\begin{itemize}
\item Nsight Compute passes at $p = 767$: Slurm 94874 on GB200, SGE 8630131 on H100.
These supply the $p = 767$ rows of \texttt{figs/data/logo\_path\_totals.csv} and nothing else.
\item Nsight Compute pass at $p = 1023$ (Figure~\ref{fig:s8}): Slurm 115157, compute node c272, commit \texttt{7004e82}, same counters and window as Slurm 83141.
Log: \texttt{figs/data/slurm-115157.out}.
\end{itemize}

\subsubsection{Instruction-pricing campaign}

Nsight Compute jobs run from one script body, clock lock disabled.

\begin{itemize}
\item GB200: Slurm 92507 ($p = 63$, 255, 511, compute node c393); 92628 (MMA-shape A/B, compute node c286); 92904 (all orders, LSU-wavefront counters added); 93312 (additional stall counters).
\item A100: TCS 9568133 ($p = 63$, 255, 511) and 9569112 (all orders), compute node wa35.
\item H100: SGE 8621140 ($p = 63$, 255, 511, compute node r18n2); 8621412 and 8621936 ($16\times 8\times 4$ shape); 8621935 (all orders at $8\times 8\times 4$); 8648736 ($p = 767$ at $16\times 8\times 4$, \texttt{FUSED\_TC}, \texttt{GEMM\_FUSED}, \texttt{GEMM}, and \texttt{GEMM\_CUTE}).
The last job supplies the preferred-shape cells of Appendix~\ref{app:crossover-map} of the main text and of Table~\ref{tab:fastest-path-preferred} at that order.
The cycles per \texttt{DMMA} instruction that Section~\ref{sec:instruction-priced-subpipe} quotes are computed as (\texttt{sm\_\_pipe\_shared\_cycles\_active} in SM cycles, less the price of Equation~\ref{eq:subpipe-prices} per warp instruction of \texttt{sm\_\_inst\_executed\_pipe\_fp64}) divided by \texttt{sm\_\_inst\_executed\_pipe\_tensor}, both per SM.
The elapsed SM cycles are \texttt{gpu\_\_time\_duration} times the \texttt{sm\_\_cycles\_elapsed.avg.per\_second} that the same launch records.
The $8\times 8\times 4$ values on the three machines and the cuBLAS $16\times 8\times 8$ values on H100 are the $p \ge 511$ rows of \texttt{figs/data/subpipe\_closure.csv} (Section~\ref{sup:closure}).
On H100, these rows come from 8621935, 8630116, 8630131, and 8631255.
The $16\times 8\times 4$ values are from 8621412 at $p = 511$ and from 8621936 over eight orders, per launch.
The recorded clock matters.
The H100 launches that issue the wide shapes run at 1858 to 1969 MHz at $p \ge 511$ for $16\times 8\times 8$, at 1846 to 1955 MHz at $p = 511$ for $16\times 8\times 4$, and at 1817 to 1990 MHz over the eight orders.
The boost clock is 1980 MHz.
At the nominal clock, the same counters would read 4.1 to 4.4 and 8.1 to 8.6 cycles.
\item The paths that the first order sweeps did not cover, \texttt{FUSED\_DFMA}, \texttt{GEMM\_CUTE}, and the Ozaki path, at all orders: Slurm 94875 (GB200), TCS 9575505 (A100), SGE 8630116 (H100).
\texttt{GEMM\_CUTE} at $p = 767$: Slurm 106536 (GB200), SGE 8631255 (H100).
\item Re-taken library paths at $p \le 31$, after the low-order flux kernel was added to the profiler kernel filter: Slurm 96222 (GB200), TCS 9579761 (A100), SGE 8631948 (H100).
These replace the same cells of the sweeps above.
\item The per-kernel counters that the model reads from these jobs are \texttt{figs/data/subpipe\_inputs.csv} (666 rows, one per machine, order, path, and kernel role).
For every kernel instance, the file holds the job, commit, kernel name, launch configuration, the requested Nsight Compute metrics under their metric names, and the derived demands, binding resource, and $\eta$.
\texttt{figs/data/subpipe\_inputs\_README.md} defines the columns and the aggregation.
The table reproduces the path totals of Table~\ref{tab:holdout-h100} and the inputs of \texttt{logo\_path\_totals.csv} and \texttt{sn\_cells.csv}.
\item The closure of the counter-defined FP64 budget per kernel role is \texttt{figs/data/subpipe\_closure.csv} (666 rows, same keys), and its summary is \texttt{figs/data/subpipe\_closure\_summary.csv}.
Both files are derived from \texttt{subpipe\_inputs.csv} alone by one script of the analysis repository (Section~\ref{sup:closure}).
\end{itemize}

\subsubsection{Run-to-run widths}

Three repetitions of the Nsight Compute pass within a single job, clock left to the driver: Slurm 95805 ($p = 63$, 511, 575) and 106648 ($p \le 31$) on GB200; SGE 8650143 ($p = 63$, 511, 575), 8650141 ($p \le 31$), and 8650255 ($p = 255$) on H100; TCS 9592751 ($p \le 31$), 9593022 ($p = 63$), and 9593023 ($p = 127$) on A100.
The earlier two-round H100 high-order job is SGE 8631255.

\subsubsection{Balance campaign (MPS)}

\begin{itemize}
\item Sweeps: Slurm 96451 on GB200 (compute node c144, commit \texttt{f3e7c12}, three rounds); TCS 9580356 on A100 (compute node wa34, commit \texttt{b5d5eff}, two rounds).
Logs under \texttt{figs/data/mps\_gb200\_96451/} and \texttt{figs/data/mps\_a100\_9580356/}.
\item Freeze of the model: \texttt{f3e7c12} carries the model form and the profiled inputs and is dated 30 minutes before the GB200 sweep began.
\texttt{b5d5eff} is a later commit that adds only the three sweep scripts and leaves the analysis directory identical.
This commit is dated two minutes before the A100 sweep began.
\item Repeat of the unrestricted A100 point: TCS 9631289 (compute node wa35, five rounds).
\item Repeat of the whole A100 sweep in one job: TCS 9631437, compute node wa35, commit \texttt{9ea83bd}, three rounds.
The A100 markers of Figure~\ref{fig:f6}(a)--(b) and the response-curve numbers of the text use the five runs of 9580356 and 9631437 together.
The per-cell ``clean'' medians (runs within 0.5\% of the fastest) are \texttt{figs/data/subpipe\_balance\_a100\_clean.csv}, a byte-identical copy of \texttt{analysis/subpipe\_balance\_a100\_clean.csv} at upstream \texttt{d2f58bb}.
The 190-cell ranking score (Figure~\ref{fig:f7}, \texttt{sn\_cells.csv}) uses 9580356 alone.
\item SM counts (\texttt{\%smid} readout): Slurm 96586 (GB200), TCS 9580732 (A100); log \texttt{figs/data/dg\_a100\_mps\_smid.9580732.out}.
\item Balance-matched point: TCS 9624726, compute node wa35, commit \texttt{725cc1c}, three rounds; logs under \texttt{figs/data/mps\_a100\_match\_9624726/}.
Repeats: TCS 9624364 (three rounds) and 9624517 (five rounds), compute node wa34, same commit.
TCS 9580806 (\texttt{figs/data/mps\_a100\_match\_9580806/}) timed only \texttt{GEMM\_FUSED} and \texttt{FUSED\_TC} at $p = 511$ and is replaced.
GB200 side of the pair: Slurm 113977, compute node c358, commit \texttt{96ea0db} (kernel sources identical to \texttt{725cc1c}), three rounds at 100\% with \texttt{\%smid} = 152 read in the same job.
The logs are under \texttt{figs/data/mps\_gb200\_match\_113977/}.
Against the sweep 96451, every step-loop path time is within 0.5\% (all faster by 0.05 to 0.52\%).
The ratio \texttt{GEMM\_FUSED/FUSED\_TC} is within 0.1\% (0.9504 against 0.9499 at $p = 511$).
Therefore, the text and Figure~\ref{fig:f6} keep the sweep as the GB200 side.
\item Wave control: Slurm 96631 (\texttt{figs/data/mps\_gb200\_waves\_96631/}); Slurm 96619 (\texttt{figs/data/mps\_gb200\_waves\_96619/}) is replaced.
\item Attempt to move balance with the clock instead: Slurm 94583.
\end{itemize}

\subsubsection{Other floats}

\begin{itemize}
\item Figure~\ref{fig:f1}(b): GB200 Slurm 109215 (\texttt{figs/data/slurm-109215-f1.out}), H100 SGE 8586793 (\texttt{figs/data/dg\_h100\_paths.o8586793}).
\item Figure~\ref{fig:a11a}(b): GB200 82944, A100 9523596.
\item Figure~\ref{fig:a11b}(a): kernel decomposition GB200 83053 and A100 9523829; Figure~\ref{fig:a11b}(b): measured ratios GB200 83462 and A100 9527108.
\item Table~\ref{tab:nesweep}: GB200 83260 and 83261, compute node c359, commit \texttt{baf69b2}.
\item Table~\ref{tab:mma}: GB200 81659 and H100 8577233, at the default locked clock of the profiler.
\end{itemize}

\subsection{Campaign checks and reproducibility}
\label{sup:data-provenance-reproducibility}
This section describes the measurement campaigns behind the figures and tables, and it reports the checks that relate them to each other.
The scheduler job, the compute node, the source revision, and the log path of every measurement, and the jobs that a later job replaces, are listed in Sections~\ref{sup:floats-main}--\ref{sup:campaigns}.
Those identifiers are not repeated here.

The balance sweeps of Section~\ref{sec:balance-controlling-variable} are one job of three rounds on GB200 and one job of two rounds on A100, each over six SM fractions, six paths, and three orders.
In the A100 sweep, the first round of the unrestricted $p = 63$ runs of \texttt{FUSED\_TC} and \texttt{FUSED} followed an idle GPU.
This round was 4.1 and 9.6\% slower than the second round.
A repeat of the whole A100 sweep on another compute node agrees with the first sweep to within 0.06\% in the median of its 102 cells.
However, 30 of its 306 runs are one-sidedly more than 1\% slower than the median of their cell.
The A100 markers of Figure~\ref{fig:f6}(a)--(b) and the response curves of Section~\ref{sec:predicting-balance-crossover} therefore use the five runs of the two jobs together.
The estimator of a cell is the median of the runs within 0.5\% of the fastest one (archived as \texttt{subpipe\_balance\_a100\_clean.csv}).
With this estimator, the $p = 511$ ratios move by less than 0.7\% from the first sweep alone.
The ranking score of Section~\ref{sec:signal-to-noise-rule} uses the device timer of the first sweep alone.
The revision at which the GB200 sweep ran carries the model form and the profiled inputs.
This revision was committed 30 minutes before that sweep began.
The profiled kernels of GB200 and A100 at $p = 63$, 255, and 511 are the inputs of this axis.
These kernels are unchanged from that revision to the revision at which the reported predictions are computed.
The later edits to the input table are H100 kernels at $p \le 31$.
The SM counts of Table~\ref{tab:balance-knob}, the supply-matched point and its repeats, the wave control, and the attempt to move balance with the clock are separate jobs that use the production timer only.
No Nsight Compute measurement was taken under MPS.
The GB200 side of the supply-matched pair was repeated on another compute node four days after the sweep.
Every path time agrees with the sweep to 0.5\%, and the ratio agrees to 0.1\%.
Therefore, the 1.6\% of Section~\ref{sec:same-reversal-one-machine} does not depend on which GB200 job is taken.
The two comma-separated files \texttt{sn\_cells.csv} and \texttt{logo\_path\_totals.csv} under \texttt{figs/data/} hold the per-cell model output behind Figures~\ref{fig:f7} and~\ref{fig:f5}.
Table~\ref{tab:fastest-path} is recomputed from \texttt{logo\_path\_totals.csv} and a third file, \texttt{sn\_cells\_all\_pairs.csv}.
The predicted and measured fastest paths and the loss come from the \texttt{pred} and \texttt{meas} columns of the rows with \texttt{ledger\_ok} equal to 1.
The smallest $S/E$ over the other paths of the cell comes from the machine-axis \texttt{sn\_led\_log} column of the third file.
The third file holds the same schema as \texttt{sn\_cells.csv} for all fifteen pairs of the six paths.
It is row-for-row identical to \texttt{sn\_cells.csv} on the seven scored ratios.
The machine-axis scores of Section~\ref{sec:signal-to-noise-rule} use the \texttt{sn\_led\_log} column, whose family error \texttt{famerr\_led\_log} is taken over the cells whose kernel lists match.
The balance-axis scores use \texttt{sn\_log}.
The production-measure scores of Section~\ref{sec:operational-predictions} (81 of 88 cells at $p \ge 63$) use the machine-axis columns \texttt{meas\_prod} and \texttt{hit\_prod}.
The column \texttt{meas\_prod} is the ratio of the production CUDA-device times per stage of the two paths.
These times are taken from the same three sweep logs as Figure~\ref{fig:f2}.
Each time is the median over the interleaved rounds of the CUDA-event sum divided by the number of stages.
The column \texttt{meas\_prod} is empty on the balance axis and wherever the sweep does not time both paths.
The preferred-shape totals of Section~\ref{sec:operational-predictions} and Table~\ref{tab:fastest-path-preferred} are in a fourth file of the same schema, \texttt{logo\_path\_totals\_16x8x4.csv}.
This file is the output of the same model with the target shape set to $16\times 8\times 4$.
The 20 H100 rows of this file, which are calibrated on GB200 and A100, give the 20 path totals and the twelve ranking cells of Section~\ref{sec:operational-predictions}.
Table~\ref{tab:fastest-path-preferred} takes the two shape-dependent paths from these rows and the shape-independent paths from \texttt{logo\_path\_totals.csv}.
The file also holds two GB200 rows at $p = 767$.
In these rows, the prediction equals the common-shape prediction, because the assembler expands the wider shape into the same instructions on GB200.
The measured value in these rows is that of a different job (Slurm 106536 in place of 94874 for \texttt{GEMM\_FUSED}) within the run-to-run width.
These two rows are calibrated on A100 and H100 and are used nowhere in this paper.
The two agree at the threshold of 1 in their counts.
A fifth file of the same schema, \texttt{logo\_path\_totals\_nofloor.csv}, holds the floor ablation of Table~\ref{tab:floor-ablation}.
This ablation uses the same model and inputs, but the four-cycle floor of Equation~\ref{eq:subpipe-prices} is removed.
The price of a \texttt{DMMA} is then $F_{\mathrm{DMMA}}/C_{\mathrm{tc},m}$ on every machine, and $\eta$ is recalibrated with that price.
The table is recomputed from the \texttt{err} and \texttt{pred} columns of that file and of \texttt{logo\_path\_totals.csv}.
The recomputed values are the medians over the $p \ge 511$ rows of each held-out machine, the three GB200 rows at $p = 767$, and the quotient of the H100 \texttt{GEMM} and \texttt{FUSED\_TC} rows at $p = 511$ and 767.
The last column of the same table comes from two further files of the same two schemas, \texttt{logo\_path\_totals\_maxform.csv} and \texttt{sn\_cells\_maxform.csv}.
The first file holds the same model and inputs, but the sum of Equation~\ref{eq:subpipe} is replaced by the larger of its two terms and $\eta$ is recalibrated.
The second file holds the ranking cells that this variant produces.
The counts of Section~\ref{sec:instruction-priced-subpipe} (56 changed path totals, one changed sign, 101 screened in, and 95 correct machine-axis cells) are recomputed by comparing these two files row by row with \texttt{logo\_path\_totals.csv} and \texttt{sn\_cells.csv}.
The columns \texttt{width}, \texttt{resolved}, and \texttt{width\_job} hold the run-to-run width of the measured ratio of each cell, in percent.
The width is the maximum minus the minimum over the rounds of the repeat job that covers both paths at that order, divided by the minimum.
The path totals are formed on the same kernel lists as the scored cell.
The column \texttt{resolved} is 1 when the measured difference from 1 exceeds the width.
The three columns are empty when no repeat job covers the cell.
The repeat jobs are GB200 106648, 95805, and 106952, H100 8650141, 8650255, 8650143, and 8650257 with 8650967 and 8652177, and A100 9592751, 9593022, 9593023, and 9592752.
On the balance axis, the width comes from the rounds of the MPS sweep itself.

The graph-enabled timings of Section~\ref{sec:timing-protocol} of the main text and Section~\ref{sup:graph-replay} are one job per machine.
The jobs are Slurm 124807 on GB200 (compute node c051, commit \texttt{ccc26f9}), SGE 8714147 on H100 (compute node r1n6) and TCS 9659039 on A100 (compute node wa34).
The last two jobs are at commit \texttt{9a7364c}.
The kernels are unchanged since \texttt{4cc3084}.
Each job runs six interleaved rounds over the six paths at $p = 7$ and 15 with the graph off and on.
The runs use the namelists of Table~\ref{tab:a-configs}, and only the graph option is changed.
On H100, the jobs also run the two CUTLASS paths at both MMA shapes.
All 480 runs completed, and the graph-on and graph-off solutions agree bit for bit except on \texttt{GEMM\_FUSED}, where they differ by at most $1.3 \times 10^{-8}$.
The per-run values are \texttt{figs/data/graph\_ab\_lowp.csv}.
The Nsight Systems version comparison of Section~\ref{sup:profiler-invocation} is Slurm 124790, 124792, 124793, and 124798 on GB200.
The graph-disabled \texttt{FUSED\_TC} trace at $p = 7$ of the same section is Slurm 45686.

Figure~\ref{fig:f2} combines the three base sweeps with the added GB200 and H100 fused-path points at $p = 767$.
These points were measured earlier at two other revisions from the same wrapper and the same committed namelist.
Between the revision of the added GB200 points and that of the base GB200 sweep, the only source files that differ are the three that hold the CUDA-event timer.
Both revisions use nvfortran 26.3 and CUDA 13.1.
The three library-path controls in the added-point jobs reproduce the base $p = 767$ values to within 0.1\% on GB200 and 0.22\% on H100 (\texttt{figs/data/h100\_p767\_fused\_8650257.csv}).
Therefore, the combination introduces no source or build change that could produce a performance difference at these points.

Figure~\ref{fig:f4} was measured on GB200 at one source revision with one frozen executable.
One job supplies the six \texttt{volume\_flux} traces and all fourteen path-timing cells.
A second job on another compute node reuses the identical executable for the fold-only axis.
In this second job, fold on and fold off are both fixed to a four-stage carrier, and overlap is disabled.
Each timing cell is the median of ten interleaved rounds, and all 120 runs returned zero.
The paired min--max ranges do not overlap at any order.
The unrounded values are in \texttt{fig4a\_uniform.csv} and \texttt{fig4b\_uniform.csv}.
Both plotted paths pass the point-varying full-field validation through $p = 255$, including a 134,217,728-point follow-up at that order with maximum absolute errors of $3.55\times10^{-15}$.

The Nsight Compute passes at $p = 767$ are one job on GB200 and one on H100, each over \texttt{FUSED\_TC}, \texttt{GEMM\_FUSED}, and \texttt{GEMM}.
They supply the $p = 767$ rows of \texttt{logo\_path\_totals.csv} and nothing else.
A100 has no pass because its device memory stops the sweep at $p = 575$.
The $p = 1023$ points of Figure~\ref{fig:s8} are a later GB200 pass with the same counters and window as the passes of Table~\ref{tab:binding}.

The run-to-run widths of Section~\ref{sup:sn-scoring} come from separate jobs.
Each of these jobs repeats the Nsight Compute pass over the paths of one machine three times within a single job at the common instruction shape.
The clock is left to the driver.
The jobs cover $p \le 31$, 63, 511, and 575 on GB200, the same orders and $p = 255$ on H100, and $p \le 31$, 63, and 127 on A100.
A width is the full spread of the ratio over the rounds of one job, relative to its smallest round.
The numerator and the denominator of a ratio are always rounds of the same job.
The balance axis needs no job of its own, because every cell of the sweeps above already carries a width.
This width is 0.1\% in the median over the 190 cells and 1.4\% at worst.

Tables~\ref{tab:a-timing-gb200}--\ref{tab:a-timing-a100} were generated from the same three raw logs as Figure~\ref{fig:f2}, with the parsing rules of \texttt{figs/f2\_landscape.py}, which is included in the archived data set \citep{thiswork-data}.
Only runs with return code zero enter the medians.
All 405 runs on GB200, 390 on H100, and 285 on A100 satisfy that condition.
Every run reports the number of tendency calls that its CUDA-event sum covers.
The maximum round-to-round median absolute deviation of the stage time is 0.2\% on GB200, 0.4\% on H100, and 0.2\% on A100.

The three campaigns share the job script, the namelists, the interleaved-round protocol, and the benchmark source.
An independent campaign on an earlier source revision, on another H100 compute node and from an independently built executable, reproduces the machine-dependent ordering in two low-order cases.
At $p = 7$, the CUDA Core path requires 0.7\% less step-loop time than the Tensor Core path.
At $p = 63$, the CUDA Core path requires 1.0\% less.
In both cases, the repeated-run ranges do not overlap.
At $p = 7$, the maximum CUDA Core time is 1042.2~$\mu$s per stage, and the minimum Tensor Core time is 1048.8.
These margins agree with the ten-round campaign of Section~\ref{sec:architecture-dependent-reversal} to within 0.1 percentage point.

A ratio between two machines therefore carries three sources of uncertainty of comparable size.
These sources are the compiler version (approximately 0.4\% in the median, Table~\ref{tab:variation}), the compute node (0.1 to 0.6\%), and the source revision (approximately 0.5\%).
This is why Section~\ref{sec:sources-variation-claim-threshold} does not interpret cross-machine differences below 1\%.

The clock-lock measurements of Section~\ref{sup:profiler-measurements-effect-clock} are two jobs on two GB200 compute nodes against frozen executables.
The six profiler-against-wall-clock comparisons are interleaved A/B runs on exclusively allocated GB200 compute nodes.
The compiler and library separation of Section~\ref{sup:separating-compiler-code-generation} is three jobs on A100.
The validation of the single-element orders above $p = 255$ on H100 and A100 in Table~\ref{tab:a-validation-coverage} comes from jobs in which timing was disabled.
These jobs ran at a source revision later than that of the timing tables.
The GB200 results at those orders were taken earlier, in the per-order studies.

The instruction-pricing campaign of Sections~\ref{sec:instruction-priced-subpipe}--\ref{sec:subpipe-model-negatives} is a separate set of Nsight Compute jobs that were run from one script body on the three machines.
The clock lock was disabled in these jobs, so the counters and the times come from the same runs.
At $p \le 31$, the library paths were re-taken after the low-order flux kernel had been added to the kernel filter of the profiler.
Those runs replace the same cells of the earlier jobs.
The H100 jobs are the comparison against which the held-out prediction is reported.
These jobs come after the model, and none of their timings enters the model.
However, this is a statement about how the model was constructed, and the job records cannot establish it on their own.

\subsection{The $p = 511$ ratio, by timer and campaign}
\label{sup:p511-ratio-provenance}
The ratio $\texttt{GEMM\_FUSED}/\texttt{FUSED\_TC}$ at $p = 511$ is quoted in the main text from several campaigns and under the three timers of Section~\ref{sec:timing-protocol}.
Therefore, the same ratio appears with slightly different values.
Table~\ref{tab:a-p511-ratio} collects every value that the main text uses, with its timer, MMA shape, and job.
Table~\ref{tab:two-shapes-p511} of the main text places the sweep values of the two comparisons side by side.
None of the values is a copy of another.
The A100 value 1.068 (dedicated job, device time) is close to the H100 sweep value 1.069 in the same timer by chance.
The A100 sweep gives 1.077, and a second dedicated A100 job gives 1.066.

\begin{apptable}
\captionof{table}{Every value of $\texttt{GEMM\_FUSED}/\texttt{FUSED\_TC}$ at $p = 511$ that the main text quotes, with its timer, MMA shape, and campaign. Device time is the CUDA-event sum per stage in the sweeps and dedicated jobs, and the sum of the profiled window per tendency call in the profiling campaign. Values in parentheses are not quoted in the main text and are listed for completeness.}
\label{tab:a-p511-ratio}
\noindent\begin{tabularx}{\tblwidth}{@{}>{\hsize=.6\hsize}L>{\hsize=.5\hsize}R>{\hsize=.7\hsize}L>{\hsize=.6\hsize}L>{\hsize=1.75\hsize}L>{\hsize=1.0\hsize}L@{}}
\toprule
Machine & Value & Timer & Shape & Campaign (job, compute node, commit) & Quoted in \\
\midrule
GB200 & 0.951, 0.950 & step loop, end-to-end & $8{\times}8{\times}4$ & base sweep, Slurm 123614, c286, \texttt{3d6ce77} & Figure~\ref{fig:f1}(a), Sections~\ref{sec:order-dependent-crossover}, \ref{sec:architecture-dependent-reversal} \\
GB200 & 0.948 & device & $8{\times}8{\times}4$ & dedicated three-timer job, Slurm 83462, c397, \texttt{3e95f7c}; the base sweep and the profiling window give the same value & Sections~\ref{sec:does-accounting-transfer-a100}, \ref{sec:logo-three-directions}, Section~\ref{sup:agreement-three-timing-measures} \\
H100 & 1.061 & step loop & $8{\times}8{\times}4$ & base sweep, SGE 8708077, r17n2, \texttt{1cfb57e} & Figure~\ref{fig:f1}(a), Section~\ref{sec:architecture-dependent-reversal} \\
H100 & 1.069 & device & $8{\times}8{\times}4$ & base sweep, SGE 8708077 & Sections~\ref{sec:does-accounting-transfer-a100}, \ref{sec:operational-predictions} \\
H100 & 1.097 & device (profiled window) & $8{\times}8{\times}4$ & profiling campaign, SGE 8621935, r22n3, \texttt{e19c834} & Table~\ref{tab:holdout-h100} \\
H100 & 0.870 & step loop & $16{\times}8{\times}4$ & base sweep, SGE 8708077 & Section~\ref{sec:architecture-dependent-reversal} \\
H100 & 0.854 & device & $16{\times}8{\times}4$ & base sweep, SGE 8708077 & Sections~\ref{sec:does-accounting-transfer-a100}, \ref{sec:operational-predictions} \\
H100 & 0.860 & device (profiled window) & $16{\times}8{\times}4$ & profiling campaign: \texttt{GEMM\_FUSED} from SGE 8621936 (r20n2) and \texttt{FUSED\_TC} from SGE 8621935 (r22n3), both \texttt{e19c834} & Table~\ref{tab:reversal-map} \\
H100 & (0.857) & device & $16{\times}8{\times}4$ & re-measurement on one source revision, SGE 8591427, r21n9, \texttt{65ada12} & --- \\
A100 & 1.069, 1.068 & step loop, end-to-end & $8{\times}8{\times}4$ & base sweep, TCS 9652100, wa35, \texttt{1cfb57e} & Figure~\ref{fig:f1}(a), Section~\ref{sec:architecture-dependent-reversal} \\
A100 & (1.077) & device & $8{\times}8{\times}4$ & base sweep, TCS 9652100 & --- \\
A100 & 1.068, 1.061, 1.061 & device, step loop, end-to-end & $8{\times}8{\times}4$ & dedicated three-timer job, eight rounds, TCS 9527108, wa35, \texttt{7a95be3} & Section~\ref{sec:does-accounting-transfer-a100}, Section~\ref{sup:agreement-three-timing-measures} \\
A100 & (1.066, 1.059, 1.059) & device, step loop, end-to-end & $8{\times}8{\times}4$ & second dedicated job, eight rounds, TCS 9555962, wa34, \texttt{e19c834} & --- \\
A100 & 1.069 & step loop & $8{\times}8{\times}4$ & balance sweep at 100\% of the SMs, TCS 9580356, wa34, \texttt{b5d5eff} & Section~\ref{sec:same-reversal-one-machine} \\
A100 & 1.079 & device (profiled window) & $8{\times}8{\times}4$ & profiling campaign, TCS 9569112, wa35, \texttt{fa6d434} & Table~\ref{tab:reversal-map} \\
\bottomrule
\end{tabularx}
\end{apptable}

On H100 at the common $8\times 8\times 4$ shape, the two device-time values 1.097 and 1.069 in Table~\ref{tab:a-p511-ratio} differ by 2.6\%.
The spread exceeds the 1\% threshold of Section~\ref{sec:sources-variation-claim-threshold}, so it is a real difference between the two measurements, which we do not explain.
Each comparison in the main text uses the value from its own measurement.
The sustained clocks of two H100 compute nodes differ by 3.8\% under the profiler (Section~\ref{sup:whether-attained-fraction-preserved}).
However, a clock difference of that size moves this ratio by less than 1\%, because only the \texttt{volume\_flux} share of \texttt{GEMM\_FUSED} is priced against DRAM.
Therefore, the clock difference does not account for the spread.

The profiled-window values that Table~\ref{tab:reversal-map} uses at $p = 511$ are 1.079 on A100 and 0.860 on H100.
They differ from the device-time values 1.077 and 0.854 in Table~\ref{tab:a-p511-ratio} by 0.2\% and 0.7\%.
No spread moves a cell across 1.

Section~\ref{sup:sensitivity-dram-saturation-threshold} also quotes a reconstruction of the same reversal from independent data.
This reconstruction uses a different input configuration, code revision, and timing scope, and it is aggregated to the whole tendency evaluation.
The reconstruction is given to two decimals (0.95 on GB200 and 1.07 on A100 against a measured A100 ratio of 1.09) and is not combined with Table~\ref{tab:a-threshold-sweep}.

\section{Order dependence of the full-fusion saving and the contraction-kernel gap}
\label{sup:low-order-fusion-axis}

\subsection{Attribution at the two orders on either side of the crossover}
The decomposition of Section~\ref{sec:why-optimum-moves-order} records the individual contractions of both paths, with the kernel structure of Section~\ref{sec:operator-implementation-paths}.
The face-flux kernel runs on a side stream in both paths, and it is outside the critical path.
At $p = 255$, the removal of \texttt{volume\_flux} saves 128.0 $\mu$s.
However, the three hand-written contractions require 78.6 $\mu$s more than the three library GEMMs.
The net contribution of these two terms is therefore $-$49.4 $\mu$s.
Relative to the 924.1 $\mu$s of CUDA-device time per stage of \texttt{GEMM\_FUSED}, these terms are a 13.9\% full-fusion saving, an 8.5\% contraction-kernel gap, and an attributed difference of $-$5.3\%.
The measured device-time difference is $-$6.4\%.
The remaining 1.1 points are the launch gaps on the main stream and the fork and join with the side stream.
These costs lie outside the counted kernels.
They are 36.3 $\mu$s in \texttt{GEMM\_FUSED} and 25.9 $\mu$s in \texttt{FUSED\_TC}.
The 10.4 $\mu$s difference is 1.1\% of the same denominator.
At $p = 511$, the full-fusion saving is 1.04 ms, the hand-written contractions require 1.71 ms more than the library GEMMs, and the net contribution is +0.67 ms.
Relative to the 12.18 ms of \texttt{GEMM\_FUSED} device time per stage, the contributions are 8.5\%, 14.0\%, and +5.5\%, against a measured difference of +5.5\%.
The shares of \texttt{volume\_flux} on the production \texttt{GEMM\_FUSED} path behind Figure~\ref{fig:f4}(a) are 12.6\% at $N_q = 256$ and 8.1\% at $N_q = 512$.
These shares agree within 1.3 and 0.4 percentage points, respectively, with this decomposition.
This decomposition records both paths and normalizes by CUDA-device time and not by the stage period.
The arithmetic lower bounds behind the attained fractions are 214 $\mu$s per contraction at $N_q = 256$ and 3.43 ms at $N_q = 512$.
Against the $2 N_q^4$ count and the 40.1 TFLOP/s roof, the three hand-written contraction kernels of \texttt{FUSED\_TC} reach an average of 75.2\% at $N_q = 256$ and 80.1\% at $N_q = 512$.
The three CUTLASS volume GEMMs of \texttt{GEMM\_FUSED} reach 83.1\% and 92.5\%, respectively.
Between the two orders, the library kernels therefore improve by 9.4 percentage points and the hand-written kernels by 4.9 points.

\subsection{The rising branch of the fusion axis at low order}
This subsection decomposes the rising branch of $\texttt{GEMM\_FUSED}/\texttt{GEMM\_CUTE}$ below $N_q = 128$ per kernel on GB200 in CUDA device time.
At these orders the carry does not absorb \texttt{volume\_flux}.
The unfused path computes volume and element-boundary flux in one kernel, and the fused path splits this kernel in two at 12--16 $\mu$s per stage.
The carry instead absorbs the lift and assembly kernel, which costs 163.9 to 154.9 $\mu$s per stage as $N_q$ grows by a factor of 16.
The extra cost of the epilogue carrier stands against this absorbed cost.
The extra cost falls from 180.8 to 34.8 $\mu$s per stage because the longer mainloop hides more of the epilogue.
Fusion loses at $N_q = 8$ because the extra cost exceeds the absorbed lift.
The plateau between $N_q = 32$ and $64$ is where the overlap advantage passes from the fused path to the unfused path.
The extra cost is measured across two different carrier kernels.
It is therefore the size of the effect at each order and not the scaling of one kernel.
The overlap contributions are inferred from kernel spans.

\subsection{The fold series: deviations from $1/N_q$ scaling and the concurrency price of fusion}
On GB200, both sides are fixed to a four-stage carrier, and overlap is disabled.
Under these conditions, the fold requires 32.4, 25.2, 14.1, 12.2, 9.5, and 5.6\% less CUDA-device time than no fold at $N_q = 128$, $256$, $512$, $576$, $768$, and $1024$.
Over the central intervals, the decrease is approximately consistent with $1/N_q$ scaling.
The two outer intervals depart from this scaling for opposite reasons.
Between $N_q = 128$ and 256 the meshes hold the volume points fixed.
Therefore the kernels with $O(1)$ work per point (34.6\% of the stage at $N_q = 128$) do not grow, and the stage cost does not scale with volume points times $N_q$.
Between $N_q = 768$ and 1024 the carrier moves the same 116.5 B per point but hides more of it behind a longer mainloop.
Its added cost per point therefore falls by 47\%, and the no-fold cost does not scale with volume points.
Between $N_q = 256$ and 512 each assumption is violated by about 20\% in opposite directions.
The central agreement is therefore a cancellation and not independent evidence.

Fusion also removes concurrency.
The three unfused volume GEMMs are independent and can execute across streams.
In contrast, the carrier depends on the other derivatives.
For $N_q \ge 64$ the fold removes the remaining overlap as well.
Against the same unfolded baseline, the lost overlap is 0.9\% at $N_q = 128$, 0.9\% at $N_q = 256$, and nothing at $N_q \ge 512$.
The fold itself saves 5.6--32.4\%, so the removed concurrency is the smaller term.

\section{Element-count sweep at fixed order}
\label{sup:element-count-sweep}
This section holds the control of Section~\ref{sec:architecture-dependent-reversal} of the main text.
This control varies the element count at fixed order, time step, step count, and timing protocol.
At $p = 255$ with $N_e = 1$, $2^3$, and $3^3$, the number of volume points rises from 16,777,216 to 452,984,832.
This is well beyond the $p = 511$, $N_e = 1$ case (Table~\ref{tab:nesweep}).
The ordering of the fused paths against the library paths is unchanged at every size.
\texttt{GEMM} and \texttt{GEMM\_CUTE} are the two sides of the library axis, and these sides are nearly identical at this order (Section~\ref{sec:one-axis-evidence}).
These two paths stay within 1.2\% of each other and exchange places at $N_e = 2^3$.
$\texttt{GEMM\_FUSED}/\texttt{FUSED\_TC}$ is 1.066, 1.058, and 1.058.
After a 27-times increase in problem size, \texttt{GEMM\_FUSED} therefore still requires approximately 5.8\% more device time than the fused kernel.
The five paths scale by 7.6 to 7.9 times for eight times the work and by 25.3 to 26.5 times for twenty-seven times the work.

At $p = 511$, $N_e = 1$ and $2^3$ take the volume points from 134,217,728 to 1,073,741,824, past the $p = 1023$, $N_e = 1$ case.
$N_e = 3^3$ would need $3.6 \times 10^{9}$ volume points, which do not fit in the 185 GiB of Table~\ref{tab:machines}.
$\texttt{GEMM\_FUSED}/\texttt{FUSED\_TC}$ is 0.948 and 0.946, a change of approximately 0.2\% for eight times the work.
All five paths scale by 8.0x to the precision that these runs resolve.
The ranking of all five paths is unchanged, and \texttt{GEMM} is ahead of \texttt{GEMM\_CUTE} by 0.6\% and 0.3\%.

\begin{table*}[tp]
\centering
\footnotesize
\caption{
  Element-count sweep at $p = 255$ on GB200, five paths in six interleaved rounds at each of the three sizes.
  CUDA device time per RK stage, and the ratio of each path to \texttt{FUSED\_TC} at the same element count.
  Only the element count changes between the columns.
  Polynomial order, time step, number of steps, CUDA-graph configuration, and timing protocol are held fixed.
  The three columns hold $256^3$, $512^3$, and $768^3$ volume points.
  The per-stage values are the CUDA-event sum divided by the $N + 1 = 58$ tendency calls that it covers (Section~\ref{sup:delayed-cuda-event-harvesting}).}
\label{tab:nesweep}
\begin{tabularx}{\tblwidth}{@{}lRRRRRR@{}}
\toprule
Path & $N_e = 1$ [$\mu$s] & $N_e = 2^3$ [$\mu$s] & $N_e = 3^3$ [$\mu$s] & ratio, $N_e = 1$ & ratio, $N_e = 2^3$ & ratio, $N_e = 3^3$ \\
\midrule
\texttt{FUSED\_TC} & 864 & 6 710 & 22 600 & 1 & 1 & 1 \\
\texttt{GEMM\_FUSED} & 921 & 7 100 & 23 900 & 1.066 & 1.058 & 1.058 \\
\texttt{GEMM\_CUTE} & 1 020 & 8 010 & 27 000 & 1.181 & 1.193 & 1.195 \\
\texttt{GEMM} & 1 030 & 7 960 & 27 200 & 1.194 & 1.186 & 1.203 \\
\texttt{FUSED} & 1 410 & 10 700 & 35 800 & 1.635 & 1.591 & 1.584 \\
\bottomrule
\end{tabularx}
\end{table*}

\section{Measurement details}
\label{sup:measurement-details}

This section reports the measurements that support the profiling rules of Section~\ref{sec:profiling-protocol} and the claim thresholds of Section~\ref{sec:sources-variation-claim-threshold} of the main text.

\subsection{Profiler measurements and the effect of the clock lock}
\label{sup:profiler-measurements-effect-clock}
Nsight Compute locks the SM clock while it profiles.
The locked clock is a different fraction of the boost clock on each machine: 1.094 GHz against 2.062 GHz on GB200, a factor of 0.53, and approximately 0.77 on A100.
Profiling with \texttt{--clock-control none} instead reproduces the production device time of the tendency kernels to within 3\%.
In contrast, the default clock policy makes the same kernels 1.82 to 1.84 times slower.

The lock does not rescale all reported quantities by a common factor.
Metrics of the form \texttt{pct\_of\_peak\_sustained\_elapsed} are normalized by an elapsed time that includes stall cycles.
A higher clock therefore lengthens the memory waits and can \emph{lower} an SM-side utilization.
For one kernel ($p = 15$, \texttt{FUSED} path), the L1/TEX throughput falls from 85.8\% at 1.08 GHz to 66.7\% at 2.04 GHz.
Across 35 profiled kernel shapes, the achieved fraction of the applicable roof falls in every case when the clock is unlocked, by between 0.1\% and 47.1\%.
The drop is larger the further the kernel operates from its roof (Spearman rank correlation +0.68).
A percentage that is measured under the clock lock therefore cannot be transferred to production clocks by any simple rule.
For the same reason, we do not compare such percentages across machines whose locked clocks are different fractions of the boost clock.

Section~\ref{sup:profiler-wallclock-bias} reports six comparisons between profiler-reported time and wall-clock time.
These comparisons show that the sign of the bias depends on which resource a change spends and which it saves.

\subsection{Separating compiler code generation from library version}
\label{sup:separating-compiler-code-generation}
\begin{widetable}
\captionof{table}{
  Independently measured sources of run-to-run, compute-node-to-compute-node, and compiler-version variation, and the basis for the two claim thresholds of Section~\ref{sec:sources-variation-claim-threshold}.
  The components are separate experiments and are not added into a single budget.}
\label{tab:variation}
\noindent\begin{tabularx}{\tblwidth}{@{}LLL@{}}
\toprule
Source & Magnitude & Measurement \\
\midrule
Compute node to compute node within one machine & 0.1--0.6\%, systematic in direction, path dependent & three compute nodes of one machine \\
Compiler version, 25.9 against 23.3 (2 years 6 months) & median approximately 0.4\%, worst 2.8--3.2\%, both on \texttt{FUSED\_DFMA} & three jobs on A100 \\
Paired within-job, round to round & median absolute deviation 0.01--0.03\%; 10 of 10 rounds clean in all six cells & one job, one compute node, one executable \\
\bottomrule
\end{tabularx}
\end{widetable}

The 0.4\% median compiler-version effect of Section~\ref{sec:sources-variation-claim-threshold} combines the code that the compiler generates with the versions of the CUDA runtime and math libraries against which the build links.
The two contributions were separated on A100.
The build with nvfortran 23.3 was run twice more while the compiler was held fixed, once against the libraries of its own toolkit and once against those of the 25.9 toolkit.
The library version alone changes the ratio by between $-$0.06 and +0.16 percentage points.
A null test between two nominally identical configurations spans $-$0.03 to +0.17\%.
The library contribution therefore lies inside the run-to-run noise.
The variation of Table~\ref{tab:variation} is therefore caused by compiler code generation.
In agreement with this, the largest compiler-version difference belongs to \texttt{FUSED\_DFMA}, a path that calls no library at all.

\subsection{SM-restriction measurements}
\label{sup:balance-knob-measurements}
Table~\ref{tab:balance-knob} lists the SM counts that the MPS restriction of Section~\ref{sec:mps-balance-knob} delivered on each machine, and the balance that results.
The SM counts are read from \texttt{\%smid} in the same job as the timings.
A restriction to a fraction $x$ of the SMs multiplies the time of a purely SM-limited path by $1/x$ and leaves a purely DRAM-limited path unchanged.
The gap below $1/x$ therefore measures the DRAM-bound share directly.
At $p = 511$ on GB200, \texttt{FUSED\_TC} takes 3.91 times longer at a quarter of the SMs against the 4.00 of a pure SM limit.
At $p = 63$ on A100, \texttt{GEMM} takes 1.74.

The SM clock was rejected before the run for the reason that Section~\ref{sec:mps-balance-knob} states.
At the base clock of GB200, the two factors are 0.528 and 0.529.
At $p = 255$, $\texttt{GEMM\_FUSED}/\texttt{FUSED\_TC}$ moves from 1.191 to 0.992 on A100 and from 1.060 to 0.975 on GB200 across the six SM fractions (Figure~\ref{fig:f6}(b)).
At that order, the ratio differs between unrestricted A100 and GB200 by 12.4\%, and between A100 at 44 SMs and GB200 by 1.1\%.
The job-to-job spread of the ratios of Section~\ref{sec:same-reversal-one-machine} is 0.5\% in the median and 1.1\% at worst.
Unrestricted A100 agrees with GB200 on the sign of eighteen of the nineteen ranking cells that the two sweeps share, and A100 at 44 SMs on sixteen.
The one unrestricted disagreement is the $p = 511$ cell that the match is for.
The three new disagreements sit 0.1 to 3.1\% from 1 on the matched A100, so a small absolute error can reverse their signs.
On the CUDA Core axis, which the restriction does not move, $\texttt{FUSED}/\texttt{FUSED\_TC}$ at $p = 255$ differs between the machines by 21.9\% unrestricted and 19.9\% at 44 SMs.
$\texttt{FUSED\_DFMA}/\texttt{FUSED\_TC}$ at $p = 511$ differs by 15.8 and 16.9\%.
In the wave control of Section~\ref{sec:wave-control}, the working sets at $N_e = 4^3$ to $8^3$ are 20 to 170 times the L2.

\begin{table}[tp]
\centering
\footnotesize
\caption{
  The SM restriction, as requested and as measured.
  SM counts are the number of distinct SMs on which blocks actually ran, which are read from \texttt{\%smid} together with the timings.
  The counts are even on both machines, so the unit that is allocated is the texture processing cluster (TPC).
  Balance is DRAM bytes per active-SM cycle, that is, the peak bandwidth divided by the product of the measured SM count and the clock.
  Unrestricted GB200 is at 25.21, and A100 reaches this value at 44 SMs.
  The six settings of each machine are one campaign.
  The A100 41\% row is the supply-matched point of Section~\ref{sec:same-reversal-one-machine}, and Section~\ref{sup:data-provenance-reproducibility} gives its provenance.}
\label{tab:balance-knob}
\begin{tabularx}{\columnwidth}{@{}LRRRR@{}}
\toprule
 & \multicolumn{2}{c}{GB200} & \multicolumn{2}{c}{A100} \\
requested & SM & B/SM-cycle & SM & B/SM-cycle \\
\midrule
100\% & 152 & 25.21 & 108 & 10.21 \\
75\%  & 114 & 33.61 &  80 & 13.78 \\
50\%  &  76 & 50.42 &  54 & 20.42 \\
41\%  & --- & ---   &  44 & 25.06 \\
40\%  &  60 & 63.87 &  42 & 26.25 \\
36\%  &  54 & 70.96 &  38 & 29.02 \\
25\%  &  38 & 100.84 & 26 & 42.41 \\
\bottomrule
\end{tabularx}
\end{table}

\subsection{Agreement of the three timing measures}
\label{sup:agreement-three-timing-measures}
Dedicated checks of the ratios that carry the claims of Sections~\ref{sec:performance-crossovers} and~\ref{sec:instruction-aware-model} measured the three timing quantities of Section~\ref{sec:timing-protocol} in the same job.
The step-loop ratios that are quoted for the main performance landscape come instead from the base sweep of Section~\ref{sup:absolute-timing-measurements}.
At $p = 511$ these ratios are 0.951 on GB200 and 1.069 on A100.
On GB200 at $p = 511$, the ratio $\texttt{GEMM\_FUSED}/\texttt{FUSED\_TC}$ is 0.948 from CUDA-device time per stage, 0.951 from step-loop time per stage, and 0.950 from end-to-end time per step.
The spread is 0.3\%.
On A100, the same three quantities give 1.068, 1.061, and 1.061, a spread of 0.7\%.
Across the H100 and GB200 sweeps, step-loop and end-to-end timing agree to within 0.2\% on both machines at every cell.
These spreads are an order of magnitude smaller than the 5 to 6.9\% differences whose signs reverse between architectures.
The reversal of Section~\ref{sec:architecture-dependent-reversal} is therefore not a result of the choice of timing quantity.
The choice still matters for the absolute values (Section~\ref{sup:difference-between-stage-device}).
For that reason, each section names the quantity that it uses.

\section{Nominal-roof two-component accounting}
\label{sup:nominal-roof-accounting}

This section gives the details behind Section~\ref{sec:does-accounting-transfer-a100}.
These details are the path-level two-roof model that the measurements reject, the kernel decomposition, the separate scaling of its two components, the component-wise failure on H100, the two explanations that the accounting replaced, and the sensitivity of its one free parameter.
The accounting was built after the A100 measurement throughout.
We report it because its failure on H100 identifies the missing coordinate and the wrong denominator that Section~\ref{sec:instruction-priced-subpipe} removes.
The two-component procedure was applied to four cross-machine ratios.
It reproduces one in both direction and magnitude.
For two, it retains the direction but overestimates the magnitude.
For one, it misses the direction.

\subsection{The path-level two-roof model}
\label{sup:path-level-two-roof}
\begin{figure*}[tp]
\centering
\includegraphics[width=\linewidth]{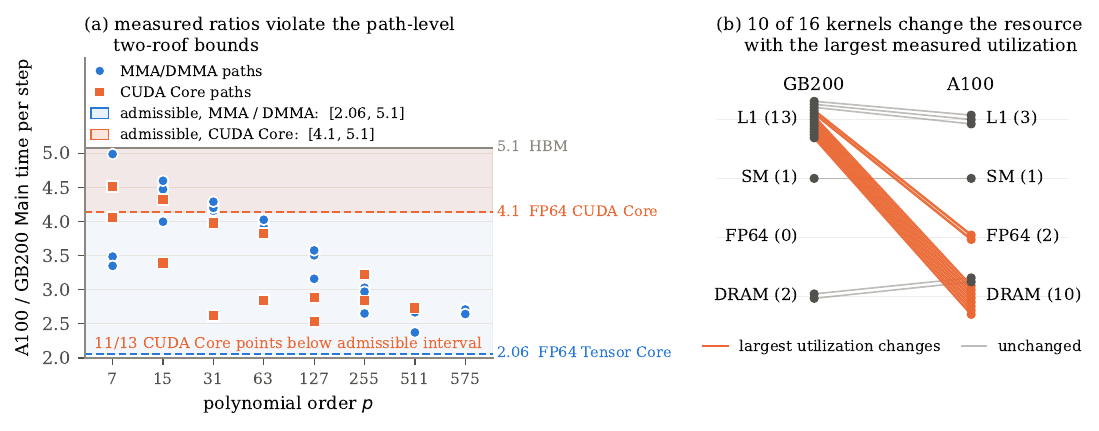}
\caption{Measured ratios violate the bounds of a path-level two-roof model, and the measured reason. (a) A100-to-GB200 time ratio for the 44 path-order combinations that were measured on both machines (median \texttt{Main} time per step), against the allowed interval of the two-roof model of Section~\ref{sup:path-level-two-roof}. The allowed interval is approximately [4.1, 5.1] for CUDA Core paths and [2.06, 5.1] for MMA/DMMA paths. The interval is a ratio of roofs and does not depend on the timing measure. The same eleven CUDA Core points fall below it in step-loop time. (b) The resource of largest measured utilization for sixteen kernels that were profiled on both machines at production clocks. The number in each category is in parentheses. This is the resource that each kernel comes closest to saturating, and it is not an established bottleneck, kernel by kernel. Section~\ref{sup:binding-resource-dominant-kernel} gives the underlying percentages. It also names the cases in which the largest percentage is itself well below peak.}
\label{fig:a11a}
\end{figure*}

A natural first model assigns each implementation path one arithmetic roof and one bandwidth roof.
This is also the form in which our own pre-measurement prediction of A100 was stated.
The arithmetic roof is selected by the instruction that the path issues.
The model combines the two roofs with two machine-independent, non-negative coefficients:
\begin{equation}
t = a\cdot W_{\mathrm{flop}}/P_m + b\cdot W_{\mathrm{bytes}}/B_m,
\end{equation}
where $t$ is the predicted path execution time, $W_{\mathrm{flop}}$ is the floating-point operation count, $W_{\mathrm{bytes}}$ is the data traffic in bytes, $P_m$ and $B_m$ are the arithmetic throughput and memory bandwidth of machine $m$, respectively, and $a$ and $b$ are machine-independent, non-negative coefficients.
That prediction was in error by as much as 43\%.
The fact that this model can exactly reproduce two machines does not validate it, because two measurements only determine its two coefficients.
Such a model constrains the time ratio between two machines to lie between the arithmetic-roof ratio and the bandwidth-roof ratio.
Figure~\ref{fig:a11a}(a) shows that eleven of the thirteen CUDA Core path measurements lie below their admissible interval.
Therefore no non-negative choice of the two coefficients can reproduce these measurements.
This result rejects the simple form, not roofline reasoning.
The model omits a roof for the resource that in fact binds.
On GB200 the CUDA Core path is L1-limited at each of the three orders that were profiled (Section~\ref{sup:binding-resource-dominant-kernel}).
The measured reason is that the resource-utilization pattern of a kernel is not preserved across architectures.
Among the sixteen kernels that were profiled on both GB200 and A100 at production clocks, the largest measured utilization falls in the same resource category for only six (Figure~\ref{fig:a11a}(b)).
The machine balance itself changes substantially between GB200 and A100.
GB200 provides 25.2 bytes of HBM bandwidth per SM cycle and A100 10.2 bytes, a factor of 2.47.
In contrast, their aggregate HBM peaks differ by a factor of 5.08.
A single roof for an entire path therefore cannot describe a path whose kernels move between resources.
We therefore classify at the level of kernels, not paths.

\subsection{The kernel decomposition}
\label{sup:kernel-decomposition}
Each kernel of one tendency evaluation at $p = 511$ was profiled on GB200 at production clocks, and Table~\ref{tab:kernels} is the per-kernel record.
A kernel is called DRAM-saturated when its DRAM column exceeds 70\% of peak.
The two groups separate clearly.
\texttt{FUSED\_TC} contains no DRAM-saturated kernel at all.
In contrast, \texttt{GEMM\_FUSED} spends 9\% of a tendency call in such kernels, and \texttt{GEMM} spends 17\%.
The kernels that lie between the two groups contribute very little to a call.
Section~\ref{sup:sensitivity-dram-saturation-threshold} gives the numbers.

\begin{table*}[tp]
\centering
\footnotesize
\caption{Decomposition of one tendency call at $p = 511$ on GB200 at production clocks, two rounds, into individual kernels. The four throughput columns are those of Table~\ref{tab:binding}, and the last column is the share of the call in kernels whose DRAM throughput exceeds 70\% of peak.}
\label{tab:kernels}
\begin{tabularx}{\tblwidth}{@{}>{\hsize=1.1\hsize}L>{\hsize=2.3\hsize}L>{\hsize=.75\hsize}R>{\hsize=.55\hsize}R>{\hsize=.55\hsize}R>{\hsize=.55\hsize}R>{\hsize=.7\hsize}R>{\hsize=1.5\hsize}R@{}}
\toprule
Path & Kernel & $\mu$s/call & SM & L1 & L2 & DRAM & Share of call \\
\midrule
\texttt{FUSED\_TC} & fused tendency z & 4 303.5 & 81.8 & 50.4 & 27.6 & 15.7 & 33.4\% \\
 & fused tendency x & 4 283.4 & 82.1 & 50.1 & 27.0 & 13.1 & 33.2\% \\
 & fused tendency y & 4 279.9 & 82.1 & 50.7 & 27.1 & 15.8 & 33.2\% \\
 & element-boundary flux (grouped) & 23.0 & 15.6 & 16.8 & 41.1 & 62.4 & 0.2\% \\
 & \textbf{total; DRAM-saturated share} & \textbf{12 889.8} &  &  &  &  & \textbf{0\%} \\
\texttt{GEMM\_FUSED} & volume GEMM z (fused epilogue) & 3 877.8 & 90.7 & 58.4 & 31.7 & 14.0 & 31.7\% \\
 & volume GEMM x & 3 630.7 & 94.7 & 45.6 & 21.3 & 11.1 & 29.7\% \\
 & volume GEMM y & 3 594.3 & 95.7 & 47.1 & 22.7 & 15.0 & 29.4\% \\
 & volume flux & 1 039.3 & 23.3 & 36.3 & 43.5 & 90.7 & 8.5\% \\
 & element-boundary flux & 74.9 & 7.9 & 16.4 & 49.2 & 74.7 & 0.6\% \\
 & \textbf{total; DRAM-saturated share} & \textbf{12 217.0} &  &  &  &  & \textbf{9.1\%} \\
\texttt{GEMM} & volume GEMMs (3 calls) & 10 830.9 & 95.3 & 33.8 & 14.1 & 7.4 & 82.7\% \\
 & separable lift and assembly & 1 156.0 & 54.0 & 48.8 & 48.6 & 82.4 & 8.8\% \\
 & volume flux & 1 041.1 & 23.3 & 36.2 & 43.4 & 90.5 & 8.0\% \\
 & element-boundary flux & 71.4 & 6.2 & 16.5 & 50.3 & 75.2 & 0.5\% \\
 & \textbf{total; DRAM-saturated share} & \textbf{13 099.3} &  &  &  &  & \textbf{17.3\%} \\
\bottomrule
\end{tabularx}
\end{table*}

\subsection{Scaling the two components separately}
\label{sup:two-component-detail}
\begin{figure*}[tp]
\centering
\includegraphics[width=\linewidth]{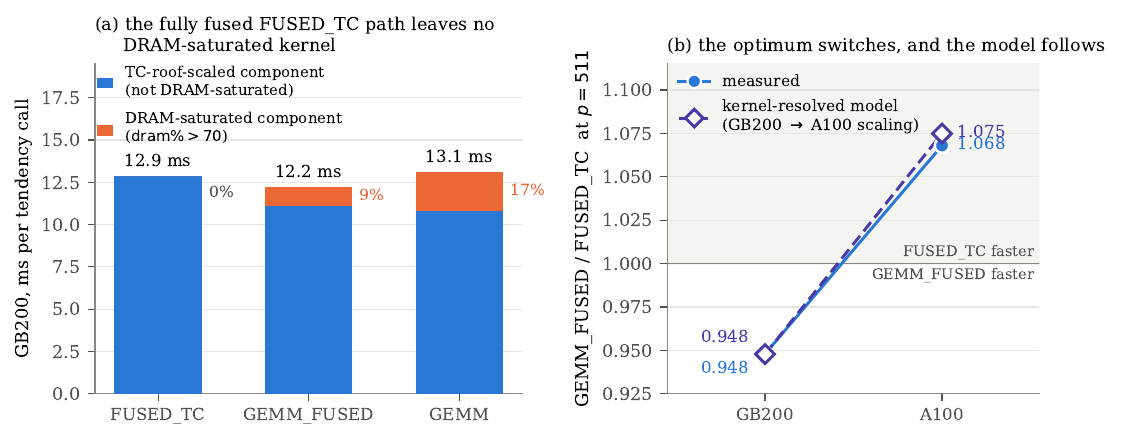}
\caption{Kernel-resolved accounting of the machine-dependent implementation crossover at $p = 511$, in CUDA device time per Runge--Kutta stage. (a) Share of each GB200 tendency call in DRAM-saturated kernels, that is, kernels with \texttt{dram\_\_throughput} $> 70\%$ in Nsight Compute at production clocks. The other component is everything else, which is dominated by the tensor-contraction mainloops but is not asserted to be compute-bound, kernel by kernel. (b) The DRAM-saturated components are scaled by the GB200-to-A100 HBM-bandwidth ratio (5.08), and the remaining components are scaled by the FP64 Tensor Core roof ratio (2.06). The scaled values are compared with the measured $\texttt{GEMM\_FUSED}/\texttt{FUSED\_TC}$ ratios of Section~\ref{sec:does-accounting-transfer-a100}. The same ordering and the same reversal are obtained from step-loop and end-to-end time. These two times differ from device time by 0.3\% on GB200 and 0.7\% on A100 (Section~\ref{sup:agreement-three-timing-measures}).}
\label{fig:a11b}
\end{figure*}

The two components are scaled separately, and no coefficient is fitted to either machine.
The DRAM-saturated component is scaled by the ratio of HBM peaks, $s_b = 5.08$.
The mainloop-dominated remainder is scaled by the ratio of nominal FP64 Tensor Core peaks, $s_a = 2.06$.
This scaling gives $\texttt{GEMM\_FUSED}/\texttt{FUSED\_TC} = 0.948$ on GB200 and 1.075 on A100, against measured CUDA-device ratios of 0.948 and 1.068 (Figure~\ref{fig:a11b}(b)).
The reversal across 1 is therefore reproduced in direction and in magnitude.
The mechanism is the relative cost of the surrounding work that fusion removes, and not the relative cost of the contraction.
The 9\% DRAM-saturated share of \texttt{GEMM\_FUSED} grows to about 20\% under the A100 scaling (18\% measured on A100 directly), and \texttt{FUSED\_TC} has no such share.

Let $C$ and $D$ be the arithmetic-classified time and the DRAM-saturated time of a path on the source machine.
Because \texttt{FUSED\_TC} has no DRAM-saturated component, the model ratio of the two paths is

\begin{equation}
R = C_g/C_f + (D_g/C_f)\cdot(s_b/s_a),
\label{eq:two-component-R}
\end{equation}
where $R$ is the predicted target-machine time ratio, and subscripts $g$ and $f$ identify \texttt{GEMM\_FUSED} and \texttt{FUSED\_TC}.
For A100, the two additive terms are 0.8614 and 0.2136.
The second term includes the attained fraction of \texttt{FUSED\_TC} as well as the DRAM-saturated component of \texttt{GEMM\_FUSED}.
The GB200 value involves no scaling, and it is the ratio of the profiled decomposition itself.
Section~\ref{sec:does-accounting-transfer-a100} does not quote $\texttt{GEMM}/\texttt{FUSED\_TC}$ for A100.
For this ratio, the same scaling gives 1.275 against a measured CUDA-device ratio of 1.167.
The scaling therefore retains the direction and the ordering but overestimates the magnitude.
Section~\ref{sup:sensitivity-dram-saturation-threshold} locates the residual.

A profile of the same decomposition on A100 shows which assumption holds.
The DRAM-saturated component scales by 5.09 against the nominal HBM ratio of 5.08.
In contrast, the contraction component scales by 2.22 for \texttt{FUSED\_TC} and 2.30 for \texttt{GEMM\_FUSED} against the nominal 2.06.
The model therefore underestimates both absolute times by 8 to 12\%.
The two errors enter the ratio with opposite signs and largely cancel.
The two terms sum to 1.074 against 1.086 for the same decomposition measured on A100.
The residual is 1.1\%, although the components are in error by 3 to 8\%.
The 0.7\% of Section~\ref{sec:does-accounting-transfer-a100} compares the model value 1.075 with the measured CUDA-device ratio 1.068.
That comparison is distinct from the pair above.
Neither is evidence for the other.

\subsection{Where the accounting fails on H100}
\label{sup:nominal-roof-failure-h100}
The GB200-to-H100 peak ratios are 3.30 for HBM bandwidth and 0.600 for FP64 Tensor Core throughput.
With these ratios, the procedure of Section~\ref{sup:two-component-detail} gives 1.34 for $\texttt{GEMM\_FUSED}/\texttt{FUSED\_TC}$ at $p = 511$, against a measured 1.069 in the common-shape comparison and 0.854 in the preferred-shape comparison.
For $\texttt{GEMM}/\texttt{FUSED\_TC}$, the procedure gives 1.812 against a measured 0.894.
H100 has 1.67 times the nominal FP64 Tensor Core peak of GB200.
However, \texttt{FUSED\_TC} requires 1.25 times as much CUDA-device time at $p = 511$, and 1.34 times as much end-to-end time.
This path has no MMA-shape parameter, so the departure is common to the same-shape and the preferred-shape comparisons.
Both terms of Equation~\ref{eq:two-component-R} can be measured on H100.
For the same-shape comparison, the second term is 0.475 in the model against 0.228 measured.
Neither the DRAM-saturated kernels nor \texttt{FUSED\_TC} has a shape parameter, so that overestimate of 0.25 is shape independent.
This overestimate alone accounts for almost all of the discrepancy of 0.27 with the measured 1.069.
The first term is preserved to 1\% at the common $8\times 8\times 4$ shape and departs by more than a third at the shape that H100 prefers.
The failure of the first term therefore belongs to the instruction shape and not to the machine (Section~\ref{sup:whether-attained-fraction-preserved}).
Both terms fail on H100 for one reason that the profile shows directly.
At production clocks, neither path is limited by arithmetic there.
A scaling by the ratio of Tensor Core peaks therefore prices work against a roof that neither path approaches.
On GB200, in contrast, the same kernels reach the fractions of Table~\ref{tab:a-attained-fraction} against a roof that equals the CUDA Core roof.
The attained fraction is not a property of the machine either.
On H100 the cuBLAS mainloop of \texttt{GEMM} reaches 84.6\% and our CUTLASS mainloop of \texttt{GEMM\_CUTE} reaches 46.8\% against the same roof, although the two kernels have equal DRAM traffic and occupancy (Table~\ref{tab:a-cublas-vs-cute}).
Neither the 92.7 and 45.2\% of Section~\ref{sec:does-accounting-transfer-a100} nor the 84.6 and 46.8\% above is a limit of roofline reasoning.
These values are surprising only if the Tensor Core roof is the capacity that the kernel can actually use.
Against the counter-defined FP64 budget, the $x$ volume GEMM of \texttt{GEMM\_FUSED} reaches 95.3\% on GB200 and 94.8\% on H100 (Section~\ref{sec:instruction-priced-subpipe}).

\subsection{Rejected hypotheses}
\label{sup:rejected-hypotheses}
Two alternative explanations were tested, and measurement rejected both.
The first explanation attributed the reversal to traffic volume.
It assumed that the difference in bytes moved by the fused path is large enough to matter on the machine with less bandwidth.
DRAM is not the binding resource of the kernels in question.
The contraction kernels reach only 7.4--15.8\% of peak DRAM throughput, so the bytes by which the two paths differ are moved by kernels that do not wait on DRAM.
In contrast, the bytes that do appear as time belong to the DRAM-saturated kernels, and \texttt{volume\_flux} alone moves more bytes than the difference between the paths.
The traffic is also a property of the implementation and not of the machine.
Over a tendency evaluation at $p = 511$, \texttt{FUSED\_TC} moves 15.29 GB and \texttt{GEMM\_FUSED} 19.69 GB, and both agree to within 0.7\% across GB200, H100, and A100.
A hypothesis stated in bytes alone therefore contains nothing machine dependent.

The second explanation attributed the reversal to the contraction mainloop.
On GB200 the CUTLASS kernels reach 90.6--95.7\% of SM throughput against 82.1\% for the hand-written kernel.
We expected that advantage to shrink or reverse on A100.
It does neither.
The difference grows from 13.6 to approximately 15--17 percentage points while the path-level ordering changes.
Both hypotheses considered only the kernels in the original profiling window.
The step that resolved the difference was to notice that \texttt{volume\_flux} and \texttt{separable\_lift\_assembly} lay outside that window.

\subsection{Sensitivity to the DRAM-saturation threshold}
\label{sup:sensitivity-dram-saturation-threshold}

The classification has one free parameter, the 70\% DRAM threshold.
The scaling is an account that was built after the measurement, so we check that the ordering that it explains does not depend on that threshold.
No additional measurements were performed.
Table~\ref{tab:a-threshold-sweep} re-aggregates the profiling data of Table~\ref{tab:kernels} and Figure~\ref{fig:a11b}(a), which were taken with \texttt{ncu --clock-control none}.
The only kernels inside the gap between the two dominant groups of Table~\ref{tab:kernels} are three element-boundary flux kernels.
These kernels reach 62.4, 74.7, and 75.2\% of peak DRAM throughput, and they take 0.2, 0.6, and 0.5\% of the tendency-call time in their respective paths.
The sweep runs from 50 to 90\% in integer steps with the scaling factors of Section~\ref{sup:two-component-detail}.
Rows combine consecutive thresholds for which the classification, and hence the prediction, is unchanged.

\begin{widetable}
\captionof{table}{Model ratios as a function of the DRAM-saturation threshold. All model ratios are computed from the same profiling data as Table~\ref{tab:kernels}, which were taken with \texttt{ncu --clock-control none}. The measured ratios against which they are compared are all CUDA-device ratios: 0.948 on GB200 and 1.068 on A100 for $\texttt{GEMM\_FUSED}/\texttt{FUSED\_TC}$, and 1.167 on A100 for $\texttt{GEMM}/\texttt{FUSED\_TC}$. The GB200 column is unscaled by definition, and therefore it is independent of the threshold.}
\label{tab:a-threshold-sweep}
\begin{tabularx}{\tblwidth}{@{}>{\hsize=.55\hsize}L>{\hsize=.65\hsize}R>{\hsize=.65\hsize}R>{\hsize=1.05\hsize}R>{\hsize=2.1\hsize}R@{}}
\toprule
 & \multicolumn{2}{c}{$\texttt{GEMM\_FUSED}/\texttt{FUSED\_TC}$} & $\texttt{GEMM}/\texttt{FUSED\_TC}$ & \\
threshold & GB200 & A100 & A100 & DRAM-saturated share, \texttt{FUSED\_TC} / \texttt{GEMM\_FUSED} / \texttt{GEMM} \\
\midrule
50--62\% & 0.948 & 1.072 & 1.272 & 0.2 / 9.1 / 17.3\% \\
63--74\% & 0.948 & 1.075 & 1.275 & 0 / 9.1 / 17.3\% \\
75\% & 0.948 & 1.066 & 1.275 & 0 / 8.5 / 17.3\% \\
76--82\% & 0.948 & 1.066 & 1.267 & 0 / 8.5 / 16.8\% \\
83--90\% & 0.948 & 1.066 & 1.135 & 0 / 8.5 / 8.0\% \\
\bottomrule
\end{tabularx}
\end{widetable}

Across the sweep, the predicted A100 $\texttt{GEMM\_FUSED}/\texttt{FUSED\_TC}$ ratio takes only three values, 1.066, 1.072, and 1.075.
This ratio stays above 1 while the GB200 ratio stays below, so the reversal is preserved throughout.
The 70\% threshold that Section~\ref{sup:kernel-decomposition} adopts is the worst of the three classifications for the model, at 0.7\% from the measured 1.068.
The \texttt{GEMM} path is more sensitive.
When the threshold passes from 82 to 83\%, the separable lift and assembly kernel moves to the arithmetic-scaled component.
This kernel is at 82.4\% of peak DRAM throughput and 8.8\% of the call.
The predicted A100 $\texttt{GEMM}/\texttt{FUSED\_TC}$ ratio then changes from 1.267 to 1.135 around the measured 1.167.
A scaling by the HBM ratio alone is therefore too strong for a kernel that is not purely bandwidth-bound.
We nevertheless retain 70\%.
A higher value that is chosen for its agreement with the measured \texttt{GEMM} ratio would tune the classification to that measurement.
The value is a convention, and the insensitivity that is established here covers the $\texttt{GEMM\_FUSED}/\texttt{FUSED\_TC}$ ordering only.
A reconstruction of the same reversal from independent data that were aggregated to the whole tendency evaluation gives 0.95 on GB200 and 1.07 on A100, against a measured A100 ratio of 1.09.
This reconstruction assigns kernels by the same 70\% criterion.
It therefore reproduces the reversal on independent data but does not test the criterion.

\section{Closure of the counter-defined FP64 budget}
\label{sup:closure}

This section tabulates how closely the counter that defines the FP64 budget, \texttt{sm\_\_pipe\_shared\_cycles\_active}, agrees with the priced demand of Equation~\ref{eq:subpipe} of the main text.
The comparison is kernel by kernel, on the three machines.
As in the main text, this is a counter-defined budget and not a claim that one physical execution unit stands behind the counter.
The measured side of every row is the utilization that the counter reports, times the elapsed SM cycles of the launch, times the SM count.
The priced side is $c_{\mathrm{tc}} N_{\mathrm{DMMA}} + c_{\mathrm{cc}} N_{\mathrm{fp64}}$ with the prices of Equation~\ref{eq:subpipe-prices} and Table~\ref{tab:dmma-prices} of the main text.
The priced side uses the counts that the same launch reports.
Both sides are per kernel role, as \texttt{figs/data/subpipe\_inputs.csv} records them (Section~\ref{sup:campaigns}).
The rows are \texttt{figs/data/subpipe\_closure.csv}, one per row of the input file (666 rows).
A row enters the table when four conditions hold: the counter reports a utilization above 5\%, the launch has at least as many thread blocks as the machine has SMs, the path is not the Ozaki path, and the priced demand is positive.
In total, 429 rows pass.
The block-count condition is needed because the utilization is reported per active SM.
Here, 14 cuBLAS launches at $p = 7$ and 15 run one to four thread blocks, and for these launches the product with the SM count overstates the measured cycles by a factor of 70 to 370.
The only other rows that this condition removes are two cuBLAS launches of 128 blocks at $p = 1023$ on GB200, at a ratio of 2.0.
The summary rows are \texttt{figs/data/subpipe\_closure\_summary.csv}.
Both files are written by one script of the analysis repository from the distributed counter table alone \citep{thiswork-code}.

Table~\ref{tab:a-closure} gives, per machine, instruction class, and order band, the number of kernels and the median and range of the ratio of the counter to the priced demand.
It also gives the median and range of the cycles that one \texttt{DMMA} contributes after the FP64 CUDA Core instructions of the kernel are subtracted at their own price.
The instruction class is fixed by the kernel.
The hand-written Tensor Core kernels, the CUTLASS kernels, and the cuBLAS tensor-op kernels issue the \texttt{DMMA} shape that their name gives, and the other kernels contain no \texttt{mma}.

Three readings follow.
In the observed high-order accuracy regime, $p \ge 511$, the counter and the priced demand agree to within 1\% in median on every machine and for every \texttt{DMMA} shape, and the range is 0.991 to 1.028.
For the $8\times 8\times 4$ shape, one \texttt{DMMA} contributes 4.00 to 4.03 cycles on GB200 and H100 and 3.96 to 4.11 cycles on A100.
For the $16\times 8\times 8$ shape of cuBLAS on H100, one \texttt{DMMA} contributes 7.93 to 8.04 cycles.
These are the values that Section~\ref{sec:instruction-priced-subpipe} of the main text and Table~\ref{tab:capacities} quote.
Below that observed regime, the counter exceeds the priced demand.
For the kernels that issue a \texttt{DMMA}, the median ratio is 1.015 to 1.065 by order group, machine, and shape, and the largest is 1.152, at $p \le 31$ on GB200.
The price of Equation~\ref{eq:subpipe-prices} is thus the minimum contribution to the counter-defined budget per instruction, which the kernels reach at $p \ge 511$.
At lower order the counter attributes up to 4.65 cycles to the same instruction.
The error is in the same direction on the three machines and is at most 15\%.
This error is therefore not the source of the low-order transfer errors of Section~\ref{sec:subpipe-model-negatives}.
Those errors are larger and are caused by kernels whose binding resource is not this budget.
The third reading concerns the kernels that contain no \texttt{mma} on A100.
For 33 of the 54 such kernels, the profiler's tensor-instruction counter reports a fixed fraction of the FP64 count of the kernel.
The balance prices those instructions as \texttt{DMMA} at four cycles, so the ratio falls to 0.66 in those rows.
However, each of those instructions contributes 0.5 cycles to the shared counter.
At that price the same rows close to 0.994 to 1.047 (column \texttt{ratio\_nonmma\_tensor\_at\_half} of the file).
The same counts occur on H100 in 18 rows of \texttt{separable\_lift}, all at a shared-counter utilization of about 3\%, which is below the 5\% selection of the table.
At the 0.5-cycle price those rows close to 1.002 to 1.011 as well.
On GB200 the counter reports no instruction for any kernel that contains no FP64 \texttt{mma}, apart from the cuBLAS kernels of the Ozaki path, which the table excludes.
The identification of the Tensor Core instruction counter with the \texttt{DMMA} count is therefore claimed for the kernels that issue an \texttt{mma}.
Section~\ref{sec:instruction-priced-subpipe} of the main text states this limit.
The two-branch price of Equation~\ref{eq:subpipe-prices} is also insensitive to its exact form.
A rounding of the FLOP price up to the next multiple of four cycles, $4\,\bigl\lceil F_{\mathrm{DMMA}}(k,m)/(4C_{\mathrm{tc},m})\bigr\rceil$, prices every DMMA shape identically to the floor, because every shape carries a power-of-two FLOP count.
Where the arithmetic is held fixed and only the pricing moves, $N_{\mathrm{DMMA}}$ is rescaled by the FLOP ratio between the two shapes.
Therefore $c_{\mathrm{tc}} N_{\mathrm{DMMA}}$ depends on the shape only through the floor.
None of the 33 kernels is bound by the FP64 budget in the model, so no prediction of the main text depends on this pricing.
We therefore leave Equation~\ref{eq:subpipe} as it is.
The per-machine medians over all 429 rows, 1.023, 1.010, and 1.021 on GB200, A100, and H100, respectively, are the FP64-budget entry of Table~\ref{tab:capacities}.

\begin{widetable}
\captionof{table}{Closure of the counter-defined FP64 budget of Equation~\ref{eq:subpipe} of the main text, from \texttt{figs/data/subpipe\_closure\_summary.csv}. Ratio is the counter \texttt{sm\_\_pipe\_shared\_cycles\_active} in SM cycles over the priced demand. The column $c_{\mathrm{tc}}$ is the measured cycles per \texttt{DMMA} after the CUDA Core instructions of the kernel are subtracted at their own price, and it is given only for the kernels that issue one. Kernels is the number of kernel roles that pass the selection of the text. The \texttt{DMMA.16x8x16} shape occurs on H100 at $p \le 31$ only. The rows labeled ``all'' pool every class of the machine.}
\label{tab:a-closure}
\begin{xltabular}{\tblwidth}{@{}l>{\hsize=1.85\hsize}L>{\hsize=1.0\hsize}L>{\hsize=.75\hsize}R>{\hsize=.7\hsize}R>{\hsize=1.0\hsize}R>{\hsize=.7\hsize}R>{\hsize=1.0\hsize}R@{}}
\toprule
Machine & Instruction class & Order band & Kernels & Ratio median & Ratio range & $c_{\mathrm{tc}}$ median & $c_{\mathrm{tc}}$ range \\
\midrule
GB200 & \texttt{DMMA.8x8x4} & $p \le 31$ & 41 & 1.042 & 1.013--1.152 & 4.18 & 4.08--4.65 \\
GB200 & \texttt{DMMA.8x8x4} & $p = 63$--127 & 22 & 1.065 & 1.034--1.095 & 4.27 & 4.15--4.39 \\
GB200 & \texttt{DMMA.8x8x4} & $p = 255$ & 12 & 1.025 & 1.018--1.054 & 4.11 & 4.08--4.22 \\
GB200 & \texttt{DMMA.8x8x4} & $p \ge 511$ & 39 & 1.001 & 0.999--1.005 & 4.01 & 4.00--4.02 \\
GB200 & \texttt{DMMA.8x8x4} & all & 114 & 1.023 & 0.999--1.152 & 4.10 & 4.00--4.65 \\
GB200 & no \texttt{mma} & $p \le 31$ & 15 & 1.027 & 1.015--1.096 & --- & --- \\
GB200 & no \texttt{mma} & $p = 63$--127 & 12 & 1.028 & 1.017--1.047 & --- & --- \\
GB200 & no \texttt{mma} & $p = 255$ & 8 & 1.015 & 1.013--1.033 & --- & --- \\
GB200 & no \texttt{mma} & $p \ge 511$ & 10 & 1.001 & 0.999--1.003 & --- & --- \\
GB200 & no \texttt{mma} & all & 45 & 1.023 & 0.999--1.096 & --- & --- \\
GB200 & all & all & 159 & 1.023 & 0.999--1.152 & --- & --- \\
\midrule
A100 & \texttt{DMMA.8x8x4} & $p \le 31$ & 28 & 1.015 & 1.000--1.041 & 4.07 & 4.00--4.18 \\
A100 & \texttt{DMMA.8x8x4} & $p = 63$--127 & 22 & 1.036 & 1.013--1.063 & 4.15 & 4.06--4.26 \\
A100 & \texttt{DMMA.8x8x4} & $p = 255$ & 12 & 1.019 & 1.010--1.055 & 4.08 & 4.04--4.22 \\
A100 & \texttt{DMMA.8x8x4} & $p \ge 511$ & 21 & 1.003 & 0.991--1.028 & 4.01 & 3.96--4.11 \\
A100 & \texttt{DMMA.8x8x4} & all & 83 & 1.016 & 0.991--1.063 & 4.07 & 3.96--4.26 \\
A100 & no \texttt{mma} & $p \le 31$ & 27 & 0.817 & 0.667--1.049 & --- & --- \\
A100 & no \texttt{mma} & $p = 63$--127 & 12 & 1.009 & 0.667--1.027 & --- & --- \\
A100 & no \texttt{mma} & $p = 255$ & 8 & 1.005 & 0.668--1.011 & --- & --- \\
A100 & no \texttt{mma} & $p \ge 511$ & 7 & 0.665 & 0.663--0.997 & --- & --- \\
A100 & no \texttt{mma} & all & 54 & 0.817 & 0.663--1.049 & --- & --- \\
A100 & all & all & 137 & 1.010 & 0.663--1.063 & --- & --- \\
\midrule
H100 & \texttt{DMMA.8x8x4} & $p \le 31$ & 34 & 1.015 & 1.006--1.044 & 4.07 & 4.04--4.19 \\
H100 & \texttt{DMMA.8x8x4} & $p = 63$--127 & 17 & 1.039 & 1.013--1.077 & 4.16 & 4.05--4.31 \\
H100 & \texttt{DMMA.8x8x4} & $p = 255$ & 9 & 1.028 & 1.016--1.057 & 4.11 & 4.07--4.23 \\
H100 & \texttt{DMMA.8x8x4} & $p \ge 511$ & 24 & 1.002 & 1.000--1.007 & 4.01 & 4.00--4.03 \\
H100 & \texttt{DMMA.8x8x4} & all & 84 & 1.014 & 1.000--1.077 & 4.07 & 4.00--4.31 \\
H100 & \texttt{DMMA.16x8x8} & $p \le 31$ & 4 & 1.033 & 1.026--1.039 & 8.28 & 8.22--8.33 \\
H100 & \texttt{DMMA.16x8x8} & $p = 63$--127 & 5 & 1.050 & 1.043--1.064 & 8.41 & 8.36--8.53 \\
H100 & \texttt{DMMA.16x8x8} & $p = 255$ & 3 & 1.045 & 1.042--1.047 & 8.36 & 8.34--8.38 \\
H100 & \texttt{DMMA.16x8x8} & $p \ge 511$ & 9 & 1.003 & 0.992--1.005 & 8.03 & 7.93--8.04 \\
H100 & \texttt{DMMA.16x8x8} & all & 21 & 1.033 & 0.992--1.064 & 8.28 & 7.93--8.53 \\
H100 & \texttt{DMMA.16x8x16} & $p \le 31$ & 3 & 1.035 & 1.029--1.040 & 16.59 & 16.50--16.67 \\
H100 & no \texttt{mma} & $p \le 31$ & 8 & 1.028 & 1.008--1.058 & --- & --- \\
H100 & no \texttt{mma} & $p = 63$--127 & 8 & 1.024 & 1.021--1.035 & --- & --- \\
H100 & no \texttt{mma} & $p = 255$ & 6 & 1.022 & 1.010--1.029 & --- & --- \\
H100 & no \texttt{mma} & $p \ge 511$ & 3 & 1.004 & 1.003--1.004 & --- & --- \\
H100 & no \texttt{mma} & all & 25 & 1.024 & 1.003--1.058 & --- & --- \\
H100 & all & all & 133 & 1.021 & 0.992--1.077 & --- & --- \\
\bottomrule
\end{xltabular}
\end{widetable}

\subsection{Per-SM-cycle capacities and their check against the counters}
\label{sup:capacities}
Table~\ref{tab:capacities} gives the source of each per-SM-cycle capacity of Equation~\ref{eq:model} of the main text and checks it against the counters.
For each resource, the implied peak is the counter divided by the product of the utilization that Nsight Compute reports for it, the elapsed SM cycles and the SM count.
The implied peak is evaluated for every profiled launch of the three order sweeps that has a utilization above 5\%.
The L2 capacity of 64 bytes per SM cycle is a reference normalization and not a sourced hardware peak.
The implied peaks of this capacity are lower bounds because the L2 throughput metric is the largest of several sub-metrics.
L2 binds 23 of the 666 kernel instances.
All of these instances belong to cuBLAS launches at $p \le 15$, the element-boundary flux kernel of \texttt{GEMM\_FUSED} at $p = 7$ and 15 on GB200, and the Ozaki path.
When the capacity is varied from 32 bytes to infinity, the above-threshold hit rate of Section~\ref{sec:signal-to-noise-rule} of the main text stays between 94.0 and 94.5\% on the machine axis and between 92.1 and 94.5\% on the balance axis.
The cell counts, however, move because the low-order cuBLAS cells cross the threshold.
The counts 94 of 100 and 156 of 167 are those at 64 bytes.
The FMA capacity is two warp instructions per SM cycle on A100 and four on the other two machines.
However, no kernel instance is bound by the FMA capacity on any machine, and the FMA demand reaches at most 28\% of the binding demand.
A capacity of two therefore changes no scored cell.

\begin{table*}[tp]
\centering
\footnotesize
\caption{
  The per-SM-cycle capacities of Equation~\ref{eq:model}, their source, and the peak that the Nsight Compute utilization counters imply on the three machines. The implied peak is the median over launches with utilization above 5\%, in the unit of the capacity, unless it is marked as a maximum.
  The FP64-budget pricing, the DRAM capacity, and the FMA capacity are per machine (Table~\ref{tab:machines-per-sm-cycle}). The rest are taken as the same on all three.}
\label{tab:capacities}
\begin{tabularx}{\tblwidth}{@{}>{\hsize=.8\hsize}L>{\hsize=.9\hsize}L>{\hsize=1.5\hsize}L>{\hsize=.8\hsize}L@{}}
\toprule
Resource & Capacity & Source & Implied peak, GB200 / A100 / H100 \\
\midrule
Counter-defined FP64 budget & $c_{\mathrm{tc}} N_{\mathrm{DMMA}} + c_{\mathrm{cc}} N_{\mathrm{fp64}}$ (Equation~\ref{eq:subpipe}) & measured: 3.96 to 4.02 cycles per DMMA instruction on the three machines; $C_{\mathrm{tc}}$ and $C_{\mathrm{cc}}$ from the datasheet & counter over priced demand 1.023 / 1.010 / 1.021 (Section~\ref{sup:closure}) \\
DRAM & 25.21 / 10.21 / 9.17 B & datasheet HBM peak over SM count times boost clock & 25.7 / 10.4 / 9.4 B \\
LSU wavefronts & 1 & the profiler's peak definition & 1.000 / 1.000 / 1.000 \\
Issue & 4 warp instructions & four SM sub-partitions, one warp instruction each & 0.97 / 0.98 / 0.98 of 4 \\
ALU & 2 warp instructions & the profiler's peak definition & 1.95 / 1.98 / 1.97 \\
FMA & 4 / 2 / 4 warp instructions (GB200 / A100 / H100) & FP32 execution width; this demand binds no kernel & 3.54 / 1.99 / 3.83 (maxima) \\
L2 & 64 B & reference normalization; results are insensitive from 32 B to infinity & 111 / 67 / 57 B (maxima; lower bounds) \\
\bottomrule
\end{tabularx}
\end{table*}

\subsection{Portability of the transferred counter numerators}
\label{sup:counter-numerator-portability}

The model transfers every non-DMMA counter numerator from the selected calibration machine.
Path-total agreement alone therefore does not test each numerator independently.
We therefore match the rows of \texttt{figs/data/subpipe\_inputs.csv} by order, path, and canonical kernel role across each unordered pair of machines and compare the counters measured on both sides.
Table~\ref{tab:a-counter-portability} reports the fold difference $\max(X_a,X_b)/\min(X_a,X_b)$ over 474 matched pairs of kernels that were compiled in this work.
These pairs include CUTLASS kernels and exclude cuBLAS roles.
The cuBLAS roles are excluded because the canonical role deliberately collapses different vendor-library kernels and dispatch decisions.
Equality of the role therefore does not identify the same generated kernel instance.
Pairs for which a numerator is zero on both sides are omitted for that numerator.
None of the remaining pairs has a zero on only one side.
The audit is reproduced by \texttt{figs/audit\_counter\_numerators.py} from the distributed counter table alone.

\begin{apptable}
\captionof{table}{Cross-machine portability of the non-DMMA counter numerators for 474 matched pairs of non-cuBLAS kernel roles. The entries are fold differences, so 1 denotes equality. The last two columns restrict the same calculation to pairs in which that resource binds on at least one of the two machines. A dash means that the resource binds no matched pair. ``FP64 instructions'' is $N_{\mathrm{fp64}}$ and does not include the separately transformed DMMA count.}
\label{tab:a-counter-portability}
\begin{tabularx}{\tblwidth}{@{}Lrrrr@{}}
\toprule
Numerator & Median & 90th percentile & Binding pairs & Binding-pair median / 90th percentile \\
\midrule
FP64 instructions & 1.000 & 1.000 & 116 & 1.000 / 1.000 \\
DRAM bytes & 1.045 & 1.228 & 344 & 1.046 / 1.302 \\
L1 LSU wavefronts & 1.016 & 1.078 & 93 & 1.021 / 1.202 \\
Issued instructions & 1.058 & 1.296 & 0 & --- \\
ALU instructions & 1.211 & 1.500 & 0 & --- \\
FMA instructions & 1.050 & 1.887 & 0 & --- \\
L2 bytes & 1.255 & 1.527 & 4 & 1.142 / 1.330 \\
\bottomrule
\end{tabularx}
\end{apptable}

The exact FP64 counts support transferring $N_{\mathrm{fp64}}$ for the kernels compiled here, and the L1 counts are close for most pairs.
The audit does not support treating all numerators as architecture invariant.
The 90th-percentile differences in issue, ALU, FMA, and L2 activity are 1.30 to 1.89-fold, and even the binding DRAM and L1 numerators reach 1.30 and 1.20-fold at the 90th percentile.
Issue, ALU, and FMA bind no matched pair in this data set, and L2 binds only four.
The disagreement of these numerators therefore has little opportunity to affect the reported path totals.
However, this disagreement cannot be inferred to be harmless on another workload.
The transfer results validate the complete model on these paths, not each numerator assumption separately.

\subsection{Check of the DMMA expansions}
\label{sup:dmma-expansion-check}
In the SASS of the binaries that are cross-compiled for the two targets, \texttt{mma.sync.m8n8k4.f64} compiles to one \texttt{DMMA.8x8x4} on \texttt{sm\_90} and on \texttt{sm\_100} alike.
In contrast, \texttt{m16n8k4}, \texttt{m16n8k8}, and \texttt{m16n8k16} compile to one wide instruction on \texttt{sm\_90} and to two, four, and eight \texttt{DMMA.8x8x4}, respectively, on \texttt{sm\_100}.
The \texttt{sm\_100} build, when it is read with \texttt{cuobjdump}, contains 46,340 matrix instructions across 807 kernels, all of them \texttt{DMMA.8x8x4}.
This static count is not an input of the model.
The instruction-side input of Section~\ref{sec:subpipe-model} of the main text is the shape.
The executed calibration counts are rescaled by the ratio of the FLOP per instruction of the calibration shape to that of the target shape.
This rescaling assumes that the same source at the same order and launch configuration executes the same FLOP through \texttt{DMMA} on every machine.
We test this assumption on executed counts.
We compile one kernel at two CUTLASS MMA shapes and compare the instruction count, the cycles per instruction, and the elapsed time.
No predicted elapsed time is involved in this test.
On \texttt{sm\_100}, the change from $8\times 8\times 4$ to $16\times 8\times 4$ must leave all three quantities unchanged.
Across eight kernel-shape pairs at $p = 511$ on GB200, a calibration machine, the three quantities agree to within 0.5\%.
On \texttt{sm\_90}, the identical source change must halve the instruction count and leave the cost of each instruction at four cycles.
Over sixteen kernels that span eight orders on H100, the ratio of \texttt{DMMA} counts is exactly 0.5 in sixteen cases out of sixteen.
After the CUDA Core share is subtracted, and at the SM clock that the profiler records, the cost per instruction is 3.95 to 4.28 (Section~\ref{sec:instruction-priced-subpipe} of the main text).
The executed counts and the cost per instruction in this check are counters that are read on the held-out H100.
They enter no parameter of the model.
We record them as target-side evidence for the expansions and for the rescaling rule, and not as target-side calibration.

\section{Calibration-side choices}
\label{sup:calibration-side-choices}
This section reports two choices that are made on the calibration machines before the target is touched.
The first choice is which calibration machine or set to prefer.
The second choice is whether the observed high-order accuracy regime can be selected from the calibration-side $\eta$ values.

\subsection{Choice of calibration set}
\label{sup:calibration-set-choice}
One candidate rule is to calibrate on the machine whose supply per unit of SM throughput is smallest.
The reason is that a kernel is more likely to sit on one of its roofs on that machine.
In addition, that machine can be identified from datasheets alone.
Under that rule, A100 is the calibration machine, because of its halved CUDA Core FP64 throughput as well as its bandwidth.
When all six single-source directions are scored, A100 is in fact the worst of the three sources.
In the observed $p \ge 511$ regime, the median path-total error is 4.9\% when this source predicts GB200 and 4.1\% when it predicts H100.
The best source in each case gives 1.5 and 1.7\%, respectively.
When the transition order is included, the same four numbers are 7.3, 4.1, 2.0, and 2.0\%.
The two supply coordinates that motivated the rule order the machines differently.
H100 has the smallest supply of DRAM bytes per SM cycle, at 9.17 against 10.21 and 25.21.
In contrast, A100 has the smallest CUDA Core FP64 throughput.
The data follow the bandwidth ordering.
H100 is the best single source for two of the three targets, and GB200 is the best for H100 itself.
However, with three machines we cannot separate that ordering from other orderings that agree with it by chance.
The margin is also small on one target.
In that regime on A100, H100 as the source gives 4.1\% and GB200 gives 4.9\%.
The two directions between A100 and H100 are equal at 4.1\%.
We therefore reject the tested form of the rule, and we leave the bandwidth version untested, not confirmed.

The remaining conclusion is about the calibration set, not about one machine.
The per-kernel choice of Section~\ref{sec:subpipe-model} takes the machine on which $\eta$ is closest to 1.
Two alternatives replace this choice by the larger of the two values of $\eta$ or by their mean.
With these alternatives, the median path-total error in the observed $p \ge 511$ regime changes from 1.3, 5.4, and 1.7\% to 4.1, 3.6, and 3.5\%, or to 2.4, 4.2, and 1.7\%, respectively.
Therefore neither alternative is uniformly worse in the median, and both are better in the A100 direction.
The worst case over all orders separates them.
The largest error grows from 30.6\% to 49.4 and 1285.3\%, respectively.
Inside the observed high-order regime, the largest error is 21.8\% for the closest rule, 29.7\% for the larger value, and 20.3\% for the mean.
We keep the rule that is best in the worst case over all orders.
We record two costs of this choice.
The first cost is a worse median in the A100 direction.
The second cost is the worst case in that regime, which the mean beats by 1.5 points.

\subsection{Can the observed high-order accuracy regime be selected from the calibration machine?}
\label{sup:validity-band-selection}
Section~\ref{sec:signal-to-noise-rule} of the main text reports an observed high-order accuracy regime, $p \ge 511$.
This regime is read from the measured transfers and is not derived from $\eta$.
This subsection gives the negative result behind that statement.
We scored three aggregations of the kernel $\eta$ values of a path against the 140 leave-one-GPU-out path totals.
These values are profiled on the calibration machine, and each aggregation can be computed on the calibration machine before the target is touched.
The three aggregations are the mean weighted by predicted time, the fraction of predicted path time on kernels with $\eta \le 1.2$, and the nearest-rank 90th percentile of the kernel $\eta$ values.
They are columns of the distributed table behind Figure~\ref{fig:f5}.
Over the 140 cells, the median absolute error is 7.7\% (Table~\ref{tab:validity-band-selection}).
The median is 1.9\% over the 32 cells of the observed high-order regime and 7.5\% over the 18 cells of the transition order.

\begin{apptable}
\captionof{table}{Attempted calibration-side selection of the observed high-order accuracy regime, scored on the 140 leave-one-GPU-out path totals. Each rule selects the cells that satisfy it. \emph{Median in} and \emph{median out} are the median absolute path-total errors of the selected and unselected cells. The last three columns say where the selected cells come from. The observed regime is 32 cells at $p \ge 511$, the transition order is 18 cells at $p = 255$, and the excluded low orders are 90 cells at $p \le 127$. A rule that recovered the observed regime would select 32, 0, and 0.}
\label{tab:validity-band-selection}
\noindent\begin{tabularx}{\tblwidth}{@{}>{\hsize=1.3\hsize}LRRRRRR@{}}
\toprule
rule & cells & median in & median out & from high order & from $p = 255$ & from $p \le 127$ \\
\midrule
weighted $\eta \le 1.20$ & 99 & 7.6\% & 8.2\% & 24 & 8 & 67 \\
weighted $\eta \le 1.10$ & 24 & 2.0\% & 9.2\% & 24 & 0 & 0 \\
90th percentile $\eta \le 1.20$ & 43 & 8.3\% & 7.2\% & 3 & 2 & 38 \\
$\ge 95\%$ of predicted time at $\eta \le 1.2$ & 67 & 5.6\% & 9.3\% & 24 & 4 & 39 \\
\bottomrule
\end{tabularx}
\end{apptable}

Neither statistic of the shape of the distribution separates the cells.
The 90th-percentile rule selects 43 cells whose median error, 8.3\%, is worse than the 7.2\% of the cells that it rejects.
The 95\%-of-time rule selects 67 cells, and 33 of these cells are at $p \le 31$.
The rank correlation of either scalar with the error is $-0.1$.
The weighted mean has the expected sign of correlation, $+0.26$, but it does not change in one direction with $p$.
The median of the weighted mean is 1.144 at $p \le 127$, 1.248 at the transition order $p = 255$, and 1.072 in the observed high-order regime.
The low orders are mispredicted although their $\eta$ is near 1.
The error comes from the drift of $\eta$ between machines and from the kernel family that the library chooses (Section~\ref{sec:subpipe-model-negatives}).
Neither cause is visible in $\eta$ on the calibration machine.

One narrow sufficient condition does hold on this population.
A weighted mean of at most 1.10 selects 24 cells, which are all within 10\% at a median of 2.0\%, against 9.2\% for the rejected cells.
This condition does not recover the observed regime.
The 24 cells are the three library paths at $p \ge 511$, and they include none of the eight fused-path cells of that regime, although seven of these fused-path cells are also within 10\%.
The condition also depends on the shape at which the target is priced.
At the $16\times 8\times 4$ shape of H100, it accepts nine of the 22 cells that this shape covers, and only three of the nine are within 10\%.
The reason is that the errors of the H100 CUTLASS paths at $p \ge 511$ move from within 2\% to 10.2 to 19.0\%, while their weighted $\eta$ stays between 1.06 and 1.08.
This error is the overestimate of the MMA-shape gain of Section~\ref{sup:model-departures}.
We therefore do not use it as a rule.

\section{Disagreements between the model and the measurement}
\label{sup:model-departures}
This section gives the mechanism of each of the six disagreements between the FP64-budget model and the measurement that Section~\ref{sec:subpipe-model-negatives} of the main text names.
At the transition order $p = 255$, the three library paths transfer within 0.1 to 4.9\%.
In contrast, the three fused paths are wrong by 6.3 to 30.6\%.
The second disagreement below gives the reason.

The element-boundary flux kernel is mispredicted by $-57$ to $+29\%$ over the three directions.
The sign is not even fixed.
The kernel runs at 84 to 88\% occupancy, with a long-scoreboard stall count of 100 to 400 cycles and $\eta$ between 1.6 and 2.9.
Therefore it is latency bound, not roof bound.
The counters that are available to us determine the two limiting cases of its machine ratio, but not the transition between them.
For that transition, a measure of memory-level parallelism is missing.
Its weight is 0.1 to 3.1\% of a tendency call at $p \ge 255$, and at most 1.2\% in the observed $p \ge 511$ regime.

The CUDA Core path \texttt{FUSED} is mispredicted by $-31$ to $+19\%$ at $p = 63$ and 255 in the three directions.
The reason is that the binding resource itself changes.
On GB200, the kernel is bound by LSU wavefronts, and on the targets, it is bound by DRAM, issue, or the FP64 budget.
When the roof changes, $\eta$ does not transfer, and the measured discrepancy in $\eta$ is 38 to 181\%.
This change of roof is the mechanism behind the low-order failures in general.
It is not a failure of the L1 capacity assumption, which we tested separately.

At $p = 63$, the error of the other paths spans $-28$ to $+11\%$ over the three directions.
The contraction kernels there are L1 bound on GB200 and bound by DRAM or the FP64 budget on the other two machines.
Therefore the roof that the model carries depends on the direction of the transfer, and $\eta$ does not carry across that change of roof.

The model predicts the direction and the roof of the MMA-shape change on H100 correctly in all sixteen cases.
However, it overestimates the resulting speedup at high order, by 17 to 30\% at $p \ge 255$, against a median of 0.6\% at $p \le 127$.
When the FP64-budget demand at $p = 511$ is halved, it falls from 95\% of elapsed cycles to 47 to 48\%.
This value is a near tie with L1 at 45.7\%.
The maximum in Equation~\ref{eq:model} handles such a near tie worst, and $\eta$ degrades in exactly this situation.
We know from the demands, before any timing on the target, that the two demands become close.
However, the size by which $\eta$ degrades is known only after the target time is measured.

A fifth disagreement is a component of $\eta$.
The stall classes \texttt{lg\_throttle} and \texttt{barrier} are together 7.1\% of the stall weight.
In these classes, the L1 queue for local and global (LG) memory operations is full, or sibling warps of the same thread block have not yet reached a barrier.
No single machine coordinate that we tried can carry these two classes.
Two of the machines order them oppositely under every candidate coordinate.
We report the weight, not a fit.

A sixth limit is information that the model does not receive as an input.
At $p \le 15$, cuBLAS does not dispatch the same kernels on the three machines, and the difference is one of instruction family.
At $p = 7$, cuBLAS selects a SIMT double-precision kernel on A100 and Tensor Core kernels on GB200 and H100.
Therefore the model has no path total to offer for three of the six paths on A100 at this order.
On the kernels that the machines do share, the library axis $\texttt{GEMM\_CUTE}/\texttt{GEMM}$ is predicted there at 0.860 against 1.022.
This is a wrong sign that is screened in at $S/E = 6.0$.
However, that quotient is not a statement about which implementation is faster.
Where the three machines make the same choice, the error of a path total at $p \le 31$ has a median of 16.6, 8.6, and 8.5\% in the three hold-out directions.
Where they do not make the same choice, the error is 45.6 to 57.5\% on the four such cells in each direction.
The choice is a rule that the library applies at run time, so it cannot be read from a cross-compiled binary.
In one direction, both calibration machines select the Tensor Core family and only the held-out machine selects SIMT.
Therefore nothing on the calibration side reveals the choice.

When those cells are set aside, only one factor per machine and order remains at low order, not a per-path error.
We remove the geometric mean of the predicted-to-measured ratio over the paths of a cell.
The residual then has a median of 1.7 to 10.3\% per cell at $p \le 127$, against 3.6 to 22.6\% before the factor is removed.
That factor divides out of a ratio.
In the observed $p \ge 511$ regime, the same residual is 0.3 to 5.4\%.
At the transition order $p = 255$, the residual is 4.7 to 9.6\%, so the transition order is not a common-factor case (Section~\ref{sec:signal-to-noise-rule}).
That factor is the drift of $\eta$ between the machines.
A way to predict that drift would be needed to identify a wider absolute-time domain prospectively.

\section{Empirical scoring of the separation-to-error screen}
\label{sup:sn-scoring}

This section holds the definitions and the sensitivity checks behind the empirical screening rule of Section~\ref{sec:signal-to-noise-rule} of the main text.
The separation-to-error ratio of a ranked pair is
\begin{equation}
S/E = \frac{|\ln r_{\mathrm{pred}}|}{\varepsilon_{\mathrm{family}}},
\label{eq:signal-to-noise}
\end{equation}
where a family is one pair of paths and $\varepsilon_{\mathrm{family}}$ is the median of $|\ln (r_{\mathrm{pred}} / r_{\mathrm{meas}})|$ for that pair on transfers in which the target machine takes no part at all.
The median is pooled over the orders of those transfers.
The denominator is a model prediction error that was observed on other transfers, and not an estimate of measurement noise.
Both quantities are quoted in percent, as $100\,|\ln r|$.
A ranking is \emph{screened in} when $S/E \ge 1$.
Its \emph{predicted sign} is the sign of $\ln r_{\mathrm{pred}}$.
The sign is \emph{correct} when it agrees with the sign of $\ln r_{\mathrm{meas}}$.
The screening rule and the closest-$\eta$ calibration choice were formulated after the three machines had been measured.
Therefore the machine-axis rate is an empirical characterization on the same machine population, and not an external validation.
The performance-model form and inputs were committed before the balance sweeps.
However, the available provenance does not establish a separate preregistration of this screening rule.
Therefore the balance-axis rate has the same descriptive status.
The threshold of 1 is the definition of the screen and was not adjusted against the observed hit rates.
An equal-count selection by $|\ln r_{\mathrm{pred}}|$ alone selects one more correct sign on the machine axis and four more on the balance axis (Section~\ref{sec:signal-to-noise-rule} of the main text).
We nevertheless keep the family-specific denominator, because it applies the same comparison to every family.
A single cut on $|\ln r_{\mathrm{pred}}|$ cannot do this when the family errors span 1.3 to 29.6\% on the machine axis.
Kernels that sit on a roof on the calibration machine transfer best (Table~\ref{tab:eta-strata}).
However, at the level of a path, no aggregation of $\eta$ separates the good transfers from the bad (Section~\ref{sup:validity-band-selection}).
The scored cells are not independent statistical samples.
Cells that share an order, path, kernel, or machine also share errors.
Therefore the counts describe the evaluation set and not an effective sample size.
Both quantities of Equation~\ref{eq:signal-to-noise} are logarithmic differences, so neither depends on whether a ratio is written as A/B or as B/A.
In percent, $100\,|\ln r|$ agrees with the percentage difference $100\,|r - 1|$ to within one point for differences up to 15\%.

Table~\ref{tab:eta-strata} groups the 609 kernel transfers of the three hold-out directions by the $\eta$ that was measured on the source.
The twelve kernels of the highest group are single cuBLAS launches at $p \le 15$ whose time is set by latency.
No capacity in the model represents this latency.
All twelve kernels lie in the twelve mismatched-list cells of Section~\ref{sec:logo-three-directions}.

\begin{widetable}
\captionof{table}{
  Kernel transfer error grouped by $\eta$ on the calibration machine. Here $\eta$ is the measured elapsed SM cycles of a kernel divided by its largest resource demand on that machine, so $\eta = 1$ means that the kernel sits on its roof.
  The transfers are every kernel of the profiled paths in all three source-to-target directions among GB200, A100, and H100 that the target also measured. Of the 613 transfers in all, 609 fall in the groups below and four have $\eta < 1$.
  The table establishes the separation at $\eta = 1.2$ and not a graded relation.
  The two highest groups are thin and dominated by single low-order cuBLAS launches.
  The 90th percentile is the nearest-rank one, and on the two groups with $n \le 12$ it has almost no information.}
\label{tab:eta-strata}
\begin{tabularx}{\tblwidth}{@{}>{\hsize=1.3\hsize}L>{\hsize=.45\hsize}R>{\hsize=.85\hsize}R>{\hsize=.9\hsize}R>{\hsize=1.5\hsize}L@{}}
\toprule
$\eta$ on the calibration machine & $n$ & Median error & 90th percentile & Interpretation \\
\midrule
1.0--1.2 & 383 & 5.3\% & 23.7\% & on a roof; transfers \\
1.2--1.6 & 206 & 11.9\% & 28.6\% & near a roof \\
1.6--2.5 & 8 & 8.4\% & 40.2\% & mixed; five fused kernels and three $p = 7$ cuBLAS launches \\
$> 2.5$ & 12 & 38.1\% & 46.2\% & $p \le 15$ cuBLAS launches, L2-bound with negligible demand; not transferable \\
\bottomrule
\end{tabularx}
\end{widetable}

A family is one of the seven ratios of Table~\ref{tab:sn-by-family}, that is, one pair of paths and not one order.
The denominator $\varepsilon_{\mathrm{family}}$ is the median over the orders of that ratio on transfers in which the target machine takes no part.
On the balance axis, this needs no extra hold-out.
A restricted A100 cell is predicted from A100 counters alone, and its family error is taken over the GB200 cells.
On the machine axis, every two-source calibration that holds out one GPU uses the other two GPUs.
Therefore the only transfers from which a given GPU is entirely absent are the single-source transfers between the other two GPUs.
We use those transfers.
The denominator is then measured on a different configuration from the scored one.
It is the larger of the two: 5.5\% in the median against 4.5\% for the two-source cells.
The value can still differ between targets, because the transfers that remain after the target is removed depend on which machine is held out.
The non-isolated denominator gives 99 of 106 on the machine axis instead of 94 of 100.

The scored population is 137 cells on the machine axis and 190 cells on the balance axis.
The machine-axis cells are seven ratios over every order in three hold-out directions, without the 18 cells at $p \le 15$ whose kernel lists do not match.
The balance-axis cells are seven ratios over three orders, two machines, and five restricted SM counts.
The 18 excluded cells are quotients of partial sums and are not scored.
Of the 16 excluded cells that reach the threshold, 14 have the sign of the partial-sum measurement.
However, a ratio of partial sums does not say which implementation is faster.
The eleven misses on the balance axis come from four combinations of machine, order, and ratio.
Five of these misses are one prediction, $\texttt{GEMM\_CUTE}/\texttt{GEMM}$ on GB200 at $p = 511$, that is repeated over the five restricted SM counts at $S/E = 1.07$ to $1.11$.

The threshold also changes the mix of families, machines, and orders between the screened-in and the below-threshold cells (Table~\ref{tab:sn-by-family}).
On the balance axis, the three families that contain \texttt{FUSED\_DFMA} or \texttt{FUSED} against a Tensor Core path are screened in for all 70 of their cells and are correct in all 70.
All 23 below-threshold cells lie in the four other families.
A cluster bootstrap with 20,000 resamples gives a 95\% interval for the above-threshold rate.
When the clusters are the seven ratio families, the interval is 84.2 to 100\% on the machine axis and 85.9 to 100\% on the balance axis.
When the clusters are machine and order, the intervals are 90.1 to 97.8\% and 86.7 to 98.2\%.
When one whole family is dropped, the rates move between 92.3 and 96.7\% and between 92.0 and 96.5\%, so no single family determines the result.
These intervals bound the resampling of the clusters that this study contains, and they say nothing about a population of GPUs or of operators.

The failures are not spread evenly over the ratios.
Fourteen of the seventeen misses above the threshold lie on the library axis $\texttt{GEMM\_CUTE}/\texttt{GEMM}$ and the epilogue axis $\texttt{GEMM\_CUTE}/\texttt{GEMM\_FUSED}$.
On these two axes, the two sides are closest to the same code (Section~\ref{sec:one-axis-evidence}).
Therefore a small absolute error is a large fraction of the ranked difference.

\begin{widetable}
\centering
\includegraphics[width=\linewidth]{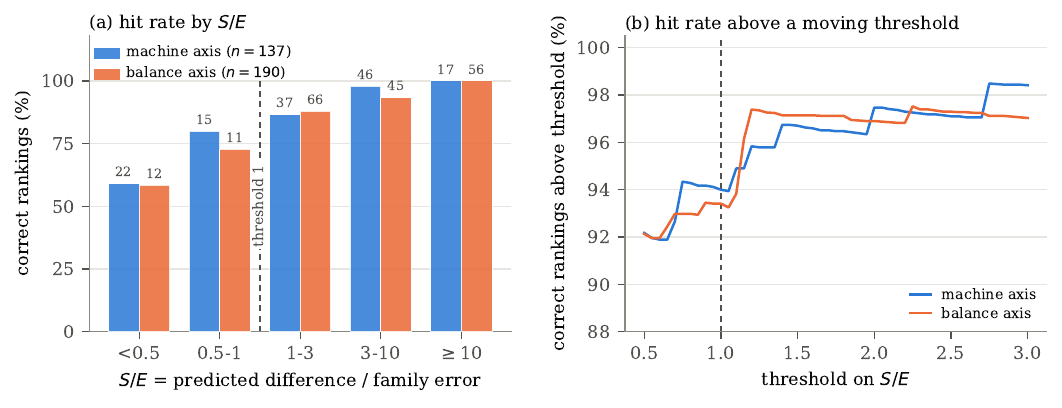}
\captionof{figure}{
  Hit rate of the empirical screening rule for a predicted ranking, on the machine axis of Section~\ref{sec:logo-three-directions} and the balance axis of Section~\ref{sec:predicting-balance-crossover}. The two axes are given equal weight.
  $S/E$ is the predicted logarithmic difference divided by the median logarithmic error of that ratio family. The median is taken leave-one-out, so both quantities are known before any target timing is inspected.
  (a) Hit rate by $S/E$, with the number of cells above each bar.
  (b) Hit rate above a moving threshold.
  Cells below the threshold are those on which the screen returns no winner.
  The rate that is plotted for them is what a forced sign would have achieved.}
\label{fig:f7}
\end{widetable}

The denominator is a median over a few cells, so we also report the hit rate by $S/E$ (Figure~\ref{fig:f7}(a)) and as a function of the threshold (Figure~\ref{fig:f7}(b)).
On the machine and balance axes, the rate above a moving threshold is 96.7 and 97.1\% at $S/E \ge 1.5$ and 97.5 and 96.9\% at $S/E \ge 2$.
At $S/E \ge 1$, the rates are 94.0 and 93.4\%.
The rise is not steady.

We call the extra time that a user pays for a wrong choice the \emph{loss}, $100\,(T_{\mathrm{chosen}}/T_{\mathrm{best}} - 1)$.
The loss is a time ratio, whereas Equation~\ref{eq:signal-to-noise} keeps the logarithmic distance $100\,|\ln r|$.
The two differ by less than 0.2 percentage points up to 6.5\% and by 4.3 points at 32\%.
In the fourteen misses on the library and epilogue axes, the measured difference is 0.2 to 6.5\%, with a median of 1.9\%.
Therefore a user who chooses the screened-in side on these two axes loses at most 6.5\% in the cells of this study.
We state no bound outside these cells.
Without these two families, the screen is correct on 66 of 69 screened-in machine-axis cells and on all 115 screened-in balance-axis cells (Table~\ref{tab:sn-by-family}).

The comparison of Section~\ref{sec:signal-to-noise-rule} of the main text with a selection by the predicted difference alone is made at equal selected-cell counts.
Therefore this comparison tests the sign hit rate of the denominator, and it does not define a rule that applies to one new cell.
The 100 machine-axis cells with the largest $|\ln r_{\mathrm{pred}}|$ are those with a predicted difference of at least 6.3\%.
The 167 balance-axis cells are those with a predicted difference of at least 1.1\%.
The two selections share 89 cells on the machine axis and 156 on the balance axis.
On the machine axis, the eleven cells that only the separation selects are correct in 9, and the eleven cells that only the criterion selects are correct in 8.
On the balance axis, the counts are 9 and 5.
The two machine-axis misses that only the separation selects are $\texttt{FUSED\_DFMA}$ ratios at $p = 15$ that the screen rejects at $S/E = 0.46$ and 0.72.
Therefore the summed loss over the wrong screened-in cells is 38 percentage points for the screen and 55 for the selection by separation alone.
On the balance axis, these losses are 25 and 24.
The three misses that only the criterion selects are $\texttt{GEMM\_CUTE}/\texttt{GEMM}$ cells whose family error of 1.3 to 4.6\% lets a predicted difference of 2.6 to 5.0\% pass.

\begin{widetable}
\captionof{table}{
  Screening and sign agreement per ratio family and axis.
  For each family, the screened-in column gives the cells at $S/E \ge 1$ with the correct sign over the screened-in cells. The below-threshold column gives the cells on which a forced sign would be correct over the cells below the threshold.
  The population is the 137 list-matched machine-axis cells and the 190 balance-axis cells defined above.
  The balance axis has no $p \le 15$ cells. The \texttt{FUSED} path is profiled through $p = 255$ only, so the two ratios that contain it have no $p = 511$ cells on either axis.}
\label{tab:sn-by-family}
\begin{tabularx}{\tblwidth}{@{}>{\hsize=2.4\hsize}L>{\hsize=.5\hsize}R>{\hsize=.9\hsize}R>{\hsize=.9\hsize}R>{\hsize=.5\hsize}R>{\hsize=.9\hsize}R>{\hsize=.9\hsize}R@{}}
\toprule
 & \multicolumn{3}{c}{Machine axis} & \multicolumn{3}{c}{Balance axis} \\
\cmidrule(lr){2-4}\cmidrule(lr){5-7}
Ratio & Cells & Screened & Below & Cells & Screened & Below \\
\midrule
$\texttt{GEMM\_FUSED}/\texttt{FUSED\_TC}$ & 20 & 18/18 & 1/2 & 30 & 24/24 & 2/6 \\
$\texttt{GEMM}/\texttt{FUSED\_TC}$ & 17 & 16/16 & 1/1 & 30 & 21/21 & 6/9 \\
$\texttt{FUSED\_DFMA}/\texttt{FUSED\_TC}$ & 21 & 14/14 & 5/7 & 30 & 30/30 & --- \\
$\texttt{GEMM\_CUTE}/\texttt{GEMM\_FUSED}$ & 23 & 22/22 & 1/1 & 30 & 22/27 & 2/3 \\
$\texttt{GEMM\_CUTE}/\texttt{GEMM}$ & 20 & 6/9 & 6/11 & 30 & 19/25 & 5/5 \\
$\texttt{FUSED\_DFMA}/\texttt{FUSED}$ & 18 & 10/13 & 4/5 & 20 & 20/20 & --- \\
$\texttt{FUSED}/\texttt{FUSED\_TC}$ & 18 & 8/8 & 7/10 & 20 & 20/20 & --- \\
\midrule
All & 137 & 94/100 & 25/37 & 190 & 156/167 & 15/23 \\
\bottomrule
\end{tabularx}
\end{widetable}

Below the threshold, the screen returns no winner.
However, a cell below the threshold is not a measured tie in general, because the family error differs between families.
On the library and epilogue axes, the family error is 0.5 to 4.6\%, and a forced sign below the threshold loses at most 5.3\%.
On the \texttt{FUSED\_DFMA} ratios, the family error is 5.9 to 29.6\% on the machine axis.
There, a below-threshold cell has a measured difference of up to 32\%, and a forced sign loses up to 15.2\%.
Therefore the choice on these ratios needs a measurement.

Run-to-run variation does not limit any of this, with one exception.
Over three rounds on each machine at $p \ge 63$, a path total reproduces to 0.1\% in the median and 1.1\% at worst.
A ratio reproduces to 0.1\% in the median and 1.3\% at worst.
At $p \le 31$, the worst ratio spread over the seven ratios is 0.9\% on GB200, 0.4\% on H100, and 2.2\% on A100.
Ten of the eleven misses on the balance axis and five of the six on the machine axis exceed their own run-to-run width.
The exception is $\texttt{GEMM\_CUTE}/\texttt{GEMM}$ on A100 at $p = 127$.
This ratio differs from 1 by 2.1\% against a width of 7.2\%, and the width comes from the A100 SM clock (1.40 to 1.13 GHz across the rounds).
Therefore the measurement does not resolve that ranking, and the miss is not evidence of a model error.
Without that cell, the machine axis is 94 of 99.
We call a cell \emph{resolved} when its measured difference from 1 exceeds its width.
Of the correct machine-axis cells above the threshold, 31 carry no width, because no repeat job covers both paths at their order.
On the resolved screened-in cells that carry a width, the sign is correct in 63 of 68 on the machine axis and 155 of 166 on the balance axis.
The width and the resolved flag of every cell are columns of \texttt{figs/data/sn\_cells.csv}.

\section{Screened rankings at low order and the choice of the fastest path}
\label{sup:low-order-and-fastest}
This section holds the cell-level tables behind Sections~\ref{sec:fastest-path-selection} and~\ref{sec:production-order-rankings} of the main text.

\subsection{Rankings in two low-order cases}
\label{sup:low-order-rankings}
Table~\ref{tab:low-order-rankings} scores the criterion of Equation~\ref{eq:signal-to-noise} at $p = 7$ and 15 on the 24 rankings of path totals that Section~\ref{sec:production-order-rankings} discusses.
The ranking $\texttt{FUSED}/\texttt{FUSED\_TC}$ reaches the threshold in none of its six cells.
On GB200 at $p = 7$, it predicts a difference of 2.7\% against a family error of 9.0\%, so $S/E$ is 0.31.
On H100, it predicts a difference of 0.6\% against a family error of 10.0\%, so $S/E$ is 0.06.
The one wrong screened-in cell is $\texttt{FUSED\_DFMA}/\texttt{FUSED}$ on H100 at $p = 15$.
This cell predicts a difference of 16.1\% against a family error of 5.9\%, so $S/E$ is 2.7.
The measured difference is 1.7\%.

\begin{table*}[tp]
\centering
\footnotesize
\caption{
  Rankings at $p = 7$ and 15 under the criterion of Equation~\ref{eq:signal-to-noise}.
  A ratio contributes one cell per order and leave-one-GPU-out direction, six in all, of which $n$ are rankings of path totals.
  The rest are quotients of partial sums, because cuBLAS does not run the same program on the two sides at these orders, and they are reported in the text.
  \emph{Screened in} counts the cells with $S/E \ge 1$ and gives the correct signs among them.
  \emph{Below thr.} counts the remainder, on which the screen returns no winner, and gives what a forced sign would have achieved there.
  True difference is the median $|\ln r_{\mathrm{meas}}|$ and error is the median $|\ln (r_{\mathrm{pred}} / r_{\mathrm{meas}})|$ over the $n$ cells, both in percent as in Table~\ref{tab:seven-ratios}. These cells are disjoint from the cells of that table.
  Only signs are scored, because these orders are outside the observed high-order accuracy regime for absolute path totals.}
\label{tab:low-order-rankings}
\begin{tabularx}{\tblwidth}{@{}>{\hsize=1.75\hsize}L>{\hsize=.45\hsize}R>{\hsize=.75\hsize}R>{\hsize=.75\hsize}R>{\hsize=.75\hsize}R>{\hsize=.60\hsize}R>{\hsize=1.95\hsize}L@{}}
\toprule
ratio & $n$ & screened in & below thr. & true diff. & error & what it compares \\
\midrule
$\texttt{FUSED\_DFMA}/\texttt{FUSED\_TC}$ & 6 & 2/2 & 2/4 & 14.7 & 15.4 & the MMA instruction (one-axis) \\
$\texttt{FUSED\_DFMA}/\texttt{FUSED}$ & 6 & 2/3 & 2/3 & 13.4 & 17.3 & MMA fragment layout \\
$\texttt{FUSED}/\texttt{FUSED\_TC}$ & 6 & --- & 4/6 & 3.3 & 4.0 & CUDA Core against Tensor Core \\
$\texttt{GEMM\_FUSED}/\texttt{FUSED\_TC}$ & 3 & 3/3 & --- & 82.0 & 6.5 & the ranking of Figure~\ref{fig:f1}(a) \\
$\texttt{GEMM\_CUTE}/\texttt{GEMM\_FUSED}$ & 3 & 3/3 & --- & 7.5 & 2.1 & epilogue fusion (one-axis) \\
$\texttt{GEMM}/\texttt{FUSED\_TC}$ & 0 & --- & --- & --- & --- & no path total at these orders \\
$\texttt{GEMM\_CUTE}/\texttt{GEMM}$ & 0 & --- & --- & --- & --- & no path total at these orders \\
\midrule
total & 24 & 10/11 & 8/13 & & & \\
\bottomrule
\end{tabularx}
\end{table*}

\subsection{Choice of the fastest path}
\label{sup:fastest-path-selection}
Table~\ref{tab:fastest-path} lists the 26 target-and-order cells behind Section~\ref{sec:fastest-path-selection}.
The simpler rule of that section takes the path that is fastest on a calibration machine at the same order.
This rule is correct in 30 of the 50 combinations of a target cell and a calibration machine.
Of its 20 wrong choices, 15 cost more than 5\%, and the largest costs 34.0\%.
The two calibration machines of a direction agree in 15 of the 24 cells that have two, and in these cells they are right in 12.
In the other nine cells, the rule gives no answer.
A cell is a target machine and an order.
The paths of a cell are those whose kernel lists match on the target (Section~\ref{sec:logo-three-directions}).
Therefore the three library paths are absent at $p = 7$, and \texttt{GEMM} is absent at $p = 15$.
The paths that were not profiled at an order on the target or on the calibration machines are absent as well.
The $S/E$ against each other path uses the family error of that pair in the direction of the cell.
Eight pairs are not among the seven scored ratios, for example, \texttt{GEMM} against \texttt{GEMM\_FUSED} and \texttt{GEMM\_CUTE} against \texttt{FUSED\_TC}.
The family errors of these eight pairs are formed in the same way as those of the seven, over the transfers in which the target machine takes no part (Section~\ref{sup:sn-scoring}).
\texttt{figs/data/sn\_cells\_all\_pairs.csv} holds every pair (Section~\ref{sup:campaigns}).
The width columns of that file cover the ratio of the claimed path to every other path of the cell in eight of the 15 claimed cells.
These cells are GB200 at $p = 31$, 575, and 767, H100 at $p = 255$, 575, and 767, and A100 at $p = 127$ and 575.
Every one of those pairs is resolved in the sense of Section~\ref{sup:sn-scoring}.
No pair is covered in four claimed cells: GB200 at $p = 127$ and 255, H100 at $p = 63$, and A100 at $p = 255$.
The other three claimed cells are GB200 at $p = 63$ and 511 and H100 at $p = 511$.
In these three cells, the pairs against \texttt{FUSED\_DFMA} have no width, and on GB200 at $p = 63$ the pairs against \texttt{FUSED} also have no width.

Table~\ref{tab:fastest-path-preferred} repeats the choice on H100 with the preferred-shape candidate set of Section~\ref{sec:fastest-path-selection}.
The totals of \texttt{GEMM\_CUTE} and \texttt{GEMM\_FUSED} are those of the $16\times 8\times 4$ binary.
These totals are predicted from GB200 and A100 and measured on H100 in the profiled window.
The totals of the other paths are those of Table~\ref{tab:fastest-path}.
At $p = 7$, the two shape-dependent paths are absent because their kernel lists do not match those of the calibration machines, as in Table~\ref{tab:fastest-path}.
No $S/E$ is listed, because no family error exists at the preferred shape (Section~\ref{sec:operational-predictions}).

\begin{table*}[tp]
\centering
\footnotesize
\caption{
  Choice of the fastest path in the 26 leave-one-GPU-out cells whose kernel lists match.
  $n$ is the number of paths in the cell.
  \emph{Loss} is the measured time of the predicted fastest path relative to the measured fastest path, in percent, so 0.0 means that the two paths are the same.
  \emph{Competitor} is the path that gives the smallest $S/E$ of Equation~\ref{eq:signal-to-noise} against the predicted fastest path among the other paths of the cell. $S/E$ is that value.
  A fastest path is \emph{claimed} when $S/E \ge 1$.}
\label{tab:fastest-path}
\begin{tabularx}{\tblwidth}{@{}>{\hsize=.7\hsize}L>{\hsize=.5\hsize}R>{\hsize=.4\hsize}R>{\hsize=1.4\hsize}L>{\hsize=1.4\hsize}L>{\hsize=.6\hsize}R>{\hsize=1.4\hsize}L>{\hsize=.9\hsize}R>{\hsize=.7\hsize}L@{}}
\toprule
target & $p$ & $n$ & predicted fastest & measured fastest & loss & competitor & $S/E$ & claimed \\
\midrule
GB200 & 7 & 3 & \texttt{FUSED\_TC} & \texttt{FUSED\_TC} & 0.0 & \texttt{FUSED} & 0.31 & no \\
GB200 & 15 & 5 & \texttt{FUSED\_TC} & \texttt{FUSED\_TC} & 0.0 & \texttt{FUSED} & 0.83 & no \\
GB200 & 31 & 6 & \texttt{FUSED\_TC} & \texttt{FUSED\_TC} & 0.0 & \texttt{FUSED} & 1.49 & yes \\
GB200 & 63 & 6 & \texttt{FUSED\_TC} & \texttt{FUSED\_TC} & 0.0 & \texttt{FUSED} & 3.81 & yes \\
GB200 & 127 & 6 & \texttt{FUSED\_TC} & \texttt{FUSED\_TC} & 0.0 & \texttt{GEMM\_FUSED} & 2.00 & yes \\
GB200 & 255 & 6 & \texttt{FUSED\_TC} & \texttt{FUSED\_TC} & 0.0 & \texttt{GEMM\_CUTE} & 1.76 & yes \\
GB200 & 511 & 5 & \texttt{GEMM\_FUSED} & \texttt{GEMM\_FUSED} & 0.0 & \texttt{FUSED\_TC} & 1.62 & yes \\
GB200 & 575 & 3 & \texttt{GEMM\_FUSED} & \texttt{GEMM\_FUSED} & 0.0 & \texttt{GEMM\_CUTE} & 2.08 & yes \\
GB200 & 767 & 4 & \texttt{GEMM\_FUSED} & \texttt{GEMM\_FUSED} & 0.0 & \texttt{GEMM\_CUTE} & 1.95 & yes \\
H100 & 7 & 3 & \texttt{FUSED\_TC} & \texttt{FUSED} & 1.0 & \texttt{FUSED\_DFMA} & 0.04 & no \\
H100 & 15 & 5 & \texttt{FUSED\_DFMA} & \texttt{FUSED\_TC} & 9.2 & \texttt{FUSED\_TC} & 0.72 & no \\
H100 & 31 & 6 & \texttt{FUSED\_TC} & \texttt{FUSED\_TC} & 0.0 & \texttt{FUSED\_DFMA} & 0.13 & no \\
H100 & 63 & 6 & \texttt{FUSED} & \texttt{FUSED} & 0.0 & \texttt{FUSED\_TC} & 1.35 & yes \\
H100 & 127 & 6 & \texttt{FUSED\_TC} & \texttt{FUSED\_TC} & 0.0 & \texttt{FUSED} & 0.37 & no \\
H100 & 255 & 6 & \texttt{FUSED\_TC} & \texttt{FUSED\_TC} & 0.0 & \texttt{GEMM} & 3.18 & yes \\
H100 & 511 & 5 & \texttt{GEMM} & \texttt{GEMM} & 0.0 & \texttt{FUSED\_TC} & 1.24 & yes \\
H100 & 575 & 3 & \texttt{GEMM} & \texttt{GEMM} & 0.0 & \texttt{GEMM\_FUSED} & 3.31 & yes \\
H100 & 767 & 4 & \texttt{GEMM} & \texttt{GEMM} & 0.0 & \texttt{FUSED\_TC} & 3.24 & yes \\
A100 & 7 & 3 & \texttt{FUSED} & \texttt{FUSED\_TC} & 0.5 & \texttt{FUSED\_TC} & 0.12 & no \\
A100 & 15 & 5 & \texttt{FUSED\_DFMA} & \texttt{FUSED\_TC} & 15.2 & \texttt{FUSED\_TC} & 0.07 & no \\
A100 & 31 & 6 & \texttt{FUSED\_TC} & \texttt{FUSED\_TC} & 0.0 & \texttt{FUSED} & 0.14 & no \\
A100 & 63 & 6 & \texttt{FUSED} & \texttt{FUSED\_TC} & 9.6 & \texttt{FUSED\_TC} & 0.34 & no \\
A100 & 127 & 6 & \texttt{FUSED\_TC} & \texttt{FUSED\_TC} & 0.0 & \texttt{FUSED} & 1.52 & yes \\
A100 & 255 & 6 & \texttt{FUSED\_TC} & \texttt{FUSED\_TC} & 0.0 & \texttt{FUSED\_DFMA} & 2.72 & yes \\
A100 & 511 & 5 & \texttt{FUSED\_TC} & \texttt{FUSED\_TC} & 0.0 & \texttt{GEMM\_FUSED} & 0.998 & no \\
A100 & 575 & 3 & \texttt{GEMM\_FUSED} & \texttt{GEMM\_FUSED} & 0.0 & \texttt{GEMM\_CUTE} & 4.06 & yes \\
\bottomrule
\end{tabularx}
\end{table*}

\begin{table}[tp]
\centering
\footnotesize
\caption{
  Choice of the fastest path on H100 with the preferred-shape candidate set, in which \texttt{GEMM\_CUTE} and \texttt{GEMM\_FUSED} use the $16\times 8\times 4$ shape.
  $n$, \emph{predicted fastest}, \emph{measured fastest} and \emph{loss} are as in Table~\ref{tab:fastest-path}.
  \emph{Margin} is the measured time of the second fastest path relative to the measured fastest path, in percent.
  The criterion is not evaluated at the preferred shape, so no path is claimed.}
\label{tab:fastest-path-preferred}
\begin{tabularx}{\tblwidth}{@{}>{\hsize=.5\hsize}R>{\hsize=.4\hsize}R>{\hsize=1.4\hsize}L>{\hsize=1.4\hsize}L>{\hsize=.6\hsize}R>{\hsize=.7\hsize}R@{}}
\toprule
$p$ & $n$ & predicted fastest & measured fastest & loss & margin \\
\midrule
7 & 3 & \texttt{FUSED\_TC} & \texttt{FUSED} & 1.0 & 1.0 \\
15 & 5 & \texttt{FUSED\_DFMA} & \texttt{FUSED\_TC} & 9.2 & 7.4 \\
31 & 6 & \texttt{FUSED\_TC} & \texttt{FUSED\_TC} & 0.0 & 2.6 \\
63 & 6 & \texttt{FUSED} & \texttt{FUSED} & 0.0 & 2.7 \\
127 & 6 & \texttt{FUSED\_TC} & \texttt{FUSED\_TC} & 0.0 & 26.9 \\
255 & 6 & \texttt{FUSED\_TC} & \texttt{FUSED\_TC} & 0.0 & 9.7 \\
511 & 5 & \texttt{GEMM\_FUSED} & \texttt{GEMM\_FUSED} & 0.0 & 6.8 \\
575 & 3 & \texttt{GEMM\_FUSED} & \texttt{GEMM\_FUSED} & 0.0 & 6.8 \\
767 & 4 & \texttt{GEMM\_FUSED} & \texttt{GEMM\_FUSED} & 0.0 & 2.8 \\
\bottomrule
\end{tabularx}
\end{table}

\section{Register-budget cost of the iso-schedule ablation}
\label{sup:register-budget-cost-iso}

\texttt{FUSED\_DFMA} and \texttt{FUSED\_TC} share one source and one \texttt{\_\_launch\_bounds\_\_} directive (Section~\ref{sec:fused-tensor-core-path}).
Therefore the DFMA control is compiled under a register budget that was selected for the Tensor Core kernels.
The DFMA form keeps more values live, and it spills where the Tensor Core form does not.
At $p = 15$ and $p = 31$, neither form spills.
At $p = 7$, 63, 127, and 255, the DFMA kernels report 32 B, 16 B, 104 B in $y$ plus 112 B in $xz$, and 56--80 B in each of four instances.
A second build lets only the DFMA version take a different second \texttt{\_\_launch\_bounds\_\_} argument.
It is outside the production dispatch.
$A_{\mathrm{iso}}$ denotes $\texttt{FUSED\_DFMA}/\texttt{FUSED\_TC}$ under the shared budget, which is the ablation of Section~\ref{sec:fused-tensor-core-path}.
$A_{\mathrm{opt}}$ denotes the ratio under the best DFMA budget of a sweep through $p = 255$.
At $p = 511$, only one budget was tested.

Table~\ref{tab:a-measured-cost} reports CUDA device time per Runge--Kutta stage.

\begin{widetable}
\captionof{table}{Cost of the shared launch-bound budget. CUDA device time per Runge--Kutta stage for \texttt{FUSED\_TC}, for \texttt{FUSED\_DFMA} under the shared production budget ($A_{\mathrm{iso}}$), and for \texttt{FUSED\_DFMA} under its own best measured budget ($A_{\mathrm{opt}}$). All variants were measured within a single job on an exclusively allocated GPU, with twelve interleaved rounds. The entries are medians. The measured ranges do not overlap except where noted. The per-stage values are the CUDA-event sum divided by the $N + 1$ calls that it covers (Section~\ref{sup:delayed-cuda-event-harvesting}).}
\label{tab:a-measured-cost}
\begin{tabularx}{\tblwidth}{@{}lRRRRRL@{}}
\toprule
$p$ & \texttt{FUSED\_TC} [$\mu$s] & \texttt{FUSED\_DFMA} iso [$\mu$s] & \texttt{FUSED\_DFMA} own budget [$\mu$s] & $A_{\mathrm{iso}}$ & $A_{\mathrm{opt}}$ & own $-$ iso \\
\midrule
7 & 250.2 & 328.5 & 316.2 & 1.313 & 1.264 & $-$3.8\% \\
15 & 268.3 & 440.2 & 440.5 & 1.641 & 1.642 & +0.07\% (overlapping) \\
31 & 356.7 & 789.7 & 789.6 & 2.214 & 2.214 & $-$0.01\% (overlapping) \\
63 & 400.9 & 822.6 & 843.6 & 2.052 & 2.104 & +2.6\% \\
127 & 590.5 & 1452.1 & 1387.3 & 2.459 & 2.349 & $-$4.5\% \\
255 & 868.6 & 1972.9 & 1780.7 & 2.271 & 2.050 & $-$9.7\% \\
511 & 12 846 & 29 716 & 27 189 & 2.313 & 2.117$\dagger$ & $-$8.5\% \\
\bottomrule
\end{tabularx}
\end{widetable}

$\dagger$ Single tested DFMA-specific budget, not the result of a sweep.

At $p = 15$ and $p = 31$, the kernels are byte-identical between the builds, so those rows measure run-to-run reproducibility.
The relaxed budget is 2.6\% slower at $p = 63$ and saves 4.5\% at $p = 127$ and 9.7\% at $p = 255$.
Therefore the register-budget penalty inside the iso-schedule ratio is negligible at some orders and approaches 10\% at $p = 255$.

A larger budget removes executed spill operations, but it can lower the resident blocks per SM.
The executed count of spill operations is what matters.
At $p = 7$ and $p = 63$, the spilling kernels have no backward branch and execute six to eight spill operations per warp.
At $p = 127$ and $p = 255$, the spills sit inside the chunk loop and execute 100 to 422 times per warp in each affected kernel (Table~\ref{tab:a-spill-residency}).

\begin{apptable}
\captionof{table}{Spill removal against resident warps. At each order at which the relaxation of the shared register budget removes spills, the table gives the spill loads and stores that are removed per warp, the accompanying change in resident blocks per SM, and the measured change in device time. Negative values denote a speedup.}
\label{tab:a-spill-residency}
\noindent\begin{tabularx}{\tblwidth}{@{}LRRL@{}}
\toprule
$p$ & spill loads/stores per warp removed & affected residency & measured change \\
\midrule
7 & 8 & 6 $\rightarrow$ 5 blocks ($-$17\%) & $-$3.8\% \\
63 & 6 & 2 $\rightarrow$ 1 block ($-$50\%) & +2.6\% \\
127 & 620 ($y$; 212 remain in $xz$) & 2 $\rightarrow$ 1 block ($-$50\%) & $-$4.5\% \\
255 & 1576 across three kernels & 3 $\rightarrow$ 2 blocks ($-$33\%) & $-$9.7\% \\
\bottomrule
\end{tabularx}
\end{apptable}

The $p = 63$ and $p = 127$ $y$ kernels take the same residency step, two blocks to one.
Their removed spill counts differ by two orders of magnitude, and their signs differ accordingly.
At $p = 7$, the shared-budget kernel is at 93.4\% of peak L1/TEX throughput with 8.4 million spill sectors.
The removal lowers the long-scoreboard stall from 4.43 to 2.80 warps per issue at unchanged issue-active (31.8\%).
At $p = 127$, spill traffic in $y$ is 4.2 times its global-load traffic.
Its removal cuts executed instructions by 22.4\%, and the lower residency raises the math-pipe-throttle stall from 0.51 to 1.41.
At $p = 255$, short-scoreboard stalls fall from 2.43 to 0.57, and FP64-pipe utilization rises from 34.0 to 37.7\%.
At the same time, achieved occupancy falls from 18.2 to 12.3\%.
At $p = 63$, the removal lowers L1/TEX utilization from 95.1 to 88.1\% and halves the mio-throttle stall.
However, the step to one resident block multiplies the long-scoreboard and lg-throttle stalls by 15 and by 7.
That exposed latency outweighs the small spill saving.

$A_{\mathrm{opt}}$ is not the instruction cost after the budget effect is removed, for three reasons.
A different budget also changes the resident-block count.
Therefore $A_{\mathrm{opt}}$ mixes the instruction difference with an occupancy difference, whereas $A_{\mathrm{iso}}$ holds the launch configuration fixed.
The ratio $A_{\mathrm{iso}}$ therefore remains the ablation that the paper reports.
The sweep does not remove all spills.
At $p = 127$, the $xz$ kernel is already at the minimum second argument and keeps 112 B.
In addition, the $p = 511$ entry is one inherited setting, not an optimum.
The thread shapes are shared as well.
Trial DFMA variants that also retune the thread shape are outside the iso-schedule ablation.
These variants are 28.3\% faster at $p = 31$, 7.9\% at $p = 63$, and 26.5\% at $p = 127$.
The $p = 31$ and $p = 127$ values are the ones that Section~\ref{sec:one-axis-evidence} quotes.
These gains appear although the budget axis alone costs nothing at $p = 31$ and the largest $p = 63$ gain comes from a kernel that does not spill.
The budget rows therefore price one identified component of the cost of the shared schedule and do not bound it.
When the same alternative thread shapes are applied to \texttt{FUSED\_TC}, they are slower in all nine variants at $p = 31$ and $p = 63$, by 2.1\% to 200\% with non-overlapping ranges.
There, the Tensor Core kernel is at only 63\% of peak L1/TEX throughput, and arithmetic-pipe and global-memory-latency stalls dominate.
Therefore trading memory traffic for resident warps buys nothing.

The own-budget and shared-budget builds produce bit-identical owned \texttt{dqdt} values at every tested order under pointwise-varying coefficients.
Both builds agree with the reference to a maximum absolute difference of $3.6\times 10^{-15}$ over 16,777,216 points.
The $N_e > 1$ check at $p = 255$ (134,217,728 points, same maximum difference) was run in the own-budget configuration only.
Therefore this check supports the reference agreement and not the agreement between builds.

\section{Absolute timing measurements}
\label{sup:absolute-timing-measurements}

This section reports the absolute timing values behind the timing definitions of Section~\ref{sec:timing-protocol}.
It also reports the measurements that support the profiling rules of Section~\ref{sec:profiling-protocol}.
The data are obtained from the same raw logs and the same aggregation procedure as Figures~\ref{fig:f1}(a), \ref{fig:f2}, \ref{fig:a11a}(a), and \ref{fig:f3}.
The two added GB200 fused-path points at $p = 767$ are described in Section~\ref{sup:data-provenance-reproducibility}.

Each cell of Tables~\ref{tab:a-timing-gb200}--\ref{tab:a-timing-a100} gives two quantities from the same run, in the form \emph{stage time / device time}.
The stage time is the wall-clock time of the time-step loop per Runge--Kutta stage.
The device time is the CUDA-event time of the tendency evaluation per stage.
The stage time includes the Runge--Kutta update, the halo exchange, and the launch gaps at stage boundaries.
The device time covers only the instrumented tendency region.
Therefore, the two must not be interchanged.
Unless marked otherwise, times are in microseconds per stage, and values are medians over interleaved measurement rounds (Section~\ref{sec:timing-protocol}).
A dash means that the path was not run at that order.
For $p \le 255$, the number of volume degrees of freedom is fixed at 16,777,216 and the number of elements changes from $32^3$ to $1^3$.
For $p \ge 511$, a single element is used and the problem size grows as $N_q^3$.
Therefore, absolute times across this boundary cannot be compared directly.
The raw logs also contain the reference CUDA Fortran baseline (\texttt{CUDAFORTRAN\_SPLIT}) and the \texttt{@16x8x4} shape variants.
The tables omit them, so that the tables use the six paths of Table~\ref{tab:paths}.

\begin{widetable}
\captionof{table}{Absolute time per Runge--Kutta stage on the GB200 system (RIKYU). Each cell gives \emph{stage time / device time}. Here and in Tables~\ref{tab:a-timing-h100} and~\ref{tab:a-timing-a100}, the device value is the CUDA-event sum divided by $N$. This divisor is the number of timed stages of Table~\ref{tab:a-configs}. It is also the number of tendency calls that the sum of these executables covers (Section~\ref{sup:delayed-cuda-event-harvesting}). The raw sums are in the archived logs \citep{thiswork-data}.}
\label{tab:a-timing-gb200}
\begin{tabularx}{\tblwidth}{@{}lRRRRRR@{}}
\toprule
$p$ & FUSED & FUSED\_DFMA & FUSED\_TC & GEMM & GEMM\_CUTE & GEMM\_FUSED \\
\midrule
7 & 374 / 287 & 415 / 329 & 336 / 249 & 1670 / 1590 & 1530 / 1440 & 1540 / 1460 \\
15 & 373 / 286 & 527 / 439 & 355 / 268 & 902 / 813 & 721 / 633 & 677 / 588 \\
31 & 509 / 421 & 872 / 785 & 443 / 355 & 697 / 609 & 683 / 594 & 633 / 544 \\
63 & 543 / 455 & 911 / 822 & 489 / 401 & 688 / 598 & 691 / 602 & 640 / 551 \\
127 & 909 / 821 & 1540 / 1450 & 679 / 590 & 801 / 711 & 811 / 721 & 723 / 634 \\
255 & 1500 / 1420 & 2060 / 1970 & 956 / 867 & 1120 / 1030 & 1110 / 1020 & 1020 / 926 \\
511 & --- & 30.4 ms / 29.7 ms & 13.5 ms / 12.8 ms & 13.7 ms / 13.0 ms & 13.7 ms / 13.1 ms & 12.8 ms / 12.2 ms \\
575 & --- & --- & --- & 21.4 ms / 20.5 ms & 21.4 ms / 20.5 ms & 20.0 ms / 19.2 ms \\
767 & --- & --- & --- & 63.5 ms / 61.4 ms & 63.8 ms / 61.7 ms & 60.5 ms / 58.4 ms \\
1023 & --- & --- & --- & 193 ms / 188 ms & 193 ms / 188 ms & 185 ms / 180 ms \\
\bottomrule
\end{tabularx}
\end{widetable}

\begin{widetable}
\captionof{table}{Absolute time per Runge--Kutta stage on the H100 SXM 94 GB system (TSUBAME 4). Each cell gives \emph{stage time / device time}. The device values are the CUDA-event sum divided by $N$, as in Table~\ref{tab:a-timing-gb200}.}
\label{tab:a-timing-h100}
\begin{tabularx}{\tblwidth}{@{}lRRRRRR@{}}
\toprule
$p$ & FUSED & FUSED\_DFMA & FUSED\_TC & GEMM & GEMM\_CUTE & GEMM\_FUSED \\
\midrule
7 & 1040 / 788 & 1180 / 927 & 1050 / 795 & 2980 / 2730 & 2880 / 2630 & 2830 / 2580 \\
15 & 1020 / 759 & 1020 / 763 & 953 / 698 & 2100 / 1840 & 1990 / 1740 & 1890 / 1630 \\
31 & 1200 / 937 & 1420 / 1170 & 1170 / 917 & 1790 / 1530 & 1790 / 1540 & 1670 / 1420 \\
63 & 1130 / 869 & 1380 / 1120 & 1140 / 880 & 1670 / 1410 & 1680 / 1430 & 1580 / 1320 \\
127 & 1540 / 1280 & 2480 / 2220 & 1210 / 951 & 1620 / 1360 & 1740 / 1480 & 1570 / 1310 \\
255 & 2920 / 2670 & 2800 / 2540 & 1520 / 1260 & 1710 / 1450 & 2040 / 1780 & 1780 / 1520 \\
511 & --- & 39.1 ms / 37.1 ms & 18.1 ms / 16.1 ms & 16.4 ms / 14.4 ms & 22.2 ms / 20.2 ms & 19.2 ms / 17.2 ms \\
575 & --- & --- & --- & 24.6 ms / 21.8 ms & 33.8 ms / 31.0 ms & 29.4 ms / 26.6 ms \\
767 & --- & --- & --- & 66.9 ms / 60.3 ms & 97.0 ms / 90.4 ms & 86.0 ms / 79.4 ms \\
\bottomrule
\end{tabularx}
\end{widetable}

\begin{widetable}
\captionof{table}{Absolute time per Runge--Kutta stage on the A100 SXM 40 GB system (Wisteria/BDEC-01 Aquarius). Each cell gives \emph{stage time / device time}. The device values are the CUDA-event sum divided by $N$, as in Table~\ref{tab:a-timing-gb200}.}
\label{tab:a-timing-a100}
\begin{tabularx}{\tblwidth}{@{}lRRRRRR@{}}
\toprule
$p$ & FUSED & FUSED\_DFMA & FUSED\_TC & GEMM & GEMM\_CUTE & GEMM\_FUSED \\
\midrule
7 & 1690 / 1290 & 1690 / 1290 & 1680 / 1280 & 5620 / 5210 & 5330 / 4930 & 5170 / 4770 \\
15 & 1620 / 1210 & 1790 / 1380 & 1600 / 1200 & 3610 / 3200 & 3220 / 2820 & 3120 / 2710 \\
31 & 2030 / 1620 & 2280 / 1880 & 1890 / 1480 & 2900 / 2490 & 2870 / 2460 & 2720 / 2310 \\
63 & 2080 / 1670 & 2590 / 2180 & 1930 / 1530 & 2750 / 2340 & 2710 / 2300 & 2580 / 2170 \\
127 & 2620 / 2220 & 3900 / 3490 & 2150 / 1730 & 2870 / 2460 & 2840 / 2430 & 2590 / 2180 \\
255 & 4850 / 4440 & 5850 / 5440 & 2540 / 2130 & 3400 / 2990 & 3370 / 2960 & 3020 / 2610 \\
511 & --- & 82.9 ms / 79.8 ms & 32.0 ms / 28.8 ms & 37.1 ms / 33.9 ms & 37.9 ms / 34.7 ms & 34.2 ms / 31.0 ms \\
575 & --- & --- & --- & 58.4 ms / 53.9 ms & 58.0 ms / 53.5 ms & 53.0 ms / 48.5 ms \\
\bottomrule
\end{tabularx}
\end{widetable}

\subsection{Effect of the CUDA-graph replay at the production orders}
\label{sup:graph-replay}
Table~\ref{tab:a-graph-replay} gives the time that the graph replay removes from the end-to-end time per step, on the three machines at $p = 7$ and 15, for the six paths.
This time is the difference between the graph-off median and the graph-on median over six interleaved rounds (Section~\ref{sup:data-provenance-reproducibility} lists the jobs).
On the fused paths, the saving is 23 to 32 $\mu$s per step on every machine and at both orders.
This saving is 8 to 11 $\mu$s per Runge--Kutta stage, which is the cost of the launches of one stage.
On the library paths, the saving is larger, because they launch more kernels per stage.
It is largest on \texttt{GEMM\_FUSED} on A100 at $p = 15$ (204 $\mu$s).
As a fraction of the time per step, the saving is 1.6 to 3.5\% on GB200 and 0.5 to 2.2\% on H100 and A100.
The rankings are read in step-loop time per stage, as in Figure~\ref{fig:f1}.
In the six cells, none of the 90 ratios of two paths at the common shape crosses 1 between the two configurations.
The ratios that involve the $16\times 8\times 4$ variants on H100 do not cross either.
The measured fastest path of every cell is the same in both configurations.
Among the ratios within 20\% of 1, the largest movement is 1.9 percentage points ($\texttt{GEMM}/\texttt{GEMM\_FUSED}$ on A100 at $p = 15$, from 1.160 to 1.179).
The ratio closest to 1 is $\texttt{FUSED\_DFMA}/\texttt{FUSED}$ on A100 at $p = 7$ at 0.9992, and it moves by less than 0.01 points.
$\texttt{FUSED\_TC}/\texttt{FUSED}$ is the ratio of Figure~\ref{fig:f1}(b).
It reads 0.899 and 0.896 on GB200, 1.007 and 1.007 on H100, and 0.993 and 0.993 on A100, where the first value is with the graph off and the second is with the graph on.

\begin{apptable}
\captionof{table}{Time that the CUDA-graph replay removes from the end-to-end time per step, in $\mu$s, as the graph-off median minus the graph-on median over six interleaved rounds. On H100 the two CUTLASS paths are at the common shape $8\times 8\times 4$. At $16\times 8\times 4$, the values for \texttt{GEMM\_CUTE} and \texttt{GEMM\_FUSED} are 73 and 143 $\mu$s, respectively, at $p = 7$, and 44 and 94 $\mu$s, respectively, at $p = 15$.}
\label{tab:a-graph-replay}
\noindent\begin{tabularx}{\tblwidth}{@{}lrRRRRRR@{}}
\toprule
Machine & $p$ & \texttt{FUSED} & \texttt{FUSED\_DFMA} & \texttt{FUSED\_TC} & \texttt{GEMM} & \texttt{GEMM\_CUTE} & \texttt{GEMM\_FUSED} \\
\midrule
GB200 & 7 & 28.4 & 28.6 & 29.0 & 79.7 & 74.6 & 82.1 \\
GB200 & 15 & 30.8 & 28.8 & 30.9 & 48.1 & 55.2 & 70.7 \\
H100 & 7 & 24.3 & 24.2 & 24.8 & 46.7 & 73.4 & 145.5 \\
H100 & 15 & 23.4 & 26.8 & 24.5 & 35.1 & 40.1 & 106.7 \\
A100 & 7 & 29.0 & 28.9 & 29.5 & 78.6 & 88.4 & 93.5 \\
A100 & 15 & 31.7 & 32.2 & 28.8 & 61.9 & 56.1 & 203.9 \\
\bottomrule
\end{tabularx}
\end{apptable}

\subsection{Difference between stage and device timing}
\label{sup:difference-between-stage-device}
The ratio of device time to stage time quantifies the fraction of the measured stage that the instrumented tendency evaluation occupies.
Across the six implementation paths of Table~\ref{tab:paths}, the ranges are those of Table~\ref{tab:a-device-stage-ratio}.
For example, when the fastest path on each machine is compared at $p = 767$, H100 requires 10.5\% more step-loop time than GB200 (Tables~\ref{tab:a-timing-gb200} and~\ref{tab:a-timing-h100}).
However, a sum of only the tendency kernels in profiling runs of the same two configurations differs by 2.6\%.
The gap between the two percentages is dominated by the Runge--Kutta update, which is a pure bandwidth kernel and lies outside the tendency evaluation.
The two percentages come from different runs and different scopes: the sweep campaign at production clocks, and separate Nsight Compute runs with unlocked clocks.

\begin{apptable}
\captionof{table}{Ratio of CUDA device time to step-loop stage time, by polynomial order and machine. Each entry is the range of that ratio over the six implementation paths of Table~\ref{tab:paths}. A dash indicates that the order was not run on that machine. The device times are those of Tables~\ref{tab:a-timing-gb200}--\ref{tab:a-timing-a100}.}
\label{tab:a-device-stage-ratio}
\begin{tabular}{lrrr}
\toprule
$p$ & GB200 & H100 & A100 \\
\midrule
7 & 0.742--0.947 & 0.756--0.914 & 0.760--0.928 \\
15 & 0.753--0.901 & 0.733--0.877 & 0.747--0.887 \\
31 & 0.801--0.900 & 0.781--0.856 & 0.784--0.860 \\
63 & 0.819--0.903 & 0.770--0.847 & 0.789--0.852 \\
127 & 0.869--0.942 & 0.784--0.895 & 0.808--0.894 \\
255 & 0.907--0.957 & 0.831--0.912 & 0.838--0.929 \\
511 & 0.951--0.979 & 0.879--0.949 & 0.901--0.962 \\
575 & 0.956--0.959 & 0.886--0.917 & 0.915--0.923 \\
767 & 0.966--0.968 & 0.901--0.932 & --- \\
1023 & 0.974--0.975 & --- & --- \\
\bottomrule
\end{tabular}
\end{apptable}

At $p = 7$, work outside the tendency evaluation is 5 to 26\% of the stage, depending on the machine and on the path.
The fraction falls with order, because the tendency grows much faster than the remainder.
At $p \ge 511$, the two measures differ by 2 to 12\%.
At a given order, the ratio is generally smaller on H100 and A100 than on GB200.

\subsection{The CUDA-event sum covers one call more than the timed stages}
\label{sup:delayed-cuda-event-harvesting}
The CUDA device timer reads back each event pair one tendency call later, so that no read-back stalls the GPU within a stage.
The accumulator is reset at the beginning of the measurement interval, when the measurement of the last warm-up call is still pending.
The first timed call reads this pending measurement back and adds it to the sum.
The measurement of the last timed call is read back at the end of the run.
The sum over $N$ timed stages therefore covers $N + 1$ intervals.
The executables of the base three-machine sweep (Section~\ref{sup:campaigns}) discard the pending measurement at the reset.
Therefore, their sum covers exactly the $N$ timed stages in the last column of Table~\ref{tab:a-configs}.
Every run of these executables reports the number of calls that it summed.
Every other CUDA-event measurement in the main text and in this document was taken with earlier executables that behave as described above.
If their sums were divided by $N$, they would be high by the factor $(N + 1)/N$.
This factor ranges from 0.17\% at $p = 31$ to 4.76\% at $p = 1023$.
Every absolute device time from those executables therefore divides the sum by $N + 1$.
Table~\ref{tab:a-event-harvesting} below is the one exception, because it is the measurement that establishes the count.
The additional interval is an ordinary stage.
Therefore, the factor is the same for every path that runs with the same configuration, and ratios between paths that are measured in the same job are unchanged.
The step-loop time, the end-to-end time, and the Nsight Compute and Nsight Systems measurements do not use this sum.

The effect was found at $p = 1023$, where $N$ is smallest.
There, the sum divided by $N$ gave a device-to-stage ratio of 1.020--1.021, which cannot be a physical decomposition of the stage.
Table~\ref{tab:a-event-harvesting} tests this with a frozen executable that behaves in that way and the published $p = 1023$ configuration.
Only \texttt{output\_interval} was changed to 100, to remove the diagnostic output.

\begin{apptable}
\captionof{table}{Effect of the measurement interval on the $p = 1023$ device time on GB200. Stage time and device time per Runge--Kutta stage over a short and a long measurement interval. The values are from a frozen executable whose sum covers $N + 1$ intervals, and the input is the published $p = 1023$ configuration without diagnostic output. The device time is the CUDA-event sum divided by the number of timed stages $N$.}
\label{tab:a-event-harvesting}
\noindent\begin{tabularx}{\tblwidth}{@{}LRRRR@{}}
\toprule
measured steps & stages & stage time & device time / stage & ratio \\
\midrule
7 & 21 & 193.1 ms & 197.2 ms & 1.021 \\
21 & 63 & 193.1 ms & 191.2 ms & 0.990 \\
\bottomrule
\end{tabularx}
\end{apptable}

The removal of the diagnostic output leaves the 21-stage ratio unchanged.
When the interval is lengthened to 63 stages, the discrepancy shrinks in proportion to $1/N$, as it must if one additional interval is included.
If the measured device average is $(N s + e)/N$, the two rows give a steady-state device time $s = 188.2$~ms per stage and an additional interval $e = 189.0$~ms.
The additional interval is therefore 1.004 times an ordinary stage, and the count alone explains the extra.
The corrected device-to-stage ratio at $p = 1023$ is 0.974--0.975.
The stage wall time is 193.1~ms in both runs, so the additional interval lies outside the measured wall-clock window.
An interleaved rerun of one $p = 7$ cell reproduces the factor directly.
This rerun uses an executable that discards the pending measurement at the reset.
The sum of the original executable is 1.0150 times the sum of the corrected one, against $70/69 = 1.0145$ for the 69 timed stages of that configuration \citep{thiswork-data}.

Every absolute device time and every fraction of a hardware roof in this paper is therefore the sum divided by the number of calls that it covers.
This number is $N$ for the base sweeps and $N + 1$ for the measurements from the earlier executables, and these measurements include Table~\ref{tab:nesweep} of the main text.
The raw sums are in the archived logs \citep{thiswork-data}.
The stage-time results of Figure~\ref{fig:f2} and Sections~\ref{sec:performance-crossovers} and~\ref{sec:why-optimum-moves-order} do not use the sum.

\section{Benchmark configurations, implementation details, validation coverage, and profiler settings}
\label{sup:benchmark-configurations}

This section lists the published benchmark configurations, the per-order kernel structure of the fused CUDA Core path and the swept choices of the library paths.
It also gives the coverage of the numerical validation of Section~\ref{sec:numerical-validation}, the definition of the point-varying coefficients that the validation uses, and the metrics that are requested from the profiler.

\subsection{Benchmark configurations}
Every timing in the main text and in this document is produced by one script body, which a per-scheduler wrapper drives on each machine.
At a given order, the script builds the configuration of every path from a single committed namelist.
It substitutes \texttt{DqdtKernel\_Type} and, where an MMA shape is swept, \texttt{CutlassMmaShape}.
Nothing else in the file changes.
The committed namelists are therefore the published configurations, and Table~\ref{tab:a-configs} lists them.
The first \texttt{WarmupStep} steps of a run are executed but not timed, because the first step of a run includes the one-time device-side initialization.
The remaining steps contribute three timed stages each, and that count is the $N$ of Section~\ref{sup:delayed-cuda-event-harvesting}.

\stripsafebreak
\begin{widetable}
\captionof{table}{Published input configuration at each order. The committed namelist from which every timing at that order is built, together with the grid, time step, and step counts that it sets.}
\label{tab:a-configs}
\begin{tabularx}{\tblwidth}{@{}l>{\hsize=2.85\hsize}L>{\hsize=.75\hsize}R>{\hsize=.7\hsize}R>{\hsize=.55\hsize}R>{\hsize=.6\hsize}R>{\hsize=.55\hsize}R@{}}
\toprule
$p$ & Namelist & Elements & \texttt{dt} & \texttt{nstep} & Warm-up steps & Timed stages \\
\midrule
7 & \texttt{namelists/perf\_p7\_gemm.conf} & $32^3$ & $10^{-5}$ & 20 & 1 & 57 \\
15 & \texttt{namelists/perf\_p15\_gemm.conf} & $16^3$ & $10^{-5}$ & 20 & 1 & 57 \\
31 & \texttt{namelists/perf\_p31\_gemm.conf} & $8^3$ & $10^{-6}$ & 200 & 1 & 597 \\
63 & \texttt{namelists/perf\_p63\_gemm.conf} & $4^3$ & $10^{-6}$ & 20 & 1 & 57 \\
127 & \texttt{namelists/perf\_p127\_gemm.conf} & $2^3$ & $10^{-6}$ & 100 & 1 & 297 \\
255 & \texttt{namelists/perf\_p255\_gemm.conf} & 1 & $10^{-7}$ & 20 & 1 & 57 \\
511 & \texttt{namelists/perf\_p511\_gemm.conf} & 1 & $10^{-8}$ & 20 & 2 & 54 \\
575 & \texttt{namelists/perf\_p575\_gemm.conf} & 1 & $10^{-8}$ & 50 & 5 & 135 \\
767 & \texttt{namelists/perf\_p767\_gemm.conf} & 1 & $10^{-8}$ & 15 & 4 & 33 \\
1023 & \texttt{namelists/perf\_p1023\_gemm.conf} & 1 & $10^{-8}$ & 10 & 3 & 21 \\
\bottomrule
\end{tabularx}
\end{widetable}

Every namelist in the repository sets the CUDA-graph option to false, so all timings and profiler measurements launch their kernels individually.
The four single-element orders from $p = 511$ upward set the warm-up count explicitly, because one warm-up step is a large fraction of a short run there.
Elsewhere it is the program default of one step.
The time step and the step count differ by order, because the orders are separate experiments at a fixed number of degrees of freedom and not points of one sweep.
They are identical across the paths that are compared at a given order, and this is what makes each comparison one-axis.
The validation runs of Section~\ref{sec:numerical-validation} are built from the same namelists by the same substitution.
The step count and the output interval are set to one, because the field is written out at the first tendency evaluation.
The reference path is selected by order, as Table~\ref{tab:a-validation-field} states.

\stripsafebreak

\subsection{Per-order structure of the fused CUDA Core path}
\label{sup:fused-per-order-structure}
At $p = 7$ and $p = 15$, one thread block owns a complete element, and a single kernel carries all three directions.
From $p = 31$, the element is tiled over several blocks.
The directional contractions are distributed over two kernels or, at $p = 255$, over three.
The grouping of the directions, the number of blocks that share an element, and the output tile that each thread owns are specialized per order.
From $p = 63$, the numerical flux is evaluated once into a face-sized array, and the volume kernels read that array.
Every kernel applies the contraction of its own direction, the metric weighting, and the lift of the corresponding pair of faces.
Therefore, no derivative is ever written to memory, at any order.

\subsection{Swept choices of the library paths}
\label{sup:swept-choices-library-paths}
The fold of Section~\ref{sec:library-paths-gemm-gemm} is enabled from $N_q = 64$ upward.
At $N_q = 64$ it applies both metric factors inside the $y$ GEMM, and for $N_q > 64$ it applies one factor in each of the two GEMMs.
Two choices were swept per order and were not fixed in advance.
The first choice is the library that runs each of the three volume GEMMs.
A GEMM keeps cuBLAS where cuBLAS was measured to be faster.
This assignment is held identical in \texttt{GEMM\_CUTE} and \texttt{GEMM\_FUSED}, so the library axis and the fusion axis do not move together.
\texttt{GEMM\_FUSED} also keeps the mainloops, tiles, MMA shapes, stage counts, swizzles, and batch partitioning of \texttt{GEMM\_CUTE}.
Therefore, the fusion pair differs in the epilogues alone.
The second choice, which applies to \texttt{GEMM\_FUSED} alone, is the GEMM that is the carrier.
Any of the three GEMMs can host the carry, because all three derivatives are read elementwise at the same node index.
However, a cuBLAS GEMM has no epilogue, so the carrier is a CUTLASS GEMM in every assignment.
Without fusion, the three volume GEMMs are independent, and they are spread over concurrent streams.
With fusion, the carrier cannot start until the derivatives that it reads are complete.
Where the fold is enabled, the two non-carriers also depend on each other.
Therefore, nothing is left to overlap.

\subsection{Numerical validation coverage}
\label{sup:numerical-validation-coverage}
\begin{widetable}
\captionof{table}{Time-integration validation on GB200. The constant-velocity problem is integrated in time and compared with its exact solution. The convergence study varies the time-step size $dt$ (\texttt{dt} in the input) at fixed final time $t_{\mathrm{end}} = 0.01$ with \texttt{FUSED\_TC} at $p = 7$. The four rows set \texttt{dt} and \texttt{nstep} to $1.0\times 10^{-4}/100$, $5.0\times 10^{-5}/200$, $2.5\times 10^{-5}/400$, and $1.0\times 10^{-5}/1000$, and the last pair is the committed value. The agreement row is the largest difference between the errors of any two of the paths that were run at one order. These paths are the six paths of Table~\ref{tab:paths}, and also the reference CUDA Fortran baseline (\texttt{CUDAFORTRAN\_SPLIT}) at $p = 7$ and $p = 63$, where it exists. This difference is a property of the implementations, not of the discretization. The relative $L^2$ error is the spatial $L^2$ norm of the solution error divided by the $L^2$ norm of the exact solution. At the committed time step, this error has reached the round-off floor, so the table does not measure spatial convergence. The inputs are the committed namelists \texttt{namelists/acc\_p\{7,63,255\}\_gemm.conf} with \texttt{DqdtKernel\_Type} substituted. The coarse-grid control takes the same file with \texttt{NeX = NeY = NeZ = 4}, \texttt{PolyOrder = 3}, \texttt{nstep = 100}, and a non-CUDA implementation path.}
\label{tab:validation}
\noindent\begin{tabularx}{\tblwidth}{@{}LL@{}}
\toprule
Quantity & Value \\
\midrule
Temporal convergence, $dt = 1.0\times 10^{-4}/5.0\times 10^{-5}/2.5\times 10^{-5}$ & relative $L^2$ error $2.63\times 10^{-11}/3.29\times 10^{-12}/4.24\times 10^{-13}$, i.e. ratios 7.99 and 7.78 for a factor of two in \texttt{dt} (third order) \\
Same, $dt = 1.0\times 10^{-5}$ & $7.58 \times 10^{-14}$, ratio 5.59: the round-off floor \\
Agreement of the paths at one order ($p = 7$, $63$, $255$) & largest difference between path errors $2.8 \times 10^{-18}$, i.e. $5 \times 10^{-6}$ of the error itself \\
Mass drift from the exact value of zero & order $10^{-18}$ at all three orders \\
Coarse-grid control ($p = 3$, $N_e = 4^3$, non-CUDA build) & relative $L^2$ error $1.94 \times 10^{-4}$: the same diagnostic returns a discretization error nine orders larger when the grid is coarsened \\
\bottomrule
\end{tabularx}
\end{widetable}

Table~\ref{tab:a-validation-field} gives the cells behind the field-to-field comparison of Section~\ref{sec:numerical-validation} at the fixed-degree-of-freedom orders.
Table~\ref{tab:a-validation-coverage} gives the coverage of the whole sweep, including the single-element orders $p \ge 511$.
Table~\ref{tab:validation} is the time-integration check of that section.
No argument in the paper reads an individual cell of these tables.
They record which path was validated on which machine at which order, and the largest deviation that was found there.
Both tables report the same quantity, the largest absolute difference over the field divided by the largest absolute value of the reference field.
That largest absolute value is 8.55 at every order, because the point-varying coefficients of Section~\ref{sup:definition-point-varying-coefficients} do not depend on the order.
Therefore, the largest absolute differences at $p \ge 511$ are the entries multiplied by 8.55 (for example $5.5 \times 10^{-14}$ on A100 at $p = 575$).

\begin{widetable}
\captionof{table}{Point-by-point comparison of one tendency evaluation against a reference implementation. The input has $u$, $v$, $w$, \texttt{Escale}, \texttt{normal\_fn}, and \texttt{Fscale} that vary from point to point on purpose. The comparison covers the complete owned field of 16,777,216 points, at $p = 7$, $15$, $31$, $63$, $127$, and $255$. Comparison is against the reference CUDA Fortran baseline (\texttt{CUDAFORTRAN\_SPLIT}) at $p \le 127$ and against \texttt{CUDAFORTRAN\_GEMM} at $p = 255$, where that baseline does not exist. Each entry is the largest absolute difference over the field divided by the largest absolute value of the reference field. It is not a pointwise relative error. An entry of $0$ means that every one of the 16,777,216 values is identical. Where a difference occurs, the largest difference and the order at which it occurs are given, and the remaining orders of that cell are stated separately. $\dagger$: at $p = 255$ \texttt{GEMM} is itself the reference, so it has no entry there. Its row therefore covers the five orders $p = 7$ to $p = 127$. Every other row covers all six.}
\label{tab:a-validation-field}
\begin{tabularx}{\tblwidth}{@{}lLLL@{}}
\toprule
Path & GB200 & H100 & A100 \\
\midrule
\texttt{GEMM}$\dagger$ (5 orders) & $4.6 \times 10^{-32}$ at $p = 15$, $0$ at the other four orders & $0$ at all five & $3.4 \times 10^{-14}$ at $p = 7$, $0$ at the other four orders \\
\texttt{GEMM\_CUTE} (6 orders) & $0$ at all six & $0$ at all six & $1.8 \times 10^{-14}$ at $p = 7$, $0$ at the other five orders \\
\texttt{GEMM\_FUSED} (6 orders) & $4.2 \times 10^{-16}$ at $p = 63$, $2.1 \times 10^{-16}$ to $4.2 \times 10^{-16}$ elsewhere & $4.2 \times 10^{-16}$ at $p = 63$, $2.1 \times 10^{-16}$ to $4.2 \times 10^{-16}$ elsewhere & $1.8 \times 10^{-14}$ at $p = 7$, $2.1 \times 10^{-16}$ to $4.2 \times 10^{-16}$ elsewhere \\
\texttt{FUSED} (6 orders) & $1.1 \times 10^{-13}$ at $p = 31$ & $1.1 \times 10^{-13}$ at $p = 31$ & $1.1 \times 10^{-13}$ at $p = 31$ \\
\texttt{FUSED\_TC} (6 orders) & $2.1 \times 10^{-13}$ at $p = 63$ & $2.1 \times 10^{-13}$ at $p = 63$ & $2.1 \times 10^{-13}$ at $p = 63$ \\
\texttt{FUSED\_DFMA} (6 orders) & $2.1 \times 10^{-13}$ at $p = 63$ & $2.1 \times 10^{-13}$ at $p = 63$ & $2.1 \times 10^{-13}$ at $p = 63$ \\
\bottomrule
\end{tabularx}
\end{widetable}

\begin{widetable}
\captionof{table}{Coverage of the point-varying full-field comparison of Table~\ref{tab:a-validation-field}, together with the single-element orders $p \ge 511$. Those orders were validated on GB200 in separate per-order campaigns, and on H100 and A100 by the same script body as Table~\ref{tab:a-validation-field}. For those orders, the entries are the same normalized quantity as in Table~\ref{tab:a-validation-field}, and they are taken against \texttt{GEMM}. Every combination that is absent from Table~\ref{tab:a-validation-coverage} cannot be run on the machine in question.}
\label{tab:a-validation-coverage}
\begin{tabularx}{\tblwidth}{@{}lRRRL@{}}
\toprule
Order & Points & GB200 & H100 & A100 \\
\midrule
$p = 7$, $15$, $31$, $63$, $127$ & 16,777,216 & 6 paths, Table~\ref{tab:a-validation-field} & 6 paths, Table~\ref{tab:a-validation-field} & 6 paths, Table~\ref{tab:a-validation-field} \\
$p = 255$ & 16,777,216 & 5 paths, Table~\ref{tab:a-validation-field} & 5 paths, Table~\ref{tab:a-validation-field} & 5 paths, Table~\ref{tab:a-validation-field} \\
$p = 511$ & 134,217,728 & 4 paths, $0$ to $4.2 \times 10^{-16}$ & 4 paths $\times$ 2 shapes, $0$ to $4.2 \times 10^{-16}$ & 4 paths, $5.3 \times 10^{-15}$ \\
$p = 575$ & 191,102,976 & 2 paths, $0$ to $4.2 \times 10^{-16}$ & 2 paths $\times$ 2 shapes, $0$ to $2.1 \times 10^{-16}$ & 2 paths, $6.4 \times 10^{-15}$ \\
$p = 767$ & 452,984,832 & 2 paths, $0$ to $4.2 \times 10^{-16}$ & 2 paths $\times$ 2 shapes, $0$ to $2.1 \times 10^{-16}$ & exceeds memory \\
$p = 1023$ & 1,073,741,824 & 2 paths, $6.2 \times 10^{-32}$ to $4.2 \times 10^{-16}$ & exceeds memory & exceeds memory \\
\bottomrule
\end{tabularx}
\end{widetable}

\subsection{Definition of the point-varying coefficients}
\label{sup:definition-point-varying-coefficients}
The validation of Section~\ref{sec:numerical-validation} replaces the constant benchmark coefficients by point-varying ones.
Let $x$, $y$, and $z$ be the coordinates of a volume node, let $i$ be the index of a face point within an element, and let $e$ be the element index.
The volume coefficients are then

$u = u_0 + 0.15\sin(2\pi x)$,\quad $v = v_0 + 0.11\cos(2\pi y)$,\quad $w = w_0 + 0.07\sin(2\pi z)$,

where $u_0 = v_0 = w_0 = 1$ are the benchmark values, and the three directional geometric factors are multiplied by

$1 + 0.02\sin(2\pi x)$,\quad $1 + 0.03\cos(2\pi y)$,\quad $1 + 0.04\sin(2\pi z)$,

respectively.
On face points, the surface factor is multiplied by $1 + 0.01\sin(i)$.
In addition, $0.01\sin(i + e)$, $0.01\cos(i + e)$, and $0.01\sin(2i + e)$ are added to the three components of the outward normal, respectively.
The perturbations are chosen to have different amplitudes and different functional forms in the three directions.
The face perturbation is a function of the face-point index and not of position.
Therefore, no two of the six faces receive the same coefficients, and no direction can be collapsed onto another.
The normals are not renormalized, so this configuration has no exact solution.
It is used only for the field-to-field comparison, and never for the time-integration test of Table~\ref{tab:validation}.

\subsection{Profiler invocation}
\label{sup:profiler-invocation}
Tables~\ref{tab:mma}, \ref{tab:binding}, and \ref{tab:kernels}, and Sections~\ref{sup:nominal-roof-accounting} and~\ref{sup:roofline-placement-using-measured}, are built from Nsight Compute counter values.
Table~\ref{tab:a-ncu-config} gives the metrics, which are requested by name and not through a metric set.
The same list is used on all three machines.

\stripsafebreak
\begin{widetable}
\captionof{table}{Nsight Compute configuration. The counters, clock policy, kernel selection, and profiling window for every profiler measurement that this paper reports.}
\label{tab:a-ncu-config}
\begin{tabularx}{\tblwidth}{@{}LL@{}}
\toprule
Item & Value \\
\midrule
Bytes & \texttt{dram\_\_bytes.sum}, \texttt{dram\_\_bytes\_read.sum}, \texttt{dram\_\_bytes\_write.sum}, \texttt{lts\_\_t\_bytes.sum}, \texttt{l1tex\_\_t\_bytes.sum} \\
Time & \texttt{gpu\_\_time\_duration.sum}, \texttt{sm\_\_cycles\_elapsed.avg.per\_second} \\
Throughputs, \texttt{pct\_of\_peak\_sustained\_elapsed} & \texttt{sm\_\_throughput.avg}, \texttt{l1tex\_\_throughput.avg}, \texttt{lts\_\_throughput.avg}, \texttt{dram\_\_throughput.avg} \\
Pipe and occupancy, \texttt{pct\_of\_peak\_sustained\_active} & \texttt{sm\_\_pipe\_fp64\_cycles\_active.avg}, \texttt{sm\_\_warps\_active.avg} \\
Instructions & \texttt{sm\_\_inst\_executed.sum}, \texttt{smsp\_\_inst\_executed.avg.per\_cycle\_active} \\
Stalls, \texttt{average\_warps\_issue\_stalled\_...\_per\_issue\_active.ratio} & \texttt{long\_scoreboard}, \texttt{short\_scoreboard}, \texttt{mio\_throttle}, \texttt{math\_pipe\_throttle} \\
Launch & \texttt{launch\_\_registers\_per\_thread}, \texttt{launch\_\_grid\_size}, \texttt{launch\_\_block\_size} \\
Clock policy & \texttt{--clock-control none} for every value that is reported as a fraction of a roof; the default lock is used only where a mechanism, not a rate, is identified \\
Kernel selection & no \texttt{-k} filter: library kernel names are not predictable, and the surrounding kernels are part of the call \\
Window & \texttt{-s <4 tendency calls worth of launches> -c 24}, which covers a whole tendency call and detects its period \\
Other flags & \texttt{--csv --kernel-name-base demangled --rename-kernels 0 --replay-mode kernel} \\
\bottomrule
\end{tabularx}
\end{widetable}

The instruction-pricing campaign of Section~\ref{sup:campaigns} supplies the model of Section~\ref{sec:subpipe-model} of the main text.
This campaign uses the same clock policy and window, but it uses a kernel filter (\ttbreak{-k regex:tendency\_|elembnd|volume\_flux|flux\_all|separable\_lift|cutlass|gemm|Kernel}) and a wider metric list in two passes.
The first pass has 43 pipe, instruction, memory, and launch metrics, which include \texttt{sm\_\_pipe\_shared\_cycles\_active}, \ttbreak{sm\_\_inst\_executed\_pipe\_\{tensor,fp64,alu,fma,lsu\}}, \texttt{l1tex\_\_data\_pipe\_lsu\_wavefronts}, \texttt{lts\_\_t\_bytes}, and \texttt{dram\_\_bytes}.
The second pass has twelve issue-stall reasons.
The complete list, under the metric names, is the column set of \texttt{figs/data/subpipe\_inputs.csv}.
The model reads twelve of the metrics and the stall pass.

Nsight Systems is invoked with \texttt{--trace=cuda --resolve-symbols=false}, and \texttt{DEBUGINFOD\_URLS} is set to the empty string.
Without these settings, the symbol resolution does not finish on the GB200 system.
Nsight Systems can trace the graph replay of this program.
With \texttt{--cuda-graph-trace=graph}, one record per graph launch gives the device time of a step.
With \texttt{--cuda-graph-trace=node}, every kernel inside the replay is recorded.
Nsight Systems 2026.1.1 is the version of every profile in this paper.
On GB200, this version traced the replay at $p = 31$, 63, 127, and 255 on \texttt{FUSED\_TC} and at $p = 7$ and 255 on \texttt{GEMM\_FUSED} in 3 to 5 s each.
For \texttt{FUSED}, \texttt{FUSED\_TC}, and \texttt{FUSED\_DFMA} at $p = 7$ and 15, neither the same version nor 2025.6.3, 2026.1.3, or 2026.2.1 finished.
From the second graph launch on, the process stays at full CPU load while the GPU is idle, and the report contains no kernel record.
A run of two steps, which captures the graph and launches it once, completes, but a run of three steps does not.
Nsight Systems 2025.1.3 traces the same runs at both levels in 3 to 4 s.
The element count and \texttt{--cuda-event-trace=false} do not change the outcome.
We did not find the cause.
In the graph-disabled configuration, the trace of \texttt{FUSED\_TC} at $p = 7$ shows a launch gap of 30.6 $\mu$s per step in the steady state.
This gap agrees with the 29 $\mu$s per step that the graph removes from that path in Section~\ref{sup:graph-replay}.
Every namelist of Table~\ref{tab:a-configs} sets the graph option to false.
Every invocation is issued from a batch job, is bounded by a \texttt{timeout}, and profiles a frozen copy of the executable.
Therefore, a concurrent rebuild cannot relink under a running profiler, and a hung invocation costs one attempt and not the allocation.
The input is the namelist of Table~\ref{tab:a-configs} with \texttt{nstep} and the output interval reduced to four, which changes no kernel and no launch order.

\subsection{Profiler-reported time against wall-clock time}
\label{sup:profiler-wallclock-bias}
Nsight Compute locks the SM clock (Section~\ref{sup:profiler-measurements-effect-clock}), and this changes the relative price of on-chip work and global-memory waiting.
For decisions about whether to adopt a change, the sign of the resulting bias depends on which resource a change spends and which it saves.
Six comparisons between profiler-reported time and wall-clock time show this.
Four changes that reduce shared-memory traffic or instruction count are rated better by the profiler than by wall-clock time.
The largest is a shared-memory conflict fix that the profiler scores as an 8.3\% reduction but that lengthens the wall-clock time by 1.0\%.
One change that spends shared-memory traffic and instructions in order to hide a global-memory latency is rated worse.
This change is a prefetch, and the profiler scores it as an 8.3\% increase.
However, the prefetch shortens the wall-clock time by 0.5\%.
In the sixth comparison, the profiler attributes 10\% of a stage to the element-boundary flux kernel.
However, the complete removal of that kernel shows that it accounts for 13.5\% of the stage.
All three groups follow from one mechanism.
At a reduced SM clock, a global-memory access costs fewer SM cycles.
As a result, shared-memory, register, and instruction work is priced relatively high, and global-memory waiting is priced relatively low.

\section{Emulated and Ozaki-scheme FP64 GEMM}
\label{sup:emulated-ozaki-scheme-fp64}

On GB200, the FP64 Tensor Core ceiling no longer exceeds the CUDA Core FP64 ceiling (Table~\ref{tab:machines} of the main text).
Therefore, lower-precision Tensor Core arithmetic with FP64 emulation is a natural alternative for the volume GEMMs of this operator.
This section quantifies the cost of this alternative at an accuracy setting that the operator can accept.
The result is negative for the matrix shapes and the accuracy requirements of this operator, and we do not generalize it.

\subsection{Accuracy is a requirement, not a result}
\label{sup:accuracy-precondition-outcome}
Performance comparisons between emulated paths are meaningful only after an accuracy configuration that reproduces the operator has been fixed.
Table~\ref{tab:a-emulation-accuracy} gives the maximum absolute difference over the owned \texttt{dqdt} values at $p = 7$ against the reference CUDA Fortran baseline (\texttt{CUDAFORTRAN\_SPLIT}).
The comparison uses the point-varying-coefficient input.
The configurations that fail do so by twelve or more orders of magnitude, so the result does not depend on the choice of an absolute or a relative measure.

\begin{apptable}
\captionof{table}{Accuracy of the emulated FP64 GEMM configurations at $p = 7$. Maximum absolute difference over the owned \texttt{dqdt} values against the reference CUDA Fortran baseline (\texttt{CUDAFORTRAN\_SPLIT}), with the point-varying-coefficient input. The table also states whether each configuration passes.}
\label{tab:a-emulation-accuracy}
\noindent\begin{tabularx}{\tblwidth}{@{}LLRL@{}}
\toprule
path & setting & max. abs. difference & result \\
\midrule
Scheme I & fixed, 7 slices & $3.55 \times 10^{-15}$ & passes \\
Scheme I & fixed, 8 slices & $3.55 \times 10^{-15}$ & passes \\
Scheme I & adaptive, 7 or 8 slices & $3.04 \times 10^{-1}$ & fails \\
Scheme II & fixed or adaptive, 7 or 14 moduli & $1.86 \times 10^{-1}$ to $2.97 \times 10^{-1}$ & fails (all four) \\
\bottomrule
\end{tabularx}
\end{apptable}

Only the fixed Scheme I configurations reproduce the operator to the required accuracy.
The adaptive Scheme I controller stops the slice expansion far too early for this operand.
The four tested configurations of Scheme II fail regardless of the number of moduli, at $p = 7$, 255, and 511 alike.
Scheme II is the integer-modular, GEMM-oriented method of \citet{ozaki2026}.
Within the passing column, the seven-slice configuration is at round-off at $p = 7$.
However, at $p = 511$ it deviates from the reference by $1.8 \times 10^{-12}$, which is 8.4 times the largest deviation of any native path in Section~\ref{sec:numerical-validation} of the main text.
In contrast, eight slices remain below that reference at every tested order.
The timings in Section~\ref{sup:cost-accuracy-passes} use the default fixed 55-bit, seven-slice configuration.
The eight-slice configuration was not timed.

\subsection{Cost at an accuracy that passes}
\label{sup:cost-accuracy-passes}
With the accuracy requirement that Section~\ref{sup:accuracy-precondition-outcome} fixes, we replace only the three volume GEMMs in the otherwise identical unfused driver with our Scheme I implementation.
Interleaved measurements on an exclusively allocated GB200 give the times of Table~\ref{tab:a-scheme-one-cost}.

\begin{apptable}
\captionof{table}{Cost of the passing Scheme I configuration on GB200. Device time per Runge--Kutta stage for the unfused driver in which only the three volume GEMMs are replaced by the fixed 55-bit, seven-slice Scheme I implementation, against the native FP64 \texttt{GEMM} path at the same order. The per-stage values are the CUDA-event sum divided by the $N + 1$ calls that it covers (Section~\ref{sup:delayed-cuda-event-harvesting}).}
\label{tab:a-scheme-one-cost}
\noindent\begin{tabularx}{\tblwidth}{@{}lLRRR@{}}
\toprule
$p$ & $K = N_q$ & native \texttt{GEMM}, microseconds per stage & Scheme I, microseconds per stage & ratio \\
\midrule
7 & 8 & 1 583.5 & 120 918.2 & 76.36 \\
63 & 64 & 597.9 & 28 048.6 & 46.91 \\
255 & 256 & 1 032.1 & 24 048.9 & 23.30 \\
511 & 512 & 13 015.1 & 171 338.5 & 13.16 \\
575 & 576 & 20 526.4 & 248 200.0 & 12.09 \\
767 & 768 & 61 393.2 & 596 794.1 & 9.72 \\
\bottomrule
\end{tabularx}
\end{apptable}

The overhead decreases steadily with the reduction depth $K = N_q$.
However, even at $K = 768$ the passing configuration is 9.72 times slower than native FP64.

\subsection{The vendor emulation path, at a fixed mantissa setting}
\label{sup:vendor-emulation-path-fixed}
With the vendor implementation and forced emulation, the outcome depends on the mantissa setting.
In the adaptive mantissa mode, the path reproduces the reference bit for bit at the four orders at which we measured accuracy.
However, it is 104.3 times slower than native FP64 at $p = 7$ and 1.25 to 1.77 times slower from $p = 255$ to $p = 1023$.
Next, the mantissa is fixed at 55 bits, and only the three volume GEMMs of the same unfused driver are replaced.
The device time per Runge--Kutta stage for this setting is that of Table~\ref{tab:a-vendor-emulation}.

\begin{apptable}
\captionof{table}{The vendor emulation path at a fixed 55-bit mantissa on GB200. Device time per Runge--Kutta stage for the same unfused driver in which the three volume GEMMs run through the vendor emulation, against the native FP64 path at the same order. The per-stage values are the CUDA-event sum divided by the $N + 1$ calls that it covers (Section~\ref{sup:delayed-cuda-event-harvesting}).}
\label{tab:a-vendor-emulation}
\noindent\begin{tabularx}{\tblwidth}{@{}lLRRR@{}}
\toprule
$p$ & $K = N_q$ & native [ms/stage] & forced emulation, fixed 55 bits [ms/stage] & ratio \\
\midrule
7 & 8 & 1.589 & 473.38 & 297.9 \\
255 & 256 & 1.035 & 2.127 & 2.055 \\
511 & 512 & 13.04 & 12.33 & 0.946 \\
575 & 576 & 20.54 & 27.38 & 1.333 \\
767 & 768 & 61.44 & 56.25 & 0.916 \\
\bottomrule
\end{tabularx}
\end{apptable}

At $p = 511$ and $p = 767$, the emulated path requires 5.4 and 8.4\% less device time than native FP64.
A later campaign extends the comparison to $p = 1023$, where the emulated path requires 30.9\% less.
However, that point is a single value with a 46\% spread over three runs.
That result must be read together with its accuracy.
Against the point-varying-coefficient reference of Section~\ref{sec:numerical-validation} of the main text, the relative difference of this configuration grows from $6.8 \times 10^{-14}$ at $p = 7$ to $1.5 \times 10^{-10}$ at $p = 767$.
This difference is up to 701 times the largest deviation of any native path ($2.1 \times 10^{-13}$).
At $p = 1023$, the accuracy cannot be measured, because the 8 GiB emulation workspace no longer fits alongside the point-varying fields.
The two mantissa settings thus separate speed from accuracy.
The fixed 55-bit setting is faster than native FP64 at $p = 511$ and $p = 767$, but its deviation is two orders of magnitude above every native path.
The adaptive setting is exact and never faster.
Neither setting is both fast and exact.

All of the above is measured on GB200, and the sign does not carry to the other machines.
On H100, the fixed 55-bit path is slower than native FP64 at every tested order (2.18 to 3.17 times at $p = 255$ to 767).
The reason is that the fused INT8 GEMM that carries the emulation is 2.87 times slower on H100, and the native FP64 GEMM that it has to beat is 1.48 times faster there (66.9 against 40.1 TFLOP/s FP64 Tensor Core roof).
On A100, the comparison could not be made, because the installed cuBLAS 12.9.1 lacks the emulation interface.
In step-loop time, our own Scheme I implementation is 9.4 times native at $p = 767$ on GB200 (9.72 in device time, Table~\ref{tab:a-scheme-one-cost}) and never below 21 times on H100 or 16 times on A100.
For comparison, a square 8192-cubed GEMM with the same library on GB200 is approximately 3.0 times faster with forced emulation than native FP64.
This is the range in which such results are usually quoted.

\subsection{Scope and relation to previous emulation results}
\label{sup:scope-relation-previous-emulation}
Recent work reports up to 2.3 times the performance of native FP64 on GB200 for Ozaki-based emulation of square GEMMs \citep{schwarz2026}.
These results do not contradict the measurements above.
The volume GEMMs of this operator have a reduction depth of only $K = N_q$, so the costs of slicing, packing, and reconstruction are spread over a shallow reduction.
A trace of the vendor emulated GEMM at $p = 511$ shows this directly.
The fused INT8 GEMM alone takes 0.614 times the time of the native FP64 GEMM, and the two packing kernels per GEMM add 0.307 back.
Therefore, at these shapes, the comparison is decided by the slicing and packing and not by the low-precision multiplication itself.
For the matrix shapes that this operator generates, and at an accuracy setting that reproduces its FP64 result, the tested emulation paths are not competitive with native FP64 on GB200.
The vendor path is also not competitive on H100 at either mantissa setting.

\section{Roofline placement using measured and algorithmic traffic}
\label{sup:roofline-placement-using-measured}
Figure~\ref{fig:s8} places every measured combination of implementation path, polynomial order, and architecture on a roofline diagram.
The figure also shows how the byte-count convention affects the placement.
Each measurement is one horizontal segment.
The vertical axis is the algorithmic operation count of one Runge--Kutta stage divided by the Nsight Compute device time of the tendency evaluation at production clocks (\texttt{ncu --clock-control none} throughout this section).
The horizontal axis uses two byte counts.
The first is the \emph{unique} bytes of the operator ($I_{\mathrm{alg}}$), which is the traffic that would result if each array were read or written once per stage.
The second is the DRAM traffic that the profiler measures for that path, order, and machine ($I_{\mathrm{meas}}$).
The roof against which a path is reported is selected by the instruction that it issues (Section~\ref{sec:roofline-its-granularity} of the main text).
This choice matters most on A100, where the \texttt{mma} paths exceed the CUDA Core peak.

\subsection{The two byte counts differ by up to a factor of three}
\label{sup:two-byte-counts-differ}
Across the 93 measured points on the three machines, the ratio of measured to unique traffic ranges from 0.98 to 2.93.
Table~\ref{tab:a-traffic-ratio} gives the GB200 values, whose range is 0.98 to 2.75.

\begin{apptable}
\captionof{table}{Ratio of measured DRAM traffic to unique traffic on GB200. The measured traffic is the traffic that \texttt{ncu --clock-control none} reports for one Runge--Kutta stage of the tendency evaluation. The unique traffic is a property of the operator, and it is independent of path and machine. Entries with -- are orders at which the path was not measured.}
\label{tab:a-traffic-ratio}
\begin{tabular}{lrrrr}
\toprule
$p$ & \texttt{FUSED} & \texttt{FUSED\_TC} & \texttt{GEMM} & \texttt{GEMM\_FUSED} \\
\midrule
7 & 0.98 & 1.00 & 2.33 & 2.18 \\
15 & 1.03 & 1.03 & 2.60 & 2.40 \\
31 & 1.73 & 1.62 & 2.75 & 2.51 \\
63 & 1.62 & 1.63 & 2.53 & 2.36 \\
127 & 1.50 & 1.49 & 2.43 & 2.26 \\
255 & 2.53 & 1.74 & 2.40 & 2.34 \\
511 & -- & 1.77 & 2.52 & 2.28 \\
575 & -- & -- & 2.65 & 2.28 \\
767 & -- & -- & 2.66 & 2.29 \\
1023 & -- & -- & 2.68 & 2.31 \\
\bottomrule
\end{tabular}
\end{apptable}

The library paths write intermediate derivatives and re-read them, so their measured traffic is 2.18--2.75 times the unique traffic.
The fused epilogue reduces the ratio at every order.
\texttt{FUSED} and \texttt{FUSED\_TC} are at the unique count at the two lowest orders, where the cache hierarchy absorbs the face-gather traffic, and at 1.49--2.53 above these orders.
For a fixed path and order, the ratios on the three machines agree to within 17.5\%, and typically within a few percent, although the L2 capacities differ substantially.
Therefore, the amplification is set by path and order and not by the machine.
This is also why the L1 capacity of Equation~\ref{eq:model} of the main text is a wavefront count and not a byte roof.
If the measured L1 bytes are divided by the peak that \texttt{l1tex\_\_throughput} implies, the resulting capacity varies by a factor of six between kernels at one order on one machine.

\subsection{Roofline crossover depends on the architecture}
\label{sup:roofline-crossover-depends-architecture}
At $I_{\mathrm{meas}}$, the fastest path at $p = 7$ reaches 81 to 89\% of the HBM roof and at most 14.3\% of its arithmetic roof on the three machines.
At the highest measured order, the fastest path reaches 92\% of the arithmetic roof on GB200 ($p = 1023$), 54\% on H100 ($p = 767$), and 71\% on A100 ($p = 575$).
These fractions are taken at $I_{\mathrm{meas}}$ and from a different campaign than the 80.0\% at $p = 7$ of Section~\ref{sec:performance-landscape} of the main text.
That value uses the unique byte count.
When we take the fastest path at each order, the arithmetic fraction first exceeds the bandwidth fraction between $p = 63$ and 127 on GB200, between 127 and 255 on A100, and only between 575 and 767 on H100.
This sequence of machines follows the Tensor Core ridge points of 5.08, 12.5, and 27.9 FLOP/byte.
Therefore, the order at which the placement crosses the ridge is architecture dependent.

\subsection{Whether the attained fraction is preserved across architectures}
\label{sup:whether-attained-fraction-preserved}
Section~\ref{sec:does-accounting-transfer-a100} of the main text quotes the contraction-only attained fraction, and the two-component scaling of Section~\ref{sup:two-component-detail} assumes that this fraction is preserved.
This fraction divides the algorithmic contraction count $3\cdot 2\,N_q^4 N_e$ of one tendency call at $p = 511$ by the summed time of its three volume contractions.
Table~\ref{tab:a-attained-fraction} gives it on all three machines.

\begin{apptable}
\captionof{table}{Fraction of the FP64 Tensor Core roof that the three volume contractions of one tendency call reach at $p = 511$. All entries are from one campaign with \texttt{ncu --clock-control none}. The CUTLASS rows use the $8\times 8\times 4$ MMA shape on GB200 and A100 and $16\times 8\times 4$ on H100, as we discuss below. The paths \texttt{FUSED\_TC} and \texttt{GEMM} have no shape parameter.}
\label{tab:a-attained-fraction}
\begin{tabular}{lrrr}
\toprule
path & GB200 & H100 & A100 \\
\midrule
\texttt{FUSED\_TC} & 79.9\% & 39.4\% & 73.8\% \\
\texttt{GEMM\_FUSED} & 92.7\% & 61.2\% & 83.0\% \\
\texttt{GEMM} & 95.0\% & 83.0\% & 92.8\% \\
\bottomrule
\end{tabular}
\end{apptable}

The \texttt{GEMM} row bounds how much of the H100 column is a property of the machine.
The cuBLAS library runs the same three contractions at 83.0\% of the same roof, against 61.2\% for our mainloop at the shape that H100 prefers.
Therefore, the contraction itself is not the part that H100 executes poorly (Table~\ref{tab:a-cublas-vs-cute} separates the two).
The CUTLASS rows of the H100 column belong to the preferred-shape comparison.
For the same-shape comparison, the three volume GEMMs of \texttt{GEMM\_CUTE} at $8\times 8\times 4$ take 13.18 ms per stage on H100 (Table~\ref{tab:a-cublas-vs-cute}).
This time corresponds to 46.8\% of the Tensor Core roof.
A separate profile of the same kernels gives 13.17 ms and an SM throughput of 92.7 to 94.7\% of peak, so these kernels are not limited by arithmetic on H100.
The \texttt{FUSED\_TC} kernels are in the same condition at 80\% of issue and 39\% of the arithmetic roof.

No path preserves its attained fraction across machines.
The ratios of the first two rows, 1.160 on GB200, 1.125 on A100, and 1.556 on H100, are the quantity that Section~\ref{sec:does-accounting-transfer-a100} of the main text and Section~\ref{sup:nominal-roof-failure-h100} use.
At the default $8\times 8\times 4$ shape, \texttt{GEMM\_FUSED} reaches 45.2\% of the roof on H100, and the ratio is 1.149, which is 0.9\% from the GB200 value.
Therefore, the departure of 34 to 38\% belongs to the instruction shape and not to the machine.
Table~\ref{tab:a-ratio-by-shape} lists each value of this ratio with its shape, job, and clock policy.

\begin{widetable}
\captionof{table}{The attained-fraction ratio $\texttt{GEMM\_FUSED}/\texttt{FUSED\_TC}$ for the volume contractions at $p = 511$, by MMA shape, job, and clock policy. The $8\times 8\times 4$ rows correspond to the same-shape comparison of Section~\ref{sec:does-accounting-transfer-a100} (measured CUDA-device ratio 1.068). The $16\times 8\times 4$ rows correspond to the preferred-shape comparison (measured 0.853). The two $16\times 8\times 4$ entries differ by 2.7\% because the clock that the two compute nodes sustained under \texttt{--clock-control none} differs by 3.8\%. The $8\times 8\times 4$ rows agree to 0.1\% in clock and shared kernels, and they are the ones used for subtractions across jobs. The fourth row is taken without a profiler and is stated in time, as whole-path event time minus the non-contraction kernels. Its ratio is 16.15 / 14.04 because an attained fraction is inversely proportional to time, and it agrees with the third row to 0.2\%.}
\label{tab:a-ratio-by-shape}
\begin{tabularx}{\tblwidth}{@{}l>{\hsize=.7\hsize}L>{\hsize=1.0\hsize}L>{\hsize=1.6\hsize}L>{\hsize=.9\hsize}R>{\hsize=.9\hsize}R>{\hsize=.9\hsize}R@{}}
\toprule
machine & MMA shape & job & clock policy & \texttt{GEMM\_FUSED} & \texttt{FUSED\_TC} & ratio \\
\midrule
GB200 & $8\times 8\times 4$ & campaign of Table~\ref{tab:a-attained-fraction} & \texttt{ncu --clock-control none} & 92.7\% & 79.9\% & 1.160 \\
A100 & $8\times 8\times 4$ & campaign of Table~\ref{tab:a-attained-fraction} & \texttt{ncu --clock-control none} & 83.0\% & 73.8\% & 1.125 \\
H100 & $8\times 8\times 4$ & later job, second H100 compute node & \texttt{ncu --clock-control none} & 45.2\% & 39.4\% & 1.149 \\
H100 & $8\times 8\times 4$ & later job, second H100 compute node & no profiler, production clocks & 14.04 ms & 16.15 ms & 1.151 \\
H100 & $16\times 8\times 4$ & campaign of Table~\ref{tab:a-attained-fraction}, first H100 compute node & \texttt{ncu --clock-control none} & 61.2\% & 39.4\% & 1.556 \\
H100 & $16\times 8\times 4$ & later job, second H100 compute node & \texttt{ncu --clock-control none} & 62.9\% & 39.4\% & 1.598 \\
\bottomrule
\end{tabularx}
\end{widetable}

The epilogue accounts for the 3.5\% between the 45.2\% of \texttt{GEMM\_FUSED} and the 46.8\% of the mainloop that it shares with \texttt{GEMM\_CUTE}.

A control profiles only the three volume contractions of one tendency call at $p = 511$ on H100, for \texttt{GEMM} and for \texttt{GEMM\_CUTE}, at both MMA shapes (Table~\ref{tab:a-cublas-vs-cute}).
This control separates our CUTLASS mainloop from the architecture.
The pair shares its surrounding kernels and differs only in the library and its mainloop.

\begin{widetable}
\captionof{table}{The three volume contractions of one tendency call at $p = 511$ on H100, through cuBLAS and through our CUTLASS mainloop, at both MMA shapes. One job per shape on one H100 compute node, both with \texttt{ncu --clock-control none}. The algorithmic operation count is identical for all four rows by design, $3 \cdot 2 N_q^4 N_e = 4.123 \times 10^{11}$ FLOP. The roof is the H100 FP64 Tensor Core peak of 66.9 TFLOP/s. \texttt{GEMM} runs through cuBLAS and has no shape parameter, so its two rows are a reproducibility control between the two jobs. They agree to 0.7\%. These attained fractions are those of the three contractions alone, and they are not the same quantity as the whole-path fractions of Tables~\ref{tab:a-attained-fraction} and~\ref{tab:a-ratio-by-shape}.}
\label{tab:a-cublas-vs-cute}
\begin{tabularx}{\tblwidth}{@{}>{\hsize=.8\hsize}L>{\hsize=1.05\hsize}L>{\hsize=.7\hsize}R>{\hsize=.9\hsize}R>{\hsize=.95\hsize}R>{\hsize=1.35\hsize}R>{\hsize=1.2\hsize}R>{\hsize=1.05\hsize}R@{}}
\toprule
MMA shape & path & time [ms] & TFLOP/s & fraction of roof & instructions & instructions per FLOP & L1 traffic [GB] \\
\midrule
$16\times 8\times 4$ & \texttt{GEMM} (cuBLAS) & 7.29 & 56.6 & 84.6\% & $1.617 \times 10^{9}$ & 0.126 & 46.2 \\
$16\times 8\times 4$ & \texttt{GEMM\_CUTE} & 8.68 & 47.5 & 71.0\% & $3.726 \times 10^{9}$ & 0.289 & 63.4 \\
$8\times 8\times 4$ & \texttt{GEMM} (cuBLAS) & 7.34 & 56.2 & 84.0\% & $1.617 \times 10^{9}$ & 0.126 & 46.2 \\
$8\times 8\times 4$ & \texttt{GEMM\_CUTE} & 13.18 & 31.3 & 46.8\% & $3.950 \times 10^{9}$ & 0.307 & 63.4 \\
\bottomrule
\end{tabularx}
\end{widetable}

Across the four rows, the measured DRAM traffic agrees to within 0.1\%, the achieved occupancy is 12.4\% in every row, and the operation count is identical by design.
However, our mainloop issues 2.31 to 2.44 times as many instructions for the same arithmetic and moves 1.37 times as many bytes through L1.
Therefore, on one machine and one problem, the attained fraction ranges from 46.8 to 84.6\% according to the implementation alone.
These measurements support the statement that attained fractions do not transfer, not the statement that the machine sets their values.
The wider shape relieves arithmetic-pipe pressure (the math-pipe stall falls from 2.37 to 0.94 cycles) and moves the dominant throughput from the SM pipe to L1.
However, it reduces the instruction count by only 5.7\%.
The residual is consistent with register blocking (246 registers per thread for cuBLAS against 156 to 168 for our mainloop).
We have not disassembled either mainloop, and we do not attribute the residual further.

\subsection{The byte count can move a point across the roofline ridge}
\label{sup:byte-count-can-move}
For 14 of the 93 measurements, $I_{\mathrm{alg}}$ and $I_{\mathrm{meas}}$ lie on opposite sides of the ridge point.
These are all four paths at $p = 63$ and \texttt{GEMM} at $p = 127$ on GB200, seven measurements between $p = 255$ and 767 on H100, and the two library paths at $p = 255$ on A100.
In every case the algorithmic count places the measurement on the compute side and the measured traffic on the bandwidth side.
Ten of the fourteen are library paths, whose intermediate arrays the algorithmic count omits.

\stripsafebreak

\subsection{Resource utilization of the dominant fused kernel}
\label{sup:binding-resource-dominant-kernel}
Table~\ref{tab:binding} is the per-kernel record behind Figure~\ref{fig:a11a}(b) and the summary that Section~\ref{sup:path-level-two-roof} quotes.
Five of the eleven CUDA Core points of Figure~\ref{fig:a11a}(a) that lie outside the interval are at the three orders that this table profiles.
At these orders, L1 is the dominant throughput of that path on GB200.

\begin{widetable}
\captionof{table}{Resource utilization of the dominant fused kernel at production clocks (\texttt{ncu --clock-control none}). For each order and fused path: the kernel with the longest single invocation in the profiling window, its throughput as a percentage of each peak, and its share of the Runge--Kutta stage. The bracketed entry is the largest of the five, which we call the dominant measured throughput. Where it is itself well below peak, as for the H100 CUDA Core path at $p = 255$, the kernel is latency-limited and the bracket marks only the closest throughput. The same source at the same order has its dominant throughput in different resources on different machines.}
\label{tab:binding}
\setlength{\tabcolsep}{3pt}
\renewcommand{\_}{\textunderscore\allowbreak}
\begin{tabularx}{\tblwidth}{@{}>{\hsize=.55\hsize}L>{\hsize=1.25\hsize}L>{\hsize=1.0\hsize}L>{\hsize=2.6\hsize}L%
                             >{\hsize=.75\hsize}R>{\hsize=.75\hsize}R>{\hsize=.75\hsize}R>{\hsize=.85\hsize}R>{\hsize=.75\hsize}R>{\hsize=.75\hsize}R@{}}
\toprule
$p$ & Path & Machine & Kernel & SM & L1 & L2 & DRAM & FP64 & Share of stage \\
\midrule
7 & \texttt{FUSED} & GB200 & \texttt{tendency\_fused\_p7\_cc} & 42.7 & \textbf{[85.0]} & 43.9 & 68.2 & 15.1 & 70.5\% \\
7 & \texttt{FUSED} & H100 & \texttt{tendency\_fused\_p7\_cc} & 17.2 & 33.6 & 68.5 & \textbf{[87.1]} & 6.5 & 72.4\% \\
7 & \texttt{FUSED} & A100 & \texttt{tendency\_fused\_p7\_cc} & 18.1 & 35.9 & 64.9 & \textbf{[84.6]} & 13.7 & 72.9\% \\
7 & \texttt{FUSED\_TC} & GB200 & \texttt{tendency\_fused\_p7\_tile} & 25.3 & 69.1 & 53.7 & \textbf{[80.9]} & 8.7 & 67.3\% \\
7 & \texttt{FUSED\_TC} & H100 & \texttt{tendency\_fused\_p7\_tile} & 9.7 & 25.0 & 68.5 & \textbf{[87.5]} & 3.2 & 72.6\% \\
7 & \texttt{FUSED\_TC} & A100 & \texttt{tendency\_fused\_p7\_tile} & 10.4 & 28.2 & 66.2 & \textbf{[87.1]} & 6.7 & 72.8\% \\
63 & \texttt{FUSED} & GB200 & \texttt{tendency\_fused\_p63\_xz\_cc} & 46.2 & \textbf{[89.1]} & 24.7 & 39.6 & 47.3 & 47.5\% \\
63 & \texttt{FUSED} & H100 & \texttt{tendency\_fused\_p63\_xz\_cc} & 34.6 & 63.5 & 52.8 & \textbf{[81.7]} & 35.9 & 37.1\% \\
63 & \texttt{FUSED} & A100 & \texttt{tendency\_fused\_p63\_xz\_cc} & 51.8 & 48.5 & 36.9 & \textbf{[57.2]} & 52.8 & 44.7\% \\
63 & \texttt{FUSED\_TC} & GB200 & \texttt{tendency\_fused\_p63\_xz} & 50.4 & \textbf{[56.0]} & 25.7 & 44.1 & 4.3 & 27.9\% \\
63 & \texttt{FUSED\_TC} & H100 & \texttt{tendency\_fused\_p63\_xz} & 33.0 & 37.5 & 52.6 & \textbf{[78.9]} & 2.7 & 37.9\% \\
63 & \texttt{FUSED\_TC} & A100 & \texttt{tendency\_fused\_p63\_xz} & 35.0 & 37.2 & 44.9 & \textbf{[69.8]} & 5.2 & 31.3\% \\
255 & \texttt{FUSED} & GB200 & \texttt{tendency\_z\_p255\_cc} & 44.9 & \textbf{[83.3]} & 23.4 & 37.2 & 45.7 & 33.1\% \\
255 & \texttt{FUSED} & H100 & \texttt{tendency\_y\_p255\_cc} & 28.0 & \textbf{[50.7]} & 26.2 & 28.6 & 28.5 & 32.6\% \\
255 & \texttt{FUSED} & A100 & \texttt{tendency\_y\_p255\_cc} & 59.3 & 53.8 & 41.3 & 28.3 & \textbf{[59.9]} & 31.6\% \\
255 & \texttt{FUSED\_TC} & GB200 & \texttt{tendency\_p255} & \textbf{[76.2]} & 50.9 & 24.4 & 24.1 & 2.5 & 30.4\% \\
255 & \texttt{FUSED\_TC} & H100 & \texttt{tendency\_p255} & 54.9 & 36.9 & 47.6 & \textbf{[55.8]} & 1.7 & 31.7\% \\
255 & \texttt{FUSED\_TC} & A100 & \texttt{tendency\_p255} & \textbf{[64.3]} & 41.7 & 52.3 & 58.4 & 3.9 & 29.0\% \\
\bottomrule
\end{tabularx}
\end{widetable}

\subsection{A path is not a point}
\label{sup:path-point}
\begin{figure*}[tp]
\centering
\includegraphics[width=\linewidth]{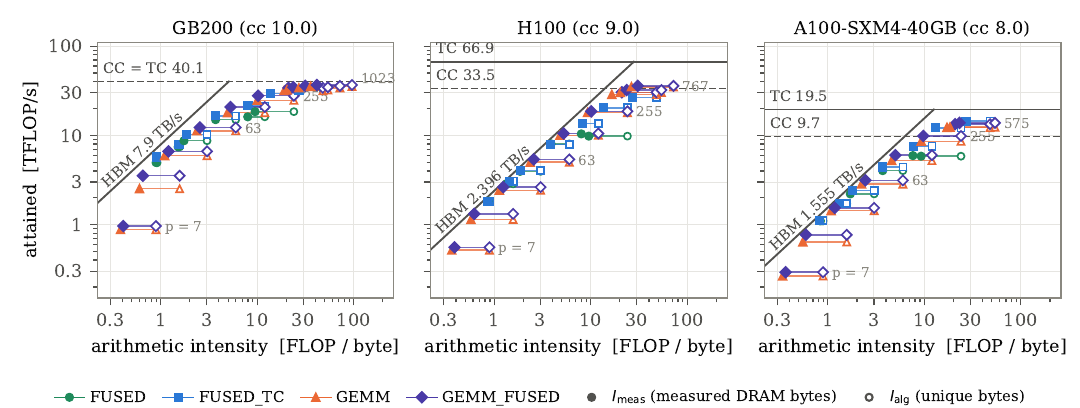}
\caption{Roofline placement of every measured implementation path, polynomial order, and architecture. Each measurement is one horizontal segment. The vertical axis is the algorithmic operation count of one Runge--Kutta stage divided by the Nsight Compute device time at production clocks (\texttt{--clock-control none}). The filled marker uses the intensity from the measured DRAM traffic, $I_{\mathrm{meas}}$. The open marker uses the unique traffic of the operator, $I_{\mathrm{alg}}$. The span of a segment is therefore the measured-to-unique traffic ratio (0.98 to 2.93 over the 93 measurements). Order increases from lower left to upper right and is labeled on the \texttt{GEMM\_FUSED} series. The CUDA Core peak (dashed) applies to \texttt{FUSED}, which issues no \texttt{mma}. The Tensor Core peak (solid) applies to the other three paths. The two coincide on GB200. For 14 of the 93 measurements, $I_{\mathrm{alg}}$ and $I_{\mathrm{meas}}$ lie on opposite sides of the ridge. The $p = 1023$ points are from a later GB200 pass (Section~\ref{sup:data-provenance-reproducibility}).}
\label{fig:s8}
\end{figure*}

A roofline reduces a path to one point.
In contrast, the accounting of Section~\ref{sec:does-accounting-transfer-a100} of the main text separates that path into kernels that different resources control.
The binding resource cannot be read from the path-level point.
For example, on GB200, L1 is the dominant throughput of the CUDA Core path at every profiled order, on both sides of the ridge (Table~\ref{tab:binding}).
Figure~\ref{fig:s8} is therefore a statement of the byte-count convention behind Section~\ref{sec:performance-landscape} of the main text and a summary of the measured traffic.
The resource-level interpretation of Section~\ref{sec:instruction-aware-model} relies on the kernel-resolved measurements.

\begin{thebibliography}{51}
\providecommand{\natexlab}[1]{#1}
\providecommand{\url}[1]{\texttt{#1}}
\providecommand{\urlprefix}{URL }
\expandafter\ifx\csname urlstyle\endcsname\relax
  \providecommand{\doi}[1]{DOI:\discretionary{}{}{}#1}\else
  \providecommand{\doi}{DOI:\discretionary{}{}{}\begingroup \urlstyle{rm}\Url}\fi

\bibitem[{Abdelfattah et~al.(2021)Abdelfattah, Costa, Dongarra, Gates, Haidar, Hammarling, Higham, Kurzak, Luszczek, Tomov and Zounon}]{abdelfattah2021}
Abdelfattah A, Costa T, Dongarra J, Gates M, Haidar A, Hammarling S, Higham NJ, Kurzak J, Luszczek P, Tomov S and Zounon M (2021) A set of batched basic linear algebra subprograms and {LAPACK} routines.
\newblock \emph{ACM Transactions on Mathematical Software} 47(3): 1--23.
\newblock \doi{10.1145/3431921}.

\bibitem[{Anderson et~al.(2021)Anderson, Andrej, Barker, Bramwell, Camier, Cerveny, Dobrev, Dudouit, Fisher, Kolev, Pazner, Stowell, Tomov et~al.}]{anderson2021}
Anderson R, Andrej J, Barker A, Bramwell J, Camier JS, Cerveny J, Dobrev V, Dudouit Y, Fisher A, Kolev T, Pazner W, Stowell M, Tomov V et~al. (2021) {MFEM}: A modular finite element methods library.
\newblock \emph{Computers \& Mathematics with Applications} 81: 42--74.
\newblock \doi{10.1016/j.camwa.2020.06.009}.

\bibitem[{Ardalani et~al.(2015)Ardalani, Lestourgeon, Sankaralingam and Zhu}]{ardalani2015}
Ardalani N, Lestourgeon C, Sankaralingam K and Zhu X (2015) Cross-architecture performance prediction ({XAPP}) using {CPU} code to predict {GPU} performance.
\newblock In: \emph{Proceedings of the 48th Annual IEEE/ACM International Symposium on Microarchitecture (MICRO-48)}. pp. 725--737.
\newblock \doi{10.1145/2830772.2830780}.

\bibitem[{Beams et~al.(2020)Beams, Abdelfattah, Tomov, Dongarra, Kolev and Dudouit}]{beams2020}
Beams N, Abdelfattah A, Tomov S, Dongarra J, Kolev T and Dudouit Y (2020) High-order finite element method using standard and device-level batch {GEMM} on {GPU}s.
\newblock In: \emph{2020 IEEE/ACM 11th Workshop on Latest Advances in Scalable Algorithms for Large-Scale Systems (ScalA)}. pp. 53--60.
\newblock \doi{10.1109/ScalA51936.2020.00012}.

\bibitem[{Bell and Garland(2009)}]{bell2009}
Bell N and Garland M (2009) Implementing sparse matrix-vector multiplication on throughput-oriented processors.
\newblock In: \emph{Proceedings of the Conference on High Performance Computing Networking, Storage and Analysis (SC09)}. pp. 1--11.
\newblock \doi{10.1145/1654059.1654078}.

\bibitem[{Brown et~al.(2021)Brown, Abdelfattah, Barra, Beams, Camier, Dobrev, Dudouit, Ghaffari, Kolev, Medina, Pazner, Ratnayaka, Thompson and Tomov}]{brown2021}
Brown J, Abdelfattah A, Barra V, Beams N, Camier JS, Dobrev V, Dudouit Y, Ghaffari L, Kolev T, Medina D, Pazner W, Ratnayaka T, Thompson J and Tomov S (2021) {libCEED}: Fast algebra for high-order element-based discretizations.
\newblock \emph{Journal of Open Source Software} 6(63): 2945.
\newblock \doi{10.21105/joss.02945}.

\bibitem[{Cao et~al.(2026)Cao, Sun, Zhang and Li}]{cao2026}
Cao Z, Sun Q, Zhang T and Li H (2026) Towards a higher roofline for matrix-vector multiplication in matrix-free {HOSFEM}.
\newblock \emph{Journal of Parallel and Distributed Computing} 216: 105309.
\newblock \doi{10.1016/j.jpdc.2026.105309}.

\bibitem[{Cui(2024)}]{cui2024}
Cui C (2024) Acceleration of tensor-product operations with tensor cores.
\newblock \emph{ACM Transactions on Parallel Computing} 11(4): 1--24.
\newblock \doi{10.1145/3695466}.

\bibitem[{Deville et~al.(2002)Deville, Fischer and Mund}]{deville2002}
Deville MO, Fischer PF and Mund EH (2002) \emph{High-Order Methods for Incompressible Fluid Flow}.
\newblock Number~9 in Cambridge Monographs on Applied and Computational Mathematics. Cambridge: Cambridge University Press.
\newblock \doi{10.1017/CBO9780511546792}.

\bibitem[{Ding and Williams(2019)}]{ding2019}
Ding N and Williams S (2019) An instruction roofline model for {GPU}s.
\newblock In: \emph{2019 IEEE/ACM Performance Modeling, Benchmarking and Simulation of High Performance Computer Systems (PMBS)}. pp. 7--18.
\newblock \doi{10.1109/PMBS49563.2019.00007}.

\bibitem[{Fischer et~al.(2022)Fischer, Kerkemeier, Min, Lan, Phillips, Rathnayake, Merzari, Tomboulides, Karakus, Chalmers and Warburton}]{fischer2022}
Fischer P, Kerkemeier S, Min M, Lan YH, Phillips M, Rathnayake T, Merzari E, Tomboulides A, Karakus A, Chalmers N and Warburton T (2022) {NekRS}, a {GPU}-accelerated spectral element {N}avier--{S}tokes solver.
\newblock \emph{Parallel Computing} 114: 102982.
\newblock \doi{10.1016/j.parco.2022.102982}.

\bibitem[{Fischer et~al.(2020)Fischer, Min, Rathnayake, Dutta, Kolev, Dobrev, Camier, Kronbichler, Warburton, {\'S}wirydowicz and Brown}]{fischer2020}
Fischer P, Min M, Rathnayake T, Dutta S, Kolev T, Dobrev V, Camier JS, Kronbichler M, Warburton T, {\'S}wirydowicz K and Brown J (2020) Scalability of high-performance {PDE} solvers.
\newblock \emph{International Journal of High Performance Computing Applications} 34(5): 562--586.
\newblock \doi{10.1177/1094342020915762}.

\bibitem[{Guo and Wang(2015)}]{guo2015}
Guo P and Wang L (2015) Accurate cross-architecture performance modeling for sparse matrix-vector multiplication ({SpMV}) on {GPUs}.
\newblock \emph{Concurrency and Computation: Practice and Experience} 27(13): 3281--3294.
\newblock \doi{10.1002/cpe.3217}.

\bibitem[{Haidar et~al.(2018)Haidar, Tomov, Dongarra and Higham}]{haidar2018}
Haidar A, Tomov S, Dongarra J and Higham NJ (2018) Harnessing {GPU} tensor cores for fast {FP16} arithmetic to speed up mixed-precision iterative refinement solvers.
\newblock In: \emph{SC18: International Conference for High Performance Computing, Networking, Storage and Analysis}. pp. 603--613.
\newblock \doi{10.1109/SC.2018.00050}.

\bibitem[{Hochgraf and Pai(2025)}]{hochgraf2025}
Hochgraf R and Pai S (2025) A few fit most: Improving performance portability of {SGEMM} on {GPU}s using multi-versioning.
\newblock arXiv preprint arXiv:2507.15277.

\bibitem[{Jarmusch and Chandrasekaran(2026)}]{jarmusch2026}
Jarmusch A and Chandrasekaran S (2026) Microbenchmark-driven analytical performance modeling across modern {GPU} architectures.
\newblock arXiv preprint arXiv:2605.04178.

\bibitem[{Kawai(2026)}]{kawai-extraction}
Kawai Y (2026) {SCALE-DG} kernel extraction.
\newblock Software, commit 283f304 (24 August 2026). Available at: \url{https://github.com/ywkawai/scale-dg-kernel-extraction} (accessed 17 September 2026).

\bibitem[{Kawai et~al.(2026)Kawai, Ren, Nishizawa, Katagiri and Tomita}]{kawai2026}
Kawai Y, Ren X, Nishizawa S, Katagiri T and Tomita H (2026) Comparison of computational advantages of high-order discontinuous {G}alerkin and conventional finite-volume dynamical cores in atmospheric turbulent simulations.
\newblock \emph{Journal of the Meteorological Society of Japan} 104(1): 20.
\newblock \doi{10.1007/s44394-026-00023-6}.

\bibitem[{Kl{\"o}ckner et~al.(2009)Kl{\"o}ckner, Warburton, Bridge and Hesthaven}]{kloeckner2009}
Kl{\"o}ckner A, Warburton T, Bridge J and Hesthaven JS (2009) Nodal discontinuous {G}alerkin methods on graphics processors.
\newblock \emph{Journal of Computational Physics} 228(21): 7863--7882.
\newblock \doi{10.1016/j.jcp.2009.06.041}.

\bibitem[{Kolev et~al.(2023)Kolev, Fischer, Abdelfattah, Atkins, Bankole, Beams, Brown, Camier, Carson, Chalmers, Dobrev et~al.}]{kolev2023}
Kolev T, Fischer P, Abdelfattah A, Atkins Z, Bankole A, Beams N, Brown J, Camier JS, Carson R, Chalmers N, Dobrev V et~al. (2023) {CEED} {ECP} milestone report: Support {ECP} applications in their exascale challenge problem runs.
\newblock Technical Report WBS 2.2.6.06, Milestone CEED-MS40, Center for Efficient Exascale Discretizations.
\newblock \doi{10.5281/zenodo.7820316}.

\bibitem[{Kolev et~al.(2021)Kolev, Fischer, Min, Dongarra, Brown, Dobrev, Warburton, Tomov, Shephard et~al.}]{kolev2021}
Kolev T, Fischer P, Min M, Dongarra J, Brown J, Dobrev V, Warburton T, Tomov S, Shephard MS et~al. (2021) Efficient exascale discretizations: High-order finite element methods.
\newblock \emph{International Journal of High Performance Computing Applications} 35(6): 527--552.
\newblock \doi{10.1177/10943420211020803}.

\bibitem[{Konstantinidis and Cotronis(2017)}]{konstantinidis2017}
Konstantinidis E and Cotronis Y (2017) A quantitative roofline model for {GPU} kernel performance estimation using micro-benchmarks and hardware metric profiling.
\newblock \emph{Journal of Parallel and Distributed Computing} 107: 37--56.
\newblock \doi{10.1016/j.jpdc.2017.04.002}.

\bibitem[{Kronbichler and Kormann(2012)}]{kronbichler2012}
Kronbichler M and Kormann K (2012) A generic interface for parallel cell-based finite element operator application.
\newblock \emph{Computers \& Fluids} 63: 135--147.
\newblock \doi{10.1016/j.compfluid.2012.04.012}.

\bibitem[{Kronbichler and Ljungkvist(2019)}]{kronbichler2019}
Kronbichler M and Ljungkvist K (2019) Multigrid for matrix-free high-order finite element computations on graphics processors.
\newblock \emph{ACM Transactions on Parallel Computing} 6(1): 1--32.
\newblock \doi{10.1145/3322813}.

\bibitem[{Lee et~al.(2025)Lee, Phanishayee and Mahajan}]{lee2025}
Lee S, Phanishayee A and Mahajan D (2025) Forecasting {GPU} performance for deep learning training and inference.
\newblock In: \emph{Proceedings of the 30th ACM International Conference on Architectural Support for Programming Languages and Operating Systems (ASPLOS 2025), Volume 1}. pp. 493--508.
\newblock \doi{10.1145/3669940.3707265}.

\bibitem[{Markidis et~al.(2018)Markidis, Chien, Laure, Peng and Vetter}]{markidis2018}
Markidis S, Chien SWD, Laure E, Peng IB and Vetter JS (2018) {NVIDIA} tensor core programmability, performance and precision.
\newblock In: \emph{2018 IEEE International Parallel and Distributed Processing Symposium Workshops (IPDPSW)}. pp. 522--531.
\newblock \doi{10.1109/IPDPSW.2018.00091}.

\bibitem[{Medina et~al.(2014)Medina, St-Cyr and Warburton}]{medina2014}
Medina DS, St-Cyr A and Warburton T (2014) {OCCA}: A unified approach to multi-threading languages.
\newblock arXiv preprint arXiv:1403.0968.

\bibitem[{Muralidharan et~al.(2016)Muralidharan, Roy, Hall, Garland and Rai}]{muralidharan2016}
Muralidharan S, Roy A, Hall M, Garland M and Rai P (2016) Architecture-adaptive code variant tuning.
\newblock In: \emph{Proceedings of the 21st International Conference on Architectural Support for Programming Languages and Operating Systems (ASPLOS 2016)}. pp. 325--338.
\newblock \doi{10.1145/2872362.2872411}.

\bibitem[{Nishizawa(2026{\natexlab{a}})}]{thiswork-code}
Nishizawa S (2026{\natexlab{a}}) {GPU} implementation paths for the {SCALE-DG} advection operator.
\newblock Software, branch feature/cuda. Available at: \url{https://github.com/seiya/scale-dg-kernel-extraction}.

\bibitem[{Nishizawa(2026{\natexlab{b}})}]{thiswork-data}
Nishizawa S (2026{\natexlab{b}}) Measurement logs and figure scripts for ``{I}nstruction-aware transfer of {GPU} implementation crossovers: a tensor-product discontinuous {G}alerkin operator on {GB200}, {H100}, and {A100}''.
\newblock Data set, Zenodo.
\newblock \doi{10.5281/zenodo.23186421}.

\bibitem[{Nishizawa et~al.(2015)Nishizawa, Yashiro, Sato, Miyamoto and Tomita}]{nishizawa2015}
Nishizawa S, Yashiro H, Sato Y, Miyamoto Y and Tomita H (2015) Influence of grid aspect ratio on planetary boundary layer turbulence in large-eddy simulations.
\newblock \emph{Geoscientific Model Development} 8(10): 3393--3419.
\newblock \doi{10.5194/gmd-8-3393-2015}.

\bibitem[{{NVIDIA Corporation}(2026{\natexlab{a}})}]{nvidia-cublas}
{NVIDIA Corporation} (2026{\natexlab{a}}) {cuBLAS} library.
\newblock Software, versions 13.2.1 (HPC SDK 26.3, CUDA 13.1 update 1) and 13.2.0 (HPC SDK 26.1, CUDA 13.1). Available at: \url{https://docs.nvidia.com/cuda/cublas/} (accessed 17 September 2026).

\bibitem[{{NVIDIA Corporation}(2026{\natexlab{b}})}]{nvidia-ncu-guide}
{NVIDIA Corporation} (2026{\natexlab{b}}) {Nsight Compute} profiling guide, version 2026.3.1.
\newblock Section ``Pipelines''. Available at: \url{https://docs.nvidia.com/nsight-compute/ProfilingGuide/} (accessed 19 September 2026).

\bibitem[{Ootomo et~al.(2024)Ootomo, Ozaki and Yokota}]{ootomo2024}
Ootomo H, Ozaki K and Yokota R (2024) {DGEMM} on integer matrix multiplication unit.
\newblock \emph{International Journal of High Performance Computing Applications} 38(4): 297--313.
\newblock \doi{10.1177/10943420241239588}.

\bibitem[{Ootomo and Yokota(2022)}]{ootomo2022}
Ootomo H and Yokota R (2022) Recovering single precision accuracy from tensor cores while surpassing the {FP32} theoretical peak performance.
\newblock \emph{International Journal of High Performance Computing Applications} 36(4): 475--491.
\newblock \doi{10.1177/10943420221090256}.

\bibitem[{Orszag(1980)}]{orszag1980}
Orszag SA (1980) Spectral methods for problems in complex geometries.
\newblock \emph{Journal of Computational Physics} 37(1): 70--92.
\newblock \doi{10.1016/0021-9991(80)90005-4}.

\bibitem[{Ozaki et~al.(2012)Ozaki, Ogita, Oishi and Rump}]{ozaki2012}
Ozaki K, Ogita T, Oishi S and Rump SM (2012) Error-free transformations of matrix multiplication by using fast routines of matrix multiplication and its applications.
\newblock \emph{Numerical Algorithms} 59(1): 95--118.
\newblock \doi{10.1007/s11075-011-9478-1}.

\bibitem[{Ozaki et~al.(2026)Ozaki, Uchino and Imamura}]{ozaki2026}
Ozaki K, Uchino Y and Imamura T (2026) {O}zaki scheme {II}: A {GEMM}-oriented emulation of floating-point matrix multiplication using an integer modular technique.
\newblock \emph{International Journal of High Performance Computing Applications}.
\newblock \doi{10.1177/10943420261467787}.

\bibitem[{Ren et~al.(2026)Ren, Kawai, Hoshino, Tomita, Katagiri, Mukunoki and Nishizawa}]{ren2026}
Ren X, Kawai Y, Hoshino T, Tomita H, Katagiri T, Mukunoki D and Nishizawa S (2026) Learning-augmented performance model for tensor product factorization in high-order {FEM}.
\newblock \emph{IEEE Access} 14: 43679--43693.
\newblock \doi{10.1109/ACCESS.2026.3675604}.

\bibitem[{Sato et~al.(2015)Sato, Nishizawa, Yashiro, Miyamoto, Kajikawa and Tomita}]{sato2015}
Sato Y, Nishizawa S, Yashiro H, Miyamoto Y, Kajikawa Y and Tomita H (2015) Impacts of cloud microphysics on trade wind cumulus: which cloud microphysics processes contribute to the diversity in a large eddy simulation?
\newblock \emph{Progress in Earth and Planetary Science} 2: 23.
\newblock \doi{10.1186/s40645-015-0053-6}.

\bibitem[{Schwarz et~al.(2026)Schwarz, Anders, Brower, Bayraktar, Gunnels, Clark, Xu, Rodriguez, Cayrols, Tabaszewski and Podlozhnyuk}]{schwarz2026}
Schwarz A, Anders A, Brower C, Bayraktar H, Gunnels J, Clark K, Xu RG, Rodriguez S, Cayrols S, Tabaszewski P and Podlozhnyuk V (2026) Guaranteed {DGEMM} accuracy while using reduced precision tensor cores through extensions of the {O}zaki scheme.
\newblock In: \emph{Proceedings of the Supercomputing Asia and International Conference on High Performance Computing in Asia-Pacific Region (SCA/HPCAsia 2026)}. pp. 91--101.
\newblock \doi{10.1145/3773656.3773670}.

\bibitem[{Sedaghati et~al.(2015)Sedaghati, Mu, Pouchet, Parthasarathy and Sadayappan}]{sedaghati2015}
Sedaghati N, Mu T, Pouchet LN, Parthasarathy S and Sadayappan P (2015) Automatic selection of sparse matrix representation on {GPU}s.
\newblock In: \emph{Proceedings of the 29th ACM International Conference on Supercomputing (ICS'15)}. pp. 99--108.
\newblock \doi{10.1145/2751205.2751244}.

\bibitem[{Shah et~al.(2024)Shah, Bikshandi, Zhang, Thakkar, Ramani and Dao}]{shah2024}
Shah J, Bikshandi G, Zhang Y, Thakkar V, Ramani P and Dao T (2024) {FlashAttention-3}: Fast and accurate attention with asynchrony and low-precision.
\newblock In: \emph{Advances in Neural Information Processing Systems 37 (NeurIPS 2024)}. pp. 68658--68685.
\newblock \doi{10.52202/079017-2193}.

\bibitem[{Stevens and Kl{\"o}ckner(2020)}]{stevens2020}
Stevens JD and Kl{\"o}ckner A (2020) A mechanism for balancing accuracy and scope in cross-machine black-box {GPU} performance modeling.
\newblock \emph{International Journal of High Performance Computing Applications} 34(6): 589--614.
\newblock \doi{10.1177/1094342020921340}.

\bibitem[{{\'S}wirydowicz et~al.(2019){\'S}wirydowicz, Chalmers, Karakus and Warburton}]{swirydowicz2019}
{\'S}wirydowicz K, Chalmers N, Karakus A and Warburton T (2019) Acceleration of tensor-product operations for high-order finite element methods.
\newblock \emph{International Journal of High Performance Computing Applications} 33(4): 735--757.
\newblock \doi{10.1177/1094342018816368}.

\bibitem[{Thakkar et~al.(2023)Thakkar, Ramani, Cecka, Shivam, Lu et~al.}]{thakkar2023}
Thakkar V, Ramani P, Cecka C, Shivam A, Lu H et~al. (2023) {CUTLASS}.
\newblock Software, version 4.7.0 plus six commits (7107b055). Available at: \url{https://github.com/NVIDIA/cutlass}.

\bibitem[{Tu et~al.(2026)Tu, Karlin, Camier, Dobrev, Kolev, Henneking and Ghattas}]{tu2026}
Tu J, Karlin I, Camier J, Dobrev V, Kolev T, Henneking S and Ghattas O (2026) Accelerating high-order finite element simulations at extreme scale with {FP64} tensor cores.
\newblock In: \emph{ISC High Performance 2026 Research Paper Proceedings (41st International Conference)}. pp. 1--12.
\newblock \doi{10.23919/ISC.2026.11520474}.

\bibitem[{Van~Lanker et~al.(2024)Van~Lanker, Taboada, Brunet and Trahay}]{vanlanker2024}
Van~Lanker L, Taboada H, Brunet E and Trahay F (2024) Predicting {GPU} kernel's performance on upcoming architectures.
\newblock In: \emph{Euro-Par 2024: Parallel Processing}, \emph{Lecture Notes in Computer Science}, volume 14801. Springer, pp. 77--90.
\newblock \doi{10.1007/978-3-031-69577-3_6}.

\bibitem[{Wang et~al.(2026)Wang, Gan, Mao, Song, Wan, Fu, Yin, Jin, Duan, Xue and Yang}]{wang2026}
Wang Y, Gan L, Mao T, Song Z, Wan W, Fu J, Yin Z, Jin Y, Duan X, Xue W and Yang G (2026) Beyond fast contractions: Attenuation and recovery of matrix-engine speedups in high-order finite elements.
\newblock arXiv preprint arXiv:2608.09175.

\bibitem[{Williams et~al.(2009)Williams, Waterman and Patterson}]{williams2009}
Williams S, Waterman A and Patterson D (2009) Roofline: An insightful visual performance model for multicore architectures.
\newblock \emph{Communications of the ACM} 52(4): 65--76.
\newblock \doi{10.1145/1498765.1498785}.

\bibitem[{Yang et~al.(2020)Yang, Kurth and Williams}]{yang2020}
Yang C, Kurth T and Williams S (2020) Hierarchical roofline analysis for {GPU}s: Accelerating performance optimization for the {NERSC-9} {P}erlmutter system.
\newblock \emph{Concurrency and Computation: Practice and Experience} 32(20): e5547.
\newblock \doi{10.1002/cpe.5547}.

\end{thebibliography}
\end{document}